\documentclass{article}
     \usepackage[final,nonatbib]{neurips_2020}

\usepackage[utf8]{inputenc} % allow utf-8 input
\usepackage[T1]{fontenc}    % use 8-bit T1 fonts
\usepackage{hyperref}       % hyperlinks
\usepackage{url}            % simple URL typesetting
\usepackage{booktabs}       % professional-quality tables
\usepackage{amsfonts}       % blackboard math symbols
\usepackage{nicefrac}       % compact symbols for 1/2, etc.
\usepackage{microtype}      % microtypography

\newcommand{\figref}[1]{Fig.~\ref{#1}}
\usepackage{adjustbox}

\usepackage{wrapfig}
\usepackage{subcaption}
\usepackage{etoolbox}
\usepackage{verbatim}

\usepackage{amssymb}
\usepackage{wrapfig}
\usepackage{mathrsfs}
\usepackage{bm}
\usepackage{algorithm}
\usepackage{algcompatible}
\usepackage[noend]{algpseudocode}
\usepackage{mathtools}
\usepackage{nccmath}
\usepackage{float}
\usepackage{times}
\usepackage{multirow,color,graphicx}
\usepackage{subcaption}
\usepackage{caption}
\usepackage{xfrac,amsbsy}
\usepackage[titletoc]{appendix}
\usepackage{xspace}
\usepackage{savesym,verbatim}
\usepackage{paralist}
\usepackage{subscript}
\usepackage{tikz}
\usepackage{amsthm}
\usepackage{bbm}

\newtheorem{definition}{Definition}

\usepackage{bbm}

\newcommand{\algoSame}{\ensuremath{\textsc{Same}}}
\newcommand{\algoTutor}{\ensuremath{\textsc{Tutor}}}
\newcommand{\algoTaskMut}{\ensuremath{\textsc{MutTask}}}

\newcommand{\algoOursNoPstar}{\ensuremath{\textsc{SynTask}}}
\usepackage{environ}
\usepackage{esvect} % declare arrow on top of character

\NewEnviron{boxcode2col}[4]{
    \begin{center}
   	 	\scalebox{#3}{
	   	 	\setlength\tabcolsep{4pt}
	   	 	\renewcommand{\arraystretch}{#4}
	   	 	\begin{tabular}{|p{#1}p{#2}|}   
    			\hline
				\BODY
   			 	[1ex]
    			\hline
    		\end{tabular}
    	}
    \end{center}
}

\NewEnviron{boxcode}[3]{
    \begin{center}
   	 	\scalebox{#2}{
	   	 	\setlength\tabcolsep{4pt}
	   	 	\renewcommand{\arraystretch}{#3}
    		\begin{tabular}{|p{#1}|}    
    			\hline
    			\vspace{0.1mm}    
				\BODY
    			\\\hline
    		\end{tabular}
    	}
    \end{center}
}
    
\newcommand{\textcode}[1]{{\fontfamily{cmtt}\selectfont #1}\xspace}

\newcommand{\task}{\text{\textcode{T}}}
\newcommand{\code}{\text{\textcode{C}}}
\newcommand{\codes}{\ensuremath{\mathcal{C}}}
\newcommand{\codesketch}{\text{\textcode{Q}}}
\newcommand{\FDissimilarity}{\ensuremath{\mathcal{F}_{\textnormal{diss}}}}
\newcommand{\FQuality}{\ensuremath{\mathcal{F}_{\textnormal{qual}}}}

\newcommand{\DSLMove}{\textcode{move}}
\newcommand{\DSLTurnL}{\textcode{turnL}}
\newcommand{\DSLTurnLeft}{\textcode{turnLeft}}
\newcommand{\DSLTurnR}{\textcode{turnR}}
\newcommand{\DSLTurnRight}{\textcode{turnRight}}
\newcommand{\DSLPickM}{\textcode{pickM}}
\newcommand{\DSLPickMarker}{\textcode{pickMarker}}
\newcommand{\DSLPutM}{\textcode{putM}}
\newcommand{\DSLPutMarker}{\textcode{putMarker}}
\newcommand{\DSLRepeat}{\textcode{\textsc{Repeat}}}
\newcommand{\DSLRepeatUntil}{\textcode{\textsc{RepeatUntil}}}
\newcommand{\DSLIf}{\textcode{\textsc{If}}}
\newcommand{\DSLIfElse}{\textcode{\textsc{IfElse}}}
\newcommand{\DSLElse}{\textcode{\textsc{Else}}}
\newcommand{\DSLdo}{\textcode{\textsc{do}}}
\newcommand{\DSLWhile}{\textcode{\textsc{While}}}
\newcommand{\DSLRun}{\textcode{\textsc{Run}}}
\newcommand{\DSLBoolGoal}{\textcode{goal}}
\newcommand{\DSLBoolPathAhead}{\textcode{pathAhead}}
\newcommand{\DSLBoolPathA}{\textcode{pathA}}

\newcommand{\DSLBoolNoPathA}{\textcode{noPathA}}

\newcommand{\DSLBoolPathLeft}{\textcode{pathLeft}}
\newcommand{\DSLBoolPathL}{\textcode{pathL}}

\newcommand{\DSLBoolNoPathL}{\textcode{noPathL}}
\newcommand{\DSLBoolPathRight}{\textcode{pathRight}}
\newcommand{\DSLBoolPathR}{\textcode{pathR}}

\newcommand{\DSLBoolNoPathR}{\textcode{noPathR}}

\newcommand{\DSLBoolMarker}{\textcode{marker}}

\newcommand{\DSLBoolNoMarker}{\textcode{noMarker}}

\newcommand{\DSLBoolTrue}{\textcode{True}}
\newcommand{\DSLBoolFalse}{\textcode{False}}
\newcommand{\hocType}{\textnormal{HOC}}
\newcommand{\hocAll}{\textnormal{H-}}
\newcommand{\hocAllBold}{\textbf{H-}}
\newcommand{\hocA}{\textnormal{H1}}
\newcommand{\hocC}{\textnormal{H2}}
\newcommand{\hocD}{\textnormal{H3}}
\newcommand{\hocF}{\textnormal{H4}}
\newcommand{\hocG}{\textnormal{H5}}
\newcommand{\hocH}{\textnormal{H6}}

\newcommand{\karelType}{\textnormal{Karel}}
\newcommand{\karelA}{\textnormal{K7}}
\newcommand{\karelC}{\textnormal{K8}}
\newcommand{\karelE}{\textnormal{K9}}
\newcommand{\karelF}{\textnormal{K10}}

\newcommand{\DSLCode}{\textnormal{code }}

\newcommand{\DSLStmtVar}{\textcode{s}}
\newcommand{\DSLAction}{\textnormal{action }}
\newcommand{\DSLActionVar}{\textcode{a}}
\newcommand{\DSLBool}{\textnormal{bool }}
\newcommand{\DSLBoolVar}{\textcode{b}}
\newcommand{\DSLIter}{\textnormal{iter }}
\newcommand{\DSLIterVar}{\textcode{x}}
\newcommand{\DSLRule}{\textnormal{rule }}
\newcommand{\DSLRuleVar}{\textcode{y}}
\newcommand{\DSLRepeatForever}{\textcode{g}}

\newcommand{\SDSLSketch}{\textnormal{sketch }}
\newcommand{\SDSLSketchVar}{\textcode{Q}}

\newcommand{\SDSLSStmtVar}{\textcode{S}}

\newcommand{\SDSLVarY}{\textcode{Y}}
\newcommand{\SDSLVarG}{\textcode{G}}

\newcommand{\SDSLBool}{\textcode{B}}
\newcommand{\SDSLIter}{\textcode{X}}
\newcommand{\SDSLAction}{\textcode{A}}

 \newcommand{\textoverline}[1]{\ensuremath{\overline{\text{#1}}}}

\newcommand{\actionseq}{\textoverline{\textcode{A}}}

\newcommand{\actionseqb}{\textcolor{blue}{\textoverline{\textcode{A}}}}

\newcommand{\codetosketch}{\ensuremath{\Omega}}
\newcommand{\sketchparams}{\ensuremath{\omega}}
\newcommand{\localblockcons}{\ensuremath{\textsc{ActionEdits}}}

\newcommand{\FScore}{\ensuremath{\mathcal{F}_{\textnormal{score}}}}
\newcommand{\FCoverage}{\ensuremath{\mathcal{F}_{\textnormal{cov}}}}
\newcommand{\FNoCrash}{\ensuremath{\mathcal{F}_{\textnormal{nocrash}}}}
\newcommand{\FNoCut}{\ensuremath{\mathcal{F}_{\textnormal{nocut}}}}
\newcommand{\FDiversity}{\ensuremath{\mathcal{F}_{\textnormal{diversity}}}}

\newcommand{\countqual}{\ensuremath{\#}}

\NewEnviron{boxcodecfg}[1]{%
\fbox{\begin{minipage}{#1}
  \begin{flalign*}
    \BODY
  \end{flalign*}
 \end{minipage}}
}

\newcommand{\sm}{\text{\textcode{M}}}
\newcommand{\tl}{\text{\textcode{L}}}
\newcommand{\tr}{\text{\textcode{R}}}
\newcommand{\pum}{\text{\textcode{puM}}}
\newcommand{\pim}{\text{\textcode{piM}}}

\newcommand{\ha}[2]{\ensuremath{\text{\SDSLAction}^{#2}_{#1}}}

\title{Synthesizing Tasks for Block-based Programming\thanks{Authors listed alphabetically; 
Correspondence to: Ahana Ghosh <\href{mailto:gahana@mpi-sws.org}{\textcode{gahana@mpi-sws.org}}>.}}
\author{
  \textbf{Umair Z. Ahmed}\textsuperscript{1} \quad 
  \textbf{Maria Christakis}\textsuperscript{2} \quad 
  \textbf{Aleksandr Efremov}\textsuperscript{2} \quad
  \textbf{Nigel Fernandez}\textsuperscript{2}\\
  \textbf{Ahana Ghosh}\textsuperscript{2} \quad
  \textbf{Abhik Roychoudhury}\textsuperscript{1} \quad
  \textbf{Adish Singla}\textsuperscript{2}\\
  \textsuperscript{1}National University of Singapore, \textcode{\{umair, abhik\}@comp.nus.edu.sg}, \\
  \textsuperscript{2}MPI-SWS, \textcode{\{maria, aefremov, nfernand, gahana, adishs\}@mpi-sws.org}\\
}
\begin{document}
\maketitle

%%%%%%%%%%%%%%%%%%%%%%%%%%%%%%%%%%%%%%%%%%%%%%%%%%%%%%%%
\newtoggle{longversion}
\settoggle{longversion}{true}

%%%%%%%%%%%%%%%%%%%%%%%%%%%%%%%%%%%%%%%%%%%%%%%%%%%%%%%%%%
% !TEX root =  main.tex
%%%%%%%%%%%%%%%%%%%%%%%%%%%%%%%%%%%%%%%%%%%%%%%%%%%%%%%%%%
%%%%%%%%%%%%%%%%%%%%%%%%%%%%%%%%%%%%%%%%%%%%%%%%%%%%%%%%%%
\begin{abstract}
%\vspace{-2mm}
Block-based visual programming environments play a critical role in introducing computing concepts to $\textnormal{K-}12$ students. One of the key pedagogical challenges in these environments is in designing new practice tasks for a student that match a desired level of difficulty and exercise specific programming concepts. In this paper, we formalize the problem of synthesizing visual programming tasks.  In particular, given a reference visual task $\task^{\textnormal{in}}$ and its solution code $\code^{\textnormal{in}}$, we propose a novel methodology to automatically generate a set $\{(\task^{\textnormal{out}}, \code^{\textnormal{out}})\}$ of new tasks along with solution codes such that tasks $\task^{\textnormal{in}}$ and $\task^{\textnormal{out}}$ are \emph{conceptually similar} but \emph{visually dissimilar}.  Our methodology is based on the realization that the mapping from the space of visual tasks to their solution codes is highly discontinuous; hence, directly mutating reference task $\task^{\textnormal{in}}$ to generate new tasks is futile. Our task synthesis algorithm operates by first mutating code $\code^{\textnormal{in}}$ to obtain a set of codes $\{\code^{\textnormal{out}}\}$. Then, the algorithm performs \emph{symbolic execution} over a code $\code^{\textnormal{out}}$ to obtain a visual task $\task^{\textnormal{out}}$; this step uses the Monte Carlo Tree Search (MCTS) procedure to guide the search in the symbolic tree. We demonstrate the effectiveness of our algorithm through an extensive empirical evaluation and user study on reference tasks taken from the \emph{Hour of Code: Classic Maze} challenge by \emph{Code.org} and the \emph{Intro to Programming with Karel} course by \emph{CodeHS.com}.
\end{abstract}
\vspace{-4mm}
\section{Introduction}\label{sec.intro}
\vspace{-2mm}
\looseness-1
%(BVPE)
Block-based visual programming environments are increasingly used nowadays to introduce computing concepts to novice programmers including children and $\textnormal{K-}12$ students. Led by the success of environments like Scratch~\cite{resnick2009scratch}, initiatives like \emph{Hour of Code} by \emph{Code.org}~\cite{hourofcode} (HOC) and online platforms like \emph{CodeHS.com}~\cite{codehscom},  block-based programming has become an integral part of introductory computer science education. Considering HOC alone, over one billion hours of block-based programming  activity has been performed so far by over 50 million unique students worldwide~\cite{hourofcode,wu2019zero}.
%%%%%%%%%%%%%%%
%offering full programming courses
%For readers unfamiliar with block-based programming, we provide two example tasks along with their solution codes in Figures~\ref{fig:intro.hoc}~and~\ref{fig:intro.karel}.
% Figure~\ref{fig:intro.karel.t0,fig:intro.karel.p0}.
% (source: HOC's classic maze challenge, problem ``16"~\cite{hourofcode_maze})
% (source: \emph{Intro to Programming with Karel} course by \emph{CodeHS.com}, problem ``Diagonal")
%%%%%%%%%%%%%%%

The societal need for enhancing $\textnormal{K-}12$ computing education has led to a surge of interest in developing AI-driven  systems for pedagogy of block-based programming~\cite{wang2017learning,price2017position,price2017isnap,weintrop2017comparing,maloney2008programming}. Existing works have studied various aspects of intelligent support, including providing real-time next-step hints when a student is stuck solving a task~\cite{piech15las,yi2017feasibility,paassen2018continuous,marwan2019impact,efremov2020zeroshot}, giving data-driven feedback about a student's misconceptions~\cite{singh2013automated,DBLP:conf/icml/PiechHNPSG15,price2017evaluation,rolim2017learning,wu2019zero}, and demonstrating a worked-out solution for a task when a student lacks the required programming concepts~\cite{zhi2019exploring}. An underlying assumption when providing such intelligent support is that afterwards the student can practice \emph{new similar tasks} to finally learn the missing concepts. However, this assumption is far from reality in existing systems---the programming tasks are typically hand-curated by experts/tutors, and the available set of tasks is limited. Consider HOC's \emph{Classic Maze} challenge~\cite{hourofcode_maze}, which provides a progression of $20$ tasks: Millions of students have attempted these tasks, yet when students fail to solve a task and receive assistance, they cannot practice similar tasks, hindering their ability to master the desired concepts. We seek to tackle this pedagogical challenge by developing techniques for synthesizing new programming tasks.

%%%%%%%%%%%%%%%%%%%%%%%%%%%%%%%%%%%%%%%%%%%%%%%%%%%%%%%%%%
%%%%%%%%%%%%%%%%%%%%%%%%%%%%%%%%%%%%%%
\begin{figure}[t!]
\centering
	%%%%%%%%%%%%%%%%%
	\begin{subfigure}[b]{.22\textwidth}
	\centering
	{
		\includegraphics[height=2.39cm]{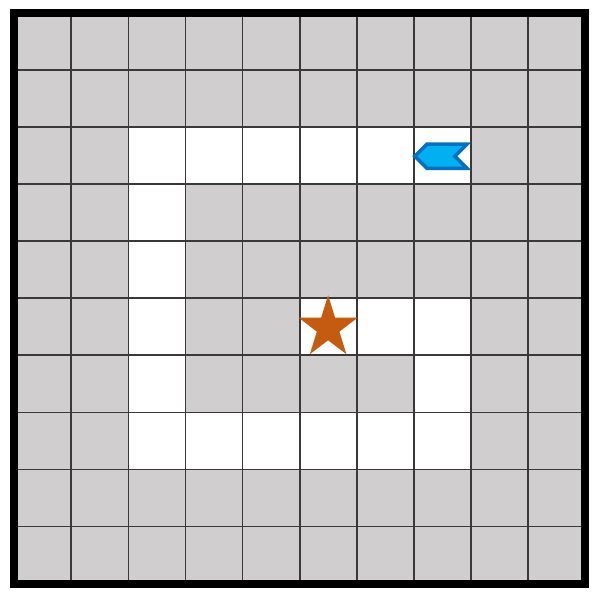}
		%\vspace{0.2mm}
		\caption{Visual puzzle for $\task^{\textnormal{in}}$}
		 \label{fig:intro.hoc.t0}
	}
	\end{subfigure}
	\
	%%%%%%%%%%%%%%%%%	
	\begin{subfigure}[b]{.22\textwidth}
	\centering
	{
		\begin{boxcode}{3.8cm}{0.75}{0.58}
				\textcode{def }\DSLRun\textcode{()\{}\\
				\quad \DSLRepeatUntil\textcode{(}\DSLBoolGoal\textcode{)\{}\\
				\quad \quad \DSLMove\\
				\quad \quad \DSLIf\textcode{(}\DSLBoolPathLeft\textcode{)\{}\\
				\quad \quad \quad \DSLTurnLeft\\
				\quad \quad \textcode{\}}\\
				\quad \textcode{\}}\\
				\textcode{\}}
				\\
				\\
		\end{boxcode}
		\vspace{-1mm}
		\caption{Solution code $\code^{\textnormal{in}}$}
		\label{fig:intro.hoc.p0}
    }
    \end{subfigure}
  	\quad \quad
	%%%%%%%%%%%%%%%%%
	\begin{subfigure}[b]{.22\textwidth}
	\centering
	{
		\includegraphics[height=2.39cm]{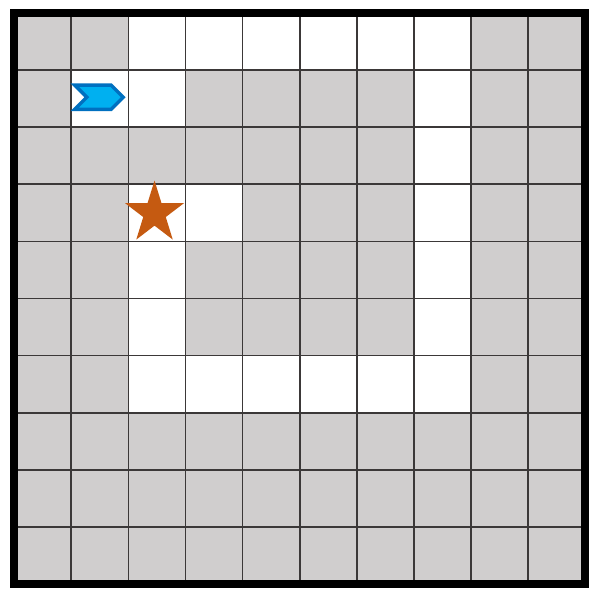}
		%\vspace{0.2mm}
		\caption{Visual puzzle for $\task^{\textnormal{out}}$}
		 \label{fig:intro.hoc.t1}
	}
	\end{subfigure}
	\
	%%%%%%%%%%%%%%%%%	
	\begin{subfigure}[b]{.22\textwidth}
	\centering
	{
		\begin{boxcode}{3.8cm}{0.75}{0.58}
			\textcode{def }\DSLRun\textcode{()\{}\\
			\quad \DSLMove\\
			\quad \DSLTurnLeft\\
			\quad \DSLRepeatUntil\textcode{(}\DSLBoolGoal\textcode{)\{}\\
			\quad \quad \DSLMove\\
			\quad \quad \DSLIf\textcode{(}\DSLBoolPathRight\textcode{)\{}\\
			\quad \quad \quad \DSLTurnRight\\
			\quad \quad \textcode{\}}\\
			\quad \textcode{\}}\\
			\textcode{\}}
		\end{boxcode}
		\vspace{-1mm}
		\caption{Solution code $\code^{\textnormal{out}}$}
		\label{fig:intro.hoc.p1}
    }
    \end{subfigure}
    %\\
    \vspace{-1.5mm}
	\caption{Illustration of our methodology for task \emph{Maze 16} from the \emph{Hour of Code: Classic Maze} challenge by \emph{Code.org}~\cite{hourofcode_maze}; the complete list of tasks with their specifications is in \figref{fig:dataset}.
	}
	\vspace{-3.5mm}
	\label{fig:intro.hoc}
\end{figure}

\begin{figure}[t!]
\centering
	%%%%%%%%%%%%%%%%%
	\begin{subfigure}[b]{.27\textwidth}
	\centering
	{
		\includegraphics[height=1.95cm]{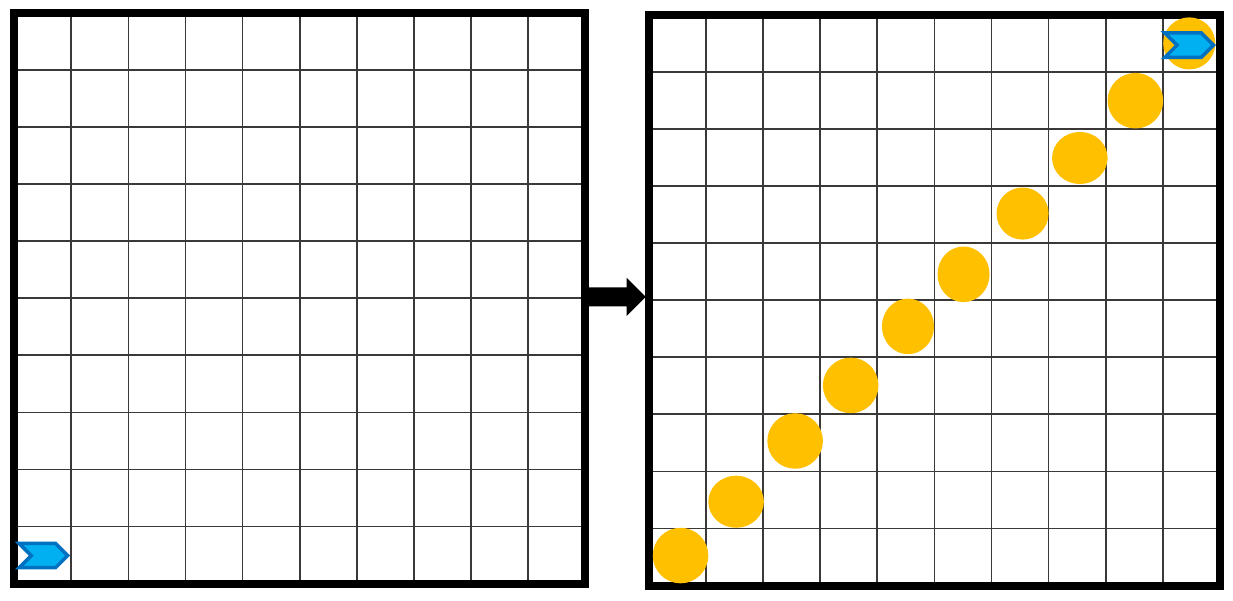}
		%\vspace{0.01mm}
		\caption{Visual puzzle for $\task^{\textnormal{in}}$}
		\label{fig:intro.karel.t0}
    }
    \end{subfigure}
  	\
  	\begin{subfigure}[b]{.195\textwidth}
  	\centering
  	 {
  	 	\begin{boxcode}{3.5cm}{0.75}{0.58}
			\textcode{def }\DSLRun\textcode{()\{}\\
			\quad \DSLPutMarker\\
			\quad \DSLWhile\textcode{(}\DSLBoolPathAhead\textcode{)\{}\\
			\quad \quad \DSLMove\\
			\quad \quad \DSLTurnLeft\\
			\quad \quad \DSLMove\\
			\quad \quad \DSLTurnRight\\
			\quad \quad \DSLPutMarker\\	
			\quad \textcode{\}}\\
			\textcode{\}}
			\\
		\end{boxcode}
		\vspace{-1.18mm}
		\caption{Solution code $\code^{\textnormal{in}}$}
		\label{fig:intro.karel.p0}
	}
	\end{subfigure}
	\quad \  
	\begin{subfigure}[b]{.27\textwidth}
	\centering
	{
		\includegraphics[height=1.95cm]{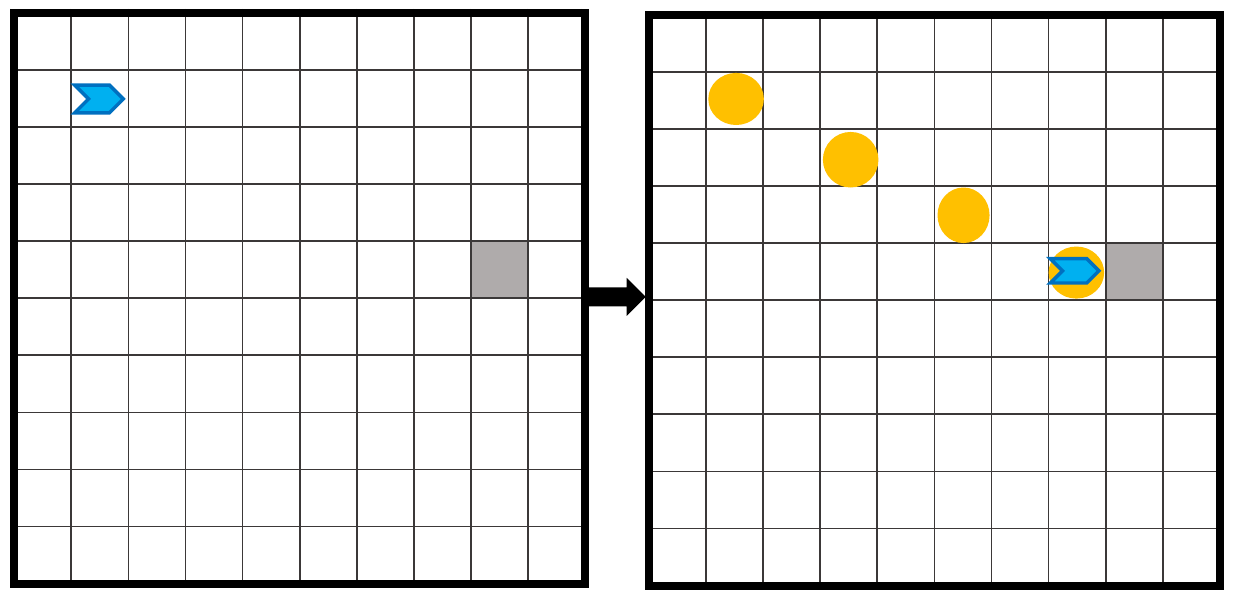}
		%\vspace{0.01mm}
		\caption{Visual puzzle for $\task^{\textnormal{out}}$}
		\label{fig:intro.karel.t1}
	}
	\end{subfigure}
	\
	\begin{subfigure}[b]{.195\textwidth}
	\centering
	{
  	 	\begin{boxcode}{3.5cm}{0.75}{0.58}
			\textcode{def }\DSLRun\textcode{()\{}\\
			\quad \DSLPutMarker\\
			\quad \DSLWhile\textcode{(}\DSLBoolPathAhead\textcode{)\{}\\
			\quad \quad \DSLMove\\
			\quad \quad \DSLMove\\
			\quad \quad \DSLTurnRight\\
			\quad \quad \DSLMove\\
			\quad \quad \DSLTurnLeft\\
			\quad \quad \DSLPutMarker\\	
			\quad \textcode{\}}\\
			\textcode{\}}
		\end{boxcode}
		\vspace{-1.18mm}
		\caption{Solution code $\code^{\textnormal{out}}$}
		\label{fig:intro.karel.p1}
	}
	\end{subfigure}  
	\vspace{-1.5mm}	
	\caption{Illustration of our methodology for task \emph{Diagonal} from the \emph{Intro to Programming with Karel} course by \emph{CodeHS.com}~\cite{intro_to_karel_codehs}; the complete list of tasks with their specifications is in \figref{fig:dataset}.
	}
	\vspace{-3.5mm}
	\label{fig:intro.karel}
\end{figure}

We formalize the problem of synthesizing visual programming tasks of the kind found in popular learning platforms like \emph{Code.org} (see \figref{fig:intro.hoc}) and \emph{CodeHS.com}  (see \figref{fig:intro.karel}). As input, we are given a reference task $\task^{\textnormal{in}}$, specified as a visual puzzle, and its solution code $\code^{\textnormal{in}}$. Our goal is to synthesize a set $\{(\task^{\textnormal{out}}, \code^{\textnormal{out}})\}$ of new tasks along with their solution codes that are \emph{conceptually similar} but \emph{visually dissimilar} to the input. This is motivated by the need for practice tasks that on one hand exercise the same concepts, while looking fresh in order to maintain student engagement.

When tackling the problem of synthesizing new tasks with the above desirable properties, three key challenges emerge. First, we are generating problems in a \emph{conceptual domain} with no \emph{well-defined procedure} that students follow to solve a task---consequently, existing work on educational problem generation in procedural domains does not apply in our setting~\cite{Andersen13,gulwani2014example}. Second, the mapping from the space of visual tasks to their solution codes is highly discontinuous; hence, template-based problem generation techniques~\cite{singh2012,polozov2015} that rely on directly mutating the input to generate new tasks is ineffective (see Section~\ref{sec.userstudy} where we use this approach as a baseline). Furthermore, such a direct task-mutation approach would require access to an automated solution synthesizer; however, state-of-the-art program synthesis techniques are not yet on par with experts and their minimal solutions~\cite{Bunel18,Devlin17,chen2018execution}. Third, the space of possible tasks and their solutions is potentially unbounded, and thus, any problem generation technique that relies on exhaustive enumeration is intractable~\cite{singh2012,Ahmed13,Alvin14}.

To overcome these challenges, we propose a novel methodology that operates by first mutating the solution code $\code^{\textnormal{in}}$ to obtain a set of codes $\{\code^{\textnormal{out}}\}$, and then performing symbolic execution over a code $\code^{\textnormal{out}}$ to obtain a visual puzzle $\task^{\textnormal{out}}$. Mutation is efficient by creating an abstract representation of $\code^{\textnormal{in}}$ along with appropriate constraints and querying an SMT solver~\cite{BarrettTinelli2018}; any solution to this query is a mutated code $\code^{\textnormal{out}}$. During symbolic execution, we use Monte Carlo Tree Search (MCTS) to guide the search over the (unbounded) symbolic execution tree. We demonstrate the effectiveness of our methodology by performing an extensive empirical evaluation and user study on a set of reference tasks from the \emph{Hour of code} challenge by \emph{Code.org} and the \emph{Intro to Programming with Karel} course by \emph{CodeHS.com}. In summary, our main contributions are:

\setlength{\leftmargini}{0.7em}
%\vspace{-1mm}
\begin{itemize}
\item We formalize the problem of synthesizing block-based visual programming tasks (Section~\ref{sec.problem}).
%\vspace{-0.5mm}
\item We present a novel approach for generating new visual tasks along with solution codes such that they are conceptually similar but visually dissimilar to a given reference task (Section~\ref{sec.approach}).
%\vspace{-0.5mm}
\item We demonstrate the effectiveness of our approach through an extensive empirical evaluation and user study on reference tasks from real-world programming platforms (Section~\ref{sec.simulations} and Section~\ref{sec.userstudy}).
\end{itemize}

%%%%%%%%%%%%%%%%%%%%%%%%%%%%%%%%%%%%%
%%%%%%%%%%%%%%%%%%%%%%%%%%%%%%%%%%%%%%%%%%%%%%%%%%%%%%%%%%

\vspace{-4mm}
\section{Problem Formulation}\label{sec.problem}
\vspace{-2mm}
%In this section, we first introduce the terminology of tasks and solution codes in block-based visual programming environments, and then formalize the problem of task synthesis.
%We begin by introducing the necessary terminology of tasks/codes in block-based visual programming.
\textbf{The space of  tasks.} We define a task as a tuple $\task := (\task_\textnormal{vis}, \task_\textnormal{store}, \task_\textnormal{size})$, where $\task_\textnormal{vis}$ denotes the visual puzzle,  $\task_\textnormal{store}$ the available block types, and $\task_\textnormal{size}$ the maximum number of blocks allowed in the solution code. For instance, considering the task $\task := \task^{\textnormal{in}}$ in \figref{fig:intro.hoc.t0}, $\task_\textnormal{vis}$ is illustrated in \figref{fig:intro.hoc.t0}, $\task_\textnormal{store} = \{\textnormal{\DSLMove, \DSLTurnL, \DSLTurnR, \DSLRepeatUntil, \DSLIf}\}$, and $\task_\textnormal{size} = 4$.

\begin{figure}[t!]
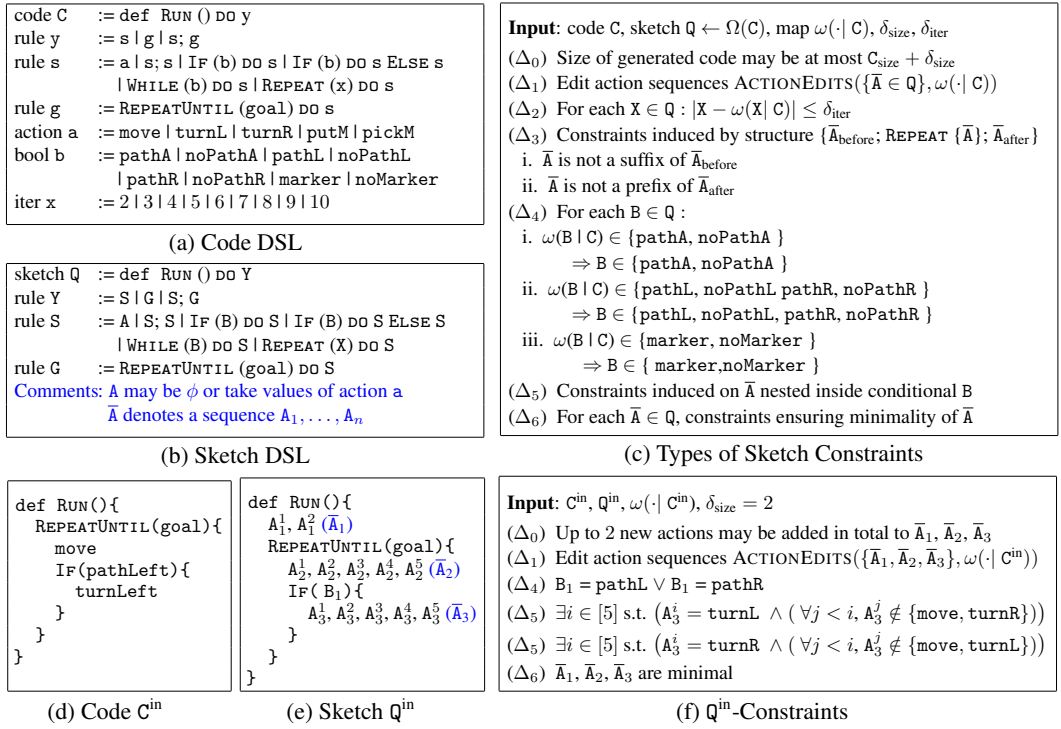

	\centering
	%%%%%%%%%%%%%%%%%%%%%%%%%%%%%%%%%%%%%%%%%%%%%%%%%%%%%%%%%%%%%%%%%%%%
	\begin{minipage}{1\textwidth}
	\begin{minipage}{0.465\textwidth}
			\begin{subfigure}[b]{1.0\textwidth}
			\centering
			{
				\begin{boxcode2col}{1.2cm}{6.7cm}{0.75}{1.08}
						\DSLCode \code &:= \textcode{def }\DSLRun() \DSLdo y \\
						\DSLRule \DSLRuleVar &:= \DSLStmtVar | \DSLRepeatForever  | \DSLStmtVar; \DSLRepeatForever \\
						% %
						\DSLRule \DSLStmtVar \hspace{1mm} &:= \DSLActionVar | $\text{\DSLStmtVar};\text{\DSLStmtVar}$ | \DSLIf(\DSLBoolVar) \DSLdo $\text{\DSLStmtVar}$ | \DSLIf(\DSLBoolVar) \DSLdo $\text{\DSLStmtVar}$ \DSLElse $\text{\DSLStmtVar}$\\
						& \quad | \DSLWhile(\DSLBoolVar) \DSLdo  $\text{\DSLStmtVar}$  | \DSLRepeat(\DSLIterVar) \DSLdo $\text{\DSLStmtVar}$ \\
						\DSLRule \DSLRepeatForever &:= \DSLRepeatUntil(\DSLBoolGoal) \DSLdo $\text{\DSLStmtVar}$\\
						%  %
						\DSLAction \DSLActionVar &:= \DSLMove| \DSLTurnL | \DSLTurnR|  \DSLPutM | \DSLPickM \\
						%   %
						\DSLBool \DSLBoolVar &:= \DSLBoolPathA | \DSLBoolNoPathA | \DSLBoolPathL | \DSLBoolNoPathL \\
						& \quad | \DSLBoolPathR | \DSLBoolNoPathR  | \DSLBoolMarker  | \DSLBoolNoMarker \\
						%    %
						\DSLIter \DSLIterVar &:= $2$ | $3$ | $4$ | $5$ | $6$ | $7$ | $8$ | $9$ | $10$\\
						%\vspace{2mm}
					\end{boxcode2col}
					\vspace{-3mm}
					\caption{Code DSL}
					\label{fig:mutation.1}
				}
			\end{subfigure}
			%\\
			%%%%%%%%%%%%%%%%%		
			\begin{subfigure}[b]{1.0\textwidth}
			\centering
			{
				\begin{boxcode2col}{1.2cm}{6.7cm}{0.75}{1.08}
					\SDSLSketch \SDSLSketchVar  &:= \textcode{def }\DSLRun() \DSLdo $\text{\SDSLVarY}$ \\
					% %
					\DSLRule \SDSLVarY & := \SDSLSStmtVar | \SDSLVarG | \SDSLSStmtVar; \SDSLVarG \\
					% %
					\DSLRule \SDSLSStmtVar &:= \SDSLAction | \SDSLSStmtVar; \SDSLSStmtVar |
																	\DSLIf(\SDSLBool) \DSLdo $\text{\SDSLSStmtVar}$ | \DSLIf(\SDSLBool) \DSLdo $\text{\SDSLSStmtVar}$ \DSLElse $\text{\SDSLSStmtVar}$ \\
					& \quad | \DSLWhile(\SDSLBool) \DSLdo $\text{\SDSLSStmtVar}$  | \DSLRepeat(\SDSLIter) \DSLdo $\text{\SDSLSStmtVar}$ \\
					% %
					\DSLRule \SDSLVarG & := \DSLRepeatUntil(\DSLBoolGoal) \DSLdo $\text{\SDSLSStmtVar}$ \\
					% %
				    \textcolor{blue}{Comments} & \textcolor{blue}{:\ \SDSLAction may be $\phi$ or take values of action \DSLActionVar}\\
				    %\textcolor{blue}{Note 2} 
				    & \textcolor{blue}{\ \ \actionseq~denotes a sequence $\text{\SDSLAction}_{1}, \ldots, \text{\SDSLAction}_{n}$}\\
				\end{boxcode2col}
				\vspace{-3mm}
				\caption{Sketch DSL}
				%\vspace{0.5mm}
				\label{fig:mutation.2}
			}
			\end{subfigure}    		
	\end{minipage}
	%\hspace{0.5em}
	%%%%%%%%%%%%%%%%%
	\begin{minipage}{0.55\textwidth}
		\begin{subfigure}[b]{1.0\textwidth}
		\centering
		{
			\begin{boxcode}{9.6cm}{0.75}{1.0}
				\textbf{Input}: code \code, sketch \SDSLSketchVar $\leftarrow$ $\codetosketch(\code)$, map $\sketchparams(\cdot|~\code)$, $\delta_\text{size}$, $\delta_\text{iter}$
				\begin{enumerate}%[\ensuremath{\Delta_{1}}]
					\item[(\ensuremath{\Delta_{0}})] Size of generated code may be at most $\code_{\textnormal{size}} + \delta_\text{size}$ 
					\item[(\ensuremath{\Delta_1})] Edit action sequences $\localblockcons(\{\actionseq \in \codesketch\},  \sketchparams(\cdot|~\code))$ 
					%
					%\item[(\ensuremath{\Delta_{1}})] $\text{For each }\text{\actionseq} \in \text{\SDSLSketchVar}$, apply \localblockcons(\actionseq, $\sketchparams(\actionseq| ~\code)$)
					%
					\item[(\ensuremath{\Delta_{2}})]  For each \SDSLIter~$ \in \text{\SDSLSketchVar}: |\text{\SDSLIter} - \sketchparams(\text{\SDSLIter}|~\code)| \leq \delta_{\textnormal{iter}}$
					\item[(\ensuremath{\Delta_{3}})] Constraints induced by structure \{\text{\actionseq}\textsubscript{before};  \DSLRepeat\{\actionseq\};  \text{\actionseq}\textsubscript{after}\}
					\begin{enumerate}[\leftmargin=0em]
						\item[i.] \actionseq~is not a suffix of \actionseq\textsubscript{before}
						\item[ii.] \actionseq~is not a prefix of \actionseq\textsubscript{after}
					\end{enumerate}
					\item[(\ensuremath{\Delta_{4}})]  For each \SDSLBool~$\in \text{\SDSLSketchVar}:$
					\begin{enumerate}[\leftmargin=0em]
						\item[i.] \sketchparams(\text{\SDSLBool}~|~\code) $\in$ \{\DSLBoolPathA, \DSLBoolNoPathA\}
							
						\item[] \qquad $\Rightarrow$ \SDSLBool $\in$ \{\DSLBoolPathA, \DSLBoolNoPathA\}							
						
						\item[ii.] \sketchparams(\text{\SDSLBool}~|~\code) $\in$ \{\DSLBoolPathL, \DSLBoolNoPathL 
							 \text{\DSLBoolPathR }, \DSLBoolNoPathR\} 
						\item[] 
							\quad \quad $\Rightarrow$ \SDSLBool $\in$ \{\DSLBoolPathL, \DSLBoolNoPathL, \DSLBoolPathR, \DSLBoolNoPathR\}
						\item[iii.] \sketchparams(\text{\SDSLBool}~|~\code) $\in$ \{\DSLBoolMarker, \DSLBoolNoMarker\} 
						\item[] \quad \quad \hspace{0.3em} $\Rightarrow$ \SDSLBool $\in$ \{ \DSLBoolMarker,\DSLBoolNoMarker\}
					\end{enumerate}
					\item[(\ensuremath{\Delta_{5}})] Constraints induced on \actionseq~nested inside conditional \SDSLBool
					\item[(\ensuremath{\Delta_{6}})] For each $\text{\actionseq} \in \text{\SDSLSketchVar}$, constraints ensuring minimality of \actionseq 
					\vspace{-0.8em}
				\end{enumerate}
			\end{boxcode}
			\vspace{-3mm}
			\caption{Types of Sketch Constraints}
			\label{fig:mutation.3}
		}
		\end{subfigure}
	\end{minipage}
	\end{minipage}
	%%%%%%%%%%%%%%%%%%%%%%%%%%%%%%%%%%%%%%%%%%%%%%%%%%%%%%%%%%%%%%%%%%%%
	%%%%%%%%%%%%%%%%%%%%%%%%%%%%%%%%%%%%%%%%%%%%%%%%%%%%%%%%%%%%%%%%%%%%
	%\begin{minipage}{1\textwidth}
    \centering
		%%%%%%%%%%%%%%%%%
		\begin{subfigure}[b]{0.215\textwidth}
		\centering
		{
			\begin{boxcode}{3.68cm}{0.75}{1.0}
				\textcode{def }\DSLRun\textcode{()\{}\\
				\quad \DSLRepeatUntil\textcode{(}\DSLBoolGoal\textcode{)\{}\\
				\quad \quad \DSLMove\\
				\quad \quad \DSLIf\textcode{(}\DSLBoolPathLeft\textcode{)\{}\\
				\quad \quad \quad \DSLTurnLeft\\
				\quad \quad \textcode{\}}\\
				\quad \textcode{\}}\\
				\textcode{\}}
				\\
				%\vspace{-1.5mm}
			\end{boxcode}
			\vspace{-3mm}
			\caption{Code \code\textsuperscript{in}}
			\label{fig:mutation.4}
			}
		\end{subfigure}
		%%%%%%%%%%%%%%%%%
		\begin{subfigure}[b]{.24\textwidth}
		\centering
		{
			\begin{boxcode}{4.05cm}{0.75}{1.0}
				\textcode{def }\DSLRun\textcode{()\{}\\
				\quad $\text{\SDSLAction}^{1}_{1}$, $\text{\SDSLAction}^{2}_{1}$ \textcolor{blue}{(\actionseqb\textsubscript{1})}\\
				\quad \DSLRepeatUntil\textcode{(}\DSLBoolGoal\textcode{)\{}\\
				\quad \quad $\text{\SDSLAction}^{1}_{2}$, $\text{\SDSLAction}^{2}_{2}$, $\text{\SDSLAction}^{3}_{2}$, 
				$\text{\SDSLAction}^{4}_{2}$, 
				$\text{\SDSLAction}^{5}_{2}$ \textcolor{blue}{(\actionseqb\textsubscript{2})}\\
				\quad \quad \DSLIf\textcode{(}$\text{\SDSLBool}_{1}$\textcode{)\{}\\
				\quad \quad \quad $\text{\SDSLAction}^{1}_{3}$, $\text{\SDSLAction}^{2}_{3}$, $\text{\SDSLAction}^{3}_{3}$, $\text{\SDSLAction}^{4}_{3}$, $\text{\SDSLAction}^{5}_{3}$ \textcolor{blue}{(\actionseqb\textsubscript{3})} \\
				\quad \quad \textcode{\}}\\
				\quad \textcode{\}}\\
				\textcode{\}}
				%\vspace{-1mm}
			\end{boxcode}
			\vspace{-3mm}
			\caption{Sketch $\text{\codesketch}^\text{in}$}
			\label{fig:mutation.5}
			}
		\end{subfigure}
		%%%%%%%%%%%%%%%%%	
			\begin{subfigure}[b]{0.53\textwidth}
			\centering
			{
				\begin{boxcode}{9.6cm}{0.75}{1.0}
					\textbf{Input}: \code\textsuperscript{in}, $\text{\codesketch}^\text{in}$, \sketchparams$(\cdot|~\text{\code}^\text{in}$), $\delta_\text{size} = 2$
					% = ($\phi$ , $\text{\DSLMove}$,  $\text{\DSLTurnL}$, \DSLBoolPathL), 									
					\begin{enumerate}
					    \item[(\ensuremath{\Delta_0})] Up to $2$ new actions may be added in total to $\text{\actionseq}_{1}$, $\actionseq_{2}$, $\actionseq_{3}$
						\item[(\ensuremath{\Delta_1})] Edit action sequences $\localblockcons(\{\actionseq_{1}, \actionseq_{2}, \actionseq_{3}\},  \sketchparams(\cdot|~\code^\text{in}))$ 
					    \item[(\ensuremath{\Delta_4})] $\text{\SDSLBool}_{1}$ = \DSLBoolPathL $\lor$ $\text{\SDSLBool}_{1}$ = \DSLBoolPathR	
					    \item[(\ensuremath{\Delta_5})] $\exists i \in [5]$ s.t. $\big(\text{\SDSLAction}^{i}_{3} =  \text{\DSLTurnL} \ \ \land$  $(\text{ } \forall j < i \text{, } \text{\SDSLAction}^{j}_{3} \notin \{\text{\DSLMove}, \text{\DSLTurnR} \})\big)$
						%Constraints if $\text{\actionseq}_{3}$ is nested in \DSLBool $\text{\SDSLBool}_{1} = \text{\DSLBoolPathL}$:
						%
						\item[(\ensuremath{\Delta_5})] $\exists i \in [5]$ s.t. $\big(\text{\SDSLAction}^{i}_{3} =  \text{\DSLTurnR} \ \ \land$  $(\text{ } \forall j<i \text{, }\text{\SDSLAction}^{j}_{3} \notin \{\text{\DSLMove}, \text{\DSLTurnL} \})\big)$
						%\item[(\ensuremath{\Delta_7})] Only one of $\actionseq_{1}, \actionseq_{2}, \actionseq_{3}$ have additional actions w.r.t. $\sketchparams(\actionseq~|~\code^\text{in})$
						%added to them.
						%
					    %\item[(\ensuremath{\Delta_6})] For $\actionseq \in \{\actionseq_{1}, \actionseq_{2}, \actionseq_{3}\}$, apply $\localblockcons(\actionseq,  \sketchparams(\actionseq~|~\code^\text{in}))$
						%\item[(\ensuremath{\Delta_7})] Only one of $\actionseq_{1}, \actionseq_{2}, \actionseq_{3}$ have additional actions w.r.t. $\sketchparams(\actionseq~|~\code^\text{in})$
						%added to them.
						%
						\item[(\ensuremath{\Delta_6})] $\text{\actionseq}_{1}$, $\actionseq_{2}$, $\actionseq_{3}$ are minimal
					\vspace{-3mm}
				\end{enumerate}	
			\end{boxcode}
			\vspace{-3mm}
			\caption{$\text{\codesketch}^\text{in}$-Constraints}
			%\caption{Constraints for Sketch $\text{\codesketch}^\text{in}$}
			\label{fig:mutation.6}
		}
		\end{subfigure}		
	%\hspace{0.5em}
	%%%%%%%%%%%%%%%%%
	%\end{minipage}
	%%%%%%%%%%%%%%%%%%%%%%%%%%%%%%%%%%%%%%%%%%%%%%%%%%%%%%%%%%%%%%%%%%%%
	\vspace{-4mm}
	\caption{Illustration of key steps in Code Mutation. \figref{fig:mutation.4} shows code $\code^{\textnormal{in}}$ from \figref{fig:intro.hoc.p0}. The code mutation stage, when applied to $\code^{\textnormal{in}}$, generates many output codes, including $\code^{\textnormal{out}}$ in \figref{fig:intro.hoc.p1}.
	}
	%; see text for details. 
	\label{fig:mutation}
	\vspace{-5mm}	
\end{figure}

\looseness-1\textbf{The space of codes.} The programming environment has a domain-specific language (DSL), which defines the set of valid codes $\codes$ and is shown in \figref{fig:mutation.1}. A code $\code \in \codes$ is characterized by several properties, such as the set $\code_{\textnormal{blocks}}$ of block types in $\code$, the number of blocks $\code_{\textnormal{size}}$, the depth $\code_{\textnormal{depth}}$ of the corresponding Abstract Syntax Tree (AST), and the nesting structure $\code_{\textnormal{struct}}$ representing programming concepts exercised by $\code$. For instance, considering the code $\code := \code^{\textnormal{in}}$ in \figref{fig:intro.hoc.p0},  $\code_{\textnormal{blocks}}=\{\textnormal{\DSLMove, \DSLTurnL, \DSLRepeatUntil, \DSLIf}\}$, $\code_{\textnormal{size}}=4$, $\code_{\textnormal{depth}}=3$, and $\code_{\textnormal{struct}} = \{\DSLRun\{\DSLRepeatUntil\{\DSLIf\}\}\}$. 

Below, we introduce two useful definitions relating the task and code space.

\looseness-1\begin{definition}[Solution code] $\code$ is a solution code for  $\task$ if the following holds: $\code$ successfully solves the visual puzzle $\task_\textnormal{vis}$, $\code_{\textnormal{blocks}}  \subseteq \task_\textnormal{store}$, and $\code_{\textnormal{size}} \leq \task_\textnormal{size}$. $\codes_{\task}$ denotes the set of all solution codes for $\task$.
\end{definition}
%%%%%%%%%%%%%%%%%%%%%%%%%%%%%%%%
%\begin{definition}[Solution code for a task] A code $\code$ is a solution code for a task $\task$ if the following conditions hold: $\code$ successfully solves the visual puzzle $\task_\textnormal{vis}$, $\code_{\textnormal{blocks}}  \subseteq T_\textnormal{store}$, and $\code_{\textnormal{size}} \leq \task_\textnormal{size}$. We denote the set of all possible solution codes for a task $\task$ as $\codes_{\task}$.
%\end{definition}
%%%%%%%%%%%%%%%%%%%%%%%%%%%%%%%%

% \begin{definition}[Solvability and $\delta$-minimality of a task] \label{def.minimality} $\task$ is solvable if $|\codes_\task| \geq 1$. Given a threshold $\delta \in \mathbb{N}$, task $\task$ is $\delta$-minimal if $\nexists \code \in \codes_\task$ such that  $\code_{\textnormal{size}} < \task_\textnormal{size} - \delta$.
% \end{definition}
\begin{definition}[Minimality of a task] \label{def.minimality} Given  a solvable task $\task$ with $|\codes_\task| \geq 1$ and a threshold $\delta \in \mathbb{N}$, the task is minimal if $\nexists \code \in \codes_\task$ such that  $\code_{\textnormal{size}} < \task_\textnormal{size} - \delta$.
\end{definition}
%$\delta_{\textnormal{mini}}$
%%%%%%%%%%%%%%%%%%%%%%%%%%%%%%%%
%\begin{definition}[Solvability and minimality] \label{def.minimality} $\task$ is solvable if $|\codes_T| \geq 1$; $\task$ is minimal if $\nexists \code \in \codes_T \textnormal{ s.t. } \code_{\textnormal{size}} < \task_\textnormal{size}$.
%\end{definition}
%%%%%%%%%%%%%%%%%%%%%%%%%%%%%%%%

%%Given a reference input $(\task^{\textnormal{in}}, \code^{\textnormal{in}})$, 
\looseness-1Next, we introduce two definitions formalizing the notion of conceptual similarity. Definition~\ref{def.conceptualsimilarity} formalizes conceptual similarity of a task $\task$ along with one solution code $\code$.  
Since a task can have multiple solution codes, Definition~\ref{def.conceptualsimilarity.task} provides a stricter notion of conceptual similarity of a task $\task$ for all its solution codes. These definitions are used in our objective of task synthesis in conditions (I) and (V) below.

\begin{definition}[Conceptual similarity of $(\task, \code)$] \label{def.conceptualsimilarity}  Given a reference  $(\task^{\textnormal{in}}, \code^{\textnormal{in}})$ and a threshold $\delta \in \mathbb{N}$, a task $\task$ along with a solution code $\code$ is conceptually similar to $(\task^{\textnormal{in}}, \code^{\textnormal{in}})$ if the following holds: $\task_\textnormal{store} = \task^{\textnormal{in}}_\textnormal{store}$, $|\task_\textnormal{size} - \task^{\textnormal{in}}_\textnormal{size}| \leq \delta$, and $\code_{\textnormal{struct}} = \code^{\textnormal{in}}_{\textnormal{struct}}$.
\end{definition}
%$\delta_{\textnormal{size}}$

%\looseness-1
\begin{definition}[Conceptual similarity of $(\task, \cdot)$] \label{def.conceptualsimilarity.task} Given a reference  $(\task^{\textnormal{in}}, \code^{\textnormal{in}})$ and a threshold $\delta \in \mathbb{N}$, a task $\task$ is conceptually similar to $(\task^{\textnormal{in}}, \code^{\textnormal{in}})$ if the following holds: $\task_\textnormal{store} = \task^{\textnormal{in}}_\textnormal{store}$, $|\task_\textnormal{size} - \task^{\textnormal{in}}_\textnormal{size}| \leq \delta$, and $\forall \code \in \codes_\task, \code_{\textnormal{struct}} = \code^{\textnormal{in}}_{\textnormal{struct}}$.
\end{definition}

%%%%%%%%%%%%%%%%%%%%%%%%%%%%%%%%%%%% 20200604-1630
%\looseness-1\begin{definition}[Conceptual similarity ] \label{def.conceptualsimilarity}  Given a threshold $\delta \in \mathbb{N}$, two tasks  with specific solution codes $(\task, \code)$ and $(\task', \code')$ are conceptually similar if $\task_\textnormal{store} = \task'_\textnormal{store}$, $|\task_\textnormal{size} - \task'_\textnormal{size}| \leq \delta$, and $\code_{\textnormal{struct}} = \code'_{\textnormal{struct}}$.
%\end{definition}

%%%%the following conditions hold:
%%%%%%%%%%%%%%%%%%%%%%%%%%%%%%%%
%%%%\begin{definition}[Conceptual similarity] \label{def.conceptualsimilarity}  Given a threshold $\delta \in \mathbb{N}$, two tasks along with specific solution codes $(\task, \code)$ and $(\task', \code')$ are conceptual similar if the following holds: $\task_\textnormal{store} = \task'_\textnormal{store}$, $|\task_\textnormal{size} - \task'_\textnormal{size}| \leq \delta$, and $\code_{\textnormal{struct}} = \code'_{\textnormal{struct}}$.
%%%%\end{definition}
%%%%%%%%%%%%%%%%%%%%%%%%%%%%%%%%

\textbf{Environment domain knowledge.} We now formalize our domain knowledge about the block-based environment to measure \emph{visual dissimilarity} of two tasks, and capture some notion of \emph{interestingness and quality} of a task. Given tasks $\task$ and $\task'$, we measure their visual dissimilarity by an environment-specific function $\FDissimilarity(\task_\textnormal{vis}, \task'_\textnormal{vis}) \in [0, 1]$. Moreover, we measure generic quality of a task with function $\FQuality(\task_\textnormal{vis},\code) \in [0, 1]$. We provide specific instantiations of $\FDissimilarity$  and $\FQuality$ in our evaluation.

\textbf{Objective of task synthesis.} Given a reference task $\task^{\textnormal{in}}$ and a solution code $\code^{\textnormal{in}} \in \codes_{\task^{\textnormal{in}}}$ as input, we seek to generate a set $\{(\task^{\textnormal{out}}, \code^{\textnormal{out}})\}$ of new tasks along with solution codes that are \emph{conceptually similar} but \emph{visually dissimilar} to the input. 
%Formally, our task synthesis problem is parameterized by $(\delta_{\textnormal{size}}, \delta_{\textnormal{diss}})$, and we seek to output new tasks with the following conditions:
Formally, given parameters $(\delta_{\textnormal{size}}, \delta_{\textnormal{diss}}, \delta_{\textnormal{qual}})$, our objective is to synthesize new tasks meeting the following conditions:
\vspace{-1.5mm}
\begin{enumerate}[(I)]
	\setlength{\itemsep}{2pt}
  	\setlength{\parskip}{0pt}
	\item $(\task^{\textnormal{out}}, \code^{\textnormal{out}})$ is conceptually similar to $(\task^{\textnormal{in}}, \code^{\textnormal{in}})$ with threshold $\delta_{\textnormal{size}}$ in Definition~\ref{def.conceptualsimilarity}.
	\item $\task^{\textnormal{out}}$ is visually dissimilar to $\task^{\textnormal{in}}$ with margin $\delta_{\textnormal{diss}}$, i.e., $\FDissimilarity(\task^{\textnormal{in}}_\textnormal{vis}, \task^{\textnormal{out}}_\textnormal{vis}) \geq \delta_{\textnormal{diss}}$.
	\item $\task^{\textnormal{out}}$ has a quality score above threshold $\delta_{\textnormal{qual}}$, i.e.,  $\FQuality(\task^{\textnormal{out}}_\textnormal{vis},\code^\textnormal{out}) \geq \delta_{\textnormal{qual}}$.
	%\maria{Do we mean to say a "high" quality score? Otherwise, isn't this property always satisfied?}	
	%\vspace{-1.5mm}
\end{enumerate}
% %%%%%%%%%%%%%%%%%%%%%%%%%%%%%%%%%%%%%%%%%%%%%%

%In addition, we seek to generate tasks with the following desirable properties:
%\looseness-1
In addition, depending on the use case, it is desirable that the new tasks satisfy the following criteria:
\vspace{-1.5mm}
\begin{enumerate}[(I)]
	\setcounter{enumi}{3}
	\setlength{\itemsep}{2pt}
  	\setlength{\parskip}{0pt}
	\item $\code^{\textnormal{out}}$ is different from the input solution code, i.e., $\code^{\textnormal{out}} \neq \code^{\textnormal{in}}$.	
	\item $\task^{\textnormal{out}}$ is conceptually similar to $(\task^{\textnormal{in}}, \code^{\textnormal{in}})$ with threshold $\delta_{\textnormal{size}}$ in Definition~\ref{def.conceptualsimilarity.task}.
	\item $\task^{\textnormal{out}}$ is minimal as per Definition~\ref{def.minimality} for a desired value of $\delta_{\textnormal{mini}}$ (e.g.,  $\delta_{\textnormal{mini}}=0$ or $\delta_{\textnormal{mini}}=1$).	
	%Either $\task^{\textnormal{out}}$ has a unique code solution, i.e., $\codes_{\task^{\textnormal{out}}} = \{\code^{\textnormal{out}}\}$, or it holds that the programming concepts exercised by all code solutions for $\task^{\textnormal{out}}$ are the same, i.e., $ \forall \code \in \codes_{\task^{\textnormal{out}}}, \code_{\textnormal{struct}} = \code^{\textnormal{out}}_{\textnormal{struct}}$.
	%a more relaxed property, that 
	%\item $\task^{\textnormal{out}}$ has a unique code solution, i.e., $\codes_{\task^{\textnormal{out}}} = \{\code^{\textnormal{out}}\}$.
%\vspace{-1.5mm}
\end{enumerate}

\vspace{-3mm}
\section{Our Task Synthesis Algorithm}\label{sec.approach}
\vspace{-2mm}
%clip,width=0.45\textwidth,
%%%%%%%%%%%%%
\begin{wrapfigure}[6]{r}{0.50\textwidth}
  \centering
  %trim={10pt 70pt 10pt 70pt},
  \includegraphics[height=1.0cm]{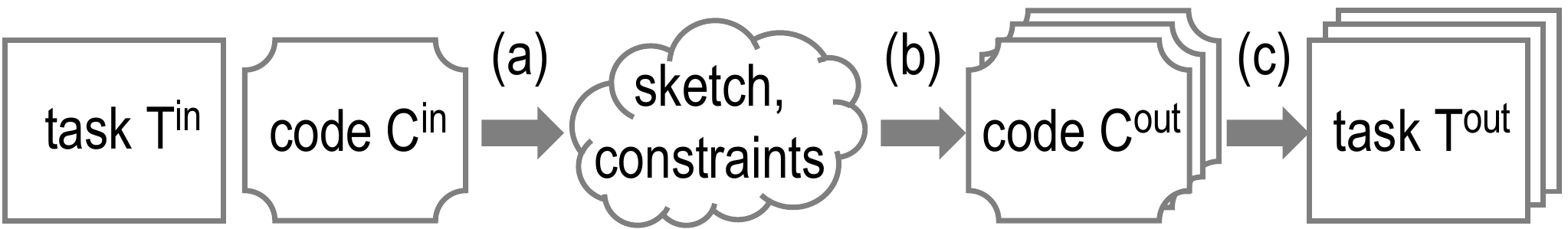}
  %\vspace{0.5mm}
  \caption{Stages in our task synthesis algorithm.}
  \label{fig:overviewofapproach}
\end{wrapfigure}
%%%%%%%%%%%%%
We now present the pipeline of our algorithm (see \figref{fig:overviewofapproach}), which takes as input a reference task $\task^{\textnormal{in}}$ and its solution code $\code^{\textnormal{in}}$, and generates a set $\{(\task^{\textnormal{out}}, \code^{\textnormal{out}})\}$ of new tasks with their solution codes. The goal is for this set to be conceptually similar to $(\task^{\textnormal{in}}, \code^{\textnormal{in}})$, but for new tasks $\{\task^{\textnormal{out}}\}$ to be visually dissimilar to $\task^{\textnormal{in}}$. This is achieved by two main stages: (1)~mutation of $\code^{\textnormal{in}}$ to obtain a set $\{\code^{\textnormal{out}}\}$, and (2)~symbolic execution of each $\code^{\textnormal{out}}$ to create a task $\task^{\textnormal{out}}$. The first stage, presented in Section~\ref{sec.approach.mutation}, converts $\code^{\textnormal{in}}$ into an abstract representation restricted by a set of constraints (\figref{fig:overviewofapproach}(a)), which must be satisfied by any generated $\code^{\textnormal{out}}$ (\figref{fig:overviewofapproach}(b)). The second stage, described in Section~\ref{sec.approach.tasksynthesis}, applies symbolic execution on each code $\code^{\textnormal{out}}$ to create a corresponding visual task $\task^{\textnormal{out}}$ (\figref{fig:overviewofapproach}(c)) while using Monte Carlo Tree Search (MCTS) to guide the search in the symbolic execution tree.

%\vspace{-10mm}
\subsection{Code Mutation}\label{sec.approach.mutation}
\vspace{-1mm}
This stage in our pipeline mutates code $\code^{\textnormal{in}}$ of task $\task^{\textnormal{in}}$ such that its conceptual elements are preserved. Our mutation procedure consists of three main steps. First, we generate an abstract representation of $\code^{\textnormal{in}}$, called \emph{sketch}. Second, we restrict the sketch with constraints that describe the space of its concrete instantiations. Although this formulation is inspired from work on generating algebra problems~\cite{singh2012}, we use it in the entirely different context of generating conceptually similar mutations of $\code^{\textnormal{in}}$. This is achieved in the last step, where we use the sketch and its constraints to query an SMT solver~\cite{BarrettTinelli2018}; the query solutions are mutated codes $\{\code^{\textnormal{out}}\}$ such that $\code^{\textnormal{out}}_{\textnormal{struct}} = \code^{\textnormal{in}}_{\textnormal{struct}}$ (see Definition~\ref{def.conceptualsimilarity}).

\textbf{Step 1: Sketch.} The sketch of code \code, denoted by  \SDSLSketchVar, is an abstraction of $\code$ capturing its skeleton and generalizing $\code$ to the space of conceptually similar codes. \SDSLSketchVar, expressed in the language of \figref{fig:mutation.2}, is generated from $\code$ with mapping $\codetosketch$. In particular, the map exploits the AST structure of the code: the AST is traversed in a depth-first manner, and all values are replaced with their corresponding sketch variables, i.e., \DSLAction \DSLActionVar, \DSLBool \DSLBoolVar, and \DSLIter \DSLIterVar are replaced with \SDSLAction, \SDSLBool, and \SDSLIter, respectively. In the following, we also use mapping $\sketchparams(\cdot|~\code)$, which takes a sketch variable in \SDSLSketchVar and returns its value in \code.

In addition to the above, we may extend a variable \SDSLAction to an \emph{action sequence} \actionseq, since any \SDSLAction is allowed to be empty ($\phi$). We may also add an action sequence of length $\delta_{\textnormal{size}}$ at the beginning and end of the obtained sketch. As an example, consider the code in \figref{fig:mutation.4} and the resulting sketch in \figref{fig:mutation.5}. Notice that, while we add an action sequence at the beginning of the sketch (\actionseq\textsubscript{1}), no action sequence is appended at the end because construct \DSLRepeatUntil renders any succeeding code unreachable.

\textbf{Step 2: Sketch constraints}. Sketch constraints restrict the possible concrete instantiations of a sketch by encoding the required semantics of the mutated codes. All constraint types are in \figref{fig:mutation.3}.

In particular, $\Delta_0$ restricts the size of the mutated code within $\delta_{\textnormal{size}}$. 
$\Delta_1$ specifies the allowed mutations to an action sequence based on its value in the code, given by $\sketchparams(\actionseq~|~\code)$. For instance, this constraint could result in converting all \DSLTurnLeft~actions of a sequence to \DSLTurnRight.
$\Delta_2$ restricts the possible values of the \DSLRepeat~counter within threshold $\delta_{\textnormal{iter}}$.
$\Delta_3$ ensures that the \DSLRepeat~counter is optimal, i.e., action subsequences before and after this construct are not nested in it.
$\Delta_4$ specifies the possible values of the \DSLIf~condition based on its value in the code, given by $\sketchparams(\textnormal{\SDSLBool}~|~\code)$.
$\Delta_5$ refers to constraints imposed on action sequences nested within conditionals. As an example, consider $\Delta_5$ in \figref{fig:mutation.6}, which states that if $\text{\SDSLBool}_{1}$ = \DSLBoolPathLeft, then the nested action sequence must have at least one \DSLTurnLeft~action, and the first occurrence of this action must not be preceded by a \DSLMove~or \DSLTurnRight, thus preventing invalid actions within the conditional.
$\Delta_6$ ensures minimality of an action sequence, i.e., optimality of the constituent actions to obtain the desired output. This constraint would, for instance, eliminate redundant sequences such as \DSLTurnLeft, \DSLTurnRight, which does not affect the output, or \DSLTurnLeft, \DSLTurnLeft, \DSLTurnLeft, whose output could be achieved by a single \DSLTurnRight. All employed elimination sequences can be found in the supplementary material.
The entire list of constraints applied on the solution code in \figref{fig:mutation.4} is shown in \figref{fig:mutation.6}.

\textbf{Step 3: SMT query.} For a sketch \SDSLSketchVar generated from code $\code$ and its constraints, we pose the following query to an SMT solver: (sketch \SDSLSketchVar, \SDSLSketchVar-constraints). As a result, the solver generates a set of instantiations, which are conceptually similar to \code. 
In our implementation, we used the Z3 solver~\cite{deMouraBjorner2008}. For the code in \figref{fig:mutation.4}, Z3 generated $66$ mutated codes in $0.8$s from an exhaustive space of $2,997$ possible codes with $\delta_{\textnormal{size}} = 2$. One such mutation is shown in \figref{fig:intro.hoc.p1}. %A detailed count of the generated mutations for all solution codes of the reference tasks used in our evaluation can be found in \figref{fig:experiments.analysis}.

%\textbf{Remark.} 
While this approach generates codes that are devoid of most semantic irregularities, it has its limitations. Certain irregularities continue to exist in some generated codes: An example of such a code included the action sequence \DSLMove, \DSLTurnLeft, \DSLMove, \DSLTurnLeft,  \DSLMove, \DSLTurnLeft,  \DSLMove, \DSLTurnLeft, which results in the agent circling back to its initial location in the task space. This kind of undesirable behaviour is eliminated in the symbolic execution stage of our pipeline.

\begin{figure}[t!]
\centering
	%%%%%%%%%%%%%%%%%
	\includegraphics[width=0.95\textwidth]{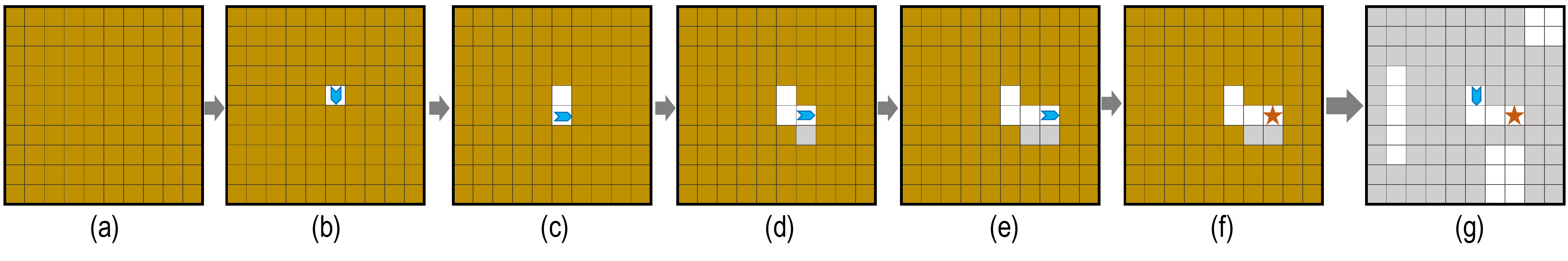}
	%\vspace{0.01mm}
	 \vspace{-1.5mm}	
	\caption{\looseness-1Illustration of symbolic execution on $\code^{\textnormal{out}}$ from \figref{fig:intro.hoc.p1}. (b) shows the initial configuration of the agent's location and orientation as well as the status of the grid cells (\emph{unknown}, \emph{free}, \emph{blocked}, \emph{goal}). (c)--(e) show the symbolic execution steps where conditions \DSLBoolGoal and \DSLBoolPathRight are \DSLBoolFalse. (f) shows the step where \DSLBoolGoal is \DSLBoolTrue. (g) shows the post-processing step where a puzzle $\task^{\textnormal{out}}_\textnormal{vis}$ is obtained.
	%which respect the status of cells.
	}
	\label{fig:symbolicexecution}
	\vspace{-3.5mm}
\end{figure}
%$\textnormal{(a)} \leftarrow \textnormal{(b)}$
%In our setting, the program inputs of $\code^{\textnormal{out}}$ are the initial location and orientation of the agent as well as the status of the grid-world cells (free, blocked, goal). Symbolic execution collects constraints over these from statements in the code. For example, the first action in Figure~\ref{fig:intro.hoc.p1} moves the agent forward, which constrains its orientation towards a cell that must be either free or the goal. If it is not the goal, the next two actions in the code impose the additional constraint that at least one cell on the agent's left must not be blocked. In this manner, symbolic execution generates a visual task for each path in $\code^{\textnormal{out}}$.
%
%However, not all of these tasks are suitable. For instance, if the goal is reached after the first \DSLMove~in Figure~\ref{fig:intro.hoc.p1}, all other statements in $\code^{\textnormal{out}}$ are not covered, rendering the task less suitable for this code. Na\"ively, symbolic execution could first enumerate all paths in $\code^{\textnormal{out}}$ and their corresponding tasks, and then rank them in terms of suitability. However, solution codes may have an unbounded number of paths, which leads to \emph{path explosion}, that is, the inability to cover all paths with tractable resources.
%%%%%%%%%%%%%%%%%%%%%%%%%%%%%%%%%%%%%
%%%%%%%%%%%%%%%%%%%%%%%%%%%%%%%%%%%%%%%%%%%%%%%%%%%%%%%%%%
\vspace{-2.5mm}
\subsection{Symbolic Execution}\label{sec.approach.tasksynthesis}
\vspace{-1mm}
%: From $\code^{\textnormal{out}}$ to $\task^{\textnormal{out}}$
Symbolic execution~\cite{King1976} is an automated test-generation technique that symbolically explores execution paths in a program. During exploration of a path, it gathers symbolic constraints over program inputs from statements along the path. These constraints are then mutated (according to a search strategy), and an SMT solver is queried to generate new inputs that explore another path.

\looseness-1\textbf{Obtaining visual tasks with symbolic execution.}
This stage in our pipeline applies symbolic execution on each generated code $\code^{\textnormal{out}}$ to obtain a suitable visual task $\task^{\textnormal{out}}$. The program inputs of $\code^{\textnormal{out}}$ are the agent's initial location/orientation and the status of the grid cells (\emph{unknown}, \emph{free}, \emph{blocked}, \emph{marker}, \emph{goal}), which is initially \emph{unknown}. Symbolic execution collects constraints over these from code statements. As in \figref{fig:symbolicexecution} for one path, symbolic execution generates a visual task for each path in $\code^{\textnormal{out}}$.

However, not all of these tasks are suitable. For instance, if the goal is reached after the first \DSLMove~in \figref{fig:intro.hoc.p1}, all other statements in $\code^{\textnormal{out}}$ are not covered, rendering the task less suitable for this code. Na\"ively, symbolic execution could first enumerate all paths in $\code^{\textnormal{out}}$ and their corresponding tasks, and then rank them in terms of suitability. However, solution codes may have an unbounded number of paths, which leads to \emph{path explosion}, that is, the inability to cover all paths with tractable resources.

\textbf{Guiding symbolic execution using Monte Carlo Tree Search (MCTS).}
To address this issue, we use MCTS~\cite{kocsis2006bandit} as a search strategy in symbolic execution with the goal of generating more suitable tasks with fewer resources---we define task suitability next. Symbolic execution has been previously combined with MCTS in order to direct the exploration towards costly paths~\cite{luckow2018monte}.
In the supplementary material, we provide an example demonstrating how MCTS could guide the symbolic execution in generating more suitable tasks.
%
%In the supplementary material, we provide an example demonstrating the combination of MCTS and symbolic execution.

\looseness-1
As previously observed~\cite{kartal2016data}, a critical component of effectively applying MCTS is to define an evaluation function that describes the desired properties of the output, i.e., the visual tasks. Tailoring the evaluation function to our unique setting is exactly what differentiates our approach from existing work. 
In particular, our evaluation function, \FScore, distinguishes suitable tasks 
% by taking into account the coverage of code $\code^{\textnormal{out}}$ by task $\task$ (\FCoverage), the dissimilarity of $\task$ to reference task $\task^{\textnormal{in}}$ (\FDissimilarity; see Section~\ref{sec.problem}), the quality of $\task$ (\FQuality; see Section~\ref{sec.problem}) as well as its validity, i.e., whether the agent crashes into a wall (\FNoCrash). 
by assigning a score ($\in [0,1]$) to them, which guides the MCTS search. A higher \FScore~indicates a more suitable task. Its constituent components are: (i) $\FCoverage(\task^\textnormal{out}_\textnormal{vis},\code^\textnormal{out}) \in \{0,1\}$, which evaluates to 1 in the event of complete coverage of code $\code^{\textnormal{out}}$ by task $\task_\textnormal{vis}^\textnormal{out}$ and 0 otherwise; (ii) $\FDissimilarity(\task_\textnormal{vis}^\textnormal{out}, \task_\textnormal{vis}^\textnormal{in}) \in [0,1]$, which evaluates the dissimilarity of $\task^\textnormal{out}$ to $\task^{\textnormal{in}}$ (see Section~\ref{sec.problem});
(iii) $\FQuality(\task_\textnormal{vis}^\textnormal{out},\code^\textnormal{out}) \in [0,1]$, which evaluates the quality and validity of $\task^\textnormal{out}$; (iv) $\FNoCrash(\task_\textnormal{vis}^\textnormal{out}, \code^\textnormal{out}) \in \{0,1\}$, which evaluates to 0 in case the agent crashes into a wall and 1 otherwise; and (v) $\FNoCut(\task_\textnormal{vis}^\textnormal{out}, \code^\textnormal{out}) \in \{0,1\}$, which evaluates to 0 if there is a \emph{shortcut sequence} of actions (\DSLActionVar in Fig.~\ref{fig:mutation.1}) smaller than $\code^\textnormal{out}_\textnormal{size}$
that solves \task\textsuperscript{out} and 1 otherwise. \FQuality~and~\FNoCut~also resolve the limitations of our mutation stage by eliminating codes and tasks that lead to undesirable agent behavior. We instantiate \FScore~in the next section.

\begin{figure}[t!]
\centering
	%%%%%%%%%%%%%%%%%
	%\scalebox{0.9}{
	\scalebox{0.85}{
	\setlength\tabcolsep{2.3pt}
	\renewcommand{\arraystretch}{1.05}
	\begin{tabular}{c|l|c|c|l}
			%Task id& \multicolumn{1}{c|}{$\task_\textnormal{store}$: Types of blocks allowed in the code} & Solution size & Solution depth & \multicolumn{1}{c}{Source} \\	
			Task $\task$& \multicolumn{1}{c|}{$\task_\textnormal{store}$} & $\task_\textnormal{size}$ (= $\code_{\textnormal{size}}$) & $\code_{\textnormal{depth}}$ & \multicolumn{1}{c}{Type: Source} \\			
			\toprule
            \hocA 			& \DSLMove, \DSLTurnL, \DSLTurnR											& $5$ &  $1$ & \hocType: \emph{Maze 4}~\cite{hourofcode_maze} \\ 
    		\hocC 			& \DSLMove, \DSLTurnL, \DSLTurnR, \DSLRepeat 								&  $3$ & $2$ & \hocType: \emph{Maze 7}~\cite{hourofcode_maze}   \\
    		\hocD 			& \DSLMove, \DSLTurnL, \DSLTurnR, \DSLRepeat 								&  $5$ & $2$ & \hocType: \emph{Maze 8}~\cite{hourofcode_maze}   \\
    		\hocF 			& \DSLMove, \DSLTurnL, \DSLTurnR, \DSLRepeatUntil 						& 5 & 2 & \hocType: \emph{Maze 12}~\cite{hourofcode_maze}  \\    	   		
            \hocG 			& \DSLMove, \DSLTurnL, \DSLTurnR, \DSLRepeatUntil, \DSLIf 			& $4$ & $3$ & \hocType: \emph{Maze 16}~\cite{hourofcode_maze} \\ 
    		\hocH 			& \DSLMove, \DSLTurnL, \DSLTurnR, \DSLRepeatUntil, \DSLIfElse 		& $4$ & $3$ & \hocType: \emph{Maze 18}~\cite{hourofcode_maze} \\ 
			%\midrule
            \karelA 		& \DSLMove, \DSLTurnL, \DSLTurnR, \DSLPickM, \DSLPutM 										& $5$ & $1$ & \karelType: \emph{Our first} \cite{intro_to_karel_codehs} \\
    		\karelC 		& \DSLMove, \DSLTurnL, \DSLTurnR, \DSLPickM, \DSLPutM, \DSLRepeat 					& $4$ &	$2$ & \karelType: \emph{Square} \cite{intro_to_karel_codehs} \\
    		\karelE 		& \DSLMove, \DSLTurnL, \DSLTurnR, \DSLPickM, \DSLPutM, \DSLRepeat, \DSLIfElse 	& $5$ &	$3$ & \karelType: \emph{One ball in each spot} \cite{intro_to_karel_codehs} \\
    		\karelF 		& \DSLMove, \DSLTurnL, \DSLTurnR, \DSLPickM, \DSLPutM, \DSLWhile 					&  $7$ & $2$ & \karelType: \emph{Diagonal} \cite{intro_to_karel_codehs} \\	
    		%%%%%%%%%%%%%%%%%%%%%%
    		%%%% add karelH and karelI in supplementary
    		%karel~\karelH 		& \DSLMove, \DSLTurnL, \DSLTurnR, \DSLPickM, \DSLPutM, \DSLWhile, \DSLIf 			& $8$ & $3$ & \emph{Stairway to heaven} \cite{intro_to_karel_codehs} \\
    		%% ``Stairway to heaven with pick"
    		%karel~\karelI 			& \DSLMove, \DSLTurnL, \DSLTurnR, \DSLPickM, \DSLPutM, \DSLWhile, \DSLIfElse 	& $5$ & $3$ & \emph{Maze 18}~\cite{hourofcode_maze} with markers\\
    		%%%%%%%%%%%%%%%%%%%%%% Ignore the tasks below
    		%%%%%\hocI 			& \DSLMove, \DSLTurnL, \DSLTurnR, \DSLRepeatUntil, \DSLIfElse 		& $6$ & $4$ & \emph{Maze 20}~\cite{hourofcode_maze} \\     		
    		%%%%%\karelJ 			& \DSLMove, \DSLTurnL, \DSLTurnR, \DSLPickM, \DSLPutM, \DSLWhile, \DSLIfElse 	& $7$ & $4$ & \emph{Maze 20}~\cite{hourofcode_maze} with markers\\
    		%	\vspace{-2mm}
    	    \bottomrule
   \end{tabular}
   }
    \vspace{-1mm}
	\caption{Datasets for HOC and Karel tasks.}
	\label{fig:dataset}
	\vspace{-3.5mm}
\end{figure}
%%%%%%%%%%%%%%%%%%%%%%%%%%%%%%%%%%%%%

%%%%%%%%%%%%%%%%%%%%%%%%%%%%%%%%%%%%%%%%%%%%%%%%%%%%%%%%%%
%%%%%%%%%%%%%%%%%%%%%%%%%%%%%%%%%%%%%%%%%%%%%%%%%%%%%%%%%%
\vspace{-2mm}
\section{Experimental Evaluation}\label{sec.simulations}
\vspace{-2mm}
In this section, we evaluate our task synthesis algorithm on \hocType~and~\karelType~tasks. Our implementation is publicly available.\footnote{\href{https://github.com/adishs/neurips2020_synthesizing-tasks_code}{https://github.com/adishs/neurips2020\_synthesizing-tasks\_code}\label{footnote.githubrepo}} While we give an overview of key results here, a detailed description of our setup and additional experiments can be found in the supplementary material.
%Our implementation will be released together with the final version of the paper. 

%%%%%%%%%%%%%%%%%%%%%%%%%%%%%%%%%%%%%%%%%%%%%%%%%%%%%%%%%%
\vspace{-2mm}
\subsection{Reference Tasks and Specifications}
\vspace{-1mm}
\textbf{Reference tasks.} We use a set of ten reference tasks from \hocType~and~\karelType, shown in \figref{fig:dataset}. The \hocType~tasks were selected from the \textit{Hour of Code: Classic Maze} challenge by \textit{Code.org}~\cite{hourofcode_maze} and the Karel tasks from the \textit{Intro to Programming with Karel} course by \textit{CodeHS.com}~\cite{intro_to_karel_codehs}. The DSL of \figref{fig:mutation.1} is generic in that it includes both HOC and Karel codes, with the following differences: (i)~construct \DSLWhile, marker-related actions \DSLPutM, \DSLPickM, and conditions \DSLBoolNoPathA, \DSLBoolNoPathL, \DSLBoolNoPathR, \DSLBoolMarker, \DSLBoolNoMarker~are specific to Karel only; (ii)~construct \DSLRepeatUntil and \DSLBoolGoal are specific to HOC only. Furthermore, the puzzles for HOC and Karel are of different styles (see \figref{fig:intro.hoc} and \figref{fig:intro.karel}). 
%However, the Karel language is more complex and 
%Our experimental results are presented in \figref{fig:experiments.analysis}.
For all tasks, the grid size of the puzzles is fixed to $10\times10$ cells (grid-size parameter $n = 10$).

\textbf{Specification of scoring functions.} $\FQuality(\task_\textnormal{vis}^\textnormal{out},\code^\textnormal{out}) \in [0,1]$ was approximated as the sum of the normalized counts of `moves', `turns', `segments', and `long-segments' in the grid; segments and long-segments are sequences of $\geq 3$ and $\geq 5$ \DSLMove actions respectively. 
More precisely, for \hocType~tasks, we used the following function where features are computed by executing $\code^\textnormal{out}$ on $\task^\textnormal{out}_\textnormal{vis}$:
\begin{align*}
    \FQuality^{\hocType}(\task_\textnormal{vis}^\textnormal{out},\code^\textnormal{out}) & = \frac{1}{4}\Big{(} \frac{\countqual\text{moves}}{2n} +  \frac{\countqual\text{turns}}{n} +  \frac{\countqual\text{segments}}{n/2} +  
    \frac{\countqual\text{long-segments}}{n/3} \Big{)}.
\end{align*}
%\looseness-1
Furthermore, in our implementation, $\FQuality(\cdot)$ value was set to $0$ when $\FNoCrash(\cdot) = 0$. 
For \karelType~tasks, \FQuality~additionally included the normalized counts of \DSLPutM and \DSLPickM, and is provided in the supplementary material.
$\FDissimilarity(\task_\textnormal{vis}^\textnormal{out}, \task_\textnormal{vis}^\textnormal{in}) \in [0,1]$ was computed based on the dissimilarity of the agent's initial location/orientation w.r.t. $\task_\textnormal{vis}^\textnormal{in}$, and the grid-cell level dissimilarity based on the Hamming distance between $\task_\textnormal{vis}^\textnormal{out}$ and $\task_\textnormal{vis}^\textnormal{in}$. More precisely, we used the following function:
%we instantiated it as a linear combination of the dissimilarity features as follows:
\begin{align*}
    \FDissimilarity(\task_\textnormal{vis}^\textnormal{out}, \task_\textnormal{vis}^\textnormal{in}) & = \frac{1}{3}\Big{(}
    \textnormal{diss}(\textnormal{loc}~|~\task_\textnormal{vis}^\textnormal{out}, \task_\textnormal{vis}^\textnormal{in}) +  \textnormal{diss}(\textnormal{dir}~|~\task_\textnormal{vis}^\textnormal{out}, \task_\textnormal{vis}^\textnormal{in}) + 
    \textnormal{diss}(\textnormal{grid-cells}~|~\task_\textnormal{vis}^\textnormal{out}, \task_\textnormal{vis}^\textnormal{in})
    \Big{)}
\end{align*}
where $\textnormal{diss}(\textnormal{loc}~|~\task_\textnormal{vis}^\textnormal{out}, \task_\textnormal{vis}^\textnormal{in}) \in \{0, 1\}$, $\textnormal{diss}(\textnormal{dir}~|~\task_\textnormal{vis}^\textnormal{out}, \task_\textnormal{vis}^\textnormal{in}) \in \{0, 1\}$, and $\textnormal{diss}(\textnormal{grid-cells}~|~\task_\textnormal{vis}^\textnormal{out}, \task_\textnormal{vis}^\textnormal{in}) \in [0, 1]$ (after the Hamming distance is normalized with a factor of $\frac{2}{n^2}$). 
%Additional details and the specific \FQuality~function used for \karelType~tasks are in the supplementary material. 
%
%

%Our evaluation function $\FScore(\task^\textnormal{out}, \code^\textnormal{out}, \task^{\textnormal{in}}, \code^{\textnormal{in}}) \in [0,1]$ measures the suitability of a generated task. A higher \FScore~indicates a more suitable task. We describe the elements of our evaluation function in greater detail here. We defined it in Section~\ref{sec.simulations} and present it here again for completeness:

Next, we define the evaluation function $\FScore(\task^\textnormal{out}, \code^\textnormal{out}, \task^{\textnormal{in}}, \code^{\textnormal{in}}) \in [0,1]$ used by MCTS:
%We define \FScore~as follows
%\vspace{-2mm}
\begin{align*}
    \FScore(\task^\textnormal{out}, \code^\textnormal{out}, \task^\textnormal{in}, \code^\textnormal{in}) & = \underbrace{\mathbbm{1}{\big(\medmath{ \FQuality(\task^\textnormal{out}_\textnormal{vis}, \code^\textnormal{out}) \geq \delta_{\textnormal{qual}}, \FNoCrash(\task^\textnormal{out}_\textnormal{vis}, \code^\textnormal{out}) = 1, \FNoCut(\task^\textnormal{out}_\textnormal{vis}, \code^\textnormal{out}) = 1 }\big)}}_\text{(i)} \cdot \\
    & \ \ \ \ \underbrace{\big[ \alpha_{1}\FCoverage(\task^\textnormal{out}_\textnormal{vis}, \code^\textnormal{out}) + \alpha_{2}\FQuality(\task^\textnormal{out}_\textnormal{vis}, \code^\textnormal{out}) + \alpha_{3} \FDissimilarity(\task^\textnormal{out}_\textnormal{vis}, \task^\textnormal{in}_\textnormal{vis}) \big]}_\text{(ii)}
%\label{evalfunc} \tag{2}
\end{align*}
where $\mathbbm{1}$ is an indicator function and each constant $\alpha = 1/3$. 
%Component (ii) in the above function supplies the gradients for guiding the search in MCTS; at the end of the MCTS run, the \textit{best} task (i.e, the one with the highest \FScore~value) is picked only from the pool of generated tasks which have $\FScore(\cdot) > 0$ and satisfy $\FCoverage(\cdot) = 1$.
%
Component (ii) in the above function supplies the gradients for guiding the search in MCTS; Component (i) is applied at the end of the MCTS run to pick the output.  More precisely, the \textit{best} task (i.e, the one with the highest \FScore~value) is picked only from the pool of generated tasks which have $\FScore(\cdot) > 0$ and satisfy $\FCoverage(\cdot) = 1$.
%%%We discuss the functions \FQuality~and~\FDissimilarity~next.
%%%$2\textnormal{b}$ in Eq.\ref{evalfunc} 

%parameters
\looseness-1\textbf{Specification of task synthesis and MCTS.}  As per Section~\ref{sec.problem}, we set the following thresholds for our algorithm: (i) $\delta_\textnormal{size} = 2$, (ii) $\delta_\textnormal{diss} = 0.33$, and (iii) $\delta_\textnormal{qual}=0.2$ for codes with \DSLWhile or \DSLRepeatUntil, and $0.05$ otherwise.  We run MCTS $10$ times per code, with each run generating one task.
We set the maximum iterations of a run to $2$ million (M) and the exploration constant to $2$~\cite{kocsis2006bandit}. Even when considering a tree depth of $2n~(= 20)$, there are millions of leaves for difficult tasks \hocG~and \hocH, reflecting the complexity of task generation.  For each code $\code^{\textnormal{out}}$, we generated $10$ different visual tasks. To ensure sufficient diversity among the tasks generated for the same code, we introduced a measure \FDiversity. This measure, not only ensures visual task dissimilarity, but also ensures sufficient diversity in entire symbolic paths during generation (for details, see supplementary material).

%%%%%%%%%%%%%%%%%%%%%%%%%%%%%%%%%%%%%%%%%%%%%%%%%%%%%%%%%%
\vspace{-2mm}
\subsection{Results}
\looseness-1\textbf{Performance of task synthesis algorithm.}
%\subsection{Evaluation of code mutation}
\figref{fig:experiments.analysis} shows the results of our algorithm. The second column illustrates the enormity of the unconstrained space of mutated codes; we only impose size constraint $\Delta_0$ from \figref{fig:mutation.3}. We then additionally impose constraint $\Delta_1$ resulting in a partially constrained space of mutated codes (column 3), and finally apply all constraints from \figref{fig:mutation.3} to obtain the final set of generated codes (column 4). This reflects the systematic reduction in the space of mutated codes by our constraints. Column 5 shows the total running time for generating the final codes, which denotes the time taken by Z3 to compute solutions to our mutation query. As discussed in Section~\ref{sec.approach.mutation}, few codes with semantic irregularities still remain after the mutation stage. The symbolic execution stage eliminates these to obtain the reduced set of valid codes (column 6). Column 7 shows the final number of generated tasks and column 8 is the average time per output task (i.e., one MCTS run).

%%%%%%%%%%%%%%%%%%%%%%%%%%%%%%%%%%%%%
%\DSLMove, \DSLTurnL, \DSLTurnR
\begin{figure}[t!]
\centering
	%%%%%%%%%%%%%%%%%
	\scalebox{0.85}{
	\setlength\tabcolsep{2pt}
	\renewcommand{\arraystretch}{1.1}
	\begin{tabular}{c||rrrr||rrr||rrr}
			\toprule
			Task & \multicolumn{4}{c||}{Code Mutation} & \multicolumn{3}{c||}{Symbolic Execution} & \multicolumn{3}{c}{Fraction of $\task^{\textnormal{out}}$ with criteria} \\
			    \multicolumn{1}{c||}{$\task^{\textnormal{in}}$} & \multicolumn{1}{r}{\textbf{2:}$\#\code^{\textnormal{out}}_{\Delta=\textnormal{0}}$} & \multicolumn{1}{r}{\textbf{3:}$\#\code^{\textnormal{out}}_{\Delta=\textnormal{0,1}}$} & \multicolumn{1}{r}{\textbf{4:}$\#\code^{\textnormal{out}}_{\Delta=\textnormal{all}}$} & \multicolumn{1}{r||}{\textbf{5:}Time} & \multicolumn{1}{r}{\textbf{6:}$\#\code^{\textnormal{out}}$} & \multicolumn{1}{r}{\textbf{7:}$\#\task^{\textnormal{out}}$} & \multicolumn{1}{r||}{\textbf{8:}Time} & \multicolumn{1}{r}{\textbf{9:}(V)} & \multicolumn{1}{r}{\textbf{10:}$\textnormal{(VI)}_{\delta_{\textnormal{mini}}=1}$} & \multicolumn{1}{r}{\textbf{11:}$\textnormal{(VI)}_{\delta_{\textnormal{mini}}=0}$} \\
			    %& $|\widetilde{\codes}^\textnormal{out}|$ & $|\overline{\codes}^\textnormal{out}|$ & $|\codes^\textnormal{out}|$ & Time & $\big|\{\code^{\textnormal{out}}\}\big|$ & $\big|\{\task^{\textnormal{out}}\}\big|$ & Time & (VI) & (V, VI) & (IV, V, VI)\\			    
			\toprule
			\hocA & $3,159$ & $112$ & $64$ & $0.6$s & $28$ & $272$ & $68$s & $1.00$ & $1.00$ & $1.00$\\
			\hocC & $8,991$ & $594$ & $138$ & $1.7$s & $48$ & $428$ & $61$s & $1.00$ & $1.00$ & $1.00$\\
			\hocD & $798,255$ & $13,122$ & $720$ & $13.3$s & $196$ & $1,126$ & $60$s & $0.90$ & $0.98$ & $0.90$\\
			\hocF & $5,913$ & $152$ & $108$ & $1.0$s & $44$ & $404$ & $167$s & $1.00$ & $1.00$ & $0.50$\\
			\hocG & $2,997$ & $294$ & $66$ & $0.8$s & $46$ & $444$ & $348$s & $0.98$ & $0.59$ & $0.27$\\
			\hocH & $1,728$ & $294$ & $54$ & $0.6$s & $48$ & $480$ & $347$s & $0.80$ & $0.45$ & $0.07$\\
			%\midrule
			\karelA & $96,875$ & $150$ & $122$ & $1.3$s & $122$ & $1,196$ & $61$s & $1.00$ & $1.00$ & $1.00$\\
			\karelC & $484,875$ & $4,506$ & $990$ & $11.6$s & $469$ & $4,506$ & $63$s & $1.00$ & $1.00$ & $1.00$\\
			\karelE & $8.595 \times 10^{6}$ & $60,768$ & $888$ & $11.3$s & $432$ & $4,258$ & $185$s & $0.92$ & $0.92$ & $0.88$\\
			%\karelE & $8.595$M
			\karelF & $132.625 \times 10^{6}$ & $19,328$ & $1,404$ & $17.1$s & $532$ & $5,032$ & $158$s & $1.00$ & $1.00$ & $1.00$\\
			%\karelF & $132.625$M 
			%%%%%%%%%%%%%%%%%%%%%%%
			%%%% add karelH and karelI in supplementary
			%%\karelH & $3,476.000$M & $198,656$ & $1,604$ & $22.5$s & $?$ & $?$ & $?$s & ? & ?\\
			%%\karelI & $4.432$M & $43,264$ & $380$ & $5.3$s & $?$ & $?$ & $?$s & ? & ?\\
			%%%%%%%%%
		\bottomrule
   \end{tabular}
   }
	\vspace{-1mm}   
	\caption{Results on \hocType~and~\karelType~tasks; details are provided in Section~\ref{sec.simulations}.}
	\label{fig:experiments.analysis}
	\vspace{-4mm}
\end{figure}
%%%%%%%%%%%%%%%%%%%%%%%%%%%%%%%%%%%%%

%%%%%%%%%%%%%%%%%%%%%%%%%%%%%%%%%%%%%
\begin{figure}[t!]
\centering
	%%%%%%%%%%%%%%%%%
	\begin{subfigure}[b]{.39\textwidth}
	\centering
	{
		\includegraphics[width=1\textwidth]{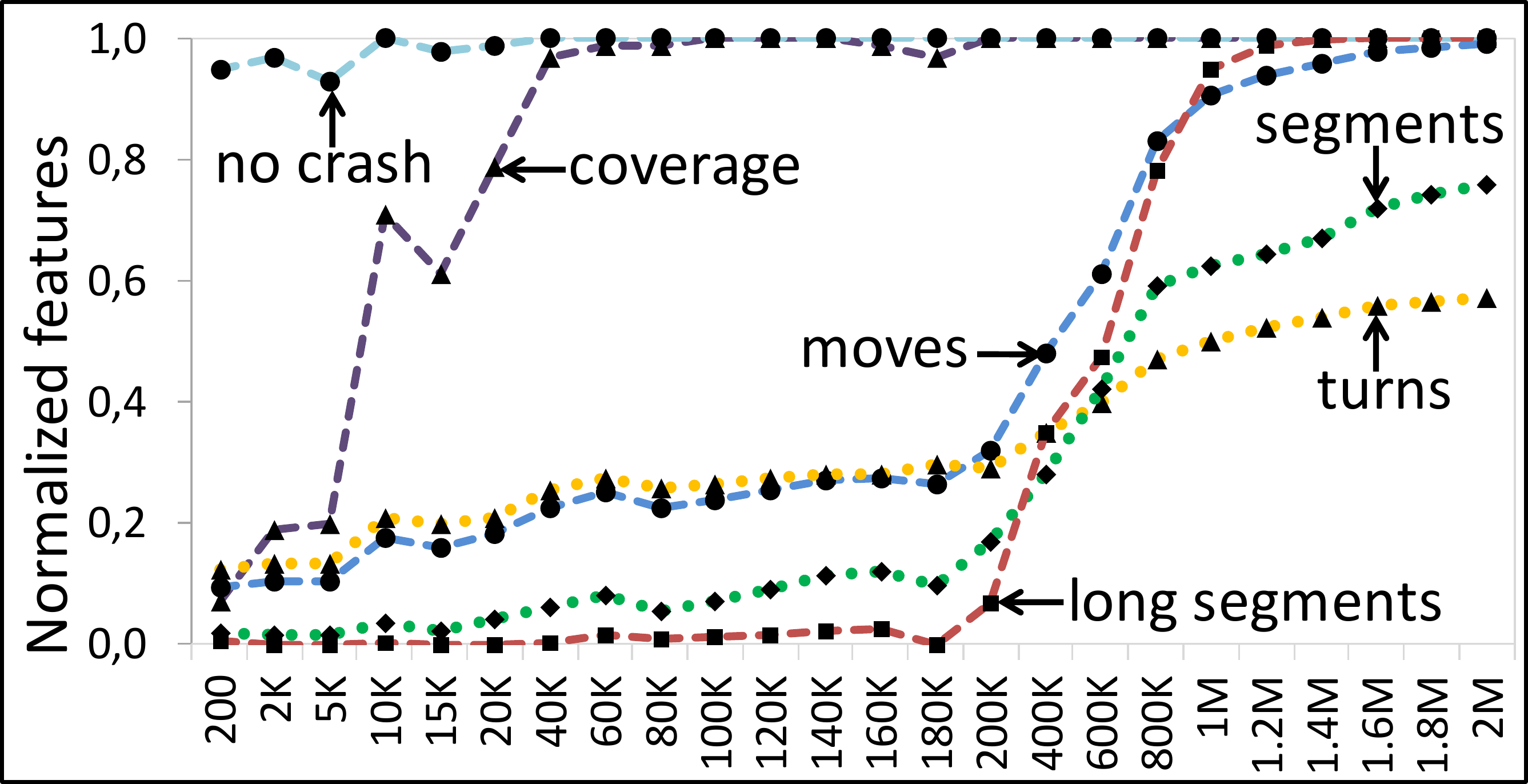}
		\caption{Trends in \FScore~features}
		%\caption{Trends in components of \FScore}
		\label{fig:experiments.mcts.varytime}
    }
    \end{subfigure}
  	\begin{subfigure}[b]{.197\textwidth}
  	\centering
  	{
		\includegraphics[height=2.84cm]{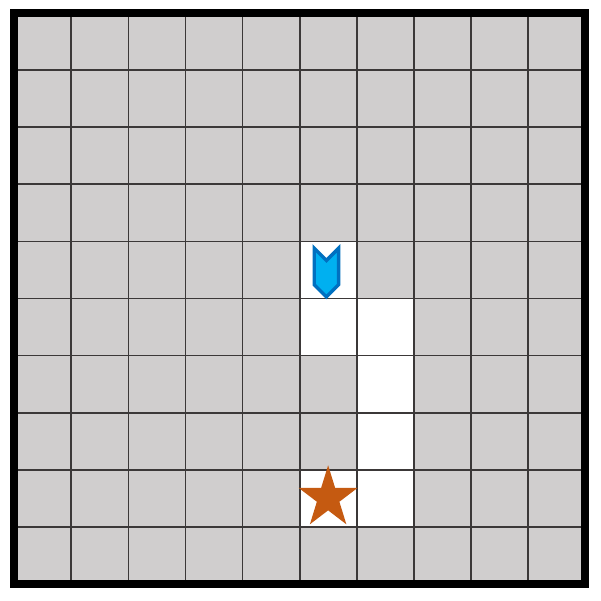}
		\caption{Best at $200$}
		\label{fig:experiments.mcts.best1}
	}
	\end{subfigure}
	\begin{subfigure}[b]{.197\textwidth}
	\centering
	{
		\includegraphics[height=2.84cm]{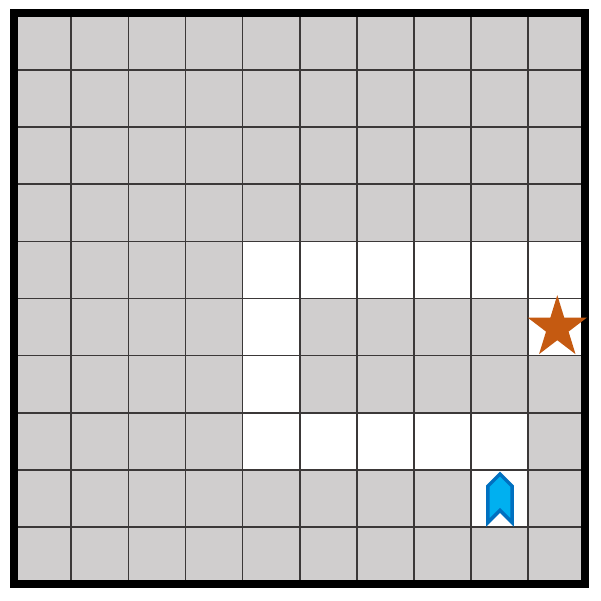}
		\caption{Best at $20$K}
		%\caption{Best at $20$K ($10^3$)}
		\label{fig:experiments.mcts.best2}
    }
    \end{subfigure}
  	\begin{subfigure}[b]{.197\textwidth}
  	\centering
  	{
		\includegraphics[height=2.84cm]{fig/mctsplots/hoc-G-mutation-6-22_run7/mctsplot_best_2M.pdf}
		\caption{Best at $2$M}
		%\caption{Best at $2$M ($10^6$)}
		\label{fig:experiments.mcts.best3}
	}
	\end{subfigure}	
	\vspace{-5mm}
	\caption{Illustration of a single MCTS run on $\code^{\textnormal{out}}$ from \figref{fig:intro.hoc.p1} obtained from solution code of task \hocG~by mutation. (a) shows the temporal trends of different feature values in \FScore~averaged over a time window of $100$ steps. (b)--(d)~show the best, i.e., highest scoring, tasks generated up to times $2 \times 10^2$, $2 \times 10^4$, and $2 \times 10^6$ respectively. $\task^{\textnormal{out}}_\textnormal{vis}$ shown in \figref{fig:intro.hoc.t1} is the puzzle produced in (d).
	}
	\vspace{-5mm}
	\label{fig:experiments}
\end{figure}
\textbf{Analyzing output tasks.}
We further analyze the generated tasks based on the objectives of Section~\ref{sec.problem}. All tasks satisfy properties (I)--(III) by design. Objective (IV) is easily achieved by excluding generated tasks for which $\code^{\textnormal{out}} = \code^{\textnormal{in}}$. For a random sample of $100$ of the generated tasks per reference task, we performed manual validation to determine whether objectives (V) and (VI) are met. The fraction of tasks that satisfy these objectives is listed in the last three columns of \figref{fig:experiments.analysis}.
%To validate task minimality (for two values of $\delta_\textnormal{mini}$), we apply delta debugging~\cite{DeltaDebugging}. 
We observe that the vast majority of tasks meet the objectives, even if not by design.
For \hocH, the fraction of tasks satisfying (VI) is low because the corresponding codes are generic enough to solve several puzzles.

%
% QUESTION: It is observed that most of our tasks that satisfied (I), satisfied objective (VI) as well. Some of the tasks had codes which were quite robust in the sense that they successfully several tasks, and violated objective (VI), particularly for the stricter variant with $\delta_\textnormal{mini}=0$. 

% QUESTION: Objective (IV) is achieved by excluding the 10 tasks generated, corresponding to the solution codes of the reference tasks.

%We compare our pipeline with alternate methods and further evaluate it through a user study, details of which we present next.

\textbf{Deep dive into an MCTS run.}
To offer more insight into the task generation process, we take a closer look at an MCTS run for task \hocG, shown in \figref{fig:experiments}. \figref{fig:experiments.mcts.varytime} illustrates the improvement in various components of \FScore~as the number of MCTS iterations increases. Best tasks at different iterations are shown in \figref{fig:experiments.mcts.best1}, \ref{fig:experiments.mcts.best2}, \ref{fig:experiments.mcts.best3}. As expected, the more the iterations, the better the tasks are.

%Comparison with enumerative methods and random search
\looseness-1\textbf{Remarks.} We also ran the mutation stage by enumerating the programs within size constraints and then post-checking other constraints without Z3. This implementation leads to a run-time increase by a factor of $10$ to $100$ for different tasks. So, Z3 seems to be very effective by jointly considering all the constraints. As a search method, although MCTS seems computationally expensive, the actual run-time and memory footprint of an MCTS run depend on the unique traces explored (i.e., unique symbolic executions done)---this number is typically much lower than the number of iterations, also see discussion in the supplementary material. Considering the MCTS output in Figs.~\ref{fig:experiments.mcts.best2}, \ref{fig:experiments.mcts.best3}, to obtain a comparable evaluation score through a random search, the corresponding number of unique symbolic executions required is at least $10$ times more than executed by MCTS. We note that while we considered one I/O pair for \karelType~tasks, our methodology can be easily extended to multiple I/O pairs by adapting techniques designed for generating diverse tasks. 
\vspace{-2.5mm}
\section{User Study and Comparison with Alternate Methods}\label{sec.userstudy}
\vspace{-2mm}
In this section, we evaluate our task synthesis algorithm with a user study focusing on tasks \hocC, \hocF, \hocG, and \hocH. We developed an online app\footnote{\href{https://www.teaching-blocks.cc/}{https://www.teaching-blocks.cc/}\label{footnote.userstudyapp}}, which uses the publicly available toolkit of Blockly Games~\cite{googleblockly} and provides an interface for a participant to practice block-based programming tasks for HOC. Each ``practice session'' of the study involves three steps: (i)~a reference task $\task^{\textnormal{in}} \in \{\text{\hocC}, \text{\hocF}, \text{\hocG}, \text{\hocH}\}$ is shown to the participant along with its solution code $\code^{\textnormal{in}}$, (ii)~a new task  $\task^{\textnormal{out}}$ is generated for which the participant has to provide a solution code, and (iii)~a post-survey asks the participant to assess the visual dissimilarity of the two tasks on a $4$-point Likert scale as used in \cite{polozov2015}. Details on the app interface and questionnaire are provided in the supplementary material. Participants for the study were recruited through Amazon Mechanical Turk. We only selected four tasks due to the high cost involved in conducting the study (about $1.8$~USD per participant). The number of participants and their performance are documented in \figref{fig:userstudy}. 
%However, we chose these tasks such that they sufficiently span the set of HOC programming concepts of varying degrees of difficulty. 
%%%
%Participants were asked to assess the conceptual similarity of generated codes to the original solution, the solvability of generated tasks, and their visual similarity to the original task. Details on the user-study interface and questionnaire are provided in the supplementary material.

%\todo{Three steps in the user study.}
%\todo{Likert scale.}
%We developed an online user study app specifically for this purpose. The App was implemented using the blockly games publicly available library, and currently supports only the HOC tasks. As part of the study, users were asked 3 questions on a forced-choice Likert scale. Further details on the interface and questionnaire are provided in the supplementary material.

%\susubsection{Experimental Setup}
%\paragraph{Online user study app.}

%%%%%%%%%%%%%%%%%%%%%%%%%%%%%%%%%%%%%%%%%%%%%%%%%%%%%%%%%%
%\vspace{-2mm}
%\subsection{Baselines and Methods Evaluated}
%\vspace{-1mm}
\textbf{Baselines and methods evaluated.}
\looseness-1 We evaluated four different methods, including three baselines (\algoSame, \algoTutor, \algoTaskMut) and our algorithm  (\algoOursNoPstar). \algoSame~generates tasks such that $\task^{\textnormal{in}} = \task^{\textnormal{out}}$. \algoTutor~produces tasks that are similar to $\task^{\textnormal{in}}$ and designed by an expert. We picked similar problems from the set of $20$ \emph{Classic Maze} challenge~\cite{hourofcode_maze} tasks exercising the same programming concepts: \emph{Maze 6, 9} for \hocC, \emph{Maze 11, 13} for \hocF, \emph{Maze 15, 17} for \hocG, and \emph{Maze 19} for \hocH.
%For each task of the study, 

%Given that the HOC \emph{Classic Maze challenge}~\cite{hourofcode_maze} has a total of $20$ tasks, each task \hocC~(\emph{Maze 7}), \hocF~(\emph{Maze 12}), \hocG~(\emph{Maze 16}), and \hocH~(\emph{Maze 18}) has a corresponding similar problem in the set of all tasks exercising the same programming concepts. Concretely, we have the following similar tasks: (i) \emph{Maze 6, Maze 9} for \hocC, (ii) \emph{Maze 11, Maze 13} for \hocF, (iii) \emph{Maze 15, Maze 17} for \hocG, and (iv) \emph{Maze 19} for \hocH.

%Our expert was the original set of 20 HOC tasks since they are already grouped by similarity. We therefore picked 1--2 similar tasks for each task in the study---the details are in the supplementary material.
%we used the following baselines: \algoSame~generates a task that is the same as the original, i.e., $\task^{\textnormal{in}} = \task^{\textnormal{out}}$. \algoTutor~produces tasks that are similar to $\task^{\textnormal{in}}$ and designed by an expert. Our expert was the original set of 20 HOC tasks since they are already grouped by similarity. We therefore picked 1--2 similar tasks for each task in the study---the details are in the supplementary material.

%In particular, the expert tasks from the 20 HOC mazes, picked were (\emph{Maze 6}, \emph{Maze 9}) for H2, ( \emph{Maze 11}, \emph{Maze 13}) for H4, ( \emph{Maze 15}) for H5 and (\emph{Maze 17}, \emph{Maze 19}) for H6. Each of the solution codes of these expert tasks adhered to code minimality, and conceptual/structural similarity of codes. 

\begin{figure}[t!]
\centering
	%%%%%%%%%%%%%%%%%
	\begin{subfigure}[b]{.1425\textwidth}
	\centering
	{
		\includegraphics[height=2.085cm]{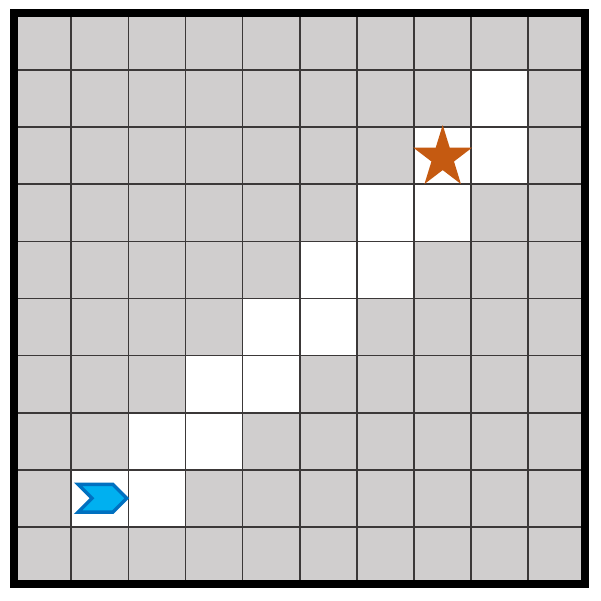}
		%\vspace{0.2mm}
		\caption{$\task^{\textnormal{in}}_\textnormal{vis}$ for \hocF}
		 \label{fig:userstudy.hoc.t0.orig}
	}
	\end{subfigure}
	%\
	%%%%%%%%%%%%%%%%%
	\begin{subfigure}[b]{.1425\textwidth}
	\centering
	{
		\includegraphics[height=2.085cm]{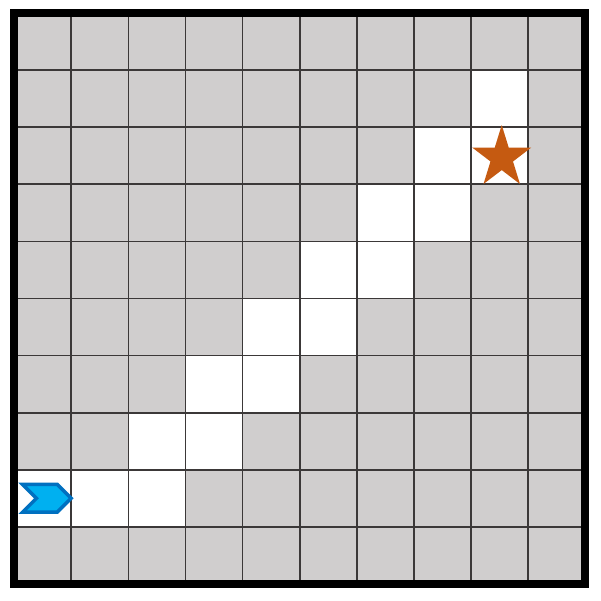}
		%\vspace{0.2mm}
		\caption{1\textsuperscript{st} $\task^{\textnormal{out}}_\textnormal{vis}$}
		 \label{fig:userstudy.hoc.t0.sim}
	}
	\end{subfigure}
	%\
	%%%%%%%%%%%%%%%%%
	\begin{subfigure}[b]{.1425\textwidth}
	\centering
	{
		\includegraphics[height=2.085cm]{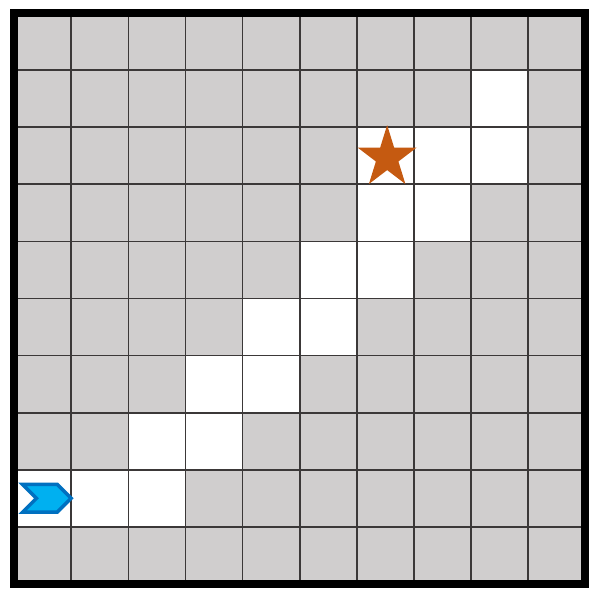}
		%\vspace{0.2mm}
		\caption{2\textsuperscript{nd} $\task^{\textnormal{out}}_\textnormal{vis}$}
		 \label{fig:userstudy.hoc.t0.diff}
	}
	\end{subfigure}
	%\
	%%%%%%%%%%%%%%%%%	
	\begin{subfigure}[b]{.197\textwidth}
	\centering
	{
		\begin{boxcode}{3.68cm}{0.7}{0.55}
				\textcode{def }\DSLRun\textcode{()\{}\\
				\quad \DSLRepeatUntil\textcode{(}\DSLBoolGoal\textcode{)\{}\\
				\quad \quad \DSLMove\\
				\quad \quad \DSLTurnLeft\\
				\quad \quad \DSLMove\\
				\quad \quad \DSLTurnRight\\				
				\quad \textcode{\}}\\
				\textcode{\}}
				%\\
				%\\	
				\\	
				\\	
		\end{boxcode}
		\vspace{-1mm}
		\caption{$\code^{\textnormal{in}}$ for \hocF}
		\label{fig:userstudy.hoc.p0.orig}
    }
    \end{subfigure}
	%\
	%%%%%%%%%%%%%%%%%	
	\begin{subfigure}[b]{.197\textwidth}
	\centering
	{
		\begin{boxcode}{3.68cm}{0.7}{0.55}
				\textcode{def }\DSLRun\textcode{()\{}\\
				\quad \DSLMove\\
				\quad \DSLMove\\				
				\quad
				\DSLRepeatUntil\textcode{(}\DSLBoolGoal\textcode{)\{}\\
				\quad \quad \DSLTurnLeft\\				
				\quad \quad \DSLMove\\
				\quad \quad \DSLTurnRight\\				
				\quad \quad \DSLMove\\				
				\quad \textcode{\}}\\
				\textcode{\}}
				%\\	
				%\\	
		\end{boxcode}
		\vspace{-1mm}
		\caption{1\textsuperscript{st} $\code^{\textnormal{out}}$}
		\label{fig:userstudy.hoc.p0.sim}
    }
    \end{subfigure}
	%\
	%%%%%%%%%%%%%%%%%	
	\begin{subfigure}[b]{.145\textwidth}
	\centering
	{
		\begin{boxcode}{2.72cm}{0.7}{0.55}
				\textcode{def }\DSLRun\textcode{()\{}\\
				\quad \DSLMove\\
				\quad \DSLMove\\
                \quad \DSLTurnLeft\\
                \quad \DSLMove\\
                \quad \DSLTurnRight\\
                \quad \DSLMove\\
                \quad \DSLTurnLeft \\
                %\quad \DSLMove\\              
                %\quad \textcode{...}\\
                \quad \textcolor{blue}{$\mathbf{15}$ \textbf{more actions}}\\
				\textcode{\}}
		\end{boxcode}
		\vspace{-0.6mm}
		\caption{2\textsuperscript{nd} $\code^{\textnormal{out}}$}
		\label{fig:userstudy.hoc.p0.diff}
    }
    \end{subfigure}
    %\\
    \vspace{-5mm}
	\caption{\algoTaskMut~applied to \hocF. $\task^{\textnormal{in}}_\textnormal{vis}$ and $\code^{\textnormal{in}}$ are shown in (a) and (d). (b)--(c) illustrate two tasks $\task^{\textnormal{out}}_\textnormal{vis}$ obtained via small mutations of $\task^{\textnormal{in}}_\textnormal{vis}$. (e) is the smallest solution code for (b) and is structurally similar to $\code^{\textnormal{in}}$. (f) is the smallest solution code for (c) and is drastically different from $\code^{\textnormal{in}}$.
	}
	\vspace{-6mm}
	\label{fig:userstudy.taskmutation}
\end{figure}

\algoTaskMut~generated tasks by directly mutating the grid-world of the original task, i.e., by moving the agent or goal by up to two cells and potentially changing the agent's orientation. A total of $18$, $20$, $15$, and $17$ tasks were generated for \hocC, \hocF, \hocG, and \hocH, respectively. \figref{fig:userstudy.taskmutation} shows two output tasks for \hocF~and illustrates the challenge in directly mutating the input task, given the high discontinuity in mapping from the space of tasks to their codes. For \hocF, a total of $14$ out of $20$ new tasks were structurally very different from the input.
%---the complete list is in the supplementary material

% As shown in Figure
%Looking at their solution codes, we noticed that, for H4, they were structurally very different from the original solution code for 14 out of 20 tasks. 

%Consider \figref{fig:userstudy.hoc.t0.diff}, where the goal is moved by a single cell in comparison to the original task in \figref{fig:userstudy.hoc.t0.orig}. However, the resulting solution code (\figref{fig:userstudy.hoc.p0.diff}), generated from $\task_{\textnormal{store}}$, significantly differs in structure from the original (\figref{fig:userstudy.hoc.p0.orig}). The remaining 6 tasks had codes that were similar to the original (one of which is shown in \figref{fig:userstudy.hoc.p0.sim}). 

%For H5, task mutation led to 5 tasks with codes that were exactly the same as the original, and 3 whose codes were structurally very different. The solution code of H6 was robust to changes in the grid-world, with 13 being exactly the same as the original and 3 being structurally very different. These results highlight the drawback of generating tasks by directly mutating the reference task; conceptual code similarity is not guaranteed.

\algoOursNoPstar~uses our algorithm to generate tasks. We picked the generated tasks from three groups based on the size of the code mutations from which they were produced, differing from the reference solution code by $+\delta_{\textnormal{size}}$ for $\delta_{\textnormal{size}} \in \{0, 1, 2\}$. For \hocC~and \hocF, we randomly selected $5$ tasks from each group, for a total of $15$ new tasks per reference task.  For \hocG~and \hocH, we selected $10$ tasks from the first group ($\delta_{\textnormal{size}}=0$) only, due to their complexity stemming from nested constructs in their codes. We observed that \algoTutor~tasks for \hocG, \hocH~were also of $\delta_{\textnormal{size}}=0$, i.e., $\code^{\textnormal{out}}_{\textnormal{size}} = \code^{\textnormal{in}}_{\textnormal{size}}$. All the generated tasks picked for \algoOursNoPstar~adhere to properties (I)--(VI) in Section~\ref{sec.problem}.

\textbf{Results on task solving.}
In terms of successfully solving the generated tasks, \algoSame~performed best (mean success = $0.94$) in comparison to \algoTutor~(mean = $0.90$), \algoOursNoPstar~(mean = $0.89$), and \algoTaskMut~(mean = $0.68$)---this is expected given the tasks generated by \algoSame. In comparison to \algoTutor, the performance of \algoOursNoPstar~was not significantly different~($\chi^2=0.04, p=0.83$); in comparison to \algoTaskMut,  \algoOursNoPstar~performed significantly better ($\chi^2=28.74,  p < e^{-8}$).
The complexity of the generated tasks is also reflected in the average time that participants spent on solving them. As shown in \figref{fig:userstudy}, they spent more time solving the tasks generated by \algoTaskMut.
%; as mentioned earlier, although the tasks are visually similar, the codes may be structurally very different.

\looseness-1
\textbf{Results on visual task dissimilarity.}
Visual dissimilarity was measured on a Likert scale ranging from 1--4, 1 being highly similar and 4 highly dissimilar.
Comparing the dissimilarity of the generated tasks w.r.t. the reference task, we found that the performance of \algoSame~was worst (mean dissimilarity = $1.07$), while that of \algoTutor~was best (mean = $2.90$). \algoOursNoPstar~(mean = $2.63$) performed significantly better than \algoTaskMut~(mean = $2.17$), yet slightly worse than \algoTutor. This is because \algoTutor~generates tasks with additional distracting paths and noise, which can also be done by our algorithm (although not done for this study). Moreover, for \hocC, which had no conditionals, the resulting codes were somewhat similar, and so were the generated puzzles. When excluding \hocC~from the analysis, the difference between \algoOursNoPstar~(mean = $2.72$) 
and \algoTutor~(mean =$2.93$) was not statistically significant. A detailed distribution of the responses can be found in the supplementary material.

\looseness-1\textbf{Remarks.} \algoSame's performance in terms of tasks solved is below $1.00$, possibly because participants overlooked the solution of Step $1$, unaware they will be receiving the same task in Step $2$, and the app did not allow them to go back to Step $1$. 
This user study provides a proof-of-concept; more elaborate studies are needed to fully reach the motivational goal of teaching $\textnormal{K-}12$ students, and evaluate the long term impact on students' concept learning. 
As additional studies, it would be important to understand the sensitivity of user study results w.r.t. the Likert scale definition; another possibility is to use pairwise comparisons in eliciting user evaluations.
\begin{figure}[t!]
\centering
	%%%%%%%%%%%%%%%%%
	\scalebox{0.85}{
	\setlength\tabcolsep{3pt}
	\renewcommand{\arraystretch}{1.1}
	\begin{tabular}{r|rrrrr|rrrrr|rrrrr|rrrrr}
			\toprule
			%Time used: mean_time_spent_on_new_task
			Method & \multicolumn{5}{c|}{Total participants} & \multicolumn{5}{c|}{Fraction of tasks solved} & \multicolumn{5}{c|}{Time spent in secs} & \multicolumn{5}{c}{Visual dissimilarity} \\
			 & \hocAllBold & \hocC & \hocF & \hocG & \hocH 	& \hocAllBold & \hocC & \hocF & \hocG & \hocH  & \hocAllBold & \hocC & \hocF & \hocG & \hocH & \hocAllBold & \hocC & \hocF & \hocG & \hocH \\ 		
			\toprule
			\algoSame & $\textbf{96}$ & $24$ & $24$ & $24$ & $24$ & $\textbf{.94}$ & $.92$ & $1.00$ & $.96$ & $.88$ & $\textbf{89}$ & $60$ & $59$ & $93$ & $145$ & $\textbf{1.07}$ & $1.12$ & $1.04$ & $1.00$ & $1.12$ \\
			\algoTutor & $\textbf{170}$ & $48$ & $48$ & $49$ & $25$ & $\textbf{.90}$ & $.90$ & $.92$ & $.88$ & $.92$ & $\textbf{121}$ & $107$ & $113$ & $118$ & $169$ & $\textbf{2.90}$ & $2.81$ & $2.79$ & $2.96$ & $3.16$ \\
			\algoTaskMut & $\textbf{278}$ & $72$ & $79$ & $60$ & $67$ & $\textbf{.68}$ & $.76$ & $.71$ & $.65$ & $.60$ & $\textbf{219}$ & $135$ & $299$ & $219$ & $215$ & $\textbf{2.17}$ & $2.36$ & $2.33$ & $1.95$ & $1.99$  \\
			%\algoOurs & $\textbf{325}$ & $83$ & $81$ & $81$ & $80$ & $\textbf{.88}$ & $.93$ & $.91$ & $.93$ & $.76$ & $\textbf{143}$ & $75$ & $165$ & $135$ & $200$ & $\textbf{2.61}$ & $2.25$ & $2.46$ & $2.54$ & $3.19$ \\
			\algoOursNoPstar & $\textbf{197}$ & $59$ & $57$ & $40$ & $41$ & $\textbf{.89}$ & $.92$ & $.89$ & $.92$ & $.83$ & $\textbf{144}$ & $85$ & $183$ & $130$ & $189$ & $\textbf{2.63}$ & $2.41$ & $2.42$ & $2.68$ & $3.20$ \\
			%\algoOurs~\textsubscript{($\text{\code}^{\textnormal{out}} \neq \text{\code}^{\textnormal{in}}$)}
			\bottomrule
   \end{tabular}
   }
   \vspace{-1mm}
	\caption{User study results for \hocType~tasks (\hocAllBold represents all tasks in the study); see Section~\ref{sec.userstudy}. }
   \vspace{-4mm}	
	\label{fig:userstudy}
\end{figure}
%%%%%%%%%%%%%%%%%%%%%%%%%%%%%%%%%%%%%
%%%%%%%%%%%%%%%%%%%%%%%%%%%%%%%%%%%%%%%%%%%%%%%%%%%%%%%%%%

%%%%%%%%%%%%%%%%%%%%%%%%%%%%%%%%%%%%%%%%%%%%%%%%%%%%%%%%%%%
%%%%%%%%%%%%%%%%%%%%%%%%%%%%%%%%%%%%%%
\vspace{-3mm}
\section{Conclusions and Outlook}\label{sec.conclusions}
\vspace{-3mm}
%\section{Concluding Discussion}\label{sec.conclusions}
\looseness-1We developed techniques for a critical aspect of pedagogy in block-based programming: Automatically generating new tasks that exercise specific programming concepts, while looking visually dissimilar to input. We demonstrated the effectiveness of our methodology through an extensive empirical evaluation and user study on reference tasks from popular programming platforms. We believe our techniques have the potential to drastically improve the success of pedagogy in block-based visual programming environments by providing tutors and students with a substantial pool of new tasks. Beyond the application domain of programming education, our methodology can be used for generating large-scale datasets consisting of tasks and solution codes with desirable characteristics---this can be potentially useful for training neural program synthesis methods.

There are several promising directions for future work, including but not limited to: Learning a policy to guide the MCTS procedure (instead of running vanilla MCTS); automatically learning the constraints and cost function from a human-generated pool of problems; and applying our methodology to other programming environments (e.g., Python problems).

\clearpage
% !TEX root =  main.tex
%%%%%%%%%%%%%%%%%%%%%%%%%%%%%%%%%%%%%%%%%%%%%%%%%%%%%%%%%%
%%%%%%%%%%%%%%%%%%%%%%%%%%%%%%%%%%%%%%%%%%%%%%%%%%%%%%%%%%
\section*{Broader Impact}\label{sec.impact}
This paper develops new techniques for improving pedagogy in block-based visual programming environments. Such programming environments are increasingly used nowadays to introduce computing concepts to novice programmers, and our work is motivated by the clear societal need of enhancing $\textnormal{K-}12$ computing education. In existing systems, the programming tasks are hand-curated by tutors, and the available set of tasks is typically very limited. This severely limits the utility of existing systems for long-term learning as students do not have access to practice tasks for mastering the programming concepts.

\looseness-1
We take a step towards tackling this challenge by developing a methodology to generate new practice tasks for a student that match a desired level of difficulty and exercise specific programming concepts. Our task synthesis algorithm is able to generate 1000's of new similar tasks for reference tasks taken from the \emph{Hour of Code: Classic Maze} challenge by \emph{Code.org} and the \emph{Intro to Programming with Karel} course by \emph{CodeHS.com}. Our extensive experiments and user study further validate the quality of the generated tasks.  Our task synthesis algorithm could be useful in many different ways in practical systems. For instance, tutors can assign new practice tasks as homework or quizzes to students to check their knowledge, students can automatically obtain new similar tasks after they failed to solve a given task and received assistance, and intelligent tutoring systems could automatically generate a personalized curriculum of problems for a student for long-term learning. 

%In a nutshell, the proposed task synthesis algorithm has the potential to drastically improve the effectiveness of pedagogy in existing block-based visual programming environments by giving tutors and students an access to a large pool of new programming tasks.

%\annotate{
%Authors are required to include a statement of the broader impact of their work, including its ethical aspects and future societal consequences. 
%Authors should discuss both positive and negative outcomes, if any. For instance, authors should discuss a)  who may benefit from this research, b) who may be put at disadvantage from this research, c) what are the consequences of failure of the system, and d) whether the task/method leverages biases in the data. If authors believe this is not applicable to them, authors can simply state this.
%}

%Use unnumbered first level headings for this section, which should go at the end of the paper. {\bf Note that this section does not count towards the eight pages of content that are allowed.}

\begin{ack}
We would like to thank the anonymous reviewers for their helpful comments. Ahana Ghosh was supported by Microsoft Research through its PhD Scholarship Programme. Umair Z. Ahmed and Abhik Roychoudhury were supported by the National Research Foundation, Singapore and National University of Singapore through its National Satellite of Excellence in Trustworthy Software Systems (NSOE-TSS) project under the National Cybersecurity R\&D (NCR) Grant award no. NRF2018NCR-NSOE003-0001. 
\end{ack}
%Nigel Fernandez did this work during an internship at MPI-SWS and implemented MCTS for symbolic execution. 

\bibliography{main}
%%%%%%%%%%%%%%%%%%%%%%%%%%%%%%%%%%%%%%%%%%%%%%%%%%%%%%%%%%
\iftoggle{longversion}{
\clearpage
\onecolumn
\appendix 
{\allowdisplaybreaks
% !TEX root =  main.tex
%%%%%%%%%%%%%%%%%%%%%%%%%%%%%%%%%%%%%%%%%%%%%%%%%%%%%%%%%%
%%%%%%%%%%%%%%%%%%%%%%%%%%%%%%%%%%%%%%%%%%%%%%%%%%%%%%%%%%
\section{List of Appendices}\label{appendix.table-of-contents}
In this section, we provide a brief description of the content provided in the appendices of the paper.   
\begin{itemize}
\item Appendix~\ref{appendix.sec.introduction} shows all the $10$ reference tasks from \figref{fig:dataset}. For each task, we also illustrate our methodology as was done in \figref{fig:intro.hoc} and \figref{fig:intro.karel}.
\item Appendix~\ref{appendix.sec.userstudy} expands on the user study analysis. (Section~\ref{sec.userstudy})
\item Appendix~\ref{appendix.sec.mutation} provides additional details on the code mutation stage. (Section~\ref{sec.approach.mutation})
\item Appendix~\ref{appendix.sec.symbolicexecution} demonstrates how MCTS could guide the symbolic execution in generating more suitable tasks. (Section~\ref{sec.approach.tasksynthesis})
\item Appendix~\ref{appendix.sec.simulations} provides additional details and results about experiments. (Section~\ref{sec.simulations})
\end{itemize}

\section{Illustration of Our Methodology for the \hocType~and \karelType~Dataset}\label{appendix.sec.introduction}
%Appendix for Introduction
In this section, we illustrate our methodology for all the $10$ reference tasks from \figref{fig:dataset}. For each reference task $(\task^{\textnormal{in}}, \code^{\textnormal{in}})$, we picked one output $(\task^{\textnormal{out}}, \code^{\textnormal{out}})$ from the entire set of generated outputs.
%\todo{replace figs}.
%is picked for each task 

%%%%%%%%%%%%%%%%%%%%%%%%%%%%%%%%%%%%%%%%%%%%%%%%%%%%%%%%%%
%%%%%%%%%%%%%%%%%%%%%%%%%%%%%%%%%%%%%%
\begin{figure}[h!]
\centering
	%%%%%%%%%%%%%%%%%
	\begin{subfigure}[b]{.22\textwidth}
	\centering
	{
		\includegraphics[height=2.39cm]{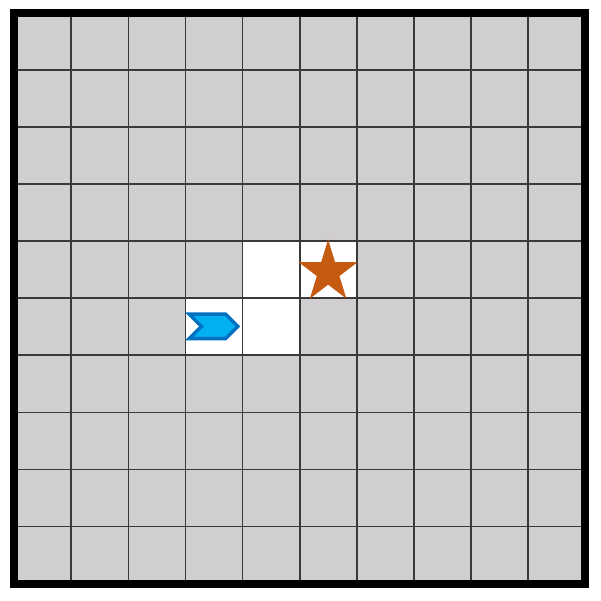}
		%\vspace{0.2mm}
		\caption{Visual puzzle for $\task^{\textnormal{in}}$}
		 \label{fig:appendix.intro.hoc.t0}
	}
	\end{subfigure}
	\
	%%%%%%%%%%%%%%%%%	
	\begin{subfigure}[b]{.22\textwidth}
	\centering
	{
		\begin{boxcode}{3.8cm}{0.75}{0.7}
			\textcode{def }\DSLRun\textcode{()\{}\\
			\quad \DSLMove\\
			\quad \DSLTurnLeft\\
			\quad \DSLMove\\
			\quad \DSLTurnRight\\
			\quad \DSLMove\\				
			\textcode{\}}
			\\
			\\
			\vspace{2.5mm}
		\end{boxcode}
		\vspace{-1mm}
		\caption{Solution code $\code^{\textnormal{in}}$}
		\label{fig:appendix.intro.hoc.p0}
    }
    \end{subfigure}
  	\quad \quad
	%%%%%%%%%%%%%%%%%
	\begin{subfigure}[b]{.22\textwidth}
	\centering
	{
		\includegraphics[height=2.39cm]{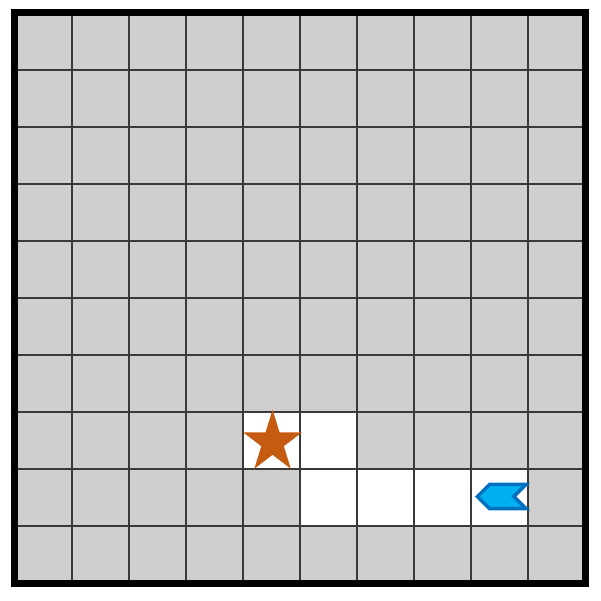}
		%\vspace{0.2mm}
		\caption{Visual puzzle for $\task^{\textnormal{out}}$}
		 \label{fig:appendix.intro.hoc.t1}
	}
	\end{subfigure}
	\
	%%%%%%%%%%%%%%%%%	
	\begin{subfigure}[b]{.22\textwidth}
	\centering
	{
		\begin{boxcode}{3.8cm}{0.75}{0.7}
			\textcode{def }\DSLRun\textcode{()\{}\\
			\quad \DSLMove\\
			\quad \DSLMove\\
			\quad \DSLMove\\			
			\quad \DSLTurnRight\\
			\quad \DSLMove\\
			\quad \DSLTurnLeft\\
			\quad \DSLMove\\
			\textcode{\}}
			\vspace{2.5mm}
		\end{boxcode}
		\vspace{-1mm}
		\caption{Solution code $\code^{\textnormal{out}}$}
		\label{fig:appendix.intro.hoc.p1}
    }
    \end{subfigure}
    %\\
    \vspace{-1.5mm}
	\caption{Task \hocA~-- Illustration of our methodology.}    
	%\caption{\todo{}Illustration of our methodology for task \emph{Maze 16} from the \emph{Hour of Code: Classic Maze} challenge by \emph{Code.org}~\cite{hourofcode_maze}; the complete list of tasks with their specifications is in \figref{fig:dataset}.}
	\vspace{-3.5mm}
	\label{fig:appendix.intro.hoc}
\end{figure}
%%%%%%%%%%%%%%%%%%%%%%%%%%%%%%%%%%%%%%%%%%%%%%%%%%%%%%%%%%

%%%%%%%%%%%%%%%%%%%%%%%%%%%%%%%%%%%%%%%%%%%%%%%%%%%%%%%%%%
%%%%%%%%%%%%%%%%%%%%%%%%%%%%%%%%%%%%%%
\begin{figure}[h!]
\centering
	%%%%%%%%%%%%%%%%%
	\begin{subfigure}[b]{.22\textwidth}
	\centering
	{
		\includegraphics[height=2.39cm]{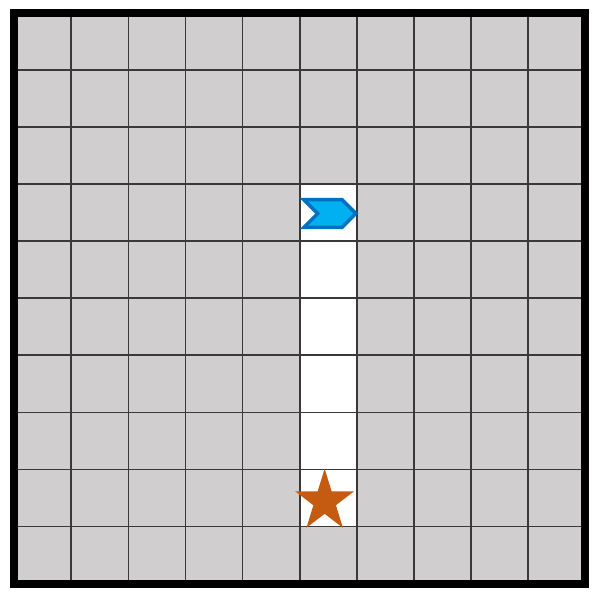}
		%\vspace{0.2mm}
		\caption{Visual puzzle for $\task^{\textnormal{in}}$}
		 \label{fig:appendix.intro.hoc.t0}
	}
	\end{subfigure}
	\
	%%%%%%%%%%%%%%%%%	
	\begin{subfigure}[b]{.22\textwidth}
	\centering
	{
		\begin{boxcode}{3.8cm}{0.75}{0.7}
			\textcode{def }\DSLRun\textcode{()\{}\\
			\quad \DSLTurnRight\\
			\quad \DSLRepeat\textcode{(5}\textcode{)\{}\\
			\quad \quad \DSLMove\\
			\quad \textcode{\}}\\
			\textcode{\}}
			\\
			\\
			\\
			\vspace{1.5mm}
		\end{boxcode}
		\vspace{-1mm}
		\caption{Solution code $\code^{\textnormal{in}}$}
		\label{fig:appendix.intro.hoc.p0}
    }
    \end{subfigure}
  	\quad \quad
	%%%%%%%%%%%%%%%%%
	\begin{subfigure}[b]{.22\textwidth}
	\centering
	{
		\includegraphics[height=2.39cm]{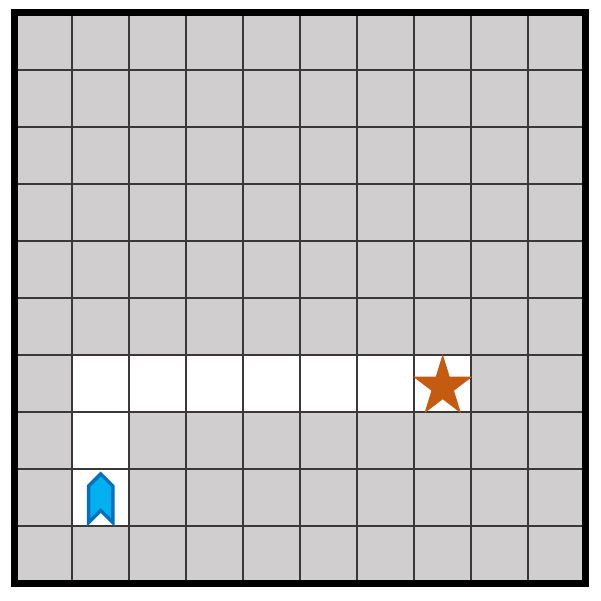}
		%\vspace{0.2mm}
		\caption{Visual puzzle for $\task^{\textnormal{out}}$}
		 \label{fig:appendix.intro.hoc.t1}
	}
	\end{subfigure}
	\
	%%%%%%%%%%%%%%%%%	
	\begin{subfigure}[b]{.22\textwidth}
	\centering
	{
		\begin{boxcode}{3.8cm}{0.75}{0.7}
			\textcode{def }\DSLRun\textcode{()\{}\\
			\quad \DSLMove\\
			\quad \DSLMove\\
			\quad \DSLTurnRight\\
			\quad \DSLRepeat\textcode{(6}\textcode{)\{}\\
			\quad \quad \DSLMove\\
			\quad \textcode{\}}\\
			\textcode{\}}
			\\
			\vspace{2mm}
		\end{boxcode}
		\vspace{-1mm}
		\caption{Solution code $\code^{\textnormal{out}}$}
		\label{fig:appendix.intro.hoc.p1}
    }
    \end{subfigure}
    %\\
    \vspace{-1.5mm}
	\caption{Task \hocC~-- Illustration of our methodology.}
	%\caption{\todo{}Illustration of our methodology for task \emph{Maze 16} from the \emph{Hour of Code: Classic Maze} challenge by \emph{Code.org}~\cite{hourofcode_maze}; the complete list of tasks with their specifications is in \figref{fig:dataset}.}
	\vspace{-3.5mm}
	\label{fig:appendix.intro.hoc}
\end{figure}
%%%%%%%%%%%%%%%%%%%%%%%%%%%%%%%%%%%%%%%%%%%%%%%%%%%%%%%%%%

%%%%%%%%%%%%%%%%%%%%%%%%%%%%%%%%%%%%%%%%%%%%%%%%%%%%%%%%%%
%%%%%%%%%%%%%%%%%%%%%%%%%%%%%%%%%%%%%%
\begin{figure}[h!]
\centering
	%%%%%%%%%%%%%%%%%
	\begin{subfigure}[b]{.22\textwidth}
	\centering
	{
		\includegraphics[height=2.39cm]{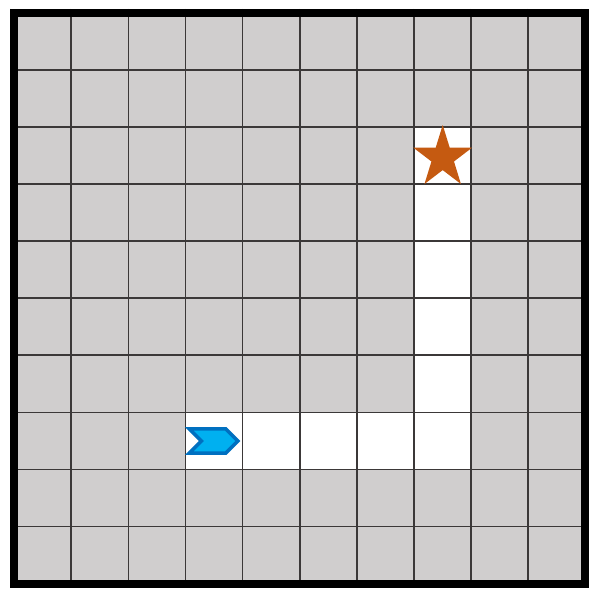}
		%\vspace{0.2mm}
		\caption{Visual puzzle for $\task^{\textnormal{in}}$}
		 \label{fig:appendix.intro.hoc.t0}
	}
	\end{subfigure}
	\
	%%%%%%%%%%%%%%%%%	
	\begin{subfigure}[b]{.22\textwidth}
	\centering
	{
		\begin{boxcode}{3.8cm}{0.75}{0.7}
			\textcode{def }\DSLRun\textcode{()\{}\\
			\quad \DSLRepeat\textcode{(4}\textcode{)\{}\\
			\quad \quad \DSLMove\\
			\quad \textcode{\}}\\
			\quad \DSLTurnLeft\\
			\quad \DSLRepeat\textcode{(5}\textcode{)\{}\\
			\quad \quad \DSLMove\\
			\quad \textcode{\}}\\
			\textcode{\}}
			\vspace{2.5mm}
		\end{boxcode}
		\vspace{-1mm}
		\caption{Solution code $\code^{\textnormal{in}}$}
		\label{fig:appendix.intro.hoc.p0}
    }
    \end{subfigure}
  	\quad \quad
	%%%%%%%%%%%%%%%%%
	\begin{subfigure}[b]{.22\textwidth}
	\centering
	{
		\includegraphics[height=2.39cm]{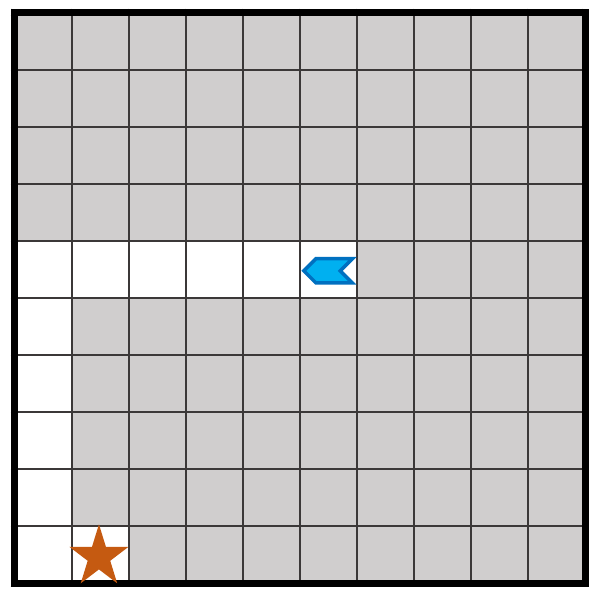}
		%\vspace{0.2mm}
		\caption{Visual puzzle for $\task^{\textnormal{out}}$}
		 \label{fig:appendix.intro.hoc.t1}
	}
	\end{subfigure}
	\
	%%%%%%%%%%%%%%%%%	
	\begin{subfigure}[b]{.22\textwidth}
	\centering
	{
		\begin{boxcode}{3.8cm}{0.75}{0.58}
			\textcode{def }\DSLRun\textcode{()\{}\\
			\quad \DSLRepeat\textcode{(5}\textcode{)\{}\\
			\quad \quad \DSLMove\\
			\quad \textcode{\}}\\
			\quad \DSLTurnLeft\\
			\quad \DSLRepeat\textcode{(5}\textcode{)\{}\\
			\quad \quad \DSLMove\\
			\quad \textcode{\}}\\
			\quad \DSLTurnLeft\\
			\quad \DSLMove\\
			\textcode{\}}
			%\vspace{1.5mm}
		\end{boxcode}
		\vspace{-1mm}
		\caption{Solution code $\code^{\textnormal{out}}$}
		\label{fig:appendix.intro.hoc.p1}
    }
    \end{subfigure}
    %\\
    \vspace{-1.5mm}
	\caption{Task \hocD~-- Illustration of our methodology.}
	%\caption{\todo{}Illustration of our methodology for task \emph{Maze 16} from the \emph{Hour of Code: Classic Maze} challenge by \emph{Code.org}~\cite{hourofcode_maze}; the complete list of tasks with their specifications is in \figref{fig:dataset}.}
	\vspace{-3.5mm}
	\label{fig:appendix.intro.hoc}
\end{figure}
%%%%%%%%%%%%%%%%%%%%%%%%%%%%%%%%%%%%%%%%%%%%%%%%%%%%%%%%%%

%%%%%%%%%%%%%%%%%%%%%%%%%%%%%%%%%%%%%%%%%%%%%%%%%%%%%%%%%%
%%%%%%%%%%%%%%%%%%%%%%%%%%%%%%%%%%%%%%
\begin{figure}[h!]
\centering
	%%%%%%%%%%%%%%%%%
	\begin{subfigure}[b]{.22\textwidth}
	\centering
	{
		\includegraphics[height=2.39cm]{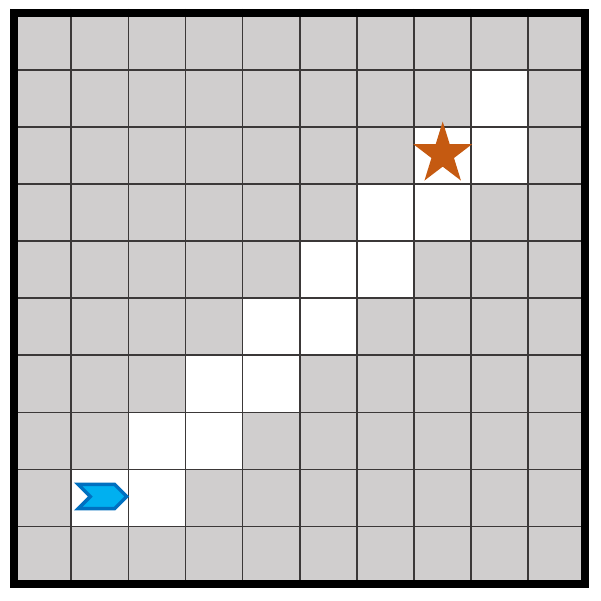}
		%\vspace{0.2mm}
		\caption{Visual puzzle for $\task^{\textnormal{in}}$}
		 \label{fig:appendix.intro.hoc.t0}
	}
	\end{subfigure}
	\
	%%%%%%%%%%%%%%%%%	
	\begin{subfigure}[b]{.22\textwidth}
	\centering
	{
		\begin{boxcode}{3.8cm}{0.75}{0.7}
			\textcode{def }\DSLRun\textcode{()\{}\\
			\quad \DSLRepeatUntil\textcode{(}\DSLBoolGoal\textcode{)\{}\\
			\quad \quad \DSLMove\\
			\quad \quad \DSLTurnLeft\\
			\quad \quad \DSLMove\\
			\quad \quad \DSLTurnRight\\			
			\quad \textcode{\}}\\
		    \textcode{\}}
			\\
		    \vspace{1.8mm}
		\end{boxcode}
		\vspace{-1mm}
		\caption{Solution code $\code^{\textnormal{in}}$}
		\label{fig:appendix.intro.hoc.p0}
    }
    \end{subfigure}
  	\quad \quad
	%%%%%%%%%%%%%%%%%
	\begin{subfigure}[b]{.22\textwidth}
	\centering
	{
		\includegraphics[height=2.39cm]{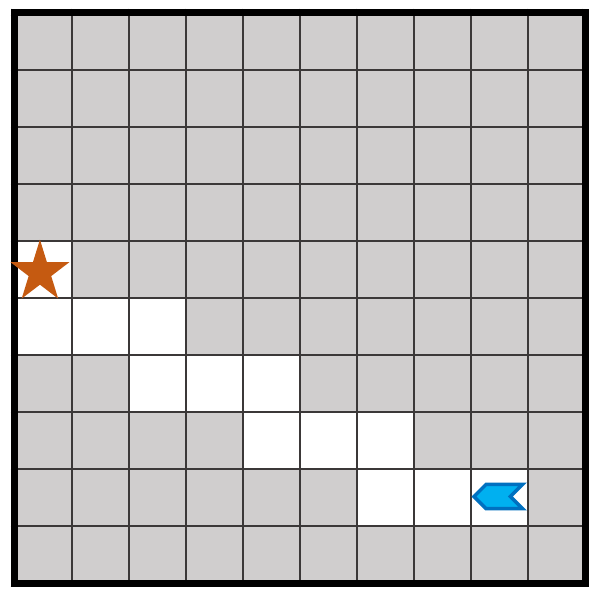}
		%\vspace{0.2mm}
		\caption{Visual puzzle for $\task^{\textnormal{out}}$}
		 \label{fig:appendix.intro.hoc.t1}
	}
	\end{subfigure}
	\
	%%%%%%%%%%%%%%%%%	
	\begin{subfigure}[b]{.22\textwidth}
	\centering
	{
		\begin{boxcode}{3.8cm}{0.75}{0.7}
			\textcode{def }\DSLRun\textcode{()\{}\\
			\quad \DSLRepeatUntil\textcode{(}\DSLBoolGoal\textcode{)\{}\\
			\quad \quad \DSLMove\\
			\quad \quad \DSLMove\\
			\quad \quad \DSLTurnRight\\
			\quad \quad \DSLMove\\
			\quad \quad \DSLTurnLeft\\			
			\quad \textcode{\}}\\
		    \textcode{\}}
		    \vspace{1.5mm}
		\end{boxcode}
		\vspace{-1mm}
		\caption{Solution code $\code^{\textnormal{out}}$}
		\label{fig:appendix.intro.hoc.p1}
    }
    \end{subfigure}
    %\\
    \vspace{-1.5mm}
	\caption{Task \hocF~-- Illustration of our methodology.}   
	%\caption{\todo{}Illustration of our methodology for task \emph{Maze 16} from the \emph{Hour of Code: Classic Maze} challenge by \emph{Code.org}~\cite{hourofcode_maze}; the complete list of tasks with their specifications is in \figref{fig:dataset}.}
	\vspace{-3.5mm}
	\label{fig:appendix.intro.hoc}
\end{figure}
%%%%%%%%%%%%%%%%%%%%%%%%%%%%%%%%%%%%%%%%%%%%%%%%%%%%%%%%%%

%%%%%%%%%%%%%%%%%%%%%%%%%%%%%%%%%%%%%%%%%%%%%%%%%%%%%%%%%%
%%%%%%%%%%%%%%%%%%%%%%%%%%%%%%%%%%%%%%
\begin{figure}[t!]
\centering
	%%%%%%%%%%%%%%%%%
	\begin{subfigure}[b]{.22\textwidth}
	\centering
	{
		\includegraphics[height=2.39cm]{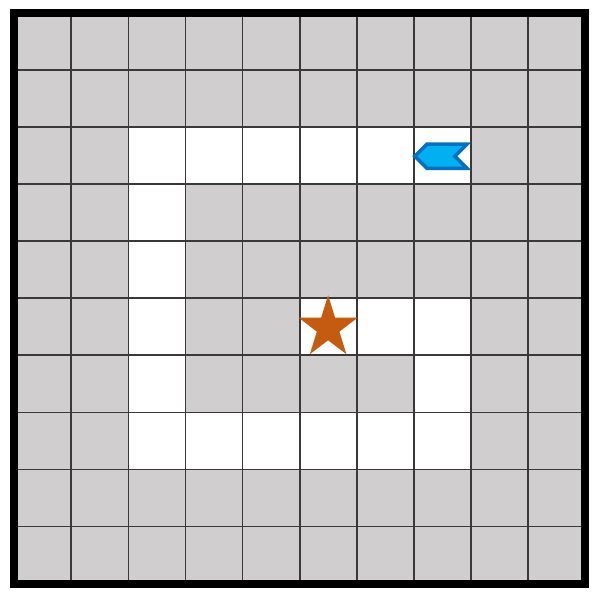}
		%\vspace{0.2mm}
		\caption{Visual puzzle for $\task^{\textnormal{in}}$}
		 \label{fig:appendix.intro.hoc.t0}
	}
	\end{subfigure}
	\
	%%%%%%%%%%%%%%%%%	
	\begin{subfigure}[b]{.22\textwidth}
	\centering
	{
		\begin{boxcode}{3.8cm}{0.75}{0.7}
				\textcode{def }\DSLRun\textcode{()\{}\\
				\quad \DSLRepeatUntil\textcode{(}\DSLBoolGoal\textcode{)\{}\\
				\quad \quad \DSLMove\\
				\quad \quad \DSLIf\textcode{(}\DSLBoolPathLeft\textcode{)\{}\\
				\quad \quad \quad \DSLTurnLeft\\
				\quad \quad \textcode{\}}\\
				\quad \textcode{\}}\\
				\textcode{\}}
				\\
				\vspace{1mm}
		\end{boxcode}
		\vspace{-1mm}
		\caption{Solution code $\code^{\textnormal{in}}$}
		\label{fig:appendix.intro.hoc.p0}
    }
    \end{subfigure}
  	\quad \quad
	%%%%%%%%%%%%%%%%%
	\begin{subfigure}[b]{.22\textwidth}
	\centering
	{
		\includegraphics[height=2.39cm]{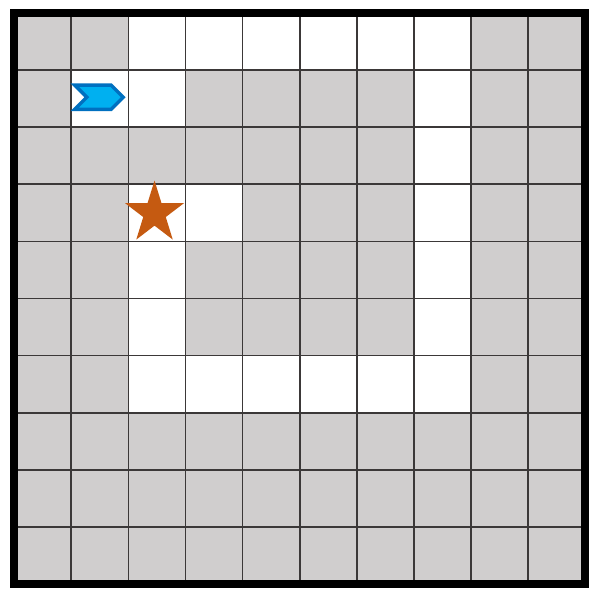}
		%\vspace{0.2mm}
		\caption{Visual puzzle for $\task^{\textnormal{out}}$}
		 \label{fig:appendix.intro.hoc.t1}
	}
	\end{subfigure}
	\
	%%%%%%%%%%%%%%%%%	
	\begin{subfigure}[b]{.22\textwidth}
	\centering
	{
		\begin{boxcode}{3.8cm}{0.75}{0.58}
			\textcode{def }\DSLRun\textcode{()\{}\\
			\quad \DSLMove\\
			\quad \DSLTurnLeft\\
			\quad \DSLRepeatUntil\textcode{(}\DSLBoolGoal\textcode{)\{}\\
			\quad \quad \DSLMove\\
			\quad \quad \DSLIf\textcode{(}\DSLBoolPathRight\textcode{)\{}\\
			\quad \quad \quad \DSLTurnRight\\
			\quad \quad \textcode{\}}\\
			\quad \textcode{\}}\\
			\textcode{\}}
		\end{boxcode}
		\vspace{-1mm}
		\caption{Solution code $\code^{\textnormal{out}}$}
		\label{fig:appendix.intro.hoc.p1}
    }
    \end{subfigure}
    %\\
    \vspace{-1.5mm}
	\caption{Task \hocG~-- Illustration of our methodology (same as \figref{fig:intro.hoc} and shown for completeness).}   
	%\caption{\todo{}Illustration of our methodology for task \emph{Maze 16} from the \emph{Hour of Code: Classic Maze} challenge by \emph{Code.org}~\cite{hourofcode_maze}; the complete list of tasks with their specifications is in \figref{fig:dataset}.}
	\vspace{-3.5mm}
	\label{fig:appendix.intro.hoc}
\end{figure}
%%%%%%%%%%%%%%%%%%%%%%%%%%%%%%%%%%%%%%%%%%%%%%%%%%%%%%%%%%

%%%%%%%%%%%%%%%%%%%%%%%%%%%%%%%%%%%%%%%%%%%%%%%%%%%%%%%%%%
%%%%%%%%%%%%%%%%%%%%%%%%%%%%%%%%%%%%%%
\begin{figure}[h!]
\centering
	%%%%%%%%%%%%%%%%%
	\begin{subfigure}[b]{.22\textwidth}
	\centering
	{
		\includegraphics[height=2.39cm]{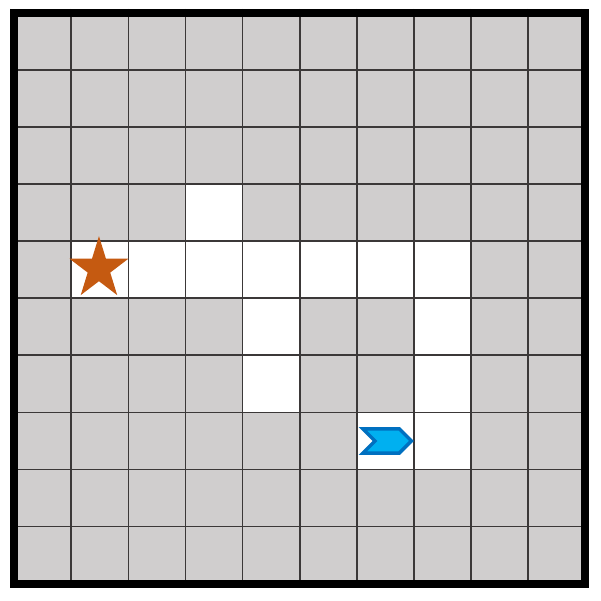}
		%\vspace{0.2mm}
		\caption{Visual puzzle for $\task^{\textnormal{in}}$}
		 \label{fig:appendix.intro.hoc.t0}
	}
	\end{subfigure}
	\
	%%%%%%%%%%%%%%%%%	
	\begin{subfigure}[b]{.22\textwidth}
	\centering
	{
		\begin{boxcode}{3.8cm}{0.75}{0.7}
				\textcode{def }\DSLRun\textcode{()\{}\\
				\quad \DSLRepeatUntil\textcode{(}\DSLBoolGoal\textcode{)\{}\\
				\quad \quad \DSLIf\textcode{(}\DSLBoolPathAhead\textcode{)\{}\\
				\quad \quad \quad \DSLMove\\
				\quad \quad \textcode{\}}\\
				\quad \quad \DSLElse\textcode{\{}\\
    			\quad \quad \quad \DSLTurnLeft\\
				\quad \quad \textcode{\}}\\
				\quad \textcode{\}}\\
				\textcode{\}}
				%\vspace{1mm}
		\end{boxcode}
		\vspace{-1mm}
		\caption{Solution code $\code^{\textnormal{in}}$}
		\label{fig:appendix.intro.hoc.p0}
    }
    \end{subfigure}
  	\quad \quad
	%%%%%%%%%%%%%%%%%
	\begin{subfigure}[b]{.22\textwidth}
	\centering
	{
		\includegraphics[height=2.39cm]{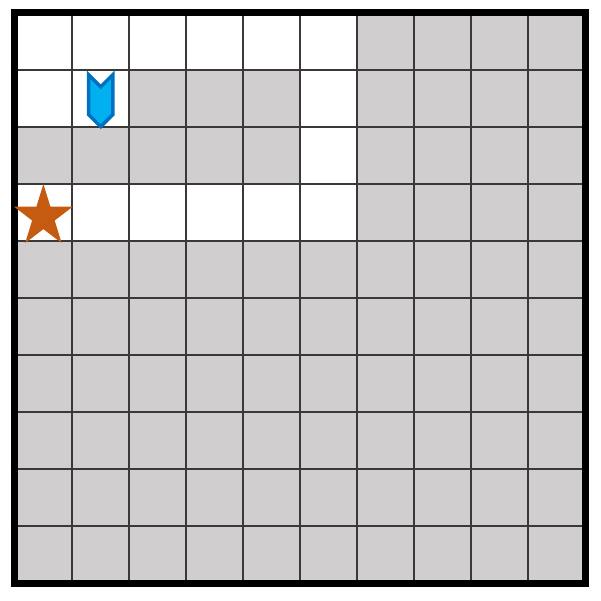}
		%\vspace{0.2mm}
		\caption{Visual puzzle for $\task^{\textnormal{out}}$}
		 \label{fig:appendix.intro.hoc.t1}
	}
	\end{subfigure}
	\
	%%%%%%%%%%%%%%%%%	
	\begin{subfigure}[b]{.22\textwidth}
	\centering
	{
		\begin{boxcode}{3.8cm}{0.75}{0.7}
				\textcode{def }\DSLRun\textcode{()\{}\\
				\quad \DSLRepeatUntil\textcode{(}\DSLBoolGoal\textcode{)\{}\\
				\quad \quad \DSLIf\textcode{(}\DSLBoolPathAhead\textcode{)\{}\\
				\quad \quad \quad \DSLMove\\
				\quad \quad \textcode{\}}\\
				\quad \quad \DSLElse\textcode{\{}\\
    			\quad \quad \quad \DSLTurnRight\\
				\quad \quad \textcode{\}}\\
				\quad \textcode{\}}\\
				\textcode{\}}
				%\vspace{1mm}
		\end{boxcode}
		\vspace{-1mm}
		\caption{Solution code $\code^{\textnormal{out}}$}
		\label{fig:appendix.intro.hoc.p1}
    }
    \end{subfigure}
    %\\
    \vspace{-1.5mm}
	\caption{Task \hocH~-- Illustration of our methodology.}
	%\caption{\todo{}Illustration of our methodology for task \emph{Maze 16} from the \emph{Hour of Code: Classic Maze} challenge by \emph{Code.org}~\cite{hourofcode_maze}; the complete list of tasks with their specifications is in \figref{fig:dataset}.}
	\vspace{-3.5mm}
	\label{fig:appendix.intro.hoc}
\end{figure}
%%%%%%%%%%%%%%%%%%%%%%%%%%%%%%%%%%%%%%%%%%%%%%%%%%%%%%%%%%
%%%%%%%%%%%%%%%%%%%%%%%%%%%%%%%%%%%%%%%%%%%%%%%%%%%%%%%%%%

%%%%%%%%%%%%%%%%%%%%%%%%%%%%%%%%%%%%%%%%%%%%%%%%%%%%%%%%%%
%%%%%%%%%%%%%%%%%%%%%%%%%%%%%%%%%%%%%%%%%%%%%%%%%%%%%%%%%%
\begin{figure}[h!]
\centering
	%%%%%%%%%%%%%%%%%
	\begin{subfigure}[b]{.27\textwidth}
	\centering
	{
		\includegraphics[height=1.95cm]{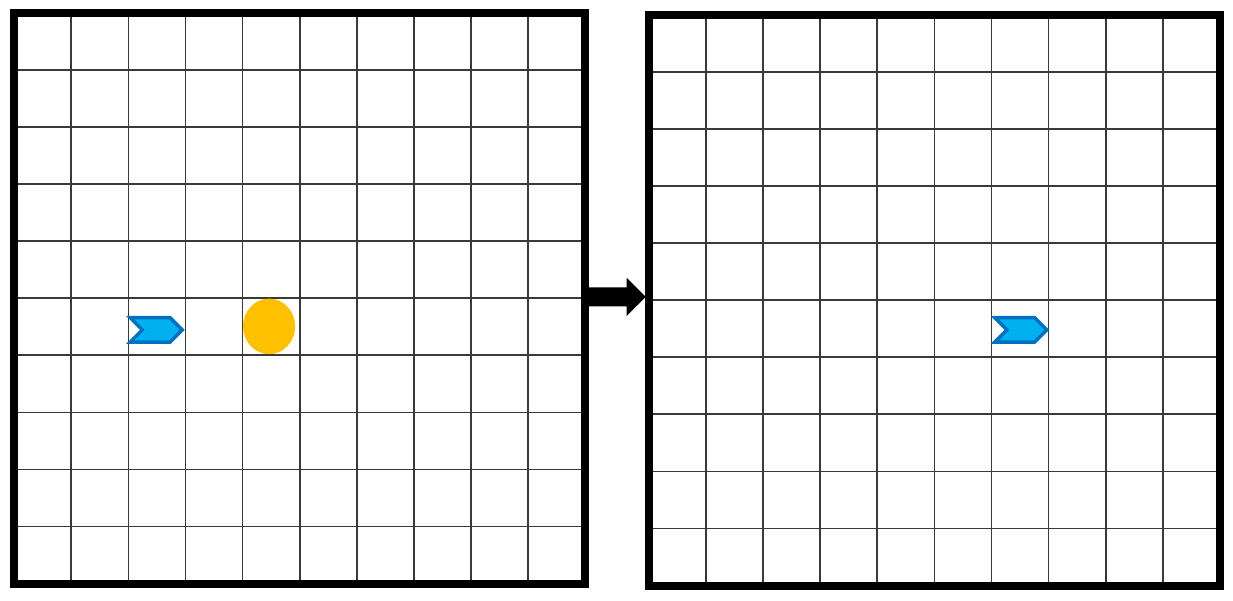}		
		%\vspace{0.01mm}
		\caption{Visual puzzle for $\task^{\textnormal{in}}$}
		\label{fig:appendix.intro.karel.t0}
    }
    \end{subfigure}
  	\
  	\begin{subfigure}[b]{.195\textwidth}
  	\centering
  	 {
  	 	\begin{boxcode}{3.5cm}{0.75}{0.7}
			\textcode{def }\DSLRun\textcode{()\{}\\
			\quad \DSLMove\\
			\quad \DSLMove\\
			\quad \DSLPickMarker\\
			\quad \DSLMove\\
			\quad \DSLMove\\			
			%\quad \textcode{\}}\\
			\textcode{\}}
			\vspace{2.5mm}
		\end{boxcode}
		\vspace{-1.18mm}
		\caption{Solution code $\code^{\textnormal{in}}$}
		\label{fig:appendix.intro.karel.p0}
	}
	\end{subfigure}
	\quad \  
	\begin{subfigure}[b]{.27\textwidth}
	\centering
	{
		\includegraphics[height=1.95cm]{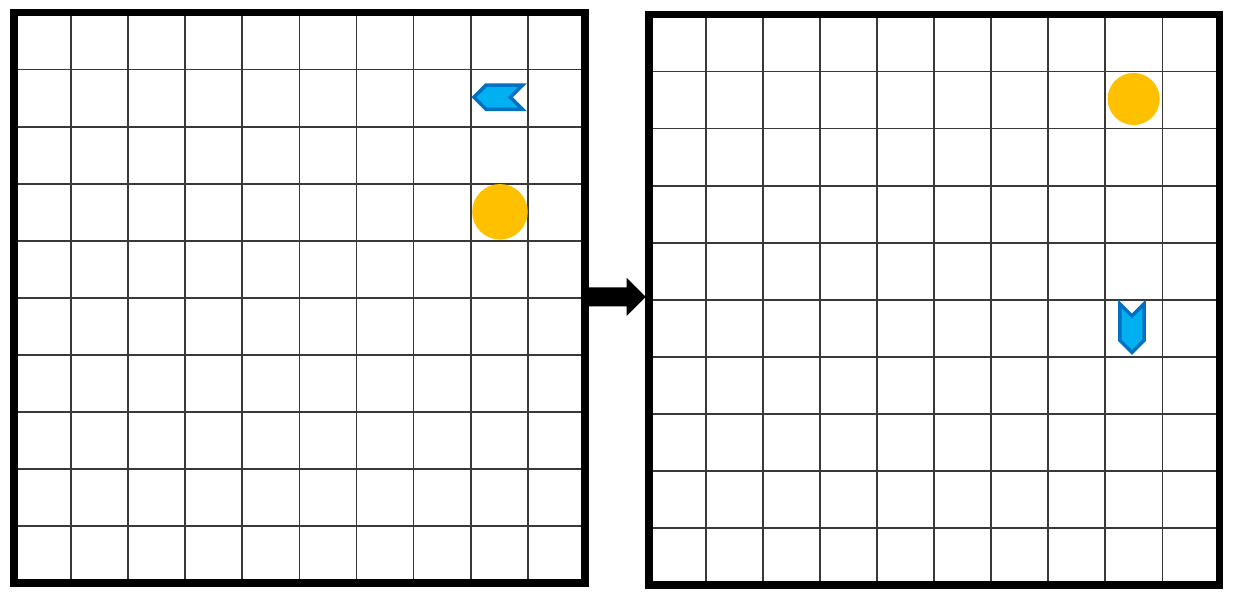}		
		%\vspace{0.01mm}
		\caption{Visual puzzle for $\task^{\textnormal{out}}$}
		\label{fig:appendix.intro.karel.t1}
	}
	\end{subfigure}
	\
	\begin{subfigure}[b]{.195\textwidth}
	\centering
	{
  	 	\begin{boxcode}{3.5cm}{0.75}{0.58}
			\textcode{def }\DSLRun\textcode{()\{}\\
			\quad \DSLPutMarker\\
			\quad \DSLTurnLeft\\
			\quad \DSLMove\\
			\quad \DSLMove\\
			\quad \DSLPickMarker\\
			\quad \DSLMove\\
			\quad \DSLMove\\			
			%\quad \textcode{\}}\\
			\textcode{\}}
		\end{boxcode}
		\vspace{-1.3mm}
		\caption{Solution code $\code^{\textnormal{out}}$}
		\label{fig:appendix.intro.karel.p1}
	}
	\end{subfigure}  
	\vspace{-1.5mm}	
	\caption{Task \karelA~-- Illustration of our methodology.}   
	%\caption{\todo{}Illustration of our methodology for task \emph{Diagonal} from the \emph{Intro to Programming with Karel} course by \emph{CodeHS.com}~\cite{intro_to_karel_codehs}; the complete list of tasks with their specifications is in \figref{fig:dataset}.}
	\vspace{-3.5mm}
	\label{fig:appendix.intro.karel1}
\end{figure}
%%%%%%%%%%%%%%%%%%%%%%%%%%%%%%%%%%%%%
%%%%%%%%%%%%%%%%%%%%%%%%%%%%%%%%%%%%%%%%%%%%%%%%%%%%%%%%%%

%%%%%%%%%%%%%%%%%%%%%%%%%%%%%%%%%%%%%%%%%%%%%%%%%%%%%%%%%%
%%%%%%%%%%%%%%%%%%%%%%%%%%%%%%%%%%%%%%%%%%%%%%%%%%%%%%%%%%
\begin{figure}[h!]
\centering
	%%%%%%%%%%%%%%%%%
	\begin{subfigure}[b]{.27\textwidth}
	\centering
	{
		\includegraphics[height=1.95cm]{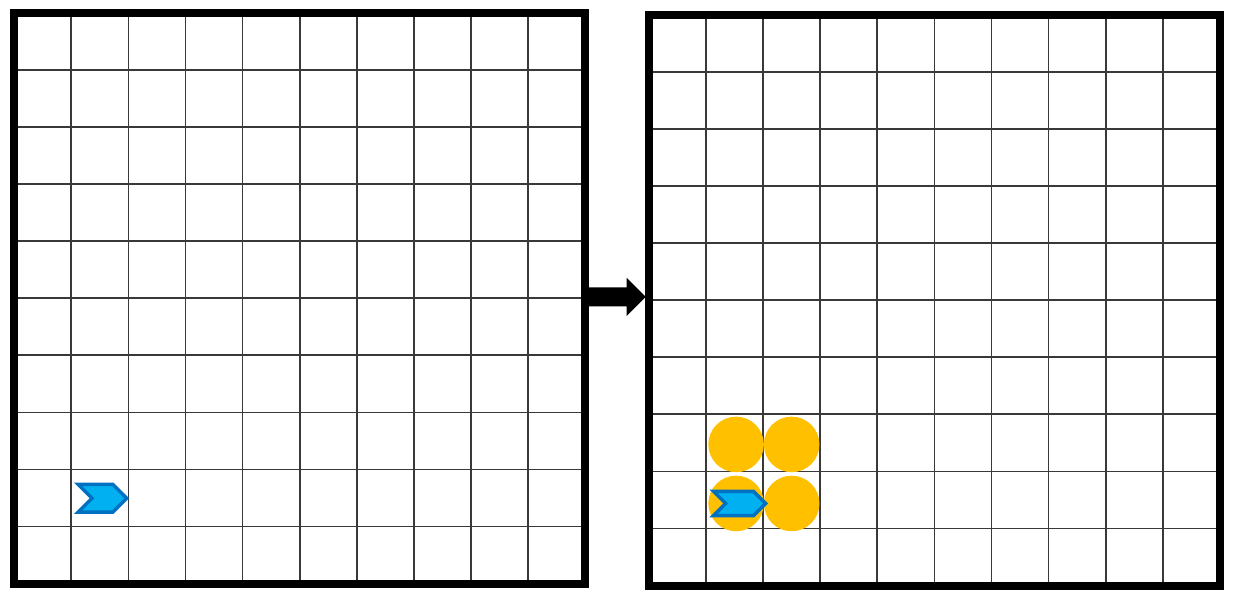}		
		%\vspace{0.01mm}
		\caption{Visual puzzle for $\task^{\textnormal{in}}$}
		\label{fig:appendix.intro.karel.t0}
    }
    \end{subfigure}
  	\
  	\begin{subfigure}[b]{.195\textwidth}
  	\centering
  	 {
  	 	\begin{boxcode}{3.5cm}{0.75}{0.7}
			\textcode{def }\DSLRun\textcode{()\{}\\
			\quad \DSLRepeat\textcode{(4}\textcode{)\{}\\
			\quad \quad \DSLPutMarker\\
			\quad \quad \DSLMove\\
			\quad \quad \DSLTurnLeft\\
			\quad \textcode{\}}\\		
			\textcode{\}}
			\vspace{3mm}
		\end{boxcode}
		\vspace{-1.18mm}
		\caption{Solution code $\code^{\textnormal{in}}$}
		\label{fig:appendix.intro.karel.p0}
	}
	\end{subfigure}
	\quad \  
	\begin{subfigure}[b]{.27\textwidth}
	\centering
	{
		\includegraphics[height=1.95cm]{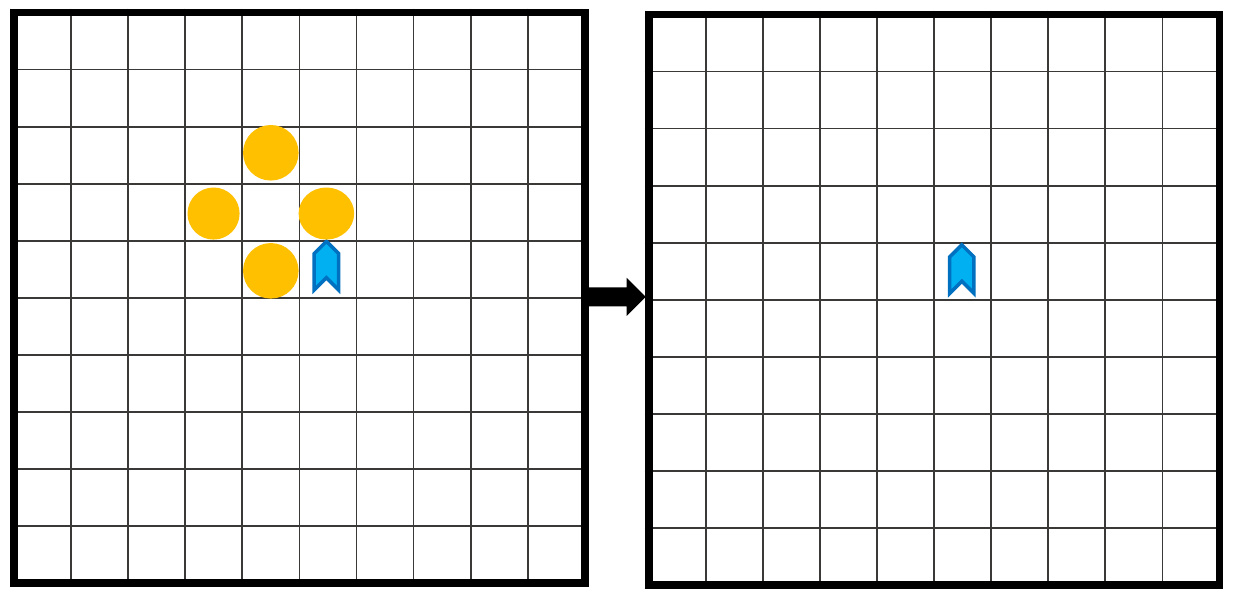}		
		%\vspace{0.01mm}
		\caption{Visual puzzle for $\task^{\textnormal{out}}$}
		\label{fig:appendix.intro.karel.t1}
	}
	\end{subfigure}
	\
	\begin{subfigure}[b]{.195\textwidth}
	\centering
	{
  	 	\begin{boxcode}{3.5cm}{0.75}{0.7}
			\textcode{def }\DSLRun\textcode{()\{}\\
			\quad \DSLRepeat\textcode{(4}\textcode{)\{}\\
			\quad \quad \DSLMove\\
			\quad \quad \DSLPickMarker\\
			\quad \quad \DSLMove\\
			\quad \quad \DSLTurnLeft\\
			\quad \textcode{\}}\\		
			\textcode{\}}
		\end{boxcode}
		\vspace{-1.18mm}
		\caption{Solution code $\code^{\textnormal{out}}$}
		\label{fig:appendix.intro.karel.p1}
	}
	\end{subfigure}  
	\vspace{-1.5mm}	
	\caption{Task \karelC~-- Illustration of our methodology.}   
	%\caption{\todo{}Illustration of our methodology for task \emph{Diagonal} from the \emph{Intro to Programming with Karel} course by \emph{CodeHS.com}~\cite{intro_to_karel_codehs}; the complete list of tasks with their specifications is in \figref{fig:dataset}.}
	\vspace{-3.5mm}
	\label{fig:appendix.intro.karel2}
\end{figure}
%%%%%%%%%%%%%%%%%%%%%%%%%%%%%%%%%%%%%
%%%%%%%%%%%%%%%%%%%%%%%%%%%%%%%%%%%%%%%%%%%%%%%%%%%%%%%%%%

%%%%%%%%%%%%%%%%%%%%%%%%%%%%%%%%%%%%%%
%%%%%%%%%%%%%%%%%%%%%%%%%%%%%%%%%%%%%%
\begin{figure}[h!]
\centering
	%%%%%%%%%%%%%%%%%
	\begin{subfigure}[b]{.27\textwidth}
	\centering
	{
		\includegraphics[height=1.95cm]{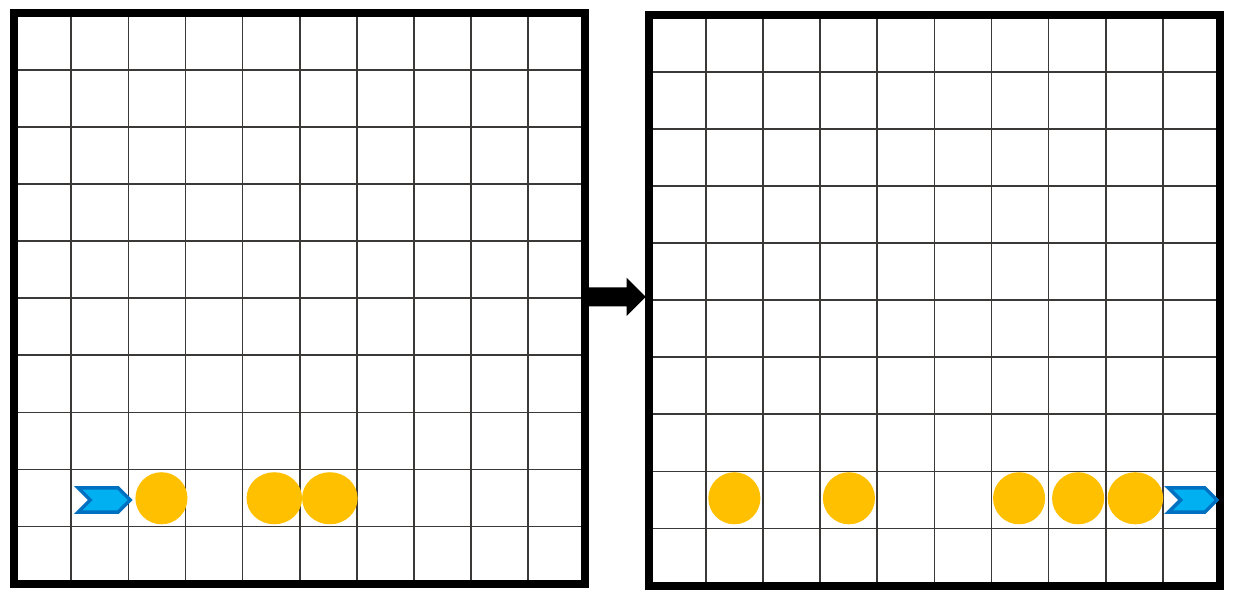}
		%\vspace{0.01mm}
		\caption{Visual puzzle for $\task^{\textnormal{in}}$}
		\label{fig:appendix.intro.karel.t0}
    }
    \end{subfigure}
  	\
  	\begin{subfigure}[b]{.195\textwidth}
  	\centering
  	 {
  	 	\begin{boxcode}{3.5cm}{0.75}{0.58}
				\textcode{def }\DSLRun\textcode{()\{}\\
				\quad \DSLRepeat\textcode{(8}\textcode{)\{}\\
				\quad \quad \DSLIf\textcode{(}\DSLBoolNoMarker\textcode{)\{}\\
				\quad \quad \quad \DSLPutMarker\\
				\quad \quad \textcode{\}}\\
				\quad \quad \DSLElse\textcode{\{}\\
    			\quad \quad \quad \DSLPickMarker\\
				\quad \quad \textcode{\}}\\
    			\quad \quad \DSLMove\\				
				\quad \textcode{\}}\\
				\textcode{\}}
				\vspace{1.5mm}
		\end{boxcode}
		\vspace{-1.18mm}
		\caption{Solution code $\code^{\textnormal{in}}$}
		\label{fig:appendix.intro.karel.p0}
	}
	\end{subfigure}
	\quad \  
	\begin{subfigure}[b]{.27\textwidth}
	\centering
	{
		\includegraphics[height=1.95cm]{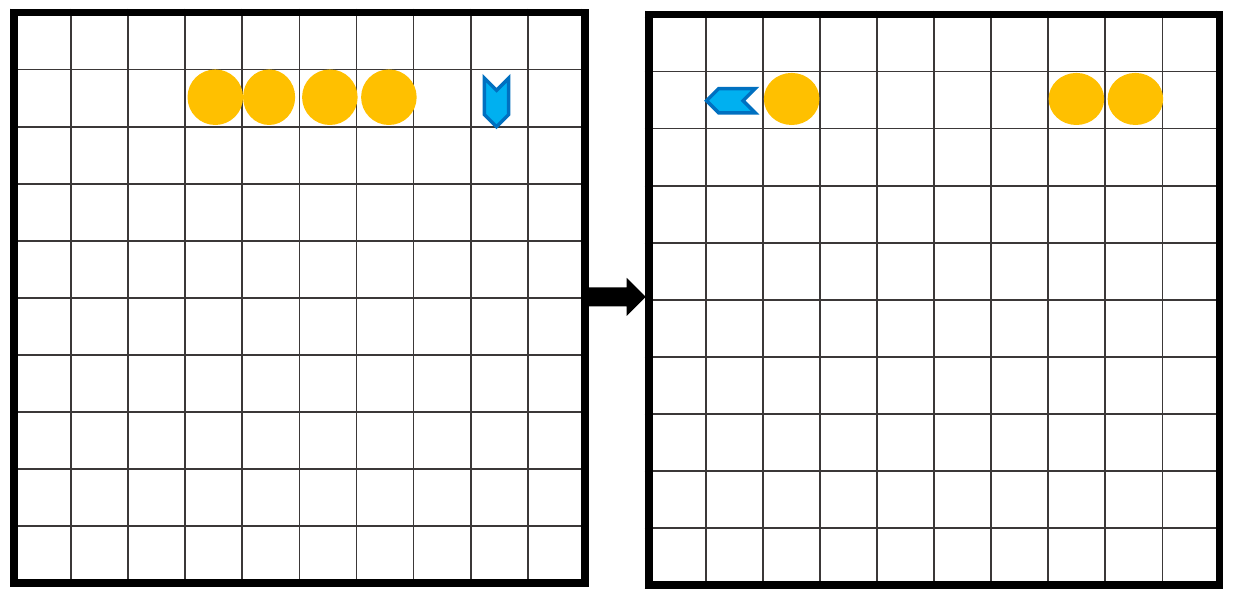}
		%\vspace{0.01mm}
		\caption{Visual puzzle for $\task^{\textnormal{out}}$}
		\label{fig:appendix.intro.karel.t1}
	}
	\end{subfigure}
	\
	\begin{subfigure}[b]{.195\textwidth}
	\centering
	{
  	 	\begin{boxcode}{3.5cm}{0.75}{0.58}
				\textcode{def }\DSLRun\textcode{()\{}\\
                \quad \DSLTurnRight\\				
				\quad \DSLRepeat\textcode{(7}\textcode{)\{}\\
				\quad \quad \DSLIf\textcode{(}\DSLBoolNoMarker\textcode{)\{}\\
				\quad \quad \quad \DSLPutMarker\\
				\quad \quad \textcode{\}}\\
				\quad \quad \DSLElse\textcode{\{}\\
    			\quad \quad \quad \DSLPickMarker\\
				\quad \quad \textcode{\}}\\
    			\quad \quad \DSLMove\\				
				\quad \textcode{\}}\\
				\textcode{\}}
				%\vspace{1mm}
		\end{boxcode}
		\vspace{-1.18mm}
		\caption{Solution code $\code^{\textnormal{out}}$}
		\label{fig:appendix.intro.karel.p1}
	}
	\end{subfigure}  
	\vspace{-1.5mm}	
	\caption{Task \karelE~-- Illustration of our methodology.}   
	%\caption{\todo{}Illustration of our methodology for task \emph{Diagonal} from the \emph{Intro to Programming with Karel} course by \emph{CodeHS.com}~\cite{intro_to_karel_codehs}; the complete list of tasks with their specifications is in \figref{fig:dataset}.}
	\vspace{-3.5mm}
	\label{fig:appendix.intro.karel3}
\end{figure}
%%%%%%%%%%%%%%%%%%%%%%%%%%%%%%%%%%%%%
%%%%%%%%%%%%%%%%%%%%%%%%%%%%%%%%%%%%%%%%%%%%%%%%%%%%%%%%%%

%%%%%%%%%%%%%%%%%%%%%%%%%%%%%%%%%%%%%%%%%%%%%%%%%%%%%%%%%%
%%%%%%%%%%%%%%%%%%%%%%%%%%%%%%%%%%%%%%%%%%%%%%%%%%%%%%%%%%
\begin{figure}[h!]
\centering
	%%%%%%%%%%%%%%%%%
	\begin{subfigure}[b]{.27\textwidth}
	\centering
	{
		\includegraphics[height=1.95cm]{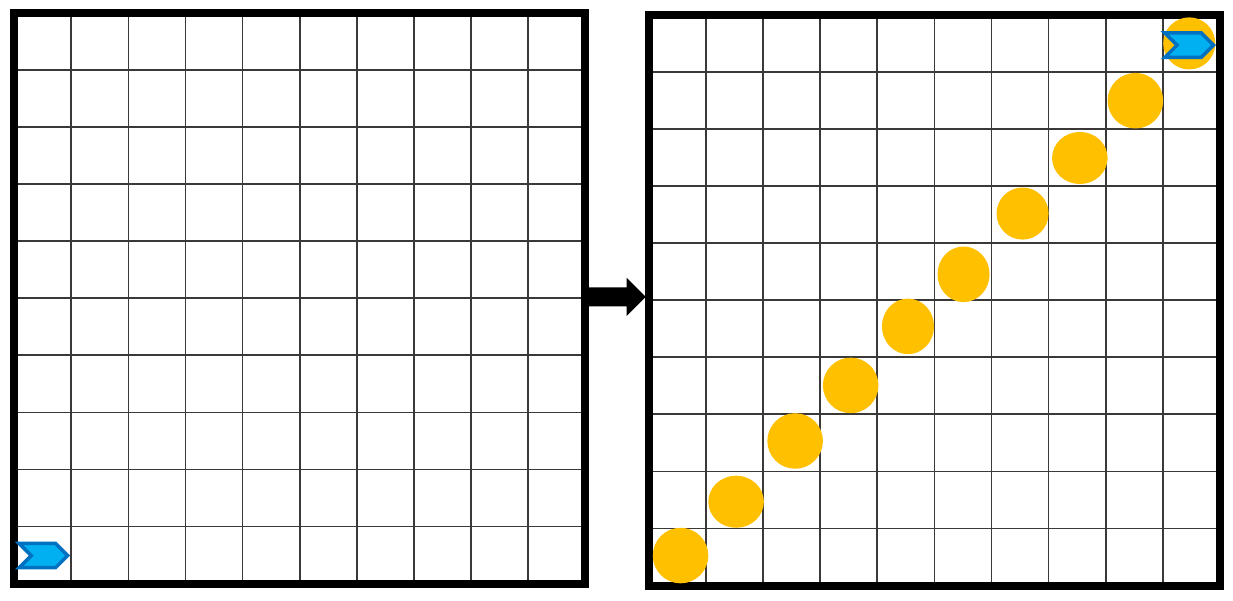}		
		%\vspace{0.01mm}
		\caption{Visual puzzle for $\task^{\textnormal{in}}$}
		\label{fig:appendix.intro.karel.t0}
    }
    \end{subfigure}
  	\
  	\begin{subfigure}[b]{.195\textwidth}
  	\centering
  	 {
  	 	\begin{boxcode}{3.5cm}{0.75}{0.58}
			\textcode{def }\DSLRun\textcode{()\{}\\
			\quad \DSLPutMarker\\
			\quad \DSLWhile\textcode{(}\DSLBoolPathAhead\textcode{)\{}\\
			\quad \quad \DSLMove\\
			\quad \quad \DSLTurnLeft\\
			\quad \quad \DSLMove\\
			\quad \quad \DSLTurnRight\\
			\quad \quad \DSLPutMarker\\	
			\quad \textcode{\}}\\
			\textcode{\}}
			\vspace{1.9mm}
		\end{boxcode}
		\vspace{-1.18mm}
		\caption{Solution code $\code^{\textnormal{in}}$}
		\label{fig:appendix.intro.karel.p0}
	}
	\end{subfigure}
	\quad \  
	\begin{subfigure}[b]{.27\textwidth}
	\centering
	{
		\includegraphics[height=1.95cm]{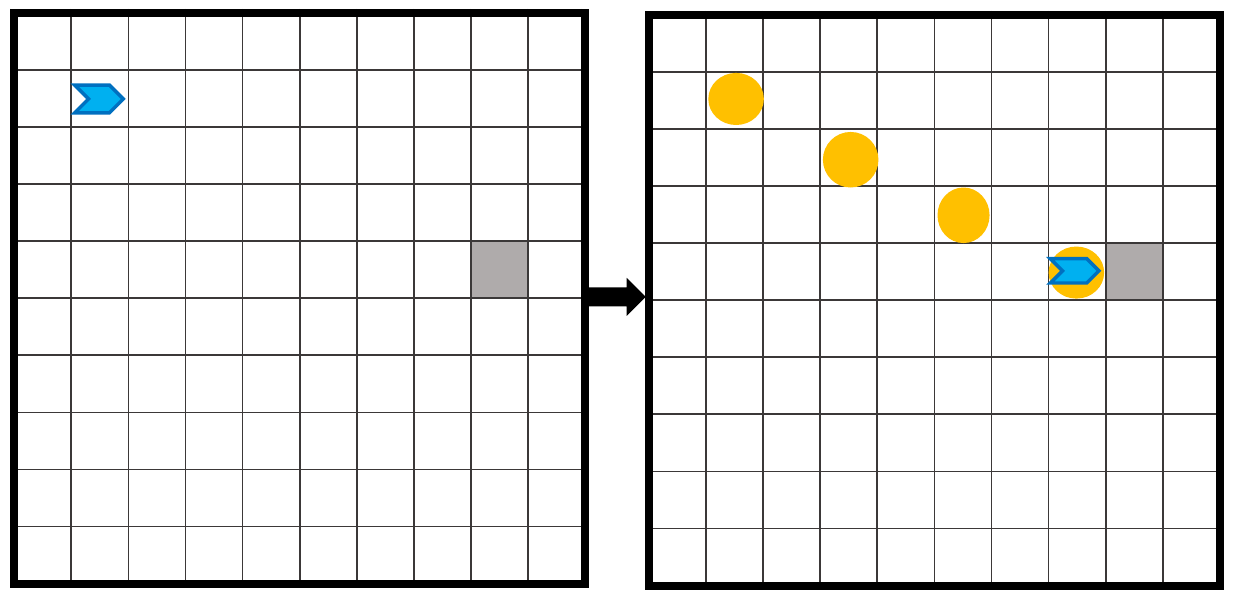}		
		%\vspace{0.01mm}
		\caption{Visual puzzle for $\task^\textnormal{out}$}   
		\label{fig:appendix.intro.karel.t1}
	}
	\end{subfigure}
	\
	\begin{subfigure}[b]{.195\textwidth}
	\centering
	{
  	 	\begin{boxcode}{3.5cm}{0.75}{0.58}
			\textcode{def }\DSLRun\textcode{()\{}\\
			\quad \DSLPutMarker\\
			\quad \DSLWhile\textcode{(}\DSLBoolPathAhead\textcode{)\{}\\
			\quad \quad \DSLMove\\
			\quad \quad \DSLMove\\
			\quad \quad \DSLTurnRight\\
			\quad \quad \DSLMove\\
			\quad \quad \DSLTurnLeft\\
			\quad \quad \DSLPutMarker\\	
			\quad \textcode{\}}\\
			\textcode{\}}
		\end{boxcode}
		\vspace{-1.18mm}
		\caption{Solution code $\code^{\textnormal{out}}$}
		\label{fig:appendix.intro.karel.p1}
	}
	\end{subfigure}  
	\vspace{-1.5mm}	
	\caption{Task \karelF~-- Illustration of our methodology (same as \figref{fig:intro.karel} and shown for completeness).}   
	%\caption{\todo{}Illustration of our methodology for task \emph{Diagonal} from the \emph{Intro to Programming with Karel} course by \emph{CodeHS.com}~\cite{intro_to_karel_codehs}; the complete list of tasks with their specifications is in \figref{fig:dataset}.}
	\vspace{-3.5mm}
	\label{fig:appendix.intro.karel4}
\end{figure}
%%%%%%%%%%%%%%%%%%%%%%%%%%%%%%%%%%%%%
%%%%%%%%%%%%%%%%%%%%%%%%%%%%%%%%%%%%%%%%%%%%%%%%%%%%%%%%%%

%\clearpage
% !TEX root =  main.tex
%%%%%%%%%%%%%%%%%%%%%%%%%%%%%%%%%%%%%%%%%%%%%%%%%%%%%%%%%%
%%%%%%%%%%%%%%%%%%%%%%%%%%%%%%%%%%%%%%%%%%%%%%%%%%%%%%%%%%
\section{User Study: Additional Details and Results}\label{appendix.sec.userstudy}
\vspace{-2mm}
In this section, we discuss additional details of our user study and expand on the key results provided in Section \ref{sec.userstudy}. For our user study, we used $4$ reference tasks (\hocC, \hocF, \hocG, and \hocH). We developed an web app where participants, recruited through Amazon Mechanical Turk, were asked to solve tasks generated by four algorithms, $\algoSame, \algoTutor, \algoTaskMut, \text{ or } \algoOursNoPstar$ (described in Section~\ref{sec.userstudy}). Next, we describe the interface of our app and details of the questionnaire. The web app is publicly available (see Footnote~\ref{footnote.userstudyapp}).
%\label{footnote.userstudyapp}}
%\textbf{Remark}: The web app will be made publicly accessible with the final version of the paper; for anonymity, we have removed the app's website URL in the submission. Furthermore, the dataset of $10$ reference tasks as well as output tasks for different methods used in the study will be released together with the final version of the paper.
%along with the output tasks from different baselines and our algorithms will be

%%%%%%%%%%%%%%%%%%%%%%%%%%%%%%%%%%%%%
\begin{figure}[h!]
\centering
	%%%%%%%%%%%%%%%%%
	\begin{subfigure}[b]{0.48\textwidth}
	\centering
	{
    	\begin{adjustbox}{varwidth=1\textwidth,fbox,center}
		    \includegraphics[trim={0.4cm 0 0.4cm 0},clip,width=1\textwidth]{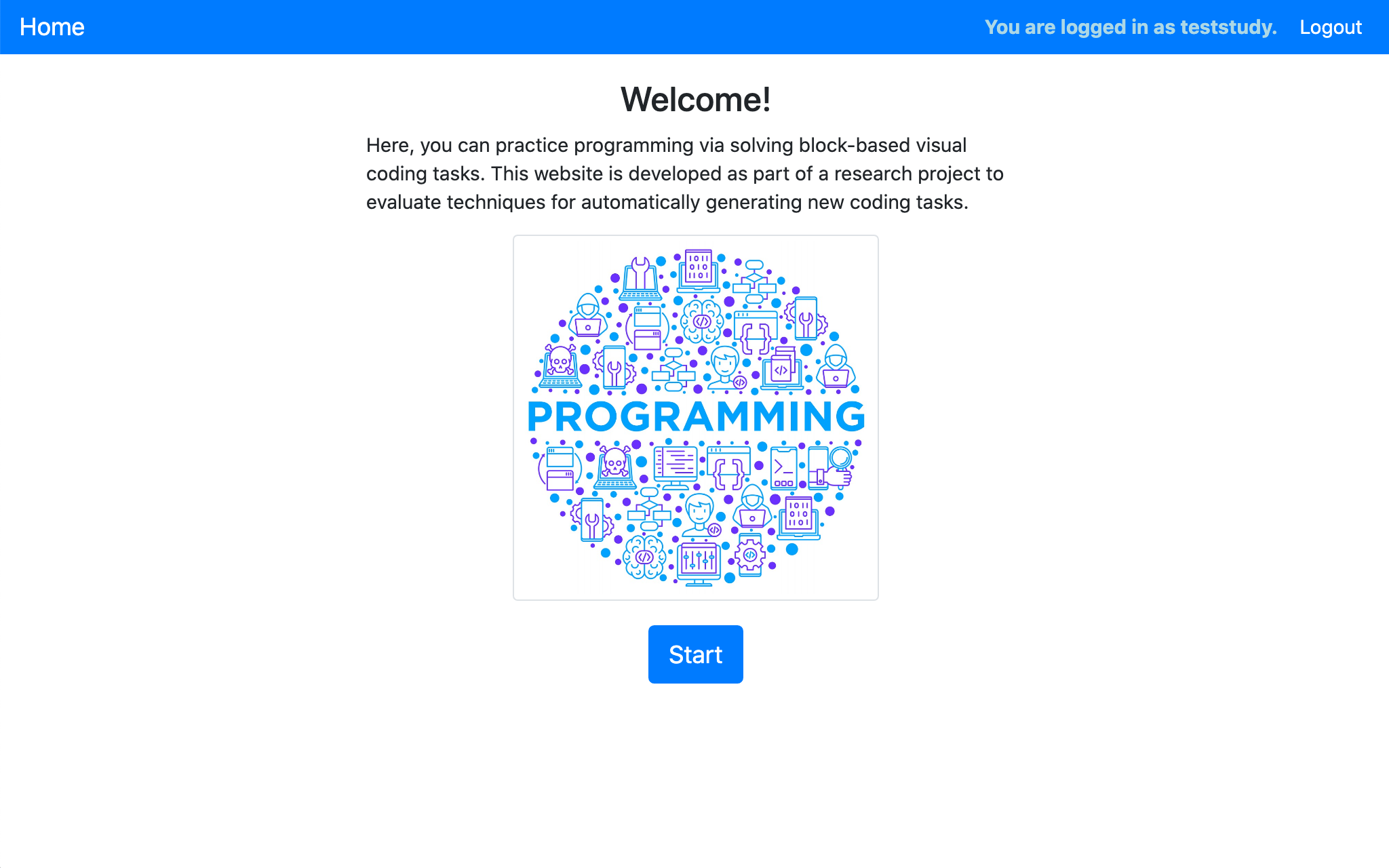}
	        %\vspace{0.2mm}
		\end{adjustbox}
		\caption{Login and welcome page}
		 \label{fig:appendix.userstudy.login}
	}
	\end{subfigure}
	\ \ 
	%%%%%%%%%%%%%%%%%
	\begin{subfigure}[b]{.48\textwidth}
	\centering
	{
    	\begin{adjustbox}{varwidth=1\textwidth,fbox,center}
		    \includegraphics[trim={0.4cm 0 0.4cm 0},clip,width=1\textwidth]{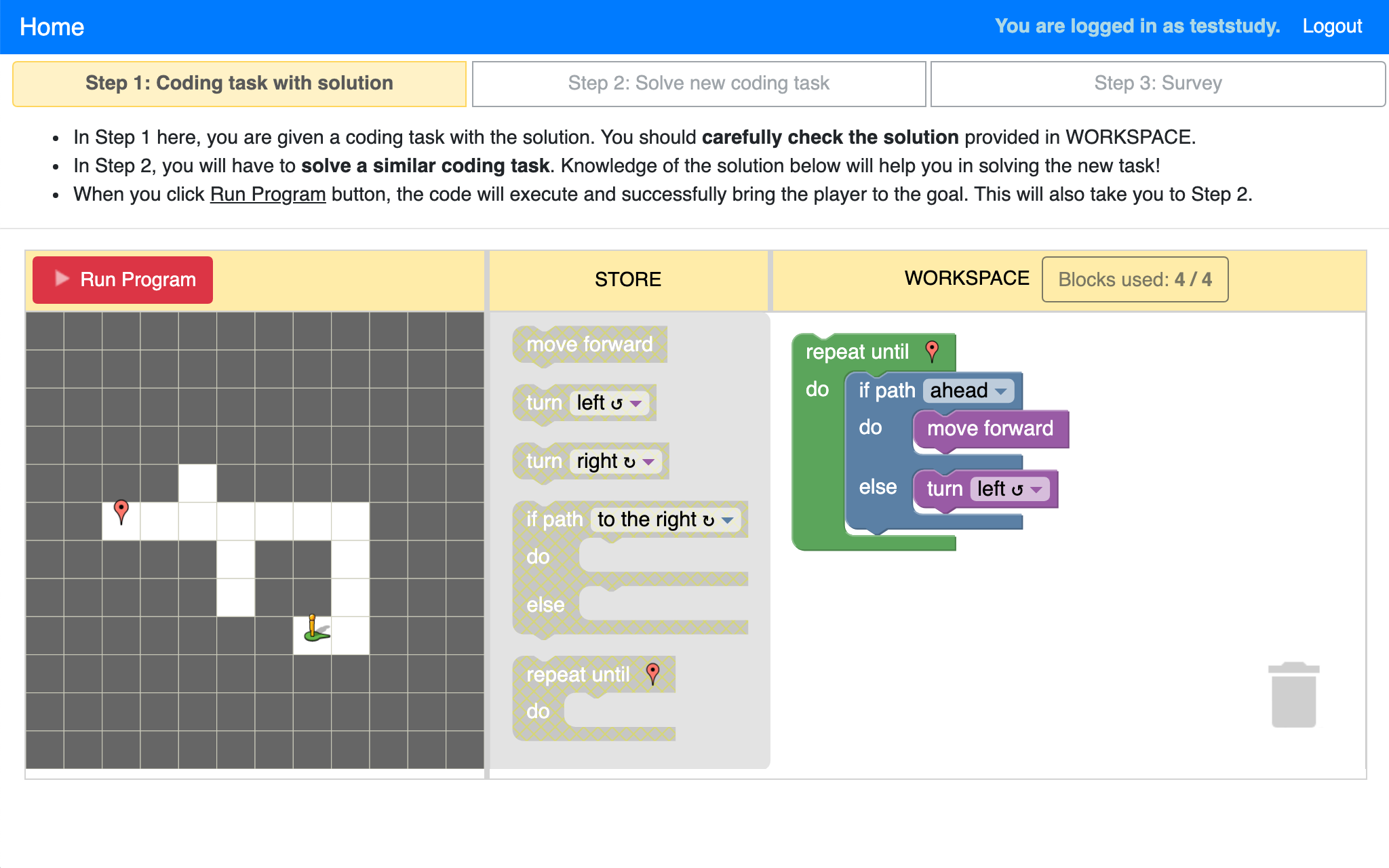}
    		%\vspace{0.2mm}
		\end{adjustbox}    		
		\caption{Step $1$: Coding task with solution}
		 \label{fig:appendix.userstudy.step1}
	}
	\end{subfigure}
	\\
	\vspace{2mm}
	%%%%%%%%%%%%%%%%%
	\begin{subfigure}[b]{.48\textwidth}
	\centering
	{
	    \begin{adjustbox}{varwidth=1\textwidth,fbox,center}
		    \includegraphics[trim={0.4cm 0 0.4cm 0},clip,width=1\textwidth]{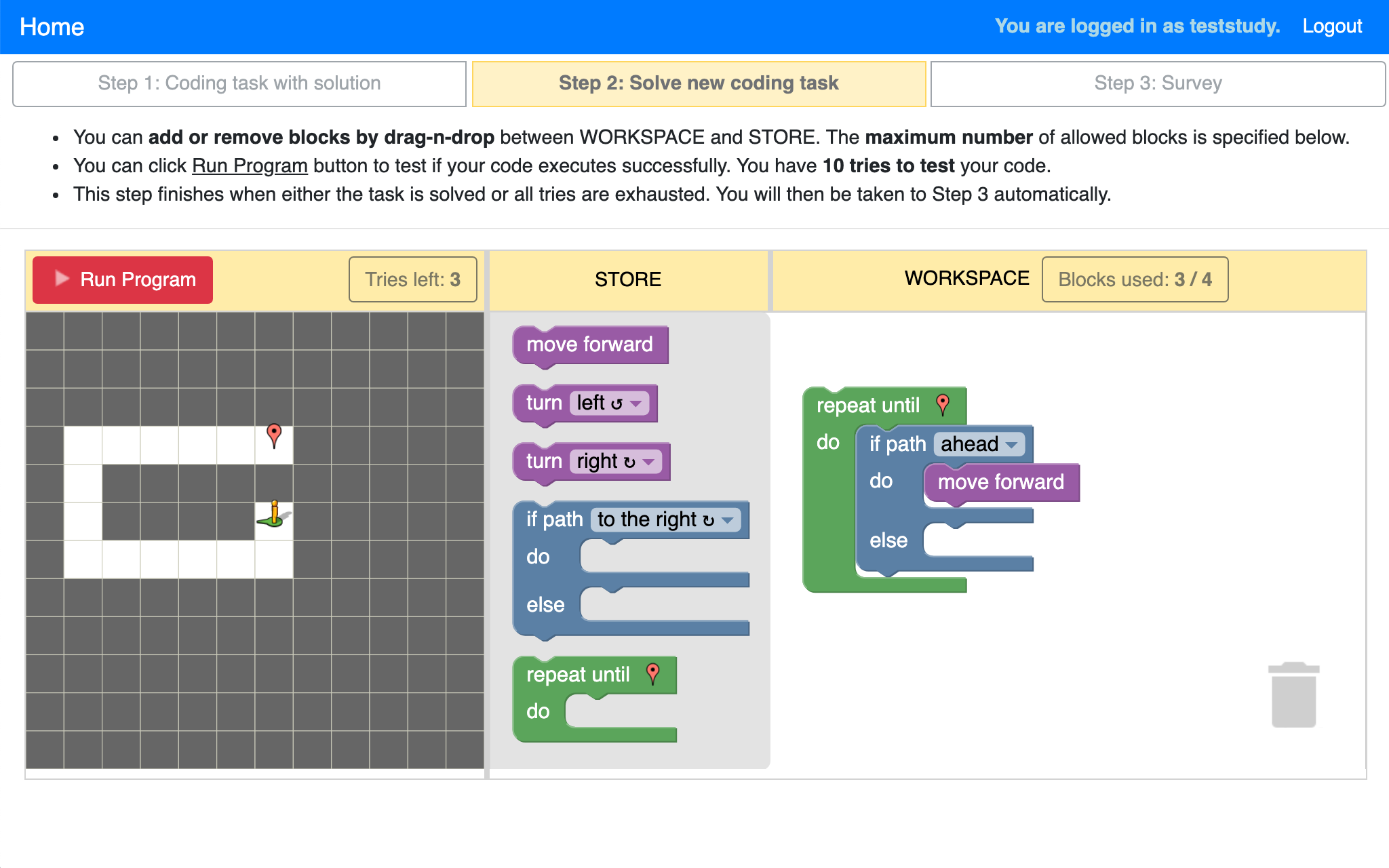}
		    %\vspace{0.2mm}
		\end{adjustbox} 		    
		\caption{Step $2$: Solve new coding task}
		 \label{fig:appendix.userstudy.step2}
	}
	\end{subfigure}
	\ \ 
	%%%%%%%%%%%%%%%%%
	\begin{subfigure}[b]{.48\textwidth}
	\centering
	{
    	\begin{adjustbox}{varwidth=1\textwidth,fbox,center}
		    \includegraphics[trim={0.4cm 0 0.4cm 0},clip,width=1\textwidth]{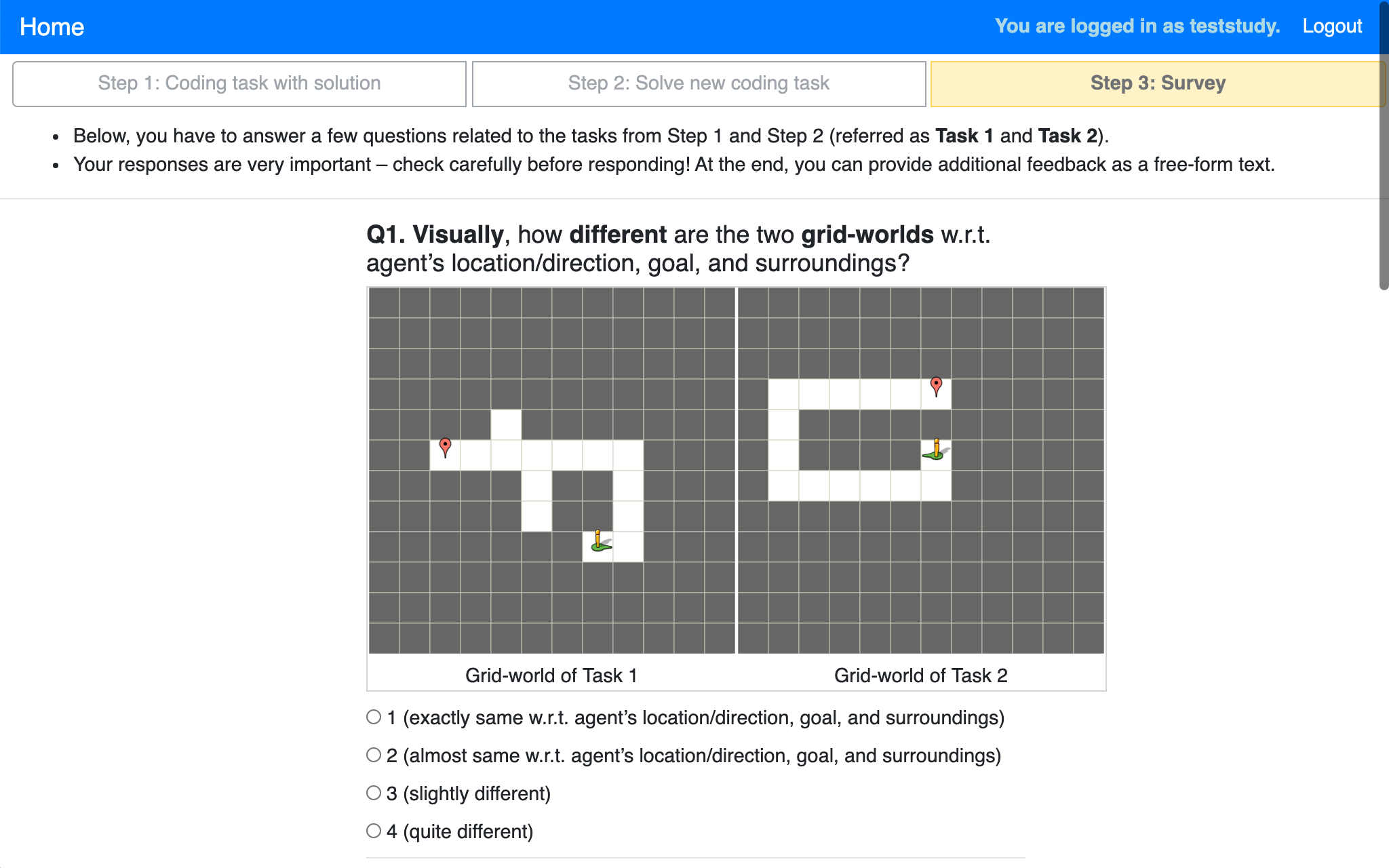}
		    %\vspace{0.2mm}
		\end{adjustbox}
		\caption{Step $3$: Survey}
		 \label{fig:appendix.userstudy.step3}
	}
	\end{subfigure}		
	%%%%%%%%%%%%%%%%%	
	\vspace{-1.5mm}
	\caption{Interface of the app.}
	% used for User Study
	%\vspace{-4mm}
	\label{fig:appendix.userstudy.app}
\end{figure}
% (\textcode{www.teaching-blocks.cc})
%%%%%%%%%%%%%%%%%%%%%%%%%%%%%%%%%%%%%
%%%%%%%%%%%%%%%%%%%%%%%%%%%%%%%%%%%%%%%%%%%%%%%%%%%%%%%%%%

\vspace{-4mm}
\subsection{App Interface and Questionnaire}
Our online app was developed using the publicly available toolkit of Blockly Games~\cite{googleblockly} and provides an interface for a participant to practice Block-based programming tasks for \hocType. Participants were familiarized with the tasks by a small tutorial given before they logged-in to the app. Each participant was randomly assigned a reference task, an algorithm (out of the four chosen for evaluation), and a particular task generated based on the chosen algorithm. These elements constituted a ``practice session" for a participant. Each session consisted of three steps. In Step $1$, the reference task along with its solution code was shown to the participant (\figref{fig:appendix.userstudy.step1}). In Step $2$, the participant was asked to solve a new task (\figref{fig:appendix.userstudy.step2}). The new task was generated by one of the four algorithms: $\algoSame, \algoTutor, \algoTaskMut, \text{ or } \algoOursNoPstar$. To solve this new task, the participant was given $10$ tries. If they successfully solved the task or ran out of tries, they were directed to Step $3$ of the practice session, the survey~(\figref{fig:appendix.userstudy.step3}). Here, they were presented with a question on the visual dissimilarity of the tasks from Step 1 and Step 2, which was to be answered on a $4$-point Likert scale as used in \cite{polozov2015}:
\begin{itemize}
\item Score $1$: indicates that the tasks are visually exactly same
\item Score $2$: indicates that the tasks are visually similar
\item Score $3$: indicates that the tasks are visually slightly different
\item Score $4$: indicates that the tasks are visually very different
\end{itemize}

%\begin{itemize}
%
%\item The first question was regarding the visual dissimilarity of the two 
%task-puzzles presented (i.e visual dissimilarity between $\task_\textnormal{vis}^\textnormal{in}, \task_\textnormal{vis}^\textnormal{out}$ ), where a score of 1 represented exactly the same, and a score of 4 represented very different. Detailed results of participant responses on this particular question, for different tasks and algorithms can be seen in \figref{fig:appendix.userstudy.q2}. 
%
%\item The second question asked participants to rate the similarity of the solution code to the reference task and to the solution code of the new task presented to them.
%
%\item The third and final question  asked participants to rate the similarity of the solution code to the task presented to them and their code to solve the task.
%
%\end{itemize}

\subsection{Results on Visual Task Dissimilarity}

%%%%%%%%%%%%%%%%%%%%%%%%%%%%%%%%%%%%%
\begin{figure}[t!]
\centering
	%%%%%%%%%%%%%%%%% 
	\begin{subfigure}[b]{.194\textwidth}
	\centering
	{
		\includegraphics[width=1\textwidth]{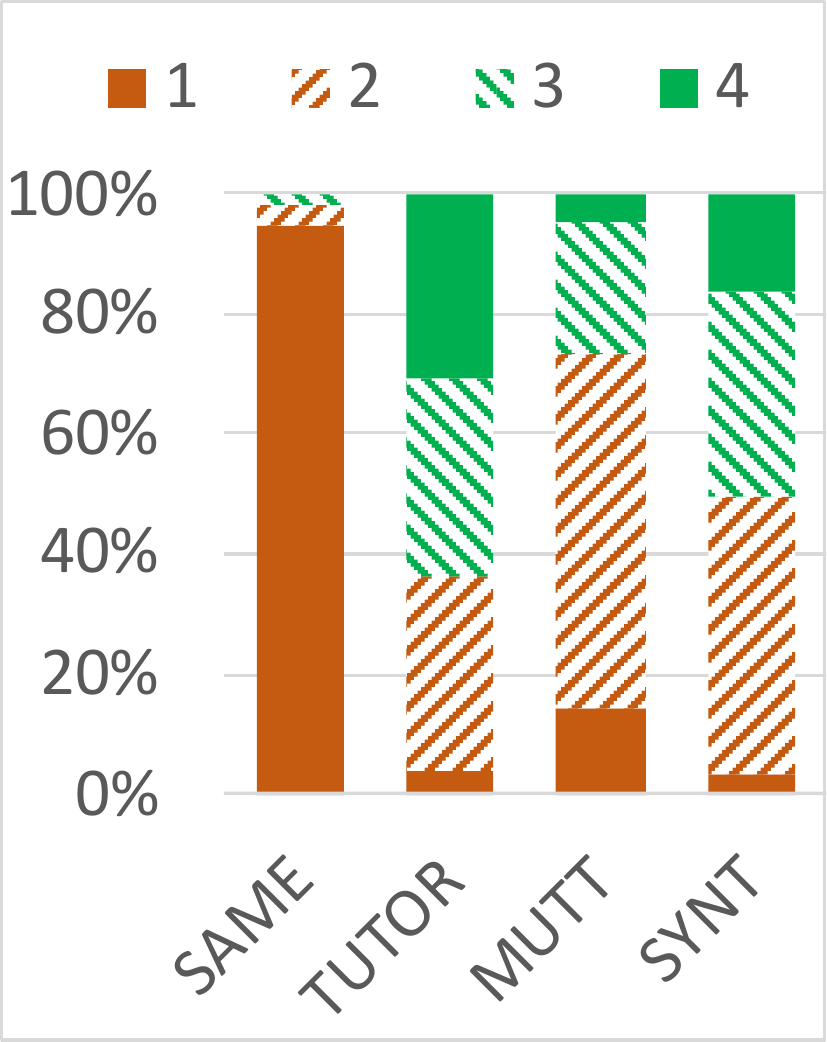}
		%\vspace{0.2mm}
		%\caption{\hocAllBold}
		\caption{\hocAll}
		 \label{fig:appendix.userstudy.q2.all}
	}
	\end{subfigure}
	%%%%%%%%%%%%%%%%%
	\begin{subfigure}[b]{.194\textwidth}
	\centering
	{
		\includegraphics[width=1\textwidth]{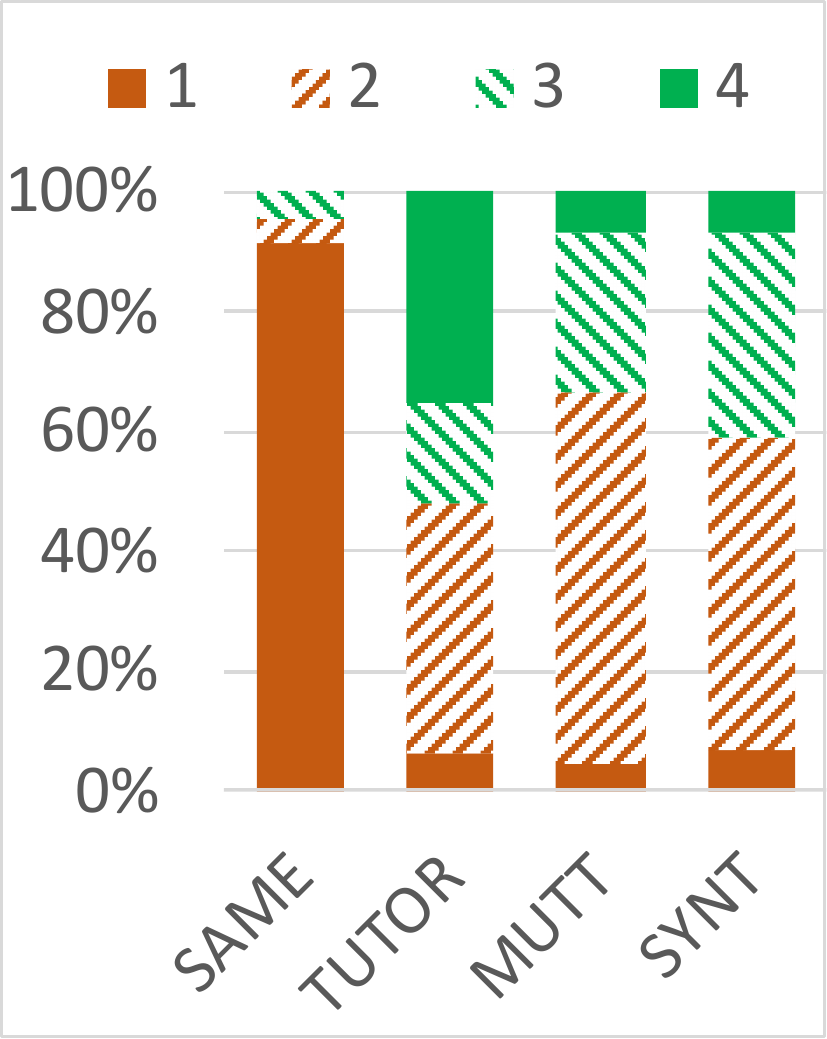}
		%\vspace{0.2mm}
		\caption{\hocC}
		 \label{fig:appendix.userstudy.q2.h2}
	}
	\end{subfigure}
	%%%%%%%%%%%%%%%%%
	\begin{subfigure}[b]{.194\textwidth}
	\centering
	{
		\includegraphics[width=1\textwidth]{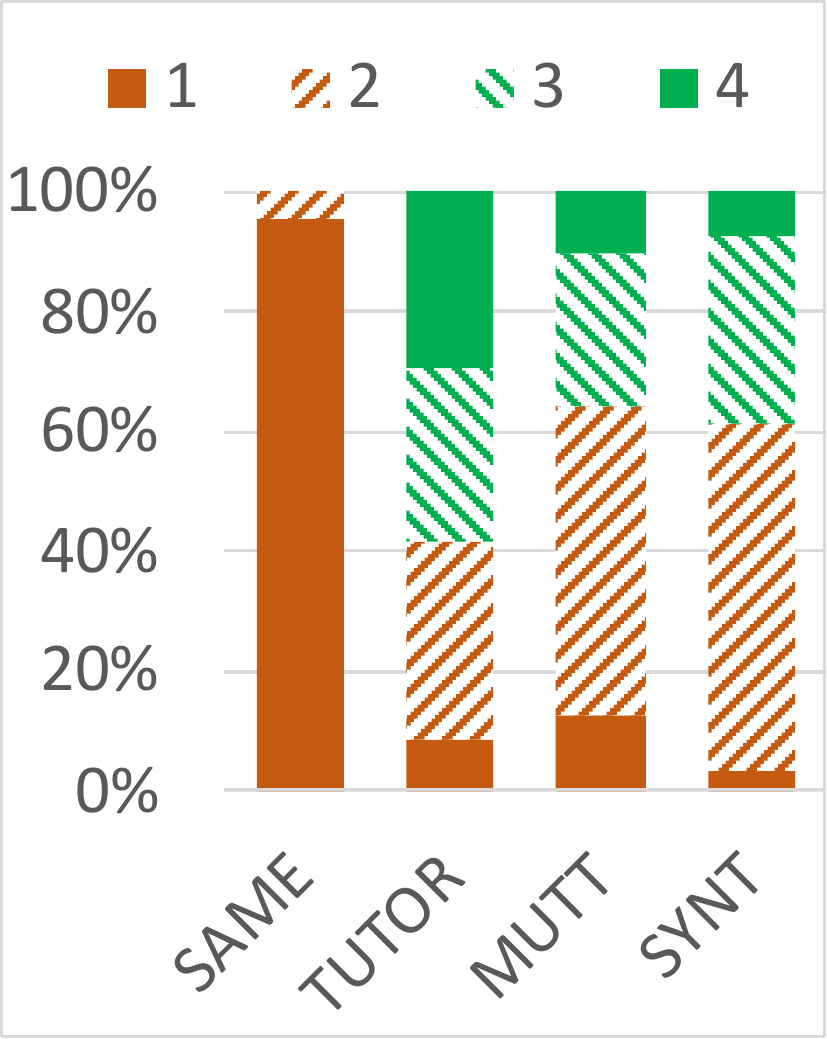}
		%\vspace{0.2mm}
		\caption{\hocF}
		 \label{fig:appendix.userstudy.q2.h4}
	}
	\end{subfigure}
	%%%%%%%%%%%%%%%%%
	\begin{subfigure}[b]{.194\textwidth}
	\centering
	{
		\includegraphics[width=1\textwidth]{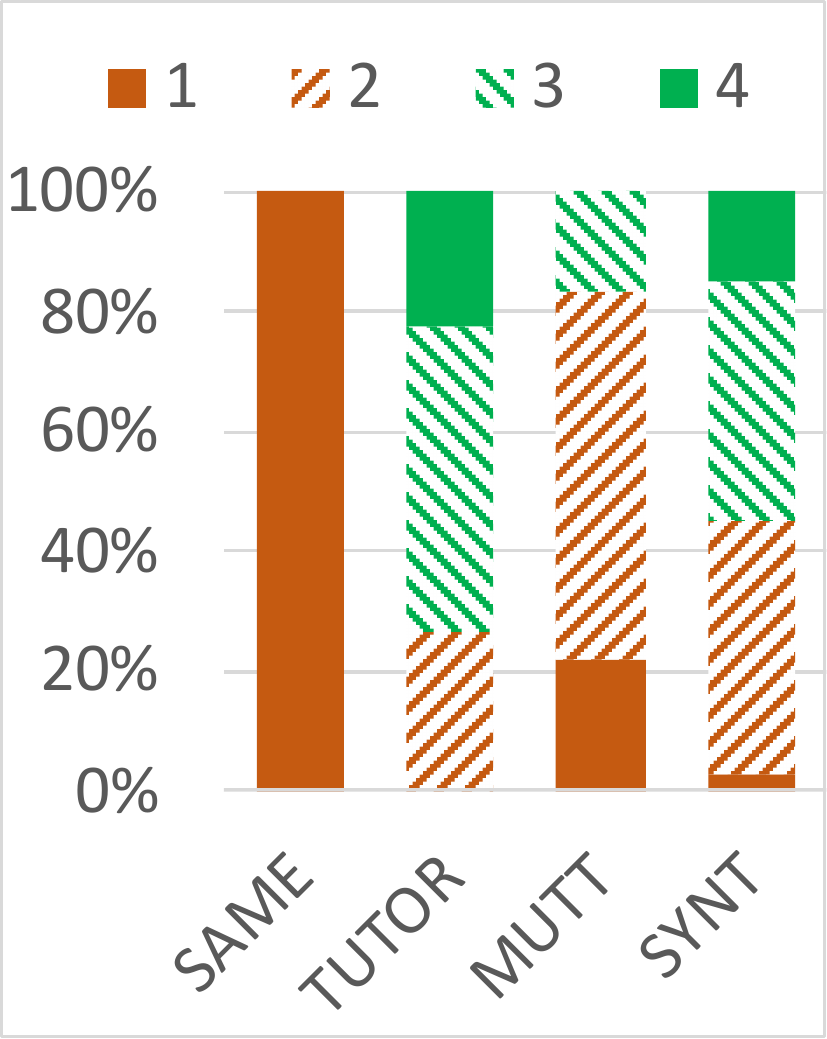}
		%\vspace{0.2mm}
		\caption{\hocG}
		 \label{fig:appendix.userstudy.q2.h5}
	}
	\end{subfigure}		
	%%%%%%%%%%%%%%%%%
	\begin{subfigure}[b]{.194\textwidth}
	\centering
	{
		\includegraphics[width=1\textwidth]{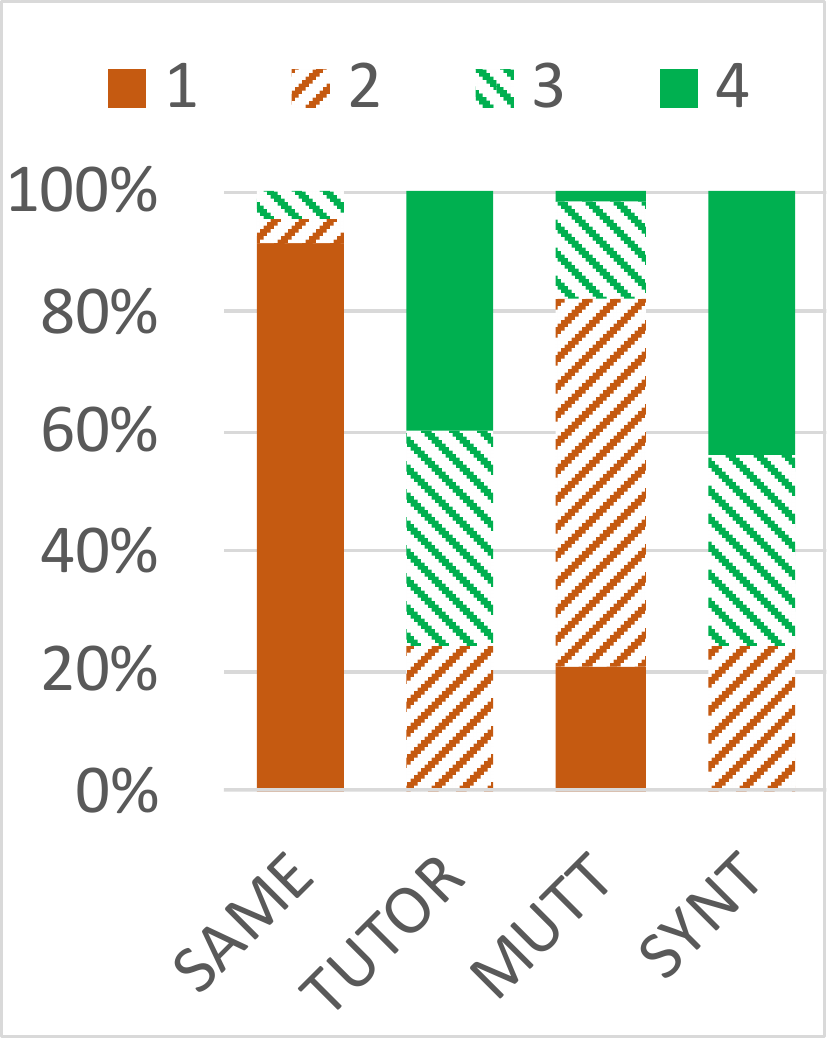}
		%\vspace{0.2mm}
		\caption{\hocH}
		 \label{fig:appendix.userstudy.q2.h6}
	}
	\end{subfigure}
	%%%%%%%%%%%%%%%%%	
	%\vspace{-4mm}
	\caption{Distribution of participant responses on visual dissimilarity of tasks, rated on a 4-point Likert scale (1: exactly same, 2: almost same, 3: slightly different, 4: very different)
	% Participants rated the tasks 
	%\loosness-1 
	}
	%\vspace{-4mm}
	\label{fig:appendix.userstudy.q2}
\end{figure}
%%%%%%%%%%%%%%%%%%%%%%%%%%%%%%%%%%%%%
%%%%%%%%%%%%%%%%%%%%%%%%%%%%%%%%%%%%%%%%%%%%%%%%%%%%%%%%%%

%First paragraph
%\begin{itemize}
%    \item 4 means --- SAME is expected to be 1. Muttask is expected to 1, 2.
%    \item ours vs. muttask SS. (similar to problem solving line)
%    \item ours vs. expert SS. (similar to problem solving line)
%\end{itemize}
%
%Second paragraph (2 tasks)
%\begin{itemize}
%    \item Explaining why Tutor is better.
%    \item Next, we looked at two tasks only. ours vs. muttask SS. ours vs. expert SS.
%    \item ours vs. taskmut SS. (similar to problem solving line)
%    \item ours vs. expert SS. (similar to problem solving line)
%\end{itemize}

%Third paragraph
%\begin{itemize}
%    \item For tasks without conditionals, we can create more powerful mutations
%    \item In general, we can do preprocessing / postprocessing
%    \item Refer to Appendix~\ref{appendix.sec.simulations.more_variability}
%\end{itemize}

The distribution of participant responses (on a 4-point scale) on the visual dissimilarity of the presented tasks is shown in \figref{fig:appendix.userstudy.q2}; the aggregate statistics are provided in \figref{fig:userstudy}.

Comparing the dissimilarity of the generated tasks w.r.t. all the reference tasks \hocAll, we found that the performance of \algoSame~was worst (mean dissimilarity = $1.07$) in comparison to \algoTaskMut~(mean = $2.17$), \algoOursNoPstar~(mean = $2.63$), and \algoTutor~(mean = $2.90$). The poor performance of \algoSame,~with the majority of scores being $1$, is expected. Furthermore, we also see in \figref{fig:appendix.userstudy.q2.all} that \algoTaskMut~has a greater percentage of lower scores of $1$ and $2$, and this is because the baseline directly mutates the task puzzle.
Comparing the performance of \algoOursNoPstar~to the baselines, we find that \algoOursNoPstar~performed significantly better than \algoTaskMut~($\chi^2 = 38.81, p < e^{-7}$). However, its performance was worse than that of \algoTutor~and this difference was significant~($\chi^2 = 12.20, p = 0.0053$).

%Comparing the dissimilarity of the generated tasks w.r.t. the reference task for all tasks denoted as \hocAll, we found that the performance of \algoSame~was worst (mean dissimilarity = $1.07$), while that of \algoTutor~was best (mean = $2.90$). The poor performance of \algoSame,~with the majority of scores being $1$, is expected. Furthermore, we also see in \figref{fig:appendix.userstudy.q2.all} that \algoTaskMut~(mean = $2.17$) has a greater percentage of lower scores of $1$ and $2$, and this is because the baseline directly mutates the task puzzle.
%Comparing the performance of \algoOursNoPstar~(mean = $2.63$) to the baselines, we find that \algoOursNoPstar~performed significantly better than \algoTaskMut ($\chi^2 = 38.81, p < e^{-8}$). However, its performance was slightly worse than that of \algoTutor~and this difference was significant ($\chi^2 = 12.20, p = 0.0053$).
%\algoTaskMut~and \algoTutor, 

When analyzing the differences in the performance of \algoOursNoPstar~w.r.t. \algoTutor, we find that \algoTutor~performs better primarily because of the following two reasons: (i) some of the tasks generated by \algoTutor~have additional distracting paths / noise, and (ii) for simpler tasks without conditionals like \hocC, more powerful code mutations were used. Next, we performed additional analysis by limiting to two complex tasks with nested conditionals, \hocG~and \hocH. On these two tasks, the mean scores of the methods \algoTaskMut, \algoOursNoPstar, and \algoTutor~were $1.97$,  $2.94$, and $3.03$ respectively. Furthermore, we find that \algoOursNoPstar's performance continued to be significantly better than \algoTaskMut~($\chi^2 = 64.33, p < e^{-13}$). But, the difference in the performance of  \algoOursNoPstar~and~\algoTutor~was not statistically significant ($\chi^2 = 2.68, p = 0.44$). 

In general, the performance of our task synthesis algorithm can be further improved by allowing for more powerful code mutations in tasks without conditionals (such as \hocC) and by adding more variability in the output tasks by incorporating techniques discussed in Appendix ~\ref{appendix.sec.simulations.more_variability}.

\section{Code Mutation: Additional Details}\label{appendix.sec.mutation}
This section describes the mutation stage of the task synthesis algorithm. In particular, we describe in detail the constraints applied on sketch \SDSLSketchVar = $\codetosketch(\code)$. Note that we denote an empty action as $\phi$.
%
%\textbf{Remark}: 
Our implementation of the code mutation stage, using the Z3 solver~\cite{deMouraBjorner2008}, is publicly available (see Footnote~\ref{footnote.githubrepo}).%\footnote{\href{https://github.com/adishs/neurips2020_synthesizing-tasks_code}{https://github.com/adishs/neurips2020\_synthesizing-tasks\_code}} 
%will be released together with the final version of the paper.
% of the algorithm

\textbf{Constraint $(\Delta_1)$: \localblockcons.}
\localblockcons~returns the values that all action sequences can take, based on their values in the reference code, $\code^\textnormal{in}$. Consider the action sequence $\actionseq$ in sketch $\text{\SDSLSketchVar}$. The function $\sketchparams(\actionseq|~\code^\textnormal{in})$ returns the value of \actionseq~in $\code^\textnormal{in}$. The types of constraints returned by \localblockcons~are:
\begin{enumerate}
    \item Local \actionseq~constraints. These constraints describe the values that one \actionseq~can take w.r.t.  $\sketchparams(\actionseq|~\code^\textnormal{in})$. It has the following rules:
    \begin{itemize}
    \item \DSLMove~action $\in \sketchparams(\actionseq|~\code^\textnormal{in})$, would imply that the corresponding action in \actionseq~must be \DSLMove.
    
    \item Set of $\text{\DSLTurnLeft}$, $\text{\DSLTurnRight}$~actions $\in \sketchparams(\actionseq|~\code^\textnormal{in})$, would imply that the corresponding actions in \actionseq~will either have all the `turn' actions changed to $\text{\DSLTurnLeft}$, or to $\text{\DSLTurnRight}$, or remain the same, or be flipped, i.e. $\text{\DSLTurnLeft} \rightarrow \text{\DSLTurnRight}$~and $\text{\DSLTurnRight} \rightarrow \text{\DSLTurnLeft}$.
    
    \item Set of $\text{\DSLPickMarker}$, $\text{\DSLPutMarker}$~actions $\in \sketchparams(\actionseq|~\code^\textnormal{in})$, would imply that the corresponding actions in \actionseq~will either have all the `marker' actions changed to $\text{\DSLPickMarker}$, or to $\text{\DSLPutMarker}$, or remain the same, or be flipped i.e. $\text{\DSLPickMarker} \rightarrow \text{\DSLPutMarker}$~and $\text{\DSLPutMarker} \rightarrow \text{\DSLPickMarker}$.
    
    \item Additional actions (up to $\delta_{\textnormal{size}}$) to \actionseq~can either be appended before the existing actions (in $\sketchparams(\actionseq|~\code^\textnormal{in})$) or after, but not both. In our experiments we set $\delta_{\textnormal{size}} = 2$.
    \end{itemize}
    
    \looseness-1These constraints are listed under ``Local \actionseq~constraints'' in \figref{fig:appendix.mutation.hoc.7} and \figref{fig:appendix.mutation.karel.7}.
    %for the example codes 
    
    \item Global \actionseq~constraints. These constraints apply on all the \actionseq's in sketch \SDSLSketchVar. They allow only one of all the \actionseq's to have additional actions added to them. These constraints are listed under ``Global \actionseq~constraints'' in \figref{fig:appendix.mutation.hoc.7} and \figref{fig:appendix.mutation.karel.7}.

\end{enumerate}

\textbf{Constraint $(\Delta_{6})$: Action sequence is minimal.}
These constraints describe the sequences that invalidate minimality of code. The constraints ensure that certain sequences of actions do not occur in \actionseq. The detailed list of sequences for the two example codes described, which invalidate \actionseq~if they occur, are given in \figref{fig:appendix.mutation.hoc.8} and \figref{fig:appendix.mutation.karel.8}.

\vspace{-1.5mm}
\subsection{Details of Code Mutation for HOC}
\vspace{-1mm}
Here, we build on the generic DSL presented in \figref{fig:mutation} for \hocType~codes.
We describe the \hocType-DSL, Sketch DSL, and constraints for \hocType~codes (shown in \figref{fig:appendix.mutation.hoc.1}, \figref{fig:appendix.mutation.hoc.2}, and \figref{fig:appendix.mutation.hoc.3}, respectively). It is to be noted that the \hocType~DSL does not contain `marker' based actions/conditionals: (action) \DSLPickMarker, (action) \DSLPutMarker, (conditional) \DSLBoolMarker, (conditional) \DSLBoolNoMarker. It does not allow few more conditionals: \DSLBoolNoPathA, \DSLBoolNoPathL~and \DSLBoolNoPathR.
It also does not contain the \DSLWhile~construct. We consider a concrete example (continued from the example presented in \figref{fig:mutation.4}), in \figref{fig:appendix.mutation.hoc.4}, illustrate its sketch in \figref{fig:appendix.mutation.hoc.5}, and its constraints in \figref{fig:appendix.mutation.hoc.6}. We elaborate its \actionseq~minimality constraints and \localblockcons~in \figref{fig:appendix.mutation.hoc.8} and \figref{fig:appendix.mutation.hoc.7}, respectively.

\textbf{A brief description of elimination sequences to ensure \actionseq-minimality.}
We identify four sequences in \hocType~codes, which must be removed to ensure code minimality. These are: \DSLTurnLeft, \DSLTurnRight~or \DSLTurnRight, \DSLTurnLeft, which do not lead to any change in the output; \DSLTurnLeft, \DSLTurnLeft, \DSLTurnLeft, which can be replaced by a single \DSLTurnRight; and finally \DSLTurnRight, \DSLTurnRight, \DSLTurnRight, which can be replaced by a single \DSLTurnLeft. It is to be noted that we also eliminate variants of the above sequences, that contain $\phi$, but effectively reduce to these four sequences only, given that $\phi$ denotes an empty action.

\vspace{-1.5mm}
\subsection{Details of Code Mutation for Karel}
\vspace{-1mm}
Here, we present a modified/detailed form of \figref{fig:mutation} for \karelType~codes in particular.
We describe the \karelType-DSL, Sketch DSL, and constraints for \karelType~codes (shown in \figref{fig:appendix.mutation.karel.1}, \figref{fig:appendix.mutation.karel.2}, and \figref{fig:appendix.mutation.karel.3}, respectively). It is to be noted that \karelType-DSL does not contain the \DSLRepeatUntil~construct. We consider a concrete example (same as the \karelType~solution code presented in \figref{fig:intro.karel.p0}), in \figref{fig:appendix.mutation.karel.4}, illustrate its sketch in \figref{fig:appendix.mutation.karel.5}, and its constraints in \figref{fig:appendix.mutation.karel.6}. We elaborate its \actionseq~minimality constraints and \localblockcons~in \figref{fig:appendix.mutation.karel.8} and \figref{fig:appendix.mutation.karel.7}, respectively.

\textbf{A brief description of elimination sequences to ensure \actionseq-minimality.} In addition to the four sequences considered in \hocType~codes, we identify sixteen more sequences in \karelType~codes, which must be removed to ensure code minimality. These are: \DSLPickMarker, \DSLPutMarker~or \DSLPutMarker, \DSLPickMarker~or \DSLPickMarker, \DSLPickMarker~or \DSLPutMarker, \DSLPutMarker, which do not lead to any change in the output or leads to a crash in the \karelType~grid (only one marker is allowed per grid-cell); \DSLTurnLeft, \DSLPickMarker, \DSLTurnRight~or  \DSLTurnRight, \DSLPickMarker, \DSLTurnLeft~or \DSLTurnLeft, \DSLPutMarker, \DSLTurnRight~or  \DSLTurnRight, \DSLPutMarker, \DSLTurnLeft, which bring about the same output without the `turn' actions; \DSLPickMarker, \DSLTurnLeft, \DSLPutMarker~or \DSLPutMarker, \DSLTurnLeft, \DSLPickMarker~or  \DSLPickMarker, \DSLTurnRight, \DSLPutMarker~or \DSLPutMarker, \DSLTurnRight, \DSLPickMarker, which bring about the same output without the `marker' actions; and finally \DSLPickMarker, \DSLTurnLeft, \DSLPickMarker~or \DSLPickMarker, \DSLTurnRight, \DSLPickMarker~or \DSLPutMarker, \DSLTurnLeft, \DSLPutMarker~or \DSLPutMarker, \DSLTurnRight, \DSLPutMarker, which leads to a crash in the \karelType~grid (only one marker is allowed per grid-cell).
It is to be noted that we also eliminate variants of the above sequences, that contain $\phi$, but effectively reduce to these basic sequences only, given that $\phi$ denotes an empty action.

% \subsection{Additional details}
% Additional details on the constraints on all $10$ code files and the generated code mutations can be found in the source code provided in the supplementary material. It contains the following files:
% \begin{itemize}
%     \item \textcode{mutations.py} - This is the main file that, when run, generates all code mutations for all $10$ reference tasks. (See \figref{fig:dataset})
%     \item \textcode{utils\_*.py} - The code files prefixed with `utils\_' contain common utility files that generate the AST structure of the code, generate constraints for any action sequence, and generate assignments from the query passed to the SMT solver.
%     \item \textcode{in\_hoc\_*.py} - The code files prefixed with `in\_hoc\_' contain all the constraints corresponding to that \hocType~reference task and generate its code mutations.
%     \item \textcode{in\_karel\_*.py} - The code files prefixed with `in\_karel\_' contain all the constraints corresponding to that \karelType~reference task and generate its code mutations.
    
% \end{itemize}
% More details are available in the \textcode{README.md} file with the code. 

%%%%%%%%%%%%%%%%%%%%%%%%%%%%%%%%%%%%%%%%%%%%%%%%%%%%%%%%%%
%%%%%%%%%%%%%%%%%%%%%%%%%%%%%%%%%%%%%%%%%%%%%%%%%%%%%%%%%%
\begin{figure}[h!]
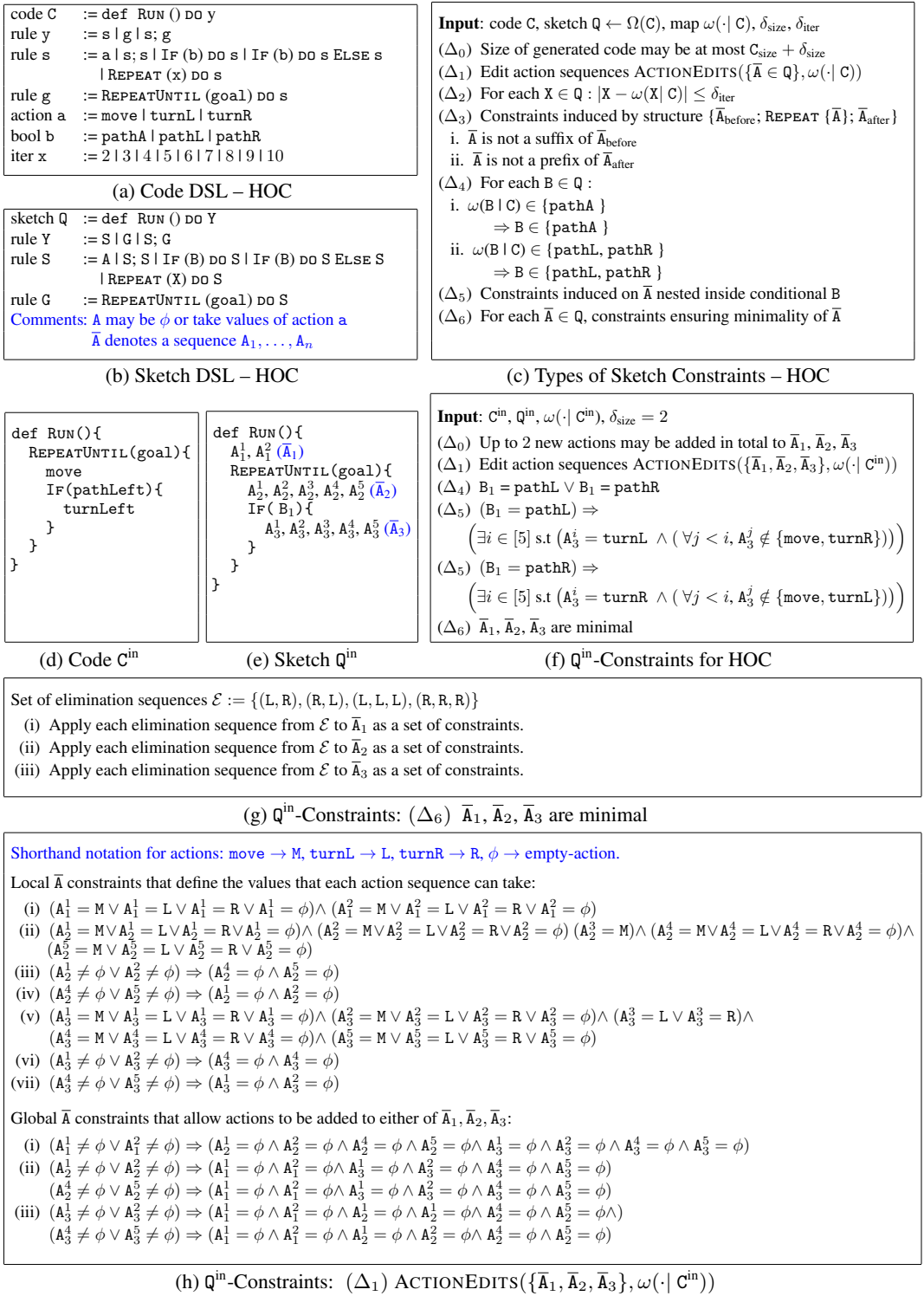

	\centering
	%%%%%%%%%%%%%%%%%%%%%%%%%%%%%%%%%%%%%%%%%%%%%%%%%%%%%%%%%%%%%%%%%%%%
	\begin{minipage}{1\textwidth}
	\begin{minipage}{0.465\textwidth}
			\begin{subfigure}[b]{1.0\textwidth}
			\centering
			{
				\begin{boxcode2col}{1.2cm}{6.7cm}{0.75}{1.08}
						\DSLCode \code &:= \textcode{def }\DSLRun() \DSLdo y \\
						\DSLRule \DSLRuleVar &:= \DSLStmtVar | \DSLRepeatForever  | \DSLStmtVar; \DSLRepeatForever \\
						% %
						\DSLRule \DSLStmtVar \hspace{1mm} &:= \DSLActionVar | $\text{\DSLStmtVar};\text{\DSLStmtVar}$ | \DSLIf(\DSLBoolVar) \DSLdo $\text{\DSLStmtVar}$ | \DSLIf(\DSLBoolVar) \DSLdo $\text{\DSLStmtVar}$ \DSLElse $\text{\DSLStmtVar}$\\
						& \quad | \DSLRepeat(\DSLIterVar) \DSLdo $\text{\DSLStmtVar}$ \\
						\DSLRule \DSLRepeatForever &:= \DSLRepeatUntil(\DSLBoolGoal) \DSLdo $\text{\DSLStmtVar}$\\
						%  %
						\DSLAction \DSLActionVar &:= \DSLMove| \DSLTurnL | \DSLTurnR \\
						%   %
						\DSLBool \DSLBoolVar &:= \DSLBoolPathA | \DSLBoolPathL | \DSLBoolPathR \\
						%    %
						\DSLIter \DSLIterVar &:= $2$ | $3$ | $4$ | $5$ | $6$ | $7$ | $8$ | $9$ | $10$\\
						%\vspace{2mm}
					\end{boxcode2col}
					\vspace{-3mm}
					\caption{Code DSL -- \hocType}
					\label{fig:appendix.mutation.hoc.1}
				}
			\end{subfigure}
			%\\
			%%%%%%%%%%%%%%%%%		
			\begin{subfigure}[b]{1.0\textwidth}
			\centering
			{
				\begin{boxcode2col}{1.2cm}{6.7cm}{0.75}{1.08}
					\SDSLSketch \SDSLSketchVar  &:= \textcode{def }\DSLRun() \DSLdo $\text{\SDSLVarY}$ \\
					% %
					\DSLRule \SDSLVarY & := \SDSLSStmtVar | \SDSLVarG | \SDSLSStmtVar; \SDSLVarG \\
					% %
					\DSLRule \SDSLSStmtVar &:= \SDSLAction | \SDSLSStmtVar; \SDSLSStmtVar |
																	\DSLIf(\SDSLBool) \DSLdo $\text{\SDSLSStmtVar}$ | \DSLIf(\SDSLBool) \DSLdo $\text{\SDSLSStmtVar}$ \DSLElse $\text{\SDSLSStmtVar}$ \\
					& \quad | \DSLRepeat(\SDSLIter) \DSLdo $\text{\SDSLSStmtVar}$ \\
					% %
					\DSLRule \SDSLVarG & := \DSLRepeatUntil(\DSLBoolGoal) \DSLdo $\text{\SDSLSStmtVar}$ \\
					% %
				    \textcolor{blue}{Comments} & \textcolor{blue}{:\ \SDSLAction may be $\phi$ or take values of action \DSLActionVar}\\
				    %\textcolor{blue}{Note 2} 
				    & \textcolor{blue}{\ \ \actionseq~denotes a sequence $\text{\SDSLAction}_{1}, \ldots, \text{\SDSLAction}_{n}$}\\
				\end{boxcode2col}
				\vspace{-3mm}
				\caption{Sketch DSL -- \hocType}
				%\vspace{0.5mm}
				\label{fig:appendix.mutation.hoc.2}
			}
			\end{subfigure}    		
	\end{minipage}
	%\hspace{0.5em}
	%%%%%%%%%%%%%%%%%
	\begin{minipage}{0.55\textwidth}
		\begin{subfigure}[b]{1.0\textwidth}
		\centering
		{
			\begin{boxcode}{9.6cm}{0.75}{1.0}
				\textbf{Input}: code \code, sketch \SDSLSketchVar $\leftarrow$ $\codetosketch(\code)$, map $\sketchparams(\cdot|~\code)$, $\delta_\text{size}$, $\delta_\text{iter}$
				\begin{enumerate}%[\ensuremath{\Delta_{1}}]
					\item[(\ensuremath{\Delta_{0}})] Size of generated code may be at most $\code_{\textnormal{size}} + \delta_\text{size}$ 
					\item[(\ensuremath{\Delta_1})] Edit action sequences $\localblockcons(\{\actionseq \in \codesketch\},  \sketchparams(\cdot|~\code))$ 
					%
					%\item[(\ensuremath{\Delta_{1}})] $\text{For each }\text{\actionseq} \in \text{\SDSLSketchVar}$, apply \localblockcons(\actionseq, $\sketchparams(\actionseq| ~\code)$)
					%
					\item[(\ensuremath{\Delta_{2}})]  For each \SDSLIter~$ \in \text{\SDSLSketchVar}: |\text{\SDSLIter} - \sketchparams(\text{\SDSLIter}|~\code)| \leq \delta_{\textnormal{iter}}$
					\item[(\ensuremath{\Delta_{3}})] Constraints induced by structure \{\text{\actionseq}\textsubscript{before};  \DSLRepeat\{\actionseq\};  \text{\actionseq}\textsubscript{after}\}
					\begin{enumerate}[\leftmargin=0em]
						\item[i.] \actionseq~is not a suffix of \actionseq\textsubscript{before}
						\item[ii.] \actionseq~is not a prefix of \actionseq\textsubscript{after}
					\end{enumerate}
					\item[(\ensuremath{\Delta_{4}})]  For each \SDSLBool~$\in \text{\SDSLSketchVar}:$
					\begin{enumerate}[\leftmargin=0em]
						\item[i.] \sketchparams(\text{\SDSLBool}~|~\code) $\in$ \{\DSLBoolPathA \}
							
						\item[] \qquad $\Rightarrow$ \SDSLBool $\in$ \{\DSLBoolPathA \}							
						
						\item[ii.] \sketchparams(\text{\SDSLBool}~|~\code) $\in$ \{\DSLBoolPathL,
							 \text{\DSLBoolPathR } \} 
						\item[] 
							\quad \quad $\Rightarrow$ \SDSLBool $\in$ \{\DSLBoolPathL, \DSLBoolPathR \}
					\end{enumerate}
					\item[(\ensuremath{\Delta_{5}})] Constraints induced on \actionseq~nested inside conditional \SDSLBool
					\item[(\ensuremath{\Delta_{6}})] For each $\text{\actionseq} \in \text{\SDSLSketchVar}$, constraints ensuring minimality of \actionseq 
					\vspace{-0.3em}
				\end{enumerate}
				\\
			\end{boxcode}
			\vspace{-3mm}
			\caption{Types of Sketch Constraints -- \hocType}
			\label{fig:appendix.mutation.hoc.3}
		}
		\end{subfigure}
	\end{minipage}
	\end{minipage}
	%%%%%%%%%%%%%%%%%%%%%%%%%%%%%%%%%%%%%%%%%%%%%%%%%%%%%%%%%%%%%%%%%%%%
	%%%%%%%%%%%%%%%%%%%%%%%%%%%%%%%%%%%%%%%%%%%%%%%%%%%%%%%%%%%%%%%%%%%%
	%\begin{minipage}{1\textwidth}
    \centering
		%%%%%%%%%%%%%%%%%
		\begin{subfigure}[b]{0.215\textwidth}
		\centering
		{
			\begin{boxcode}{3.68cm}{0.75}{1.0}
				\textcode{def }\DSLRun\textcode{()\{}\\
				\quad \DSLRepeatUntil\textcode{(}\DSLBoolGoal\textcode{)\{}\\
				\quad \quad \DSLMove\\
				\quad \quad \DSLIf\textcode{(}\DSLBoolPathLeft\textcode{)\{}\\
				\quad \quad \quad \DSLTurnLeft\\
				\quad \quad \textcode{\}}\\
				\quad \textcode{\}}\\
				\textcode{\}}
				\\
				\\
				\\
				\vspace{2mm}
				%\vspace{-1.5mm}
			\end{boxcode}
			\vspace{-3mm}
			\caption{Code \code\textsuperscript{in}}
			\label{fig:appendix.mutation.hoc.4}
			}
		\end{subfigure}
		%%%%%%%%%%%%%%%%%
		\begin{subfigure}[b]{.24\textwidth}
		\centering
		{
			\begin{boxcode}{4.05cm}{0.75}{1.0}
				\textcode{def }\DSLRun\textcode{()\{}\\
				\quad $\text{\SDSLAction}^{1}_{1}$, $\text{\SDSLAction}^{2}_{1}$ \textcolor{blue}{(\actionseqb\textsubscript{1})}\\
				\quad \DSLRepeatUntil\textcode{(}\DSLBoolGoal\textcode{)\{}\\
				\quad \quad $\text{\SDSLAction}^{1}_{2}$, $\text{\SDSLAction}^{2}_{2}$, $\text{\SDSLAction}^{3}_{2}$, 
				$\text{\SDSLAction}^{4}_{2}$, 
				$\text{\SDSLAction}^{5}_{2}$ \textcolor{blue}{(\actionseqb\textsubscript{2})}\\
				\quad \quad \DSLIf\textcode{(}$\text{\SDSLBool}_{1}$\textcode{)\{}\\
				\quad \quad \quad $\text{\SDSLAction}^{1}_{3}$, $\text{\SDSLAction}^{2}_{3}$, $\text{\SDSLAction}^{3}_{3}$, $\text{\SDSLAction}^{4}_{3}$, $\text{\SDSLAction}^{5}_{3}$ \textcolor{blue}{(\actionseqb\textsubscript{3})} \\
				\quad \quad \textcode{\}}\\
				\quad \textcode{\}}\\
				\textcode{\}}
				\\
				\\
				\vspace{1.75mm}
			\end{boxcode}
			\vspace{-3mm}
			\caption{Sketch $\text{\codesketch}^\text{in}$}
			\label{fig:appendix.mutation.hoc.5}
			}
		\end{subfigure}
		%%%%%%%%%%%%%%%%%	
			\begin{subfigure}[b]{0.53\textwidth}
			\centering
			{
				\begin{boxcode}{9.6cm}{0.75}{1.0}
					\textbf{Input}: \code\textsuperscript{in}, $\text{\codesketch}^\text{in}$, \sketchparams$(\cdot|~\text{\code}^\text{in}$), $\delta_\text{size} = 2$
					% = ($\phi$ , $\text{\DSLMove}$,  $\text{\DSLTurnL}$, \DSLBoolPathL), 									
					\begin{enumerate}
					    \item[(\ensuremath{\Delta_0})] Up to $2$ new actions may be added in total to $\text{\actionseq}_{1}$, $\actionseq_{2}$, $\actionseq_{3}$
						\item[(\ensuremath{\Delta_1})] Edit action sequences $\localblockcons(\{\actionseq_{1}, \actionseq_{2}, \actionseq_{3}\},  \sketchparams(\cdot|~\code^\text{in}))$ 
					    \item[(\ensuremath{\Delta_4})] $\text{\SDSLBool}_{1}$ = \DSLBoolPathL $\lor$ $\text{\SDSLBool}_{1}$ = \DSLBoolPathR	
					    \item[(\ensuremath{\Delta_5})] $(\text{\SDSLBool}_{1} = \text{\DSLBoolPathL}) \Rightarrow$
					    \item[] \quad $\Big{(}\exists i \in [5]$ s.t $\big(\text{\SDSLAction}^{i}_{3} =  \text{\DSLTurnL} \ \ \land$  $(\text{ } \forall j < i \text{, } \text{\SDSLAction}^{j}_{3} \notin \{\text{\DSLMove}, \text{\DSLTurnR} \})\big) \Big{)}$
						%Constraints if $\text{\actionseq}_{3}$ is nested in \DSLBool $\text{\SDSLBool}_{1} = \text{\DSLBoolPathL}$:
						%
						\item[(\ensuremath{\Delta_5})] $(\text{\SDSLBool}_{1} = \text{\DSLBoolPathR}) \Rightarrow$
						\item[] \quad $\Big{(} \exists i \in [5]$ s.t $\big(\text{\SDSLAction}^{i}_{3} =  \text{\DSLTurnR} \ \ \land$  $(\text{ } \forall j<i \text{, }\text{\SDSLAction}^{j}_{3} \notin \{\text{\DSLMove}, \text{\DSLTurnL} \})\big) \Big{)}$
						%\item[(\ensuremath{\Delta_7})] Only one of $\actionseq_{1}, \actionseq_{2}, \actionseq_{3}$ have additional actions w.r.t. $\sketchparams(\actionseq~|~\code^\text{in})$
						%added to them.
						%
					    %\item[(\ensuremath{\Delta_6})] For $\actionseq \in \{\actionseq_{1}, \actionseq_{2}, \actionseq_{3}\}$, apply $\localblockcons(\actionseq,  \sketchparams(\actionseq~|~\code^\text{in}))$
						%\item[(\ensuremath{\Delta_7})] Only one of $\actionseq_{1}, \actionseq_{2}, \actionseq_{3}$ have additional actions w.r.t. $\sketchparams(\actionseq~|~\code^\text{in})$
						%added to them.
						%
						\item[(\ensuremath{\Delta_6})] $\text{\actionseq}_{1}$, $\actionseq_{2}$, $\actionseq_{3}$ are minimal
				\vspace{-3mm}
				\end{enumerate}
			\end{boxcode}
			\vspace{-3mm}
			\caption{$\text{\codesketch}^\text{in}$-Constraints for \hocType}
			%\caption{Constraints for Sketch $\text{\codesketch}^\text{in}$}
			\label{fig:appendix.mutation.hoc.6}
		}
		\end{subfigure}		
	%\hspace{0.5em}
	%%%%%%%%%%%%%%%%%
	%\end{minipage}
	%%%%%%%%%%%%%%%%%%%%%%%%%%%%%%%%%%%%%%%%%%%%%%%%%%%%%%%%%%%%%%%%%%%%
% \\
    %%%%%%%%%%%%%%
    %\hspace{0.5em}
    \begin{subfigure}[b]{1\textwidth}
    \centering
			{
				\begin{boxcode}{18.5cm}{0.75}{1.0}
				    Set of elimination sequences $\mathcal{E} := \{(\tl,\tr), (\tr,\tl), (\tl, \tl, \tl), (\tr, \tr, \tr)\}$
				     \begin{enumerate}[(i)]
    				    \item Apply each elimination sequence from $\mathcal{E}$ to $\actionseq_{1}$ as a set of constraints.
    				    \item Apply each elimination sequence from $\mathcal{E}$ to $\actionseq_{2}$ as a set of constraints.
    				    \item Apply each elimination sequence from $\mathcal{E}$ to $\actionseq_{3}$ as a set of constraints.			    
				    \end{enumerate}
			\vspace{-2mm}
			\end{boxcode}
			\vspace{-3mm}
			\caption{$\text{\codesketch}^\text{in}$-Constraints: $(\Delta_{6})~\text{ \actionseq}_{1}$, $\actionseq_{2}$, $\actionseq_{3}$ are minimal}
			%\caption{Constraints for Sketch $\text{\codesketch}^\text{in}$}
			\label{fig:appendix.mutation.hoc.8}
		}
    \end{subfigure}
	%\vspace{-4mm}
	%%%%%%%%%%%%%%%%
    \begin{subfigure}[b]{1\textwidth}
    \centering
			{
				\begin{boxcode}{18.5cm}{0.75}{1.0}
    				%For brevity, denote actions
					\textcolor{blue}{Shorthand notation for actions: $\text{\DSLMove} \rightarrow \sm$, $\text{\DSLTurnL} \rightarrow \tl$, $\text{\DSLTurnR} \rightarrow \tr$, $\phi \rightarrow \text{empty-action}$.}
					% Constraints are joined by '$\wedge$' operator.
				   \vspace{2mm}
				   \\
				    Local \actionseq~constraints that define the values that each action sequence can take:
				    %\vspace{-0.5mm}
                    \begin{enumerate}[(i)]
					\item $(\ha{1}{1} =\sm \vee \ha{1}{1} = \tl
					\vee \ha{1}{1} = \tr \vee \ha{1}{1} = \phi) \wedge $ $(\ha{1}{2} =\sm \vee \ha{1}{2} = \tl \vee \ha{1}{2} = \tr \vee \ha{1}{2} = \phi) $
		    		%%%%%%%%% end of A1--values
		    		
					\item $(\ha{2}{1} =\sm \vee \ha{2}{1} = \tl \vee \ha{2}{1} = \tr \vee \ha{2}{1} = \phi) \wedge $ $(\ha{2}{2} =\sm \vee \ha{2}{2} = \tl \vee \ha{2}{2} = \tr \vee \ha{2}{2} = \phi) $
					$(\ha{2}{3} =\sm) \wedge $ $(\ha{2}{4} =\sm \vee \ha{2}{4} = \tl \vee \ha{2}{4} = \tr \vee \ha{2}{4} = \phi) \wedge $ $(\ha{2}{5} =\sm \vee \ha{2}{5} = \tl \vee \ha{2}{5} = \tr \vee \ha{2}{5} = \phi) $
					
					\item $(\ha{2}{1} \neq \phi \vee \ha{2}{2} \neq \phi) \Rightarrow (\ha{2}{4} = \phi \wedge \ha{2}{5} = \phi)$
					
					\item $(\ha{2}{4} \neq \phi \vee \ha{2}{5} \neq \phi) \Rightarrow (\ha{2}{1} = \phi \wedge \ha{2}{2} = \phi)$
					%%%%%%%%% end of A2--values
					
					\item $(\ha{3}{1} =\sm \vee \ha{3}{1} = \tl \vee \ha{3}{1} = \tr \vee \ha{3}{1} = \phi) \wedge $ $(\ha{3}{2} =\sm \vee \ha{3}{2} = \tl \vee \ha{3}{2} = \tr \vee \ha{3}{2} = \phi) \wedge $ $(\ha{3}{3} = \tl \vee \ha{3}{3} = \tr) \wedge $
				    \item[] $(\ha{3}{4} =\sm \vee \ha{3}{4} = \tl \vee \ha{3}{4} = \tr \vee \ha{3}{4} = \phi) \wedge $ 
				    $(\ha{3}{5} =\sm \vee \ha{3}{5} = \tl \vee \ha{3}{5} = \tr \vee \ha{3}{5} = \phi) $
				    
				    \item $(\ha{3}{1} \neq \phi \vee \ha{3}{2} \neq \phi) \Rightarrow (\ha{3}{4} = \phi \wedge \ha{3}{4} = \phi)$
				    
				    \item $(\ha{3}{4} \neq \phi \vee \ha{3}{5} \neq \phi) \Rightarrow (\ha{3}{1} = \phi \wedge \ha{3}{2} = \phi)$
				    %%%%%%%%% end of A3--values
				    \end{enumerate}
				    \vspace{2mm}
				     Global \actionseq~constraints that allow actions to be added to either of $\actionseq_{1}, \actionseq_{2}, \actionseq_{3}$:
				    
				    %\vspace{-0.5mm}
				    \begin{enumerate}[(i)]

				    \item $(\ha{1}{1} \neq \phi \vee \ha{1}{2} \neq \phi) \Rightarrow (\ha{2}{1} = \phi \wedge \ha{2}{2} = \phi \wedge \ha{2}{4} = \phi \wedge \ha{2}{5} = \phi \wedge$ $\ha{3}{1} = \phi \wedge \ha{3}{2} = \phi \wedge \ha{3}{4} = \phi \wedge \ha{3}{5} = \phi)$
				    %%%%%%%%% end of A1---phi-cons
				    
				    \item $(\ha{2}{1} \neq \phi \vee \ha{2}{2} \neq \phi) \Rightarrow (\ha{1}{1} = \phi \wedge \ha{1}{2} = \phi \wedge$ $\ha{3}{1} = \phi \wedge \ha{3}{2} = \phi \wedge \ha{3}{4} = \phi \wedge \ha{3}{5} = \phi)$
				    \item[] $(\ha{2}{4} \neq \phi \vee \ha{2}{5} \neq \phi) \Rightarrow (\ha{1}{1} = \phi \wedge \ha{1}{2} = \phi \wedge$
			        $\ha{3}{1} = \phi \wedge \ha{3}{2} = \phi \wedge \ha{3}{4} = \phi \wedge \ha{3}{5} = \phi)$
				     %%%%%%%%% end of A2---phi-cons
				    
				    \item $(\ha{3}{1} \neq \phi \vee \ha{3}{2} \neq \phi) \Rightarrow (\ha{1}{1} = \phi \wedge \ha{1}{2} = \phi \wedge \ha{2}{1} = \phi \wedge \ha{2}{1} = \phi \wedge$ $\ha{2}{4} = \phi \wedge \ha{2}{5} = \phi \wedge )$
				    \item[]$(\ha{3}{4} \neq \phi \vee \ha{3}{5} \neq \phi) \Rightarrow (\ha{1}{1} = \phi \wedge \ha{1}{2} = \phi \wedge \ha{2}{1} = \phi \wedge \ha{2}{2} = \phi \wedge$ $\ha{2}{4} = \phi \wedge \ha{2}{5} = \phi)$
				    %%%%%%%%% end of A3---phi-cons
				    \end{enumerate}
				  
			\end{boxcode}
			\vspace{-3mm}
			\caption{$\text{\codesketch}^\text{in}$-Constraints: ~$(\Delta_{1})~\localblockcons(\{\actionseq_{1}, \actionseq_{2}, \actionseq_{3}\},  \sketchparams(\cdot|~\code^\text{in}))$}
			%\caption{Constraints for Sketch $\text{\codesketch}^\text{in}$}
			\label{fig:appendix.mutation.hoc.7}
		}
    \end{subfigure}
    \caption{Illustration of Code Mutation for \hocType~on the solution code for task \emph{Maze 16} from the \emph{Hour of Code: Classic Maze} challenge by \emph{Code.org}~\cite{hourofcode_maze}; We build on \figref{fig:mutation} here.
	}
	%; see text for details. 
	\label{fig:appendix.mutation.hoc}
	\vspace{-5mm}	
\end{figure}
%%%%%%%%%%%%%%%%%%%%%%%%%%%%%%%%%%%%%%%%%%%%%%%%%%%%%%%%%%
%%%%%%%%%%%%%%%%%%%%%%%%%%%%%%%%%%%%%%%%%%%%%%%%%%%%%%%%%%

%%%%%%%%%%%%%%%%%%%%%%%%%%%%%%%%%%%%%%%%%%%%%%%%%%%%%%%%%%
%%%%%%%%%%%%%%%%%%%%%%%%%%%%%%%%%%%%%%%%%%%%%%%%%%%%%%%%%%
\begin{figure}[h!]
	\centering
	%%%%%%%%%%%%%%%%%%%%%%%%%%%%%%%%%%%%%%%%%%%%%%%%%%%%%%%%%%%%%%%%%%%%
	\begin{minipage}{1\textwidth}
	\begin{minipage}{0.465\textwidth}
			\begin{subfigure}[b]{1.0\textwidth}
			\centering
			{
				\begin{boxcode2col}{1.2cm}{6.7cm}{0.75}{1.15}
						\DSLCode \code &:= \textcode{def }\DSLRun() \DSLdo~\DSLStmtVar \\
						% %
						\DSLRule \DSLStmtVar \hspace{1mm} &:= \DSLActionVar | $\text{\DSLStmtVar};\text{\DSLStmtVar}$ | \DSLIf(\DSLBoolVar) \DSLdo $\text{\DSLStmtVar}$ | \DSLIf(\DSLBoolVar) \DSLdo $\text{\DSLStmtVar}$ \DSLElse $\text{\DSLStmtVar}$\\
						& \quad | \DSLWhile (\DSLBoolVar) \DSLdo $\text{\DSLStmtVar}$ | \DSLRepeat(\DSLIterVar) \DSLdo $\text{\DSLStmtVar}$ \\
						%  %
						\DSLAction \DSLActionVar &:= \DSLMove| \DSLTurnL | \DSLTurnR|  \DSLPutM | \DSLPickM \\
						%   %
						\DSLBool \DSLBoolVar &:= \DSLBoolPathA | \DSLBoolNoPathA | \DSLBoolPathL | \DSLBoolNoPathL \\
						& \quad | \DSLBoolPathR | \DSLBoolNoPathR  | \DSLBoolMarker  | \DSLBoolNoMarker \\
						%    %
						\DSLIter \DSLIterVar &:= $2$ | $3$ | $4$ | $5$ | $6$ | $7$ | $8$ | $9$ | $10$\\
						%\vspace{2mm}
					\end{boxcode2col}
					\vspace{-3mm}
					\caption{Code DSL -- Karel}
					\label{fig:appendix.mutation.karel.1}
				}
			\end{subfigure}
			\\
			\\
			\\
			%%%%%%%%%%%%%%%%%		
			\begin{subfigure}[b]{1.0\textwidth}
			\centering
			{
				\begin{boxcode2col}{1.2cm}{6.7cm}{0.75}{1.15}
					\SDSLSketch \SDSLSketchVar  &:= \textcode{def }\DSLRun() \DSLdo $\text{\SDSLVarY}$ \\
					% %
					\DSLRule \SDSLVarY & := \SDSLSStmtVar \\
					% %
					\DSLRule \SDSLSStmtVar &:= \SDSLAction | \SDSLSStmtVar; \SDSLSStmtVar |
					\DSLIf(\SDSLBool) \DSLdo $\text{\SDSLSStmtVar}$ | \DSLIf(\SDSLBool) \DSLdo $\text{\SDSLSStmtVar}$ \DSLElse $\text{\SDSLSStmtVar}$ \\
					& \quad | \DSLWhile(\SDSLBool) \DSLdo $\text{\SDSLSStmtVar}$  | \DSLRepeat(\SDSLIter) \DSLdo $\text{\SDSLSStmtVar}$ \\
					% %
				    \textcolor{blue}{Comments} & \textcolor{blue}{:\ \SDSLAction may be $\phi$ or take values of action \DSLActionVar}\\
				    %\textcolor{blue}{Note 2} 
				    & \textcolor{blue}{\ \ \actionseq~denotes a sequence $\text{\SDSLAction}_{1}, \ldots, \text{\SDSLAction}_{n}$}\\
				\end{boxcode2col}
				\vspace{-3mm}
				\caption{Sketch DSL -- Karel}
				%\vspace{0.5mm}
				\label{fig:appendix.mutation.karel.2}
			}
			\end{subfigure}    		
	\end{minipage}
	%\hspace{0.5em}
	%%%%%%%%%%%%%%%%%
	\begin{minipage}{0.55\textwidth}
		\begin{subfigure}[b]{1.0\textwidth}
		\centering
		{
			\begin{boxcode}{9.6cm}{0.75}{1.0}
				\textbf{Input}: code \code, sketch \SDSLSketchVar $\leftarrow$ $\codetosketch(\code)$, map $\sketchparams(\cdot|~\code)$, $\delta_\text{size}$, $\delta_\text{iter}$
				\begin{enumerate}%[\ensuremath{\Delta_{1}}]
					\item[(\ensuremath{\Delta_{0}})] Size of generated code may be at most $\code_{\textnormal{size}} + \delta_\text{size}$ 
					\item[(\ensuremath{\Delta_1})] Edit action sequences $\localblockcons(\{\actionseq \in \codesketch\},  \sketchparams(\cdot|~\code))$ 
					%
					%\item[(\ensuremath{\Delta_{1}})] $\text{For each }\text{\actionseq} \in \text{\SDSLSketchVar}$, apply \localblockcons(\actionseq, $\sketchparams(\actionseq| ~\code)$)
					%
					\item[(\ensuremath{\Delta_{2}})]  For each \SDSLIter~$ \in \text{\SDSLSketchVar}: |\text{\SDSLIter} - \sketchparams(\text{\SDSLIter}|~\code)| \leq \delta_{\textnormal{iter}}$
					\item[(\ensuremath{\Delta_{3}})] Constraints induced by structure \{\text{\actionseq}\textsubscript{before};  \DSLRepeat\{\actionseq\};  \text{\actionseq}\textsubscript{after}\}
					\begin{enumerate}[\leftmargin=0em]
						\item[i.] \actionseq~is not a suffix of \actionseq\textsubscript{before}
						\item[ii.] \actionseq~is not a prefix of \actionseq\textsubscript{after}
					\end{enumerate}
					\item[(\ensuremath{\Delta_{4}})]  For each \SDSLBool~$\in \text{\SDSLSketchVar}:$
					\begin{enumerate}[\leftmargin=0em]
						\item[i.] \ \ \sketchparams(\text{\SDSLBool}~|~\code) $\in$ \{\DSLBoolPathA, \DSLBoolNoPathA\}
							
						\item[] \qquad \ $\Rightarrow$ \SDSLBool $\in$ \{\DSLBoolPathA, \DSLBoolNoPathA\}							
						
						\item[ii.] \ \sketchparams(\text{\SDSLBool}~|~\code) $\in$ \{\DSLBoolPathL, \DSLBoolNoPathL 
							 \text{\DSLBoolPathR }, \DSLBoolNoPathR\} 
						\item[] 
							\quad \quad \ $\Rightarrow$ \SDSLBool $\in$ \{\DSLBoolPathL, \DSLBoolNoPathL, \DSLBoolPathR, \DSLBoolNoPathR\}
						\item[iii.] \sketchparams(\text{\SDSLBool}~|~\code) $\in$ \{\DSLBoolMarker, \DSLBoolNoMarker\} 
						\item[] \quad \quad \hspace{0.1em} $\Rightarrow$ \SDSLBool $\in$ \{ \DSLBoolMarker,\DSLBoolNoMarker\}
					\end{enumerate}
					\item[(\ensuremath{\Delta_{5}})] Constraints induced on \actionseq~nested inside conditional \SDSLBool
					\item[(\ensuremath{\Delta_{6}})] For each $\text{\actionseq} \in \text{\SDSLSketchVar}$, constraints ensuring minimality of \actionseq 
					\vspace{-0.3em}
				\end{enumerate}
			\end{boxcode}
			\vspace{-3mm}
			\caption{Types of Sketch Constraints -- Karel}
			\label{fig:appendix.mutation.karel.3}
		}
		\end{subfigure}
	\end{minipage}
	\end{minipage}
	%%%%%%%%%%%%%%%%%%%%%%%%%%%%%%%%%%%%%%%%%%%%%%%%%%%%%%%%%%%%%%%%%%%%
	%%%%%%%%%%%%%%%%%%%%%%%%%%%%%%%%%%%%%%%%%%%%%%%%%%%%%%%%%%%%%%%%%%%%
	%\begin{minipage}{1\textwidth}
    \centering
		%%%%%%%%%%%%%%%%%
		\begin{subfigure}[b]{0.215\textwidth}
		\centering
		{
			\begin{boxcode}{3.68cm}{0.75}{1.0}
				\textcode{def }\DSLRun\textcode{()\{}\\
			\quad \DSLPutMarker\\
			\quad \DSLWhile\textcode{(}\DSLBoolPathAhead\textcode{)\{}\\
			\quad \quad \DSLMove\\
			\quad \quad \DSLTurnLeft\\
			\quad \quad \DSLMove\\
			\quad \quad \DSLTurnRight\\
			\quad \quad \DSLPutMarker\\	
			\quad \textcode{\}}\\
			\textcode{\}}
			\\
			\\
			\vspace{2.5mm}
			\end{boxcode}
			\vspace{-3mm}
			\caption{Code \code\textsuperscript{in}}
			\label{fig:appendix.mutation.karel.4}
			}
		\end{subfigure}
		%%%%%%%%%%%%%%%%%
		\begin{subfigure}[b]{.24\textwidth}
		\centering
		{
			\begin{boxcode}{4.05cm}{0.75}{1.0}
				\textcode{def }\DSLRun\textcode{()\{}\\
				\quad $\text{\SDSLAction}^{1}_{1}$, $\text{\SDSLAction}^{2}_{1}$, $\text{\SDSLAction}^{3}_{1}$, $\text{\SDSLAction}^{4}_{1}$,
				$\text{\SDSLAction}^{5}_{1}$,
				\textcolor{blue}{(\actionseqb\textsubscript{1})}\\
				\quad \DSLWhile\textcode{(}$\text{\SDSLBool}_{1}$\textcode{)\{}\\
				\quad \quad $\text{\SDSLAction}^{1}_{2}$, $\text{\SDSLAction}^{2}_{2}$,\\ \quad \quad $\text{\SDSLAction}^{3}_{2}$, 
				$\text{\SDSLAction}^{4}_{2}$, 
				$\text{\SDSLAction}^{5}_{2}$,\\ \quad \quad
				$\text{\SDSLAction}^{6}_{2}$, 
				$\text{\SDSLAction}^{7}_{2}$,
				$\text{\SDSLAction}^{8}_{2}$, \quad \quad \textcolor{blue}{(\actionseqb\textsubscript{2})}
				\\ 
				\quad \quad
				$\text{\SDSLAction}^{9}_{2}$, 
				$\text{\SDSLAction}^{10}_{2}$, 
				$\text{\SDSLAction}^{11}_{2}$,\\ \quad \quad
				$\text{\SDSLAction}^{12}_{2}$, 
				$\text{\SDSLAction}^{13}_{2}$ \\
				\quad \textcode{\}}\\
				\quad $\text{\SDSLAction}^{1}_{3}$, $\text{\SDSLAction}^{2}_{3}$ \textcolor{blue}{(\actionseqb\textsubscript{3})}\\
				\textcode{\}}
				\\
				\vspace{1.5mm}
			\end{boxcode}
			\vspace{-3mm}
			\caption{Sketch $\text{\codesketch}^\text{in}$}
			\label{fig:appendix.mutation.karel.5}
			}
		\end{subfigure}
		%%%%%%%%%%%%%%%%%	
			\begin{subfigure}[b]{0.53\textwidth}
			\centering
			{
				\begin{boxcode}{9.9cm}{0.75}{1.0}
					\textbf{Input}: \code\textsuperscript{in}, $\text{\codesketch}^\text{in}$, \sketchparams$(\cdot|~\text{\code}^\text{in}$), $\delta_\text{size} = 2$
					% = ($\phi$ , $\text{\DSLMove}$,  $\text{\DSLTurnL}$, \DSLBoolPathL), 									
					\begin{enumerate}
					    \item[(\ensuremath{\Delta_0})] Up to $2$ new actions may be added in total to $\text{\actionseq}_{1}$, $\actionseq_{2}$, $\actionseq_{3}$
						\item[(\ensuremath{\Delta_1})] Edit action sequences $\localblockcons(\{\actionseq_{1}, \actionseq_{2}, \actionseq_{3}\},  \sketchparams(\cdot|~\code^\text{in}))$ 
					    \item[(\ensuremath{\Delta_4})] $\text{\SDSLBool}_{1}$ = \DSLBoolPathA $\lor$ $\text{\SDSLBool}_{1}$ = \DSLBoolNoPathA
					    \item[(\ensuremath{\Delta_5})] $(\text{\SDSLBool}_{1} = \text{\DSLBoolPathA} ) \Rightarrow$ 
					    \item[] \quad $\Big{(} \exists i \in [13]$ s.t $\big(\text{\SDSLAction}^{i}_{2} =  \text{\DSLMove} \ \ \land$  $(\text{ } \forall j < i \text{, } \text{\SDSLAction}^{j}_{2} \notin \{\text{\DSLTurnL}, \text{\DSLTurnR} \})\big) \Big{)}$
						%Constraints if $\text{\actionseq}_{3}$ is nested in \DSLBool $\text{\SDSLBool}_{1} = \text{\DSLBoolPathL}$:
						%
						\item[(\ensuremath{\Delta_5})] $(\text{\SDSLBool}_{1} = \text{\DSLBoolNoPathA}) \Rightarrow $ 
						\item[] \quad $\Big{(} \exists i \in [13]$ s.t $\big{(}\text{\SDSLAction}^{i}_{2} =  \text{\DSLMove}$ $\Rightarrow$  $(\text{ } \exists j<i \text{, }\text{\SDSLAction}^{j}_{2} \in \{\text{\DSLTurnL}, \text{\DSLTurnR} \}) \big{)} \Big{)}$
						%\item[(\ensuremath{\Delta_7})] Only one of $\actionseq_{1}, \actionseq_{2}, \actionseq_{3}$ have additional actions w.r.t. $\sketchparams(\actionseq~|~\code^\text{in})$
						%added to them.
						%
					    %\item[(\ensuremath{\Delta_6})] For $\actionseq \in \{\actionseq_{1}, \actionseq_{2}, \actionseq_{3}\}$, apply $\localblockcons(\actionseq,  \sketchparams(\actionseq~|~\code^\text{in}))$
						%\item[(\ensuremath{\Delta_7})] Only one of $\actionseq_{1}, \actionseq_{2}, \actionseq_{3}$ have additional actions w.r.t. $\sketchparams(\actionseq~|~\code^\text{in})$
						%added to them.
						%
						\item[(\ensuremath{\Delta_6})] $\text{\actionseq}_{1}$, $\actionseq_{2}$, $\actionseq_{3}$ are minimal
				\vspace{-2.5mm}
				\end{enumerate}
				%\vspace{0.5mm}
			\end{boxcode}
			\vspace{-3mm}
			\caption{$\text{\codesketch}^\text{in}$-Constraints for \karelType}
			%\caption{Constraints for Sketch $\text{\codesketch}^\text{in}$}
			\label{fig:appendix.mutation.karel.6}
		}
		\end{subfigure}		
	%\hspace{0.5em}
	%%%%%%%%%%%%%%%%%
	%\end{minipage}
	%%%%%%%%%%%%%%%%%%%%%%%%%%%%%%%%%%%%%%%%%%%%%%%%%%%%%%%%%%%%%%%%%%%%
% \\
	%%%%%%%%%%%%%%%%
%\hspace{0.5em}

    \begin{subfigure}[b]{1\textwidth}
    \centering
			{
				\begin{boxcode}{18.5cm}{0.75}{1.0}
				    Set of elimination sequences $\mathcal{E} := \{(\tl,\tr), (\tr,\tl), (\tl, \tl, \tl), (\tr, \tr, \tr), (\pim, \pum), (\pum, \pim), (\pim, \pim), (\pum, \pum), (\tl, \pim, \tr),$ \\
				    \qquad \qquad \qquad \qquad \qquad \qquad \qquad $(\tr, \pim, \tl), (\tl, \pum, \tr), (\tr,\pum, \tl), (\pim, \tl, \pum), (\pum, \tl, \pim), (\pim, \tr, \pum), (\pum, \tr, \pim),$\\
				    \qquad \qquad \qquad \qquad \qquad \qquad \qquad
				    $(\pim, \tl, \pim), (\pim, \tr, \pim), (\pum, \tl, \pum), (\pum, \tr, \pum)\}$
				     \begin{enumerate}[(i)]
    				    \item Apply each elimination sequence from $\mathcal{E}$ to $\actionseq_{1}$ as a set of constraints.
    				    \item Apply each elimination sequence from $\mathcal{E}$ to $\actionseq_{2}$ as a set of constraints.
    				    \item Apply each elimination sequence from $\mathcal{E}$ to $\actionseq_{3}$ as a set of constraints.			    
				    \end{enumerate}
			\vspace{-2mm}
			\end{boxcode}
			\vspace{-3mm}
			\caption{$\text{\codesketch}^\text{in}$-Constraints:~$(\Delta_{6})\text{ \actionseq}_{1}$, $\actionseq_{2}$, $\actionseq_{3}$ are minimal}
			%\caption{Constraints for Sketch $\text{\codesketch}^\text{in}$}
			\label{fig:appendix.mutation.karel.8}
		}
    \end{subfigure}
	%\vspace{-4mm}

%%%%%%%%%%%%%%

	\caption{\textbf{(a)--(g)}: Illustration of Code Mutation for \karelType~task \emph{Diagonal} from the \emph{Intro to Programming with Karel} course by \emph{CodeHS.com}~\cite{intro_to_karel_codehs}; We present the \karelType~variant of \figref{fig:mutation} here.
	}
	%; see text for details. 
	\label{fig:appendix.mutation.karel}
	\vspace{-5mm}	
\end{figure}
\begin{figure}[h!]
\ContinuedFloat
\begin{subfigure}[b]{1\textwidth}
    \centering
			{
				\begin{boxcode}{18.5cm}{0.75}{1.0}
				        %Shorthand actions
						\textcolor{blue}{Shorthand notation for actions: $\text{\DSLMove} \rightarrow \sm$, $\text{\DSLTurnL} \rightarrow \tl$, $\text{\DSLTurnR} \rightarrow \tr$,
						$\text{\DSLPickM} \rightarrow \pim$,
						$\text{\DSLPutM} \rightarrow \pum$,
						$\phi \rightarrow \text{empty-action}$.}
						%Constraints are joined by '$\wedge$' operator.
				% 	\vspace{-0.8mm}
				    \\
				    \\
				    %\vspace{4mm}
				    Local \actionseq~constraints that define the values that each action sequence can take:
				    \vspace{-0.5mm}
                    \begin{enumerate}[(i)]
					\item \hspace{0.25em} $(\ha{1}{1} =\sm \vee \ha{1}{1} = \tl
					\vee \ha{1}{1} = \tr \vee \ha{1}{1} = \pim \vee \ha{1}{1} = \pum \vee \ha{1}{1} = \phi)$ 
					\item[] $\wedge (\ha{1}{2} =\sm \vee \ha{1}{2} = \tl
					\vee \ha{1}{2} = \tr \vee \ha{1}{2} = \pim \vee \ha{1}{2} = \pum \vee \ha{1}{2} = \phi) $ 
					\item[] $\wedge ( \ha{1}{3} =\pim \vee \ha{1}{3} = \pum )$ 
					\item[] $\wedge(\ha{1}{4} =\sm \vee \ha{1}{4} = \tl
					\vee \ha{1}{4} = \tr \vee \ha{1}{4} = \pim \vee \ha{1}{4} = \pum \vee \ha{1}{4} = \phi)$ 
					\item[] $\wedge (\ha{1}{5} =\sm \vee \ha{1}{5} = \tl
					\vee \ha{1}{5} = \tr \vee \ha{1}{5} = \pim \vee \ha{1}{5} = \pum \vee \ha{1}{5} = \phi)$ 
					
					\item $\wedge (\ha{1}{1} \neq \phi \vee \ha{1}{2} \neq \phi) \Rightarrow (\ha{1}{4} = \phi \wedge \ha{1}{5} = \phi) \wedge$
					
					\item[] $\wedge (\ha{1}{4} \neq \phi \vee \ha{1}{5} \neq \phi) \Rightarrow (\ha{1}{1} = \phi \wedge \ha{1}{2} = \phi)$
					%%%%%%%%% end of A1--values

					\item \hspace{0.25em} $(\ha{2}{1} =\sm \vee \ha{2}{1} = \tl
					\vee \ha{2}{1} = \tr \vee \ha{2}{1} = \pim \vee \ha{2}{1} = \pum \vee \ha{2}{1} = \phi) $ 
					
					\item[] $\wedge(\ha{2}{2} =\sm \vee \ha{2}{2} = \tl
					\vee \ha{2}{2} = \tr \vee \ha{2}{2} = \pim \vee \ha{2}{2} = \pum \vee \ha{2}{2} = \phi)$ 
					
					\item[] $\wedge (\ha{2}{3} = \sm )$ 
					
					\item[] $\wedge (\ha{2}{4} =\sm \vee \ha{2}{4} = \tl
					\vee \ha{2}{4} = \tr \vee \ha{2}{4} = \pim \vee \ha{2}{4} = \pum \vee \ha{2}{4} = \phi)$ 
					
					\item[] $\wedge \big{(}(\ha{2}{5} = \tl \wedge \ha{2}{9} = \tl) \vee (\ha{2}{5} = \tr \wedge \ha{2}{9} = \tr) \vee (\ha{2}{5} = \tl \wedge \ha{2}{9} = \tr) \vee (\ha{2}{5} = \tr \wedge \ha{2}{9} = \tl) \big{)}$ 
					
					\item[] $\wedge (\ha{2}{6} =\sm \vee \ha{2}{6} = \tl
					\vee \ha{2}{6} = \tr \vee \ha{2}{6} = \pim \vee \ha{2}{6} = \pum \vee \ha{2}{6} = \phi)$ 
					
					\item[] $\wedge (\ha{2}{7} = \sm)$
					
					\item[] $ \wedge (\ha{2}{8} =\sm \vee \ha{2}{8} = \tl
					\vee \ha{2}{8} = \tr \vee \ha{2}{8} = \pim \vee \ha{2}{8} = \pum \vee \ha{2}{8} = \phi)$ 
					
					\item[] $\wedge (\ha{2}{10} =\sm \vee \ha{2}{10} = \tl
					\vee \ha{2}{10} = \tr \vee \ha{2}{10} = \pim \vee \ha{2}{10} = \pum \vee \ha{2}{10} = \phi)$ 
					
					\item[] $\wedge (\ha{2}{11} = \pim \vee \ha{2}{11} = \pum)$ 
					
					\item[] $\wedge (\ha{2}{12} =\sm \vee \ha{2}{12} = \tl
					\vee \ha{2}{12} = \tr \vee \ha{2}{12} = \pim \vee \ha{2}{12} = \pum \vee \ha{2}{12} = \phi)$ 
					
					\item[] $\wedge (\ha{2}{13} =\sm \vee \ha{2}{13} = \tl
					\vee \ha{2}{13} = \tr \vee \ha{2}{13} = \pim \vee \ha{2}{13} = \pum \vee \ha{2}{13} = \phi) $ 
					
					\item \hspace{0.25em} $ \big{(}(\ha{2}{1} \neq \phi \vee \ha{2}{2} \neq \phi) \Rightarrow (\ha{2}{4} = \phi \wedge \ha{2}{6} = \phi \wedge \ha{2}{8} = \phi \wedge \ha{2}{10} = \phi \wedge \ha{2}{12} = \phi \wedge \ha{2}{13} = \phi)\big{)}$
				    
				    \item[] $ \wedge \big{(}(\ha{2}{4} \neq \phi) \Rightarrow (\ha{2}{1} = \phi \wedge \ha{2}{2} = \phi \wedge  \ha{2}{6} = \phi \wedge \ha{2}{8} = \phi \wedge \ha{2}{10} = \phi \wedge \ha{2}{12} = \phi \wedge \ha{2}{13} = \phi) \big{)}$
				    
				    \item[] $ \wedge \big{(}(\ha{2}{6} \neq \phi) \Rightarrow (\ha{2}{1} = \phi \wedge \ha{2}{2} = \phi \wedge  \ha{2}{4} = \phi \wedge \ha{2}{8} = \phi \wedge \ha{2}{10} = \phi \wedge \ha{2}{12} = \phi \wedge \ha{2}{13} = \phi) \big{)}$
				     
				    \item[] $ \wedge \big{(}(\ha{2}{8} \neq \phi) \Rightarrow (\ha{2}{1} = \phi \wedge \ha{2}{2} = \phi \wedge  \ha{2}{4} = \phi \wedge \ha{2}{6} = \phi \wedge \ha{2}{10} = \phi \wedge \ha{2}{12} = \phi \wedge \ha{2}{13} = \phi) \big{)}$
				    
				    \item[] $ \wedge \big{(}(\ha{2}{10} \neq \phi) \Rightarrow (\ha{2}{1} = \phi \wedge \ha{2}{2} = \phi \wedge  \ha{2}{4} = \phi \wedge \ha{2}{6} = \phi \wedge \ha{2}{8} = \phi \wedge \ha{2}{12} = \phi \wedge \ha{2}{13} = \phi) \big{)}$
				    
				    \item[] $ \wedge \big{(}(\ha{2}{12} \neq \phi \vee \ha{2}{13} \neq \phi) \Rightarrow (\ha{2}{1} = \phi \wedge \ha{2}{2} = \phi \wedge \ha{2}{4} = \phi \wedge \ha{2}{6} = \phi \wedge \ha{2}{8} = \phi \wedge \ha{2}{10} = \phi) \big{)}$
					%%%%%%%%% end of A2--values
					
					\item \hspace{0.25em} $(\ha{3}{1} =\sm \vee \ha{3}{1} = \tl \vee \ha{3}{1} = \tr \vee \ha{3}{1} = \pim \vee \ha{3}{1} = \pum \vee \ha{3}{1} = \phi)$ 
					\item[] $\wedge  (\ha{3}{2} =\sm \vee \ha{3}{2} = \tl \vee \ha{3}{2} = \tr \vee \ha{3}{2} = \pim \vee \ha{3}{2} = \pum \vee \ha{3}{2} = \phi)$ 
				    %%%%%%%%% end of A3--values
				    \end{enumerate}
				%\vspace{4mm}
				\\
				 Global \actionseq~constraints that allow actions to be added to either of $\actionseq_{1}, \actionseq_{2}, \actionseq_{3}$:
				    
				    \vspace{-0.5mm}
				    \begin{enumerate}[(i)]
				    \item $(\ha{1}{1} \neq \phi \vee \ha{1}{2} \neq \phi \vee \ha{1}{3} \neq \phi \vee \ha{1}{4} \neq \phi)$ 
				    \item[] \qquad $\Rightarrow (\ha{2}{1} = \phi \wedge \ha{2}{2} = \phi \wedge \ha{2}{4} = \phi \wedge \ha{2}{6} = \phi \wedge \ha{2}{8} = \phi \wedge \ha{2}{10} = \phi \wedge \ha{2}{12} = \phi \wedge \ha{2}{13} = \phi$ $\wedge$ $\ha{3}{1} = \phi \wedge \ha{3}{2} = \phi)$
				    
				    % \item[] \qquad \qquad \qquad \qquad \qquad \qquad \qquad \qquad \qquad 
				    
				    %%%%%%%%% end of A1---phi-cons
				    
				     \item $(\ha{2}{1} \neq \phi \vee \ha{2}{2} \neq \phi \vee \ha{2}{4} \neq \phi \wedge \ha{2}{6} \neq \phi \wedge \ha{2}{8} \neq \phi \wedge \ha{2}{10} \neq \phi \wedge \ha{2}{12} \neq \phi \wedge \ha{2}{13} \neq \phi )$ 
				     \item[] \qquad $\Rightarrow (\ha{1}{1} = \phi \wedge \ha{1}{2} = \phi \wedge \ha{1}{4} = \phi \wedge \ha{1}{5} = \phi$ $\wedge$ $\ha{3}{1} = \phi \wedge \ha{3}{2} = \phi)$
				    %  \item[] \qquad \qquad \qquad \qquad \qquad \qquad \qquad \qquad \qquad \qquad \qquad \qquad \qquad \qquad \qquad \qquad \qquad \quad 
				    %%%%%%%%% end of A2---phi-cons
				    
				     \item $(\ha{3}{1} \neq \phi \vee \ha{3}{2} \neq \phi)$
				     \item[] \qquad $\Rightarrow (\ha{1}{1} = \phi \wedge \ha{1}{2} = \phi \wedge \ha{1}{4} = \phi \wedge \ha{1}{5} = \phi \wedge \ha{2}{1} = \phi \wedge \ha{2}{2} = \phi \wedge \ha{2}{4} = \phi \wedge \ha{2}{6} = \phi \wedge \ha{2}{8} = \phi \wedge \ha{2}{10} = \phi \wedge$ $\ha{2}{12} = \phi \wedge \ha{2}{13} = \phi)$
				     %\item[] \qquad \qquad \qquad \qquad \qquad 
				     
				    %%%%%%%%% end of A3---phi-cons
				    \end{enumerate}
				  
			\end{boxcode}
			\vspace{-3mm}
			\caption{$\text{\codesketch}^\text{in}$-Constraints:~$(\Delta_{1})~\localblockcons(\{\actionseq_{1}, \actionseq_{2}, \actionseq_{3}\},  \sketchparams(\cdot|~\code^\text{in}))$}
			%\caption{Constraints for Sketch $\text{\codesketch}^\text{in}$}
			\label{fig:appendix.mutation.karel.7}
		}
\end{subfigure}
\caption{\textbf{(h)}: Illustration of Code Mutation for \karelType~task \emph{Diagonal} from the \emph{Intro to Programming with Karel} course by \emph{CodeHS.com}~\cite{intro_to_karel_codehs}; We present the \karelType~variant of \figref{fig:mutation} here.}
\end{figure}

\clearpage
% !TEX root =  main.tex
%%%%%%%%%%%%%%%%%%%%%%%%%%%%%%%%%%%%%%%%%%%%%%%%%%%%%%%%%%
%%%%%%%%%%%%%%%%%%%%%%%%%%%%%%%%%%%%%%%%%%%%%%%%%%%%%%%%%%
\section{Symbolic Execution: Additional Details}\label{appendix.sec.symbolicexecution}
%%%%%%%%%%%%%%%%%%%%%%%%%%%%%%%%%%%%%%
%%%%%%%%%%%%%%%%%%%%%%%%%%%%%%%%%%%%%%
%%%%%%%%%%%%%%%%%%%%%%%%%%%%%%%%%%%%%
In this section, we provide an example demonstrating how MCTS could guide the symbolic
execution in generating more suitable tasks, see Fig.~\ref{fig:appendix.mcts.symbolicexecution}.

%we provide an example demonstrating the combination of MCTS and symbolic execution, see Fig.~\ref{fig:appendix.mcts.symbolicexecution}.

% we limited the unrolling of the loops to a maximum of $2n (=20)$.

\begin{figure}[h!]
\centering
	\begin{subfigure}[b]{.17\textwidth}
	\centering
	{
		\includegraphics[height=2.3cm]{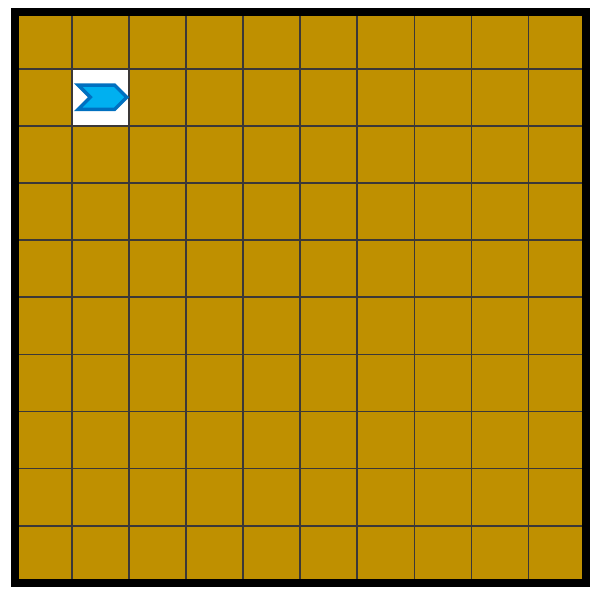}
		\caption{Initialization}
		%\caption{Trends in components of \FScore}
		\label{fig:appendix.mcts.symbolicexecution.init}
    }
    \end{subfigure}
	%%%%%%%%%%%%%%%%%
  	\begin{subfigure}[b]{.27\textwidth}
  	\centering
  	{
		\includegraphics[height=4.7cm]{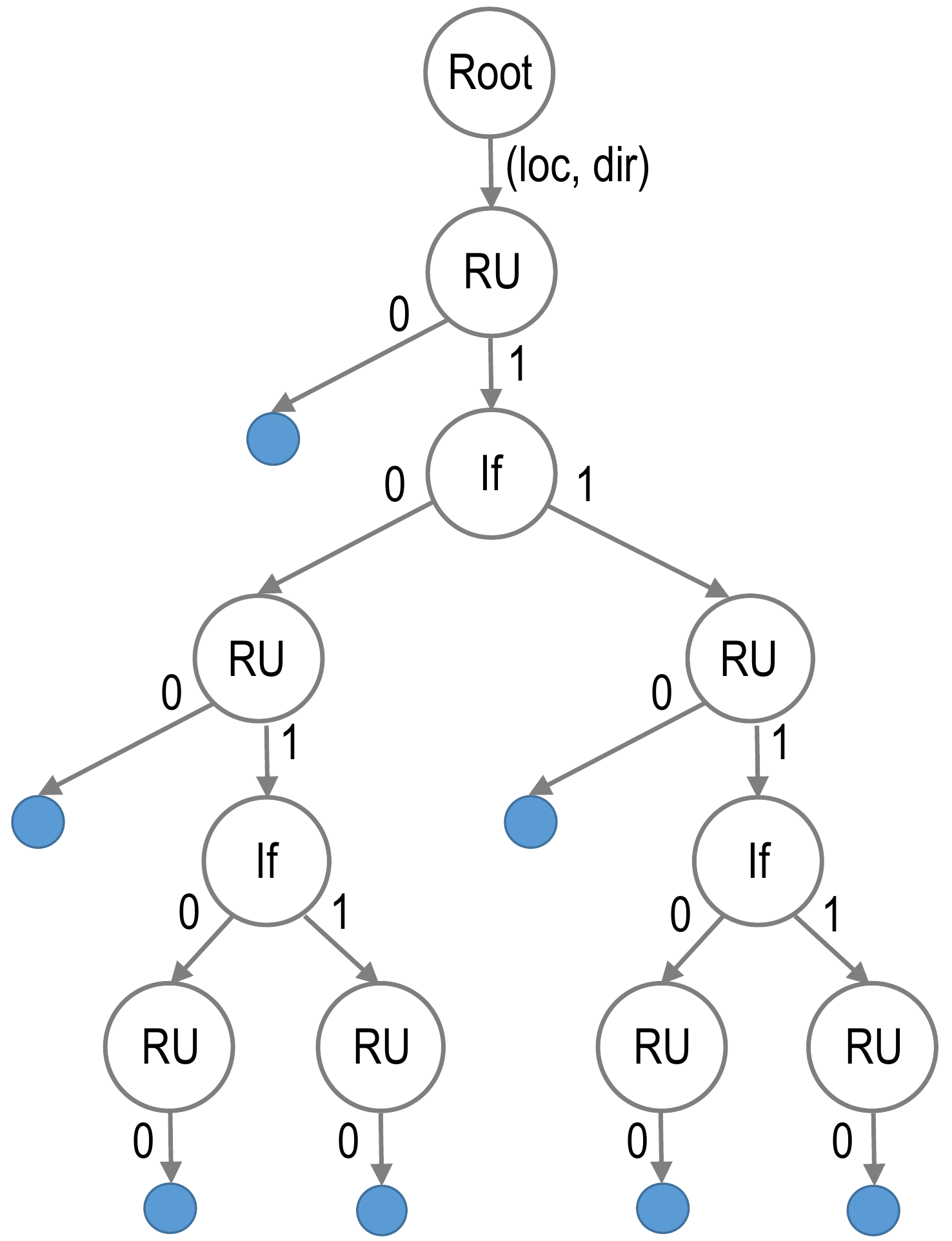}
		\caption{Search tree}
		%\caption{Best at $2$M ($10^6$)}
		\label{fig:appendix.mcts.symbolicexecution.tree}
	}
	\end{subfigure}	    
	%%%%%%%%%%%%%%%%%
  	\begin{subfigure}[b]{.17\textwidth}
  	\centering
  	{
		\includegraphics[height=2.3cm]{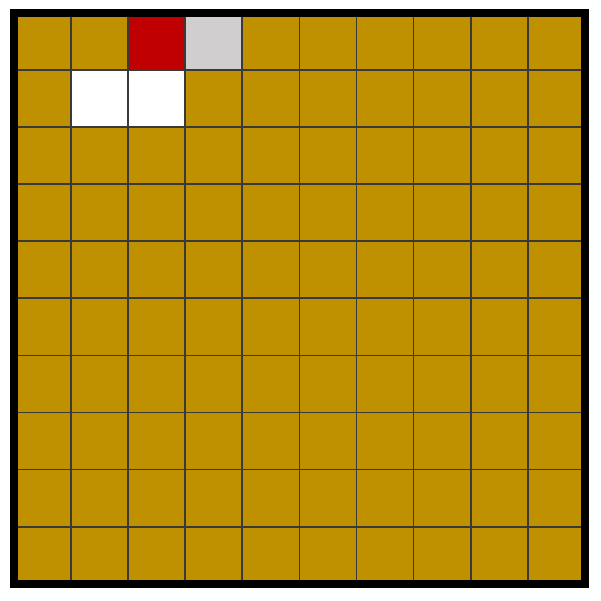}
		\caption{(1, 0, 1)} %RU-1; If-0; RU-1
		\label{fig:appendix.mcts.symbolicexecution.path1}
	}
	\end{subfigure}
	\begin{subfigure}[b]{.17\textwidth}
	\centering
	{
		\includegraphics[height=2.3cm]{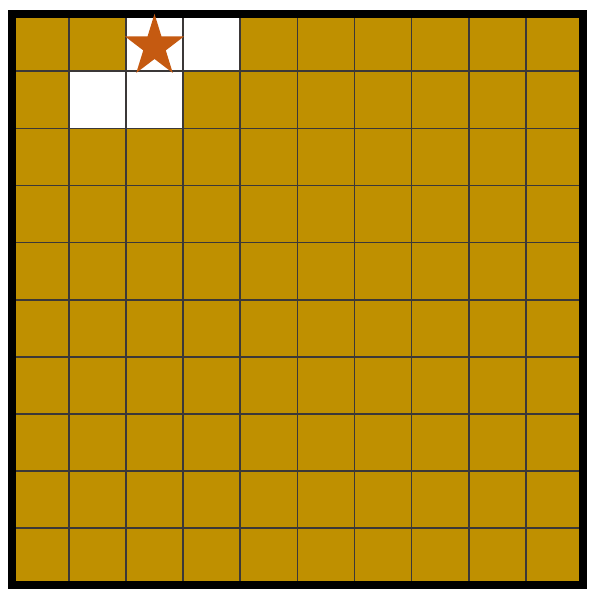}
		\caption{(1, 1, 0)} %RU-1; If-1; RU-0
		%\caption{Best at $20$K ($10^3$)}
		\label{fig:appendix.mcts.symbolicexecution.path2}
    }
    \end{subfigure}
  	\begin{subfigure}[b]{.17\textwidth}
  	\centering
  	{
		\includegraphics[height=2.3cm]{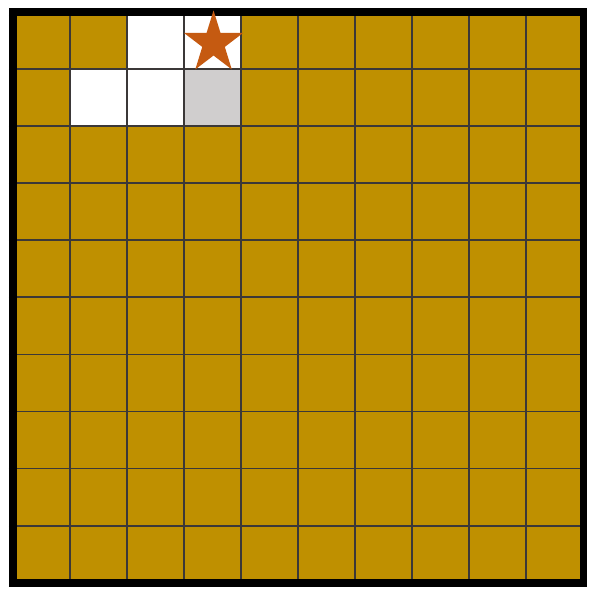}
		\caption{(1, 1, 1, 0, 0)} %RU-1; If-1; RU-1; If-0; RU-0
		%\caption{Best at $2$M ($10^6$)}
		\label{fig:appendix.mcts.symbolicexecution.path3}
	}
	\end{subfigure}
	%\vspace{-5mm}
	\caption{Search strategy for guiding the symbolic execution; see the text below for details.
	}
	%\vspace{-5mm}
	\label{fig:appendix.mcts.symbolicexecution}
\end{figure}
%Illustration of a single MCTS run on $\code^{\textnormal{out}}$ from \figref{fig:intro.hoc.p1} obtained from solution code of task \hocG~by mutation. (a) shows the temporal trends of different feature values in \FScore~averaged over a time window of $100$ steps. (b)--(d)~show the best, i.e., highest scoring, tasks generated up to times $2 \times 10^2$, $2 \times 10^4$, and $2 \times 10^6$ respectively. $\task^{\textnormal{out}}_\textnormal{vis}$ shown in \figref{fig:intro.hoc.t1} is the puzzle produced in (d)
%%%%%%%%%%%%%%%%%%%%%%%%%%%%%%%%%%%%%
%%%%%%%%%%%%%%%%%%%%%%%%%%%%%%%%%%%%%%%%%%%%%%%%%%%%%%%%%%

In Fig.~\ref{fig:appendix.mcts.symbolicexecution.tree}, we consider the MCTS search tree for the code $\code^{\textnormal{out}}$ from \figref{fig:intro.hoc.p1} obtained from the solution code of task \hocG~by mutation.  At the top of the tree, we have a ``Root'' node where the initial location/direction of the agent is picked from among the available choices. Node ``RU'' corresponds to the block \DSLRepeatUntil; the child ``1'' from a node ``RU'' corresponds to unrolling of the loop (i.e., \DSLBoolGoal is false).  Node ``If'' corresponds to the block \DSLIf; the child ``1'' from a node ``If'' corresponds to executing the code inside (i.e., \DSLBoolPathR is true). For the purpose of this demonstration, we limited the unrolling of the loops to a maximum of $3$; for our experiments, this depth is $2n (=20)$ as discussed in Appendix~\ref{appendix.sec.simulations}. Furthermore, in this demonstration, the initial location/direction for the agent are fixed as shown in  Fig.~\ref{fig:appendix.mcts.symbolicexecution.init}; for our experiments, we consider $5\times4 (=20)$ initial configurations which are picked by MCTS resulting in a large branching factor at the root of the tree.

Figs.~\ref{fig:appendix.mcts.symbolicexecution.path1},~\ref{fig:appendix.mcts.symbolicexecution.path2},~and~\ref{fig:appendix.mcts.symbolicexecution.path3} illustrate the symbolic execution output for three different paths in the search tree as discussed below:
\begin{itemize}
\item Fig.~\ref{fig:appendix.mcts.symbolicexecution.path1} corresponds to the path (``RU'':1, ``If'':0, ``RU'': 1). This path results in agent crashing the wall. As discussed in Section~\ref{sec.simulations}, $\FQuality(\cdot)$ score is set to $0$ when $\FNoCrash(\cdot) = 0$ and hence this tree path evaluates to low $\FScore(\cdot)$ score. 
\item \looseness-1Fig.~\ref{fig:appendix.mcts.symbolicexecution.path2} corresponds to the path (``RU'':1, ``If'':1, ``RU'': 0). This path results in two moves and two turns.
\item Fig.~\ref{fig:appendix.mcts.symbolicexecution.path3} corresponds to the path (``RU'':1, ``If'':1, ``RU'': 1, ``If'':0, ``RU'': 0). This path results in three moves and two turns, and is the higher scoring path compared to other two paths.
\end{itemize}

MCTS is extremely effective as a search strategy and quickly learns to pick paths with high scores. For the same code $\code^{\textnormal{out}}$ used in Fig.~\ref{fig:appendix.mcts.symbolicexecution} above, Fig.~\ref{fig:experiments.mcts.varytime} shows the temporal trends of different feature values in \FScore~averaged over a time window of $100$ steps in our experiments. For further details, we refer the reader to \cite{kocsis2006bandit} for an overview of the MCTS procedure, and to \cite{luckow2018monte} where MCTS is used with symbolic execution to direct the exploration towards costly paths.
%(maximum unrolling of the loops was set to $2n$ and there were $5\times4 (=20)$ initial configurations)
\clearpage
% !TEX root =  main.tex
%%%%%%%%%%%%%%%%%%%%%%%%%%%%%%%%%%%%%%%%%%%%%%%%%%%%%%%%%%
%%%%%%%%%%%%%%%%%%%%%%%%%%%%%%%%%%%%%%%%%%%%%%%%%%%%%%%%%%
\section{Experimental Evaluation: Additional Details and Results}\label{appendix.sec.simulations}
This section elaborates the experimental setup and results described in Section~\ref{sec.simulations}. We begin by providing details on the MCTS procedure in the symbolic execution stage of our task synthesis algorithm. Furthermore, we provide insights on generating multiple tasks for a single code. We also illustrate some example output tasks which violate criteria (V) and (VI), and mutated codes which got pruned in the symbolic execution stage (see Figure~\ref{fig:experiments.analysis}).
%
%\textbf{Remark}: 
Our implementation of the symbolic execution stage with MCTS procedure  is publicly available (see Footnote~\ref{footnote.githubrepo}).
%will be released together with the final version of the paper.  
%Our implementation of the code mutation stage, using the Z3 solver~\cite{deMouraBjorner2008}, is publicly available (see Footnote~\ref{footnote.githubrepo}).%\footnote{\href{https://github.com/adishs/neurips2020_synthesizing-tasks_code}{https://github.com/adishs/neurips2020\_synthesizing-tasks\_code}} 
%will be released together with the final version of the paper.
% of the algorithm

%generated by our algorithm, and the task-synthesis objectives (defined in Section~\ref{sec.problem}) that they satisfy.

%We also discuss the final tasks generated by our algorithm, and the task-synthesis objectives (defined in Section~\ref{sec.problem}) that they satisfy.
%, and on procedures used to add more variability to the output tasks

\subsection{Specification of MCTS for Single Run and Additional Results}
%Reference Tasks and Specifications
%\textbf{Specification of MCTS}
In this section, we elaborate on the specification of MCTS, briefly discussed in Section~\ref{sec.simulations}. We begin by discussing choices that effect the scalability and run-time of MCTS procedure.

%We begin by discussing its initial orientation and subsequent tree-depth.

\textbf{Choice of initial location and direction.} When doing symbolic execution using MCTS, the procedure begins by picking an initial location and direction for the agent (see \figref{fig:symbolicexecution}(b)). Given a grid-size $n$, and four initial directions (\emph{north}, \emph{east}, \emph{south}, \emph{west}) of the agent to choose from, we get a total of $4n^2$ choices for our initial configuration of the grid-puzzle. In the implementation used for generating the results, we restricted the set of initial choices to $20$ (by choosing only $5$ initial locations of the grid including four corners and centre, and $4$ directions)---this aids in exploration by limiting the branching factor at the root of the MCTS's tree.

\textbf{Tree depth for symbolic execution.} The depth of the  symbolic tree depends on the nature of the corresponding solution code.  For codes without the \DSLRepeatUntil~or \DSLWhile~constructs (\hocA, \hocC, \hocD,  \karelA, \karelC~and \karelE), the tree-depth is bounded. But, for more complex codes (\hocF, \hocG, \hocH, and \karelF), the tree depth is unbounded. In our implementation,  we limited the unrolling of the loops to a maximum of $2n (=20)$. To get an insight into the complexity of the problem, we note that with $20$ initial choices and a depth of $2n$, there are over $40$ million leaves in the symbolic tree for codes \hocG~and \hocH~which contain conditionals nested inside \DSLRepeatUntil~or \DSLWhile~constructs.\footnote{The actual memory footprint of the MCTS procedure is small given that the symbolic tree is dynamically constructed during a run and the actual number of leaves explored is much lesser. In fact, the average runtime per output task (i.e., one MCTS run) as reported in Column $8$ of \figref{fig:experiments.analysis} is achieved on a laptop machine with $2.8$ GHz Quad-Core Intel Core i$7$ processor and $16$ GB RAM.} Next, we describe the details of our evaluation function for MCTS.

%With this approximation, for each initial orientation, these codes yield more than $2$ million leaves in the MCTS-tree. Hence, with even a choice of $20$ initial orientations, we get over $40$ million leaves in the tree. This clearly illustrates the complexity of the task synthesis problem. We describe the details of our evaluation function, that guides the MCTS-search, next.

%For codes without the \DSLRepeatUntil~or \DSLWhile~constructs, the tree-depth is bounded. However for codes containing them, the tree depth is unbounded. In particular for codes corresponding to \hocA, \hocC, \hocD, \karelA, \karelC~and \karelE~the symbolic tree depths are $1$. But, for the more complex codes in \hocF, \hocG, and \karelF~the tree depth is unbounded. To successfully deal with this nature of the symbolic tree, we restrict the set of initial orientations to $20$ (by choosing only 5 initial locations of the grid--4 corners and centre, and 4 initial directions), and for codes with \DSLRepeatUntil~or \DSLWhile~constructs, we limit the unrolling of the loop to a maximum of $20$ iterations. With this approximation, for each initial orientation, these codes yield more than $2$ million leaves in the MCTS-tree. Hence, with even a choice of $20$ initial orientations, we get over $40$ million leaves in the tree. This clearly illustrates the complexity of the task synthesis problem. We describe the details of our evaluation function, that guides the MCTS-search, next.

\textbf{Details on the evaluation function \FScore.}
Our evaluation function $\FScore(\task^\textnormal{out}, \code^\textnormal{out}, \task^{\textnormal{in}}, \code^{\textnormal{in}}) \in [0,1]$ measures the suitability of a generated task. A higher \FScore~indicates a more suitable task. We describe the elements of our evaluation function in greater detail here. We defined it in Section~\ref{sec.simulations} and present it here again for completeness:
\begin{align*}
    \FScore(\task^\textnormal{out}, \code^\textnormal{out}, \task^\textnormal{in}, \code^\textnormal{in}) & = \underbrace{\mathbbm{1}{\big(\medmath{ \FQuality(\task^\textnormal{out}_\textnormal{vis}, \code^\textnormal{out}) \geq \delta_{\textnormal{qual}}, \FNoCrash(\task^\textnormal{out}_\textnormal{vis}, \code^\textnormal{out}) = 1, \FNoCut(\task^\textnormal{out}_\textnormal{vis}, \code^\textnormal{out}) = 1 }\big)}}_\text{2a} \cdot \\
    & \ \ \ \ \underbrace{\big[ \alpha_{1}\FCoverage(\task^\textnormal{out}_\textnormal{vis}, \code^\textnormal{out}) + \alpha_{2}\FQuality(\task^\textnormal{out}_\textnormal{vis}, \code^\textnormal{out}) + \alpha_{3} \FDissimilarity(\task^\textnormal{out}_\textnormal{vis}, \task^\textnormal{in}_\textnormal{vis}) \big]}_\text{2b} \label{evalfunc2} \tag{2}
\end{align*}
where $\mathbbm{1}$ is an indicator function and each constant $\alpha = 1/3$.
It is to be noted that component $2\textnormal{b}$ in Eq.\ref{evalfunc2} supplies the gradients for guiding the search in MCTS. At the end of the MCTS run (containing $2$ million iterations), the \textit{best} task (i.e, the one with the highest \FScore~value) is picked only from the pool of generated tasks which satisfy $\FCoverage(\cdot) = 1, \FScore(\cdot) > 0$. We discuss each constituent function of \FScore~next.

\textbf{Task quality component of evaluation function.} $\FQuality(\task_\textnormal{vis}^\textnormal{out},\code^\textnormal{out}) \in [0,1]$ evaluates the quality and validity of $\task^\textnormal{out}$. Its is defined as a linear combination of the normalized counts of certain features of $\task^\textnormal{out}_\textnormal{vis}$ when $\code^\textnormal{out}$ is executed. As certain elements differ in the two task types, \hocType~and \karelType, we define the features differently for each. More precisely, for \hocType~tasks, we have:
\begin{align*}
    \FQuality^{\hocType}(\task_\textnormal{vis}^\textnormal{out},\code^\textnormal{out}) & = \frac{1}{4}\Big{(} \frac{\countqual\text{moves}}{2n} +  \frac{\countqual\text{turns}}{n} +  \frac{\countqual\text{segments}}{n/2} +  
    \frac{\countqual\text{long-segments}}{n/3} \Big{)}
%\tag{3}
%\label{qualfunc}
\end{align*}
where the individual features are defined as 
\vspace{4mm}
\begin{itemize}
    \item \countqual moves: This refers to the count of `moves'.
    %in $\task^\textnormal{out}_\textnormal{vis}$
    \item \countqual turns: This refers to the count of `turns'.
    %in $\task^\textnormal{out}_\textnormal{vis}$
    \item \countqual segments: This refers to the number of consecutive sequence ($\geq 3$) of `moves'.% in $\task^\textnormal{out}_\textnormal{vis}$
    \item \countqual long-segments:  This refers to the number of longer consecutive sequence ($\geq 5$) of `moves'. %The count of longer-segments is also included in the count of segments.% in $\task^\textnormal{out}_\textnormal{vis}$.
\end{itemize}
For \karelType~tasks, we additionally have two marker based features to define the quality of the task i.e,
\begin{align*}
    \FQuality^{\karelType}(\task_\textnormal{vis}^\textnormal{out},\code^\textnormal{out}) & = \frac{3}{4} \cdot \frac{1}{4}\Big{(} \frac{\countqual\text{moves}}{2n} +  \frac{\countqual\text{turns}}{n} +  \frac{\countqual\text{segments}}{n/2} +  
    \frac{\countqual\text{long-segments}}{n/3} \Big{)}
    \\ & + \frac{1}{4} \cdot \frac{1}{2} \Big{(}\frac{\countqual\text{pick-markers}}{n} + \frac{\countqual\text{put-markers}}{n}\Big{)}
%\tag{4}
%\label{qualfunc}
\end{align*}
where the additional features are defined as 
\begin{itemize}
    \item \countqual pick-markers: This refers to the count of `pick-marker' activity.% in $\task^\textnormal{out}_\textnormal{vis}$
    \item \countqual put-markers: This refers to the count of `put-marker' activity.% in $\task^\textnormal{out}_\textnormal{vis}$.
\end{itemize}
\looseness-1 While we have used high-level features of  $\task^\textnormal{out}_\textnormal{vis}$ to define the quality of a task, one could also embed more specific domain knowledge in defining these features to obtain more interesting/complex tasks.

\textbf{Task dissimilarity component of the evaluation function.}
$\FDissimilarity(\task_\textnormal{vis}^\textnormal{out}, \task_\textnormal{vis}^\textnormal{in}) \in [0,1]$ evaluates the visual dissimilarity of $\task^\textnormal{out}_\textnormal{vis}$ w.r.t. $\task^{\textnormal{in}}_\textnormal{vis}$. We define it as a linear combination of the dissimilarity features as follows:
\begin{align*}
    \FDissimilarity(\task_\textnormal{vis}^\textnormal{out}, \task_\textnormal{vis}^\textnormal{in}) & = \frac{1}{3}\Big{(}
    %\text{dissimilarity between agent's initial location}
    \textnormal{diss}(\textnormal{loc}~|~\task_\textnormal{vis}^\textnormal{out}, \task_\textnormal{vis}^\textnormal{in}) +  \textnormal{diss}(\textnormal{dir}~|~\task_\textnormal{vis}^\textnormal{out}, \task_\textnormal{vis}^\textnormal{in}) + 
    \textnormal{diss}(\textnormal{grid-cells}~|~\task_\textnormal{vis}^\textnormal{out}, \task_\textnormal{vis}^\textnormal{in})
    \Big{)}
%\tag{5}
%\label{dissfunc}
\end{align*}
where the individual features are defined as
\begin{itemize}
    \item $\textnormal{diss}(\textnormal{loc}~|~\task_\textnormal{vis}^\textnormal{out}, \task_\textnormal{vis}^\textnormal{in}) \in \{0, 1\}$ measures the dissimilarity in the agent's initial location in the task-puzzles $\task^\textnormal{out}_\textnormal{vis}$ and $\task^\textnormal{in}_\textnormal{vis}$.

    \item $\textnormal{diss}(\textnormal{dir}~|~\task_\textnormal{vis}^\textnormal{out}, \task_\textnormal{vis}^\textnormal{in}) \in \{0, 1\}$ measures the dissimilarity in the agent's initial direction in the task-puzzles $\task^\textnormal{out}_\textnormal{vis}$ and $\task^\textnormal{in}_\textnormal{vis}$.

    \item $\textnormal{diss}(\textnormal{grid-cells}~|~\task_\textnormal{vis}^\textnormal{out}, \task_\textnormal{vis}^\textnormal{in}) \in [0, 1]$ measures the grid-cell level dissimilarity in the task-puzzles $\task^\textnormal{out}_\textnormal{vis}$ and $\task^\textnormal{in}_\textnormal{vis}$. This is computed as the normalized Hamming distance w.r.t. the two grid-worlds (i.e., number of cells which are different, multiplied with a normalization factor of $\frac{2}{n^2}$).
\end{itemize}

% \begin{align*}
%     \FDissimilarity(\task_\textnormal{vis}^\textnormal{out}, \task_\textnormal{vis}^\textnormal{in}) & = \frac{1}{3}\big{(}
%     \text{dissimilarity between agent's initial location}
%     \\
%     & + \text{dissimilarity between agent's initial orientation} +
%     \text{Hamming distance}(\task^{\textnormal{out}}_\textnormal{vis},\task^{\textnormal{in}}_\textnormal{vis})
%     \big{)}
% \tag{5}
% \label{dissfunc}
% \end{align*}

%We get an insight into an MCTS run for a \karelType~task next.

\textbf{Deep dive into an MCTS run for Karel.} Analogous to the example provided in Section~\ref{sec.simulations}, we take a closer look at an MCTS run for the \karelType~task~\karelF, shown in \figref{fig:appendix.experiments.mcts}. \figref{fig:appendix.experiments.mcts.varytime.1} and \figref{fig:appendix.experiments.mcts.varytime.2} illustrate the improvement in various components of \FScore~as the number of MCTS iterations increases. Best tasks at different iterations are shown in \figref{fig:appendix.experiments.mcts.best1} and   \figref{fig:appendix.experiments.mcts.best3}. As expected, the more the iterations, the better the tasks which are generated.
%\figref{fig:appendix.experiments.mcts.best2},
%%%%%%%%%%%%%%%%%%%%%%%%%%%%%%%%%%%%%
\begin{figure}[h!]
\centering
	%%%%%%%%%%%%%%%%%
	\begin{subfigure}[b]{.45\textwidth}
	\centering
	{
		\includegraphics[width=1\textwidth]{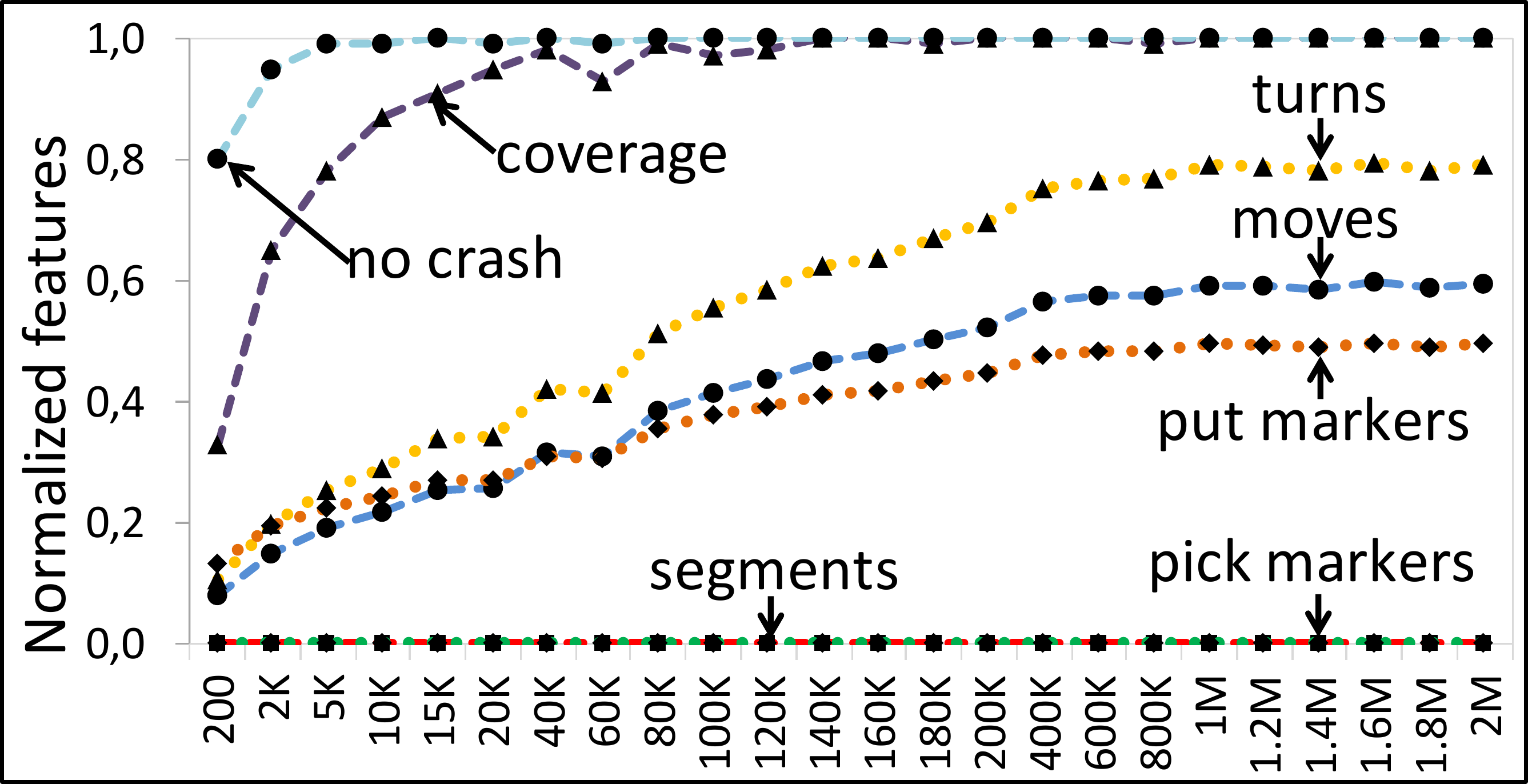}
		\caption{Trends in \FScore~features}
		%\caption{Trends in components of \FScore}
		\label{fig:appendix.experiments.mcts.varytime.1}
    }
    \end{subfigure}
    \quad
	%%%%%%%%%%%%%%%%%
	\begin{subfigure}[b]{.45\textwidth}
	\centering
	{
		\includegraphics[width=1\textwidth]{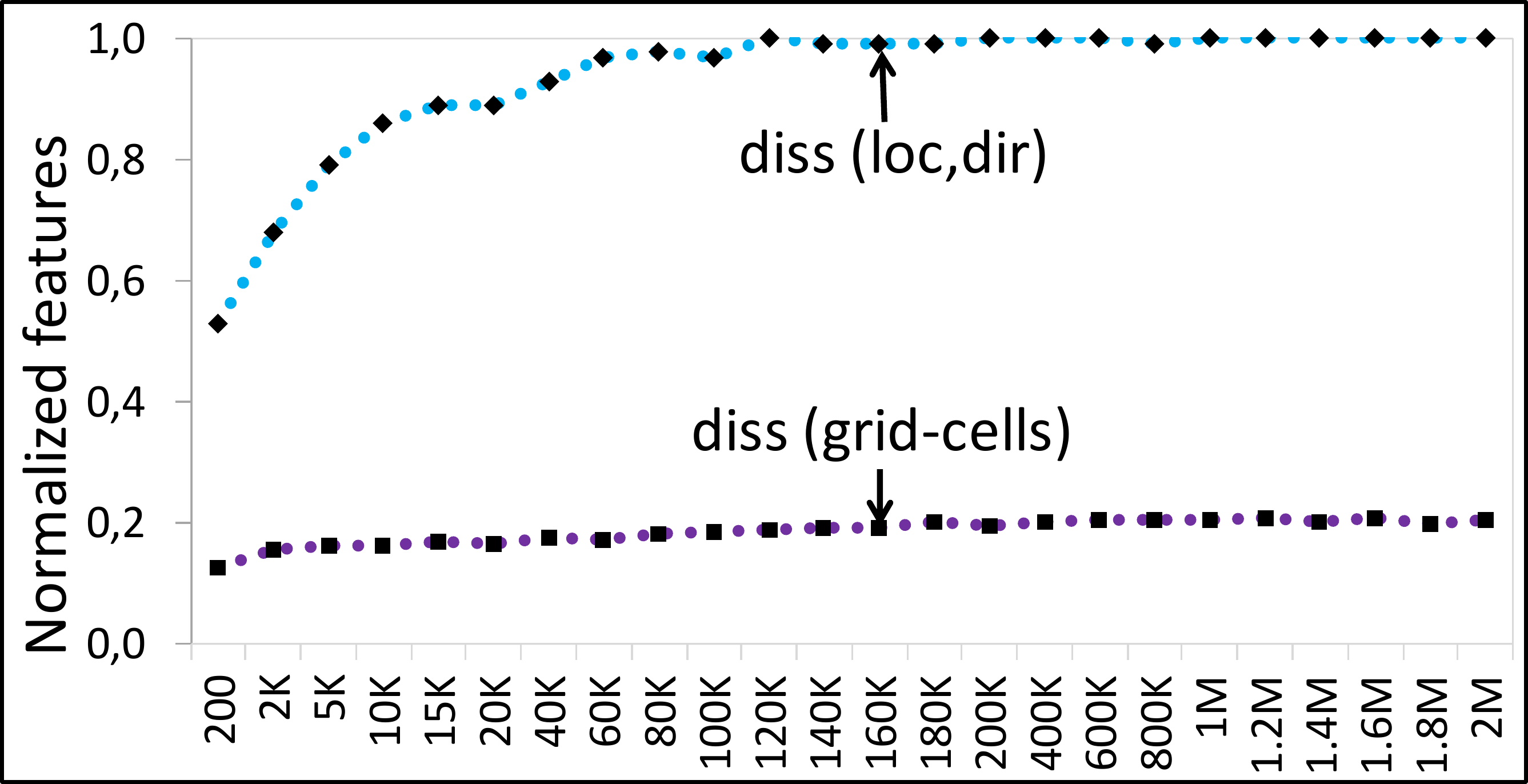}
		\caption{Trends in \FScore~features capturing dissimilarity}
		%\caption{Trends in components of \FScore}
		\label{fig:appendix.experiments.mcts.varytime.2}
    }
    \end{subfigure}
    \\
    \vspace{2mm}
	%%%%%%%%%%%%%%%%%
  	\begin{subfigure}[b]{.45\textwidth}
  	\centering
  	{
		\includegraphics[height=2.00cm]{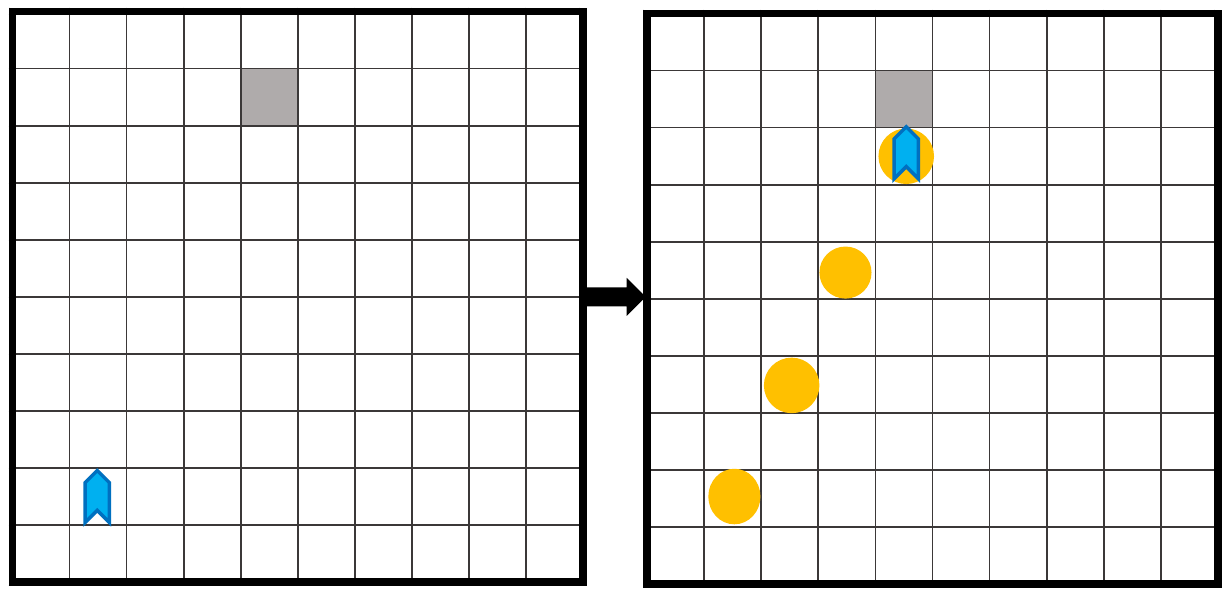}
		\caption{Best at $20$}
		\label{fig:appendix.experiments.mcts.best1}
	}
	\end{subfigure}
	\quad
% 	\begin{subfigure}[b]{.32\textwidth}
% 	\centering
% 	{
% 		\includegraphics[height=2.00cm]{fig/appendix_mctsplots_karel/20200607/karel-F_mutation-8-252_run4/mctsplot_best_200.pdf}
% 		\caption{Best at $200$}
% 		%\caption{Best at $20$K ($10^3$)}
% 		\label{fig:appendix.experiments.mcts.best2}
%     }
%     \end{subfigure}
  	\begin{subfigure}[b]{.45\textwidth}
  	\centering
  	{
		\includegraphics[height=2.00cm]{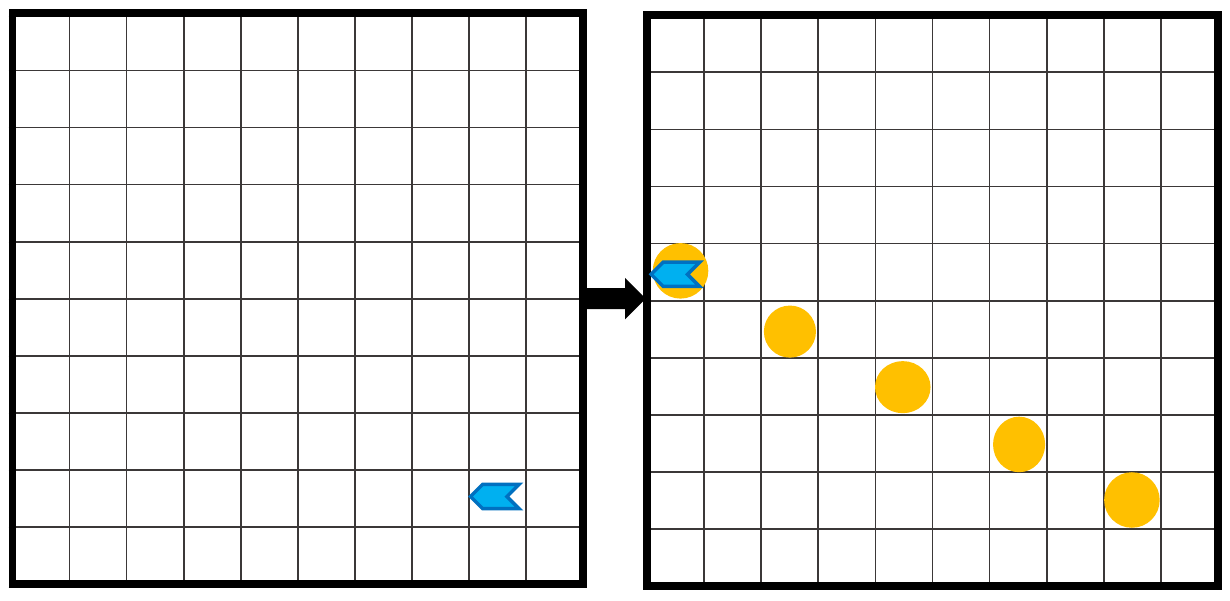}
		\caption{Best at $2$M}
		%\caption{Best at $2000$}
		%\caption{Best at $2$M ($10^6$)}
		\label{fig:appendix.experiments.mcts.best3}
	}
	\end{subfigure}	
	%\vspace{-4mm}
	\caption{Illustration of a single MCTS run on $\code^{\textnormal{out}}$ from \figref{fig:intro.karel.p1} obtained from solution code of task \karelF~by mutation. (a, b) show the temporal trends of different feature values in \FScore~averaged over a time window of $100$ steps. (c, d)~show the best, i.e., highest scoring, tasks generated up to times $2 \times 10^1$ and $2 \times 10^6$ respectively. %$\task^{\textnormal{out}}_\textnormal{vis}$ shown in \figref{fig:intro.karel.t1} is the puzzle produced in (d).
	}
	\vspace{-10mm}
	\label{fig:appendix.experiments.mcts}
\end{figure}
%%%%%%%%%%%%%%%%%%%%%%%%%%%%%%%%%%%%%
%%%%%%%%%%%%%%%%%%%%%%%%%%%%%%%%%%%%%%%%%%%%%%%%%%%%%%%%%%

\subsection{Specification of MCTS for Multiple Runs with Diversity and Additional Results}
Our task synthesis algorithm can also generate multiple tasks for a \emph{single} code, with sufficient diversity. To achieve this, we modify the evaluation function \FScore~guiding the MCTS search. We introduce a diversity measure \FDiversity~which measures the diversity between the generated tasks. More concretely, when generating a new $(k+1)^{\textsuperscript{th}}$ task, we capture the diversity score w.r.t the tasks generated in the previous k-runs of MCTS $\{ \task^\textnormal{out}_{\textnormal{vis},1}, \ldots, \task^\textnormal{out}_{\textnormal{vis},k} \}$.

% \textbf{Modified evaluation function \FScore.}
% Our evaluation function with the new diversity component is given below where  
% \begin{align*}
% &\FScore(\task^\textnormal{out}, \code^\textnormal{out}, \task^\textnormal{in}, \code^\textnormal{in}, \{ \task^\textnormal{out}_{\textnormal{vis},1}, \ldots, \task^\textnormal{out}_{\textnormal{vis},k} \})=  \underbrace{\mathbbm{1}{\big(\medmath{ \FQuality(\task^\textnormal{out}_\textnormal{vis}, \code^\textnormal{out}) \geq \delta_{\textnormal{qual}}, \FNoCrash(\task^\textnormal{out}_\textnormal{vis}, \code^\textnormal{out}) = 1, \FNoCut(\task^\textnormal{out}_\textnormal{vis}, \code^\textnormal{out}) = 1 }\big)}}_\text{3a} \cdot \\
%      &\quad  \underbrace{\big[ \alpha_{1}\FCoverage(\task^\textnormal{out}_\textnormal{vis}, \code^\textnormal{out}) + \alpha_{2}\FQuality(\task^\textnormal{out}_\textnormal{vis}, \code^\textnormal{out}) + \alpha_{3} \FDissimilarity(\task^\textnormal{out}_\textnormal{vis}, \task^\textnormal{in}_\textnormal{vis}) +
%     \alpha_{4} \FDiversity(\task^\textnormal{out}_\textnormal{vis}, \code^\textnormal{out}~|~\{ \task^\textnormal{out}_{\textnormal{vis},1} \ldots \task^\textnormal{out}_{\textnormal{vis},k} \})
%     \big]}_\text{3b} 
% \label{evalfunc3} 
% \tag{3}
% \end{align*}

\textbf{Modified evaluation function \FScore.}
Our evaluation function with the new diversity component is given below with each $\alpha = 1/4$. We have $\FScore(\task^\textnormal{out}, \code^\textnormal{out}, \task^\textnormal{in}, \code^\textnormal{in}, \{ \task^\textnormal{out}_{\textnormal{vis},1}, \ldots, \task^\textnormal{out}_{\textnormal{vis},k} \}):=$
\begin{align*}
    &\quad \underbrace{\mathbbm{1}{\big(\medmath{ \FQuality(\task^\textnormal{out}_\textnormal{vis}, \code^\textnormal{out}) \geq \delta_{\textnormal{qual}}, \FNoCrash(\task^\textnormal{out}_\textnormal{vis}, \code^\textnormal{out}) = 1, \FNoCut(\task^\textnormal{out}_\textnormal{vis}, \code^\textnormal{out}) = 1 }\big)}}_\text{3a} \cdot 
    \\
     &\quad  \underbrace{\big[ \alpha_{1}\FCoverage(\task^\textnormal{out}_\textnormal{vis}, \code^\textnormal{out}) + \alpha_{2}\FQuality(\task^\textnormal{out}_\textnormal{vis}, \code^\textnormal{out}) + \alpha_{3} \FDissimilarity(\task^\textnormal{out}_\textnormal{vis}, \task^\textnormal{in}_\textnormal{vis}) +
    \alpha_{4} \FDiversity(\task^\textnormal{out}_\textnormal{vis}, \code^\textnormal{out}~|~\{ \task^\textnormal{out}_{\textnormal{vis},1} \ldots \task^\textnormal{out}_{\textnormal{vis},k} \})
    \big]}_\text{3b} 
\label{evalfunc3} 
\tag{3}
\end{align*}

% \begin{align*}
%     &\FScore(\task^\textnormal{out}, \code^\textnormal{out}, \task^\textnormal{in}, \code^\textnormal{in}, \{ \task^\textnormal{out}_{\textnormal{vis},1}, \ldots, \task^\textnormal{out}_{\textnormal{vis},k} \}) = \underbrace{\mathbbm{1}{(\medmath{ \FQuality(\task^\textnormal{out}_\textnormal{vis}, \code^\textnormal{out}) \geq \delta_{\textnormal{qual}}, \FNoCrash(\task^\textnormal{out}_\textnormal{vis}, \code^\textnormal{out}) = 1, \FNoCut(\task^\textnormal{out}_\textnormal{vis}, \code^\textnormal{out}) = 1 })}}_\text{3a} \cdot 
%     \\
%      &\ \ \underbrace{\Big[ \alpha_{1}\FCoverage(\task^\textnormal{out}_\textnormal{vis}, \code^\textnormal{out}) + \alpha_{2}\FQuality(\task^\textnormal{out}_\textnormal{vis}, \code^\textnormal{out}) + \alpha_{3} \FDissimilarity(\task^\textnormal{out}_\textnormal{vis}, \task^\textnormal{in}_\textnormal{vis}) +
%     \alpha_{4} \FDiversity(\task^\textnormal{out}_\textnormal{vis}, \code^\textnormal{out}~|~\{ \task^\textnormal{out}_{\textnormal{vis},1} \ldots \task^\textnormal{out}_{\textnormal{vis},k} \})
%     \Big]}_\text{3b} 
% \label{evalfunc3} 
% \tag{3}
% \end{align*}

%We describe the diversity component of the \FScore~function next.

\textbf{Diversity score of tasks \FDiversity.} Here, we describe the diversity component of the evaluation function. $\FDiversity(\task^\textnormal{out}_\textnormal{vis}, \code^\textnormal{out}~|~\{ \task^\textnormal{out}_{\textnormal{vis},1},\ldots, \task^\textnormal{out}_{\textnormal{vis},k} \}) \in [0,1]$ operates on a pool of generated tasks, and computes a diversity score for the new task w.r.t the tasks in the pool. Initially, the pool of tasks generated is empty. In this case, we have $\FDiversity(\task^\textnormal{out}_\textnormal{vis}, \code^\textnormal{out}~|~\{\}) = 1$.
% \begin{align*}

% %\tag{7}
% \label{divfunc1}
% \end{align*}
% For a subsequent task $\task^\textnormal{out}_\textnormal{vis}$, if $\task^\textnormal{out}_\textnormal{vis}$ = $\task^\textnormal{out}_{\textnormal{vis},1}$, then its diversity score is $0$. 
% For a subsequent task $\task^\textnormal{out}_\textnormal{vis}$,  When,  then its diversity score is $0$. 
% Otherwise, its diversity score is defined as,

After one run of MCTS, we have one task in the task pool $\{\task^\textnormal{out}_{\textnormal{vis},1}$\}. We define the diversity score for a subsequent task $\task^\textnormal{out}_\textnormal{vis}$ as follows. First, if $\task^\textnormal{out}_\textnormal{vis} = \task^\textnormal{out}_{\textnormal{vis},1}$ then we set $\FDiversity(\task^\textnormal{out}_\textnormal{vis}, \code^\textnormal{out}~|~ \{\task^\textnormal{out}_{\textnormal{vis},1} \}) = 0$. Otherwise, the diversity score is given by 
\begin{align*}
\FDiversity(\task^\textnormal{out}_\textnormal{vis}, \code^\textnormal{out}~|~ \{\task^\textnormal{out}_{\textnormal{vis},1} \}) & = \frac{1}{4} \Big{(}
    %\text{dissimilarity between agent's initial location}
    \textnormal{diss}(\textnormal{loc}~|~\task_\textnormal{vis}^\textnormal{out}, \task_\textnormal{vis,1}^\textnormal{out}) 
    +  \textnormal{diss}(\textnormal{dir}~|~\task_\textnormal{vis}^\textnormal{out}, \task_\textnormal{vis,1}^\textnormal{out}) \\
   & + \textnormal{diss}(\textnormal{grid-cells}~|~\task_\textnormal{vis}^\textnormal{out}, \task_\textnormal{vis,1}^\textnormal{out})
    +  \textnormal{diss}(\textnormal{symbolic-paths}~|~\task_\textnormal{vis}^\textnormal{out}, \task_\textnormal{vis,1}^\textnormal{out})
    \Big{)}
% %\tag{8}
\end{align*}
where the individual features are defined as 
\begin{itemize}
    \item $\textnormal{diss}(\textnormal{loc}~|~\task_\textnormal{vis}^\textnormal{out}, \task_\textnormal{vis,1}^\textnormal{out}) \in \{0,1\}$ measures the dissimilarity in the agent's initial location in the task-puzzles $\task^\textnormal{out}_\textnormal{vis}$ and $\task^\textnormal{out}_{\textnormal{vis},1}$.
    \item $\textnormal{diss}(\textnormal{dir}~|~\task_\textnormal{vis}^\textnormal{out}, \task_\textnormal{vis,1}^\textnormal{out}) \in \{0,1\}$ measures the dissimilarity in the agent's initial direction in the task-puzzles $\task^\textnormal{out}_\textnormal{vis}$ and $\task^\textnormal{out}_{\textnormal{vis},1}$.
    \item $\textnormal{diss}(\textnormal{grid-cells}~|~\task_\textnormal{vis}^\textnormal{out}, \task_{\textnormal{vis},1}^\textnormal{out}) \in [0,1]$ measures the grid-cell level dissimilarity in the task-puzzles $\task^\textnormal{out}_\textnormal{vis}$ and $\task^\textnormal{out}_{\textnormal{vis},1}$.  This is computed as the normalized Hamming distance w.r.t. the two grid-worlds (i.e., number of cells which are different, multiplied with a normalization factor of $\frac{2}{n^2}$).
     \item \looseness-1 $\textnormal{diss}(\textnormal{symbolic-paths}~|~\task_\textnormal{vis}^\textnormal{out}, \task_{\textnormal{vis},1}^\textnormal{out}) \in [0,1]$ measures the dissimilarity in the symbolic-paths used in the generation of  $\task^\textnormal{out}_\textnormal{vis}$ and $\task^\textnormal{out}_{\textnormal{vis},1}$. This is computed as the normalized ``edit distance" between these paths.
     %two symbolic 
\end{itemize}

After we have run $k$ MCTS runs, the pool of tasks generated is populated ($\{\task^\textnormal{out}_{\textnormal{vis},1} \ldots \task^\textnormal{out}_{\textnormal{vis},k}\} $)
and the diversity score of the subsequent task is computed as follows:
\begin{align*}
    \FDiversity(\task^\textnormal{out}_\textnormal{vis}, \code^\textnormal{out}~|~\{\task^\textnormal{out}_{\textnormal{vis},1}, \ldots, \task^\textnormal{out}_{\textnormal{vis},k}\} ) & = \min_{i \in [k]} \FDiversity(\task^\textnormal{out}_\textnormal{vis}, \code^\textnormal{out}~|~\{\task^\textnormal{out}_{\textnormal{vis},i}\} )
\end{align*}

% \begin{align*}
%     \FDiversity(\task^\textnormal{out}, \code^\textnormal{out}~|~ \{\task^\textnormal{out}_{i=\{1,\ldots, k \}}\} ) & = \min_{i \in k} \FDiversity(\task^\textnormal{out}, \code^\textnormal{out}| \task^\textnormal{out}_{i} )
% \tag{9}
% \label{dicfunc3}
% \end{align*}
%\vspace{-2mm}
As the pool of tasks grow, it becomes more constrained to generate a task that is diverse from all the tasks in the pool. In general, there is a limit to the number of tasks per code that can be generated---eventually, the new task will have \FDiversity~to be $0$ or \FScore~to be $0$. As stated in Section~\ref{sec.simulations}, for each code $\code^{\textnormal{out}}$, we generated up to $10$ different visual tasks using this process.

%The process is terminated when the diversity score of a new task becomes $0$ (i.or the evaluation score evaluates to $0$. Using this procedure for each code in our set of reference tasks, we were able to generate a maximum $10$ diverse tasks, before the diversity score reached $0$. 
 
\textbf{Illustration of output tasks using diversity.} We illustrate our diverse task generation process on both \hocType~and \karelType~tasks. \figref{fig:appendix.experiments.mcts.hoc.diversity} shows the $10$ diverse tasks generated for code shown in \figref{fig:intro.hoc.p1}, while \figref{fig:appendix.experiments.mcts.karel.diversity} shows the $6$ diverse tasks generated for the code shown in \figref{fig:intro.karel.p1}.

%%%%%%%%%%%%%%%%%%%%%%%%%%%%%%%%%%%%%
\begin{figure}[h!]
\centering
	%%%%%%%%%%%%%%%%%
	\begin{subfigure}[b]{.18\textwidth}
	\centering
	{
		\includegraphics[height=2.39cm]{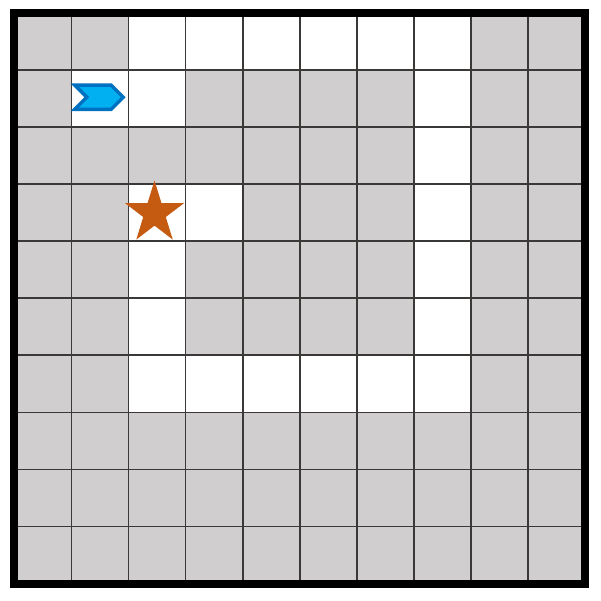}
		%\vspace{0.2mm}
		\caption{$\task^\textnormal{out}_{\textnormal{vis},1}$}
		 \label{fig:appendix.experiments.mcts.hoc.diversity.1}
	}
	\end{subfigure}
	%%%%%%%%%%%%%%%%%
	\begin{subfigure}[b]{.18\textwidth}
	\centering
	{
		\includegraphics[height=2.39cm]{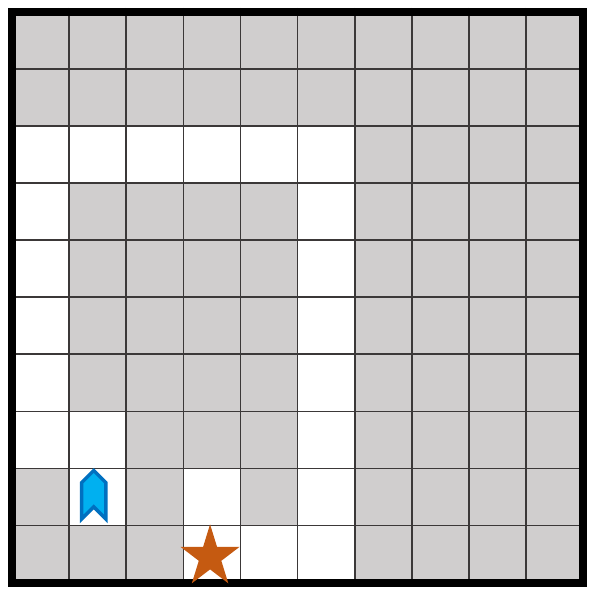}
		%\vspace{0.2mm}
		\caption{$\task^\textnormal{out}_{\textnormal{vis},2}$}
		 \label{fig:appendix.experiments.mcts.hoc.diversity.2}
	}
	\end{subfigure}
	%%%%%%%%%%%%%%%%%
	\begin{subfigure}[b]{.18\textwidth}
	\centering
	{
		\includegraphics[height=2.39cm]{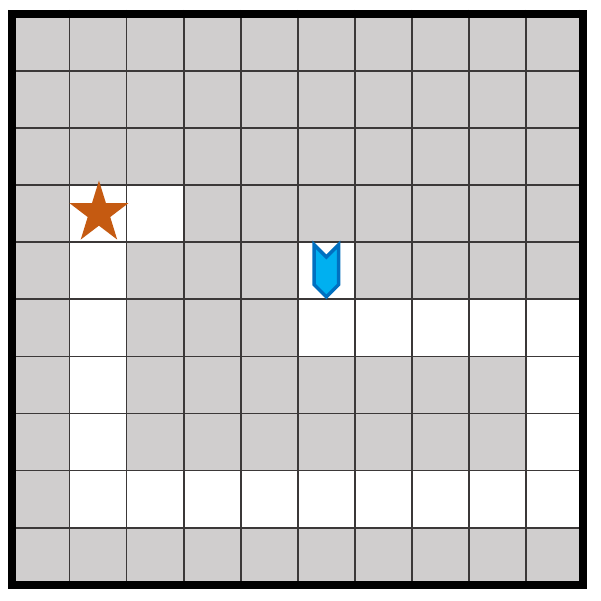}
		%\vspace{0.2mm}
		\caption{$\task^\textnormal{out}_{\textnormal{vis},3}$}
		 \label{fig:appendix.experiments.mcts.hoc.diversity.3}
	}
	\end{subfigure}
	%%%%%%%%%%%%%%%%%
	\begin{subfigure}[b]{.18\textwidth}
	\centering
	{
		\includegraphics[height=2.39cm]{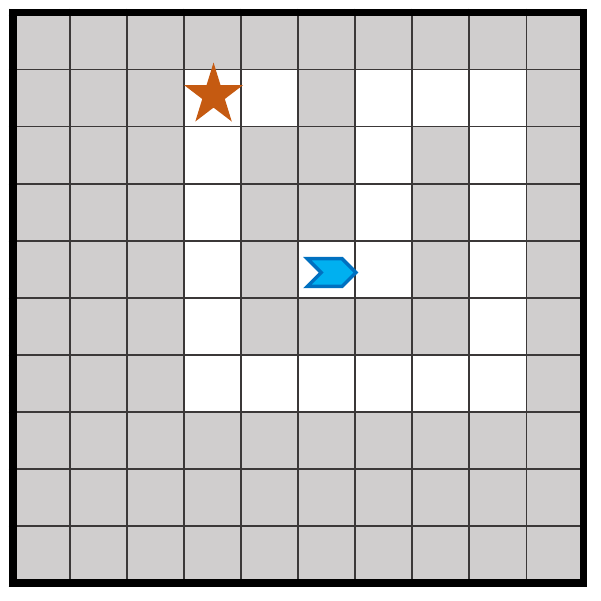}
		%\vspace{0.2mm}
		\caption{$\task^\textnormal{out}_{\textnormal{vis},4}$}
		 \label{fig:appendix.experiments.mcts.hoc.diversity.4}
	}
	\end{subfigure}	
	%%%%%%%%%%%%%%%%%
	\begin{subfigure}[b]{.18\textwidth}
	\centering
	{
		\includegraphics[height=2.39cm]{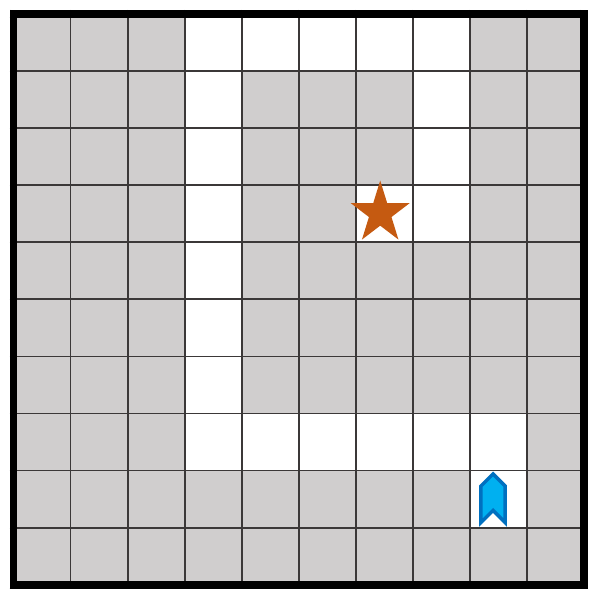}
		%\vspace{0.2mm}
		\caption{$\task^\textnormal{out}_{\textnormal{vis}, 5}$}
		 \label{fig:appendix.experiments.mcts.hoc.diversity.5}
	}
	\end{subfigure}		
	%%%%%%%%%%%%%%%%%
	\begin{subfigure}[b]{.18\textwidth}
	\centering
	{
		\includegraphics[height=2.39cm]{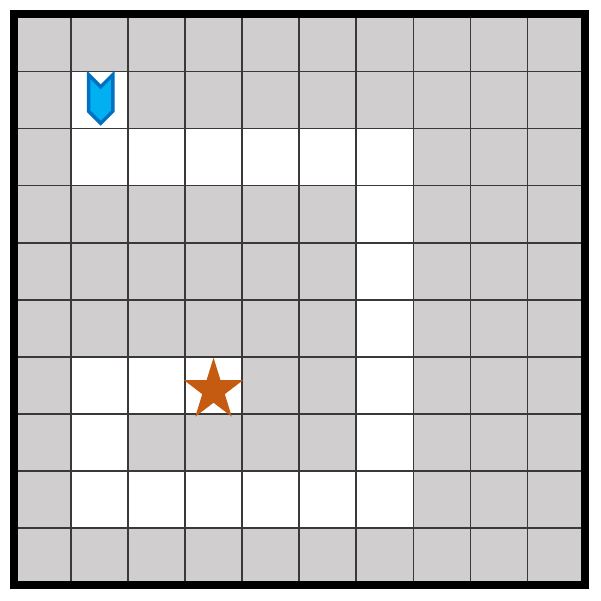}
		%\vspace{0.2mm}
		\caption{$\task^\textnormal{out}_{\textnormal{vis}, 6}$}
		 \label{fig:appendix.experiments.mcts.hoc.diversity.6}
	}
	\end{subfigure}
	%%%%%%%%%%%%%%%%%
	\begin{subfigure}[b]{.18\textwidth}
	\centering
	{
		\includegraphics[height=2.39cm]{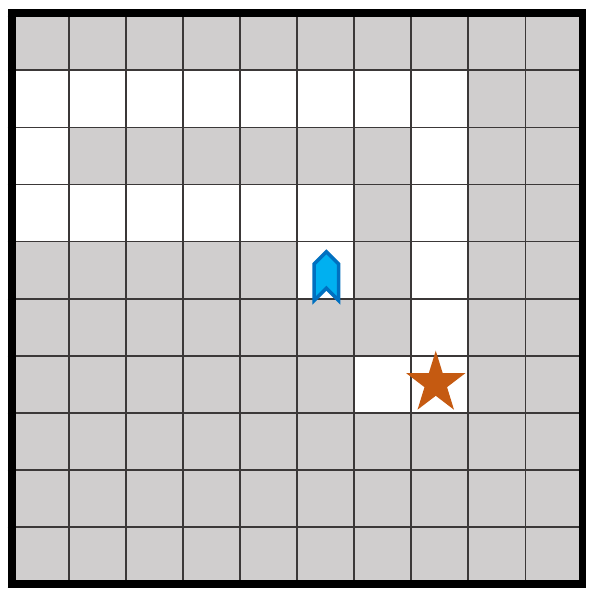}
		%\vspace{0.2mm}
		\caption{$\task^\textnormal{out}_{\textnormal{vis}, 7}$}
		 \label{fig:appendix.experiments.mcts.hoc.diversity.7}
	}
	\end{subfigure}
	%%%%%%%%%%%%%%%%%
	\begin{subfigure}[b]{.18\textwidth}
	\centering
	{
		\includegraphics[height=2.39cm]{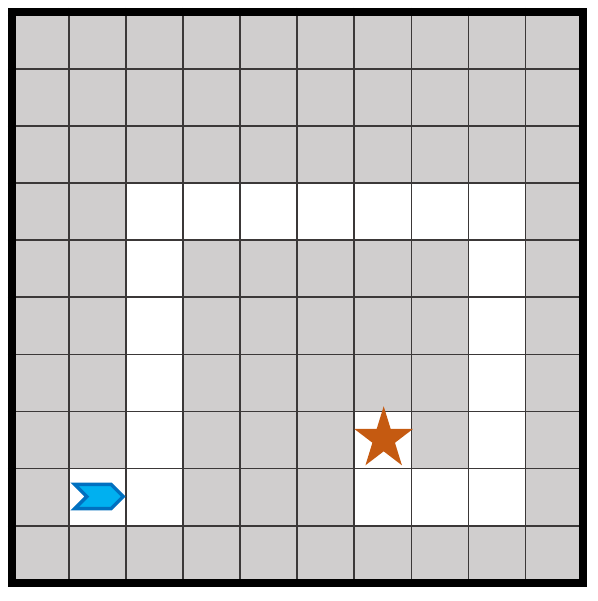}
		%\vspace{0.2mm}
		\caption{$\task^\textnormal{out}_{\textnormal{vis}, 8}$}
		 \label{fig:appendix.experiments.mcts.hoc.diversity.8}
	}
	\end{subfigure}
	%%%%%%%%%%%%%%%%%
	\begin{subfigure}[b]{.18\textwidth}
	\centering
	{
		\includegraphics[height=2.39cm]{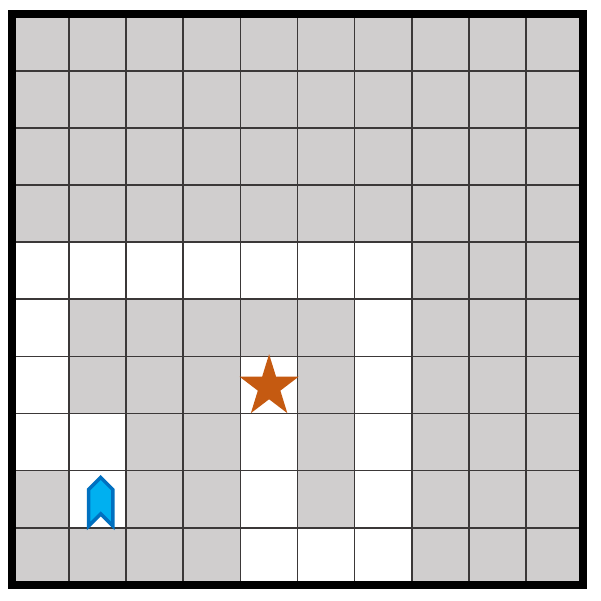}
		%\vspace{0.2mm}
		\caption{$\task^\textnormal{out}_{\textnormal{vis}, 9}$}
		 \label{fig:appendix.experiments.mcts.hoc.diversity.9}
	}
	\end{subfigure}	
	%%%%%%%%%%%%%%%%%
	\begin{subfigure}[b]{.18\textwidth}
	\centering
	{
		\includegraphics[height=2.39cm]{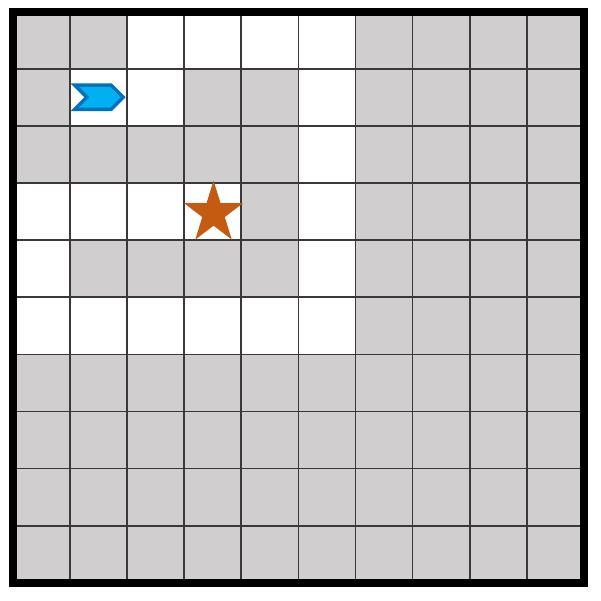}
		%\vspace{0.2mm}
		\caption{$\task^\textnormal{out}_{\textnormal{vis}, 10}$}
		 \label{fig:appendix.experiments.mcts.hoc.diversity.10}
	}
	\end{subfigure}		
	%%%%%%%%%%%%%%%%%
	%%%%%%%%%%%%%%%%%	
	%\vspace{-4mm}
	\caption{Illustration of task diversity on $\code^{\textnormal{out}}$ from \figref{fig:intro.hoc.p1} which was obtained from solution code of task \hocG~by mutation. All the $10$ diverse tasks generated are shown here.
	%using the modified MCTS evaluation function 
	}
	\vspace{-4mm}
	\label{fig:appendix.experiments.mcts.hoc.diversity}
\end{figure}
%%%%%%%%%%%%%%%%%%%%%%%%%%%%%%%%%%%%%
%%%%%%%%%%%%%%%%%%%%%%%%%%%%%%%%%%%%%%%%%%%%%%%%%%%%%%%%%%

%%%%%%%%%%%%%%%%%%%%%%%%%%%%%%%%%%%%%
\begin{figure}[h!]
\centering
	%%%%%%%%%%%%%%%%%
	\begin{subfigure}[b]{.32\textwidth}
	\centering
	{
		\includegraphics[height=2.00cm]{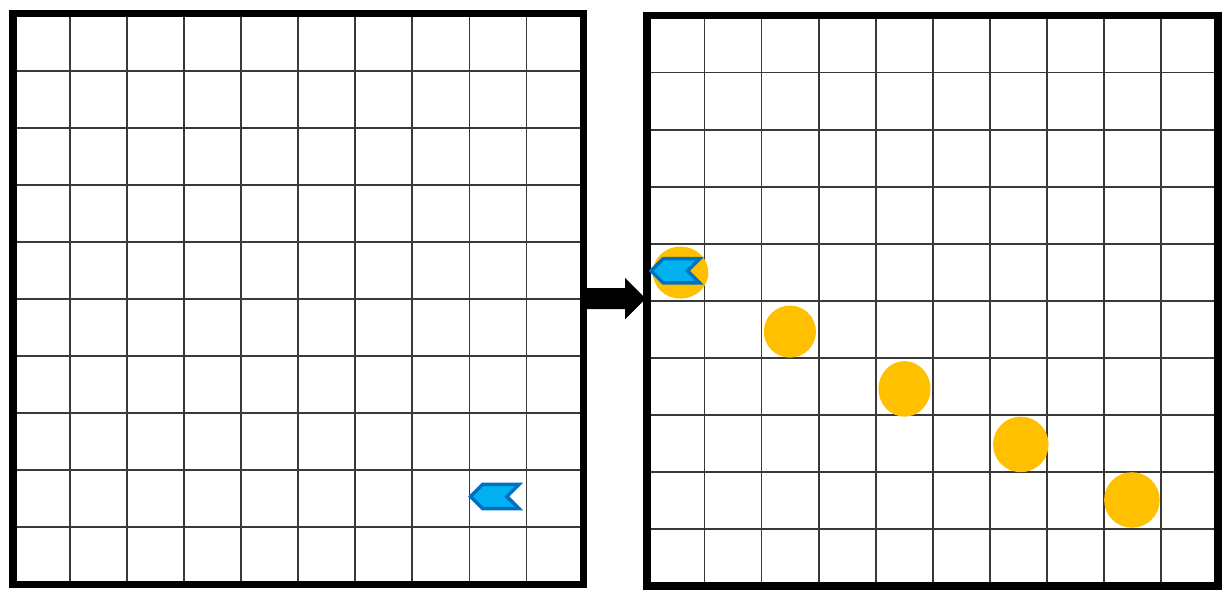}
		%\vspace{0.2mm}
		\caption{$\task^\textnormal{out}_{\textnormal{vis}, 1}$}
		 \label{fig:appendix.experiments.mcts.karel.diversity.1}
	}
	\end{subfigure}
	%%%%%%%%%%%%%%%%%
	\begin{subfigure}[b]{.32\textwidth}
	\centering
	{
		\includegraphics[height=2.00cm]{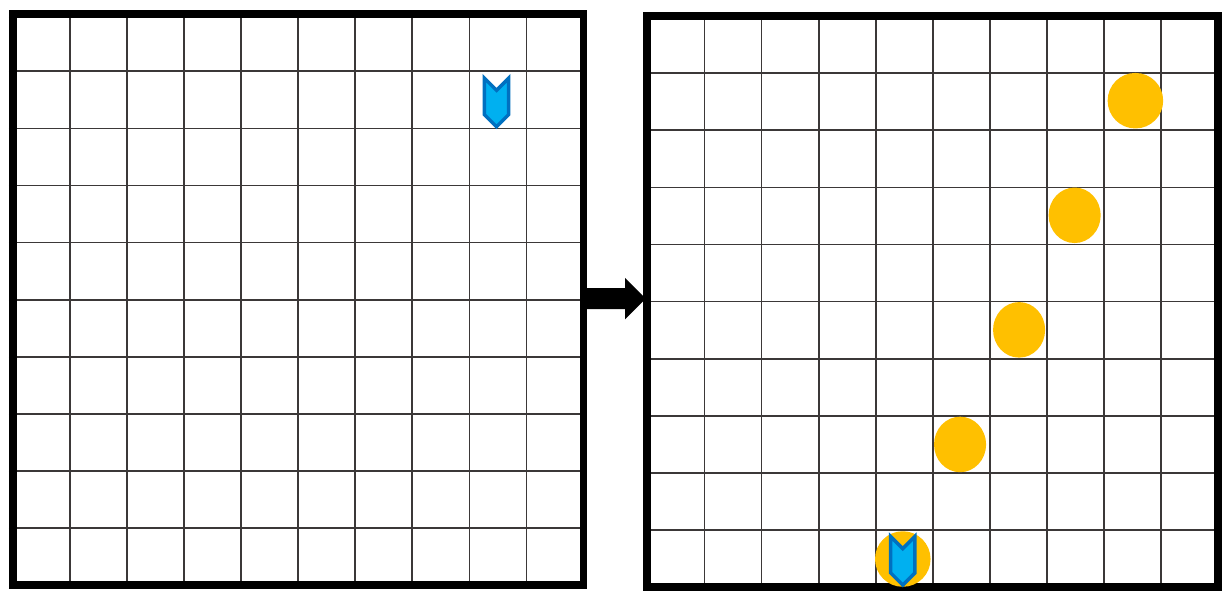}
		%\vspace{0.2mm}
		\caption{$\task^\textnormal{out}_{\textnormal{vis}, 2}$}
		 \label{fig:appendix.experiments.mcts.karel.diversity.2}
	}
	\end{subfigure}
	%%%%%%%%%%%%%%%%%
	\begin{subfigure}[b]{.32\textwidth}
	\centering
	{
		\includegraphics[height=2.00cm]{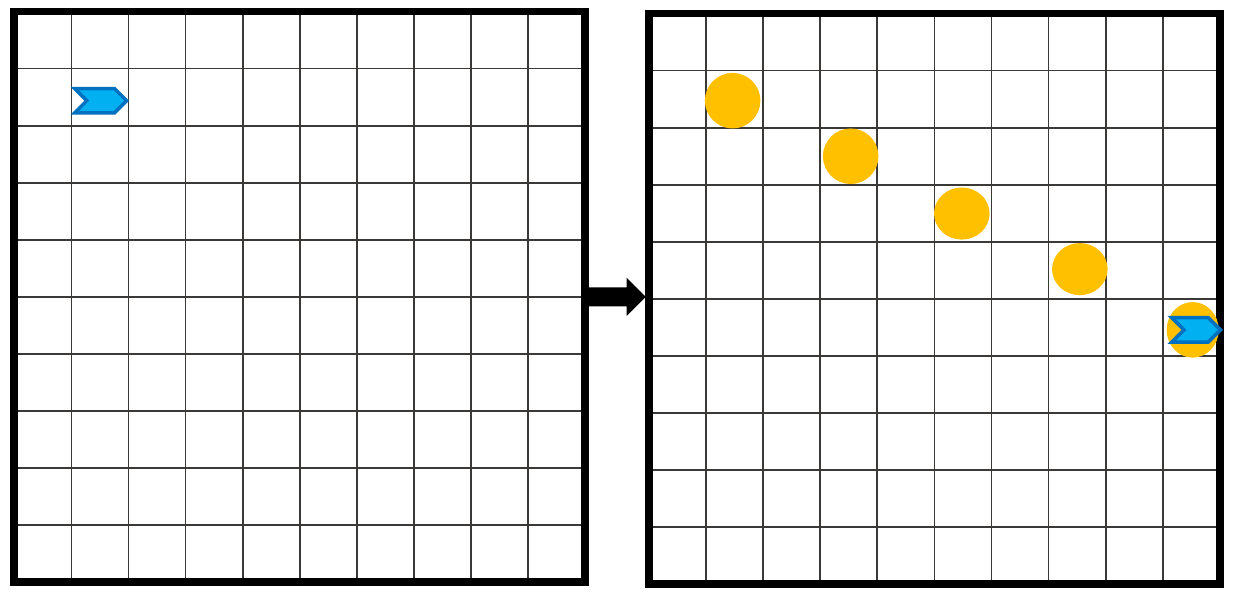}
		%\vspace{0.2mm}
		\caption{$\task^\textnormal{out}_{\textnormal{vis}, 3}$}
		 \label{fig:appendix.experiments.mcts.karel.diversity.3}
	}
	\end{subfigure}
	%%%%%%%%%%%%%%%%%
	\begin{subfigure}[b]{.32\textwidth}
	\centering
	{
		\includegraphics[height=2.00cm]{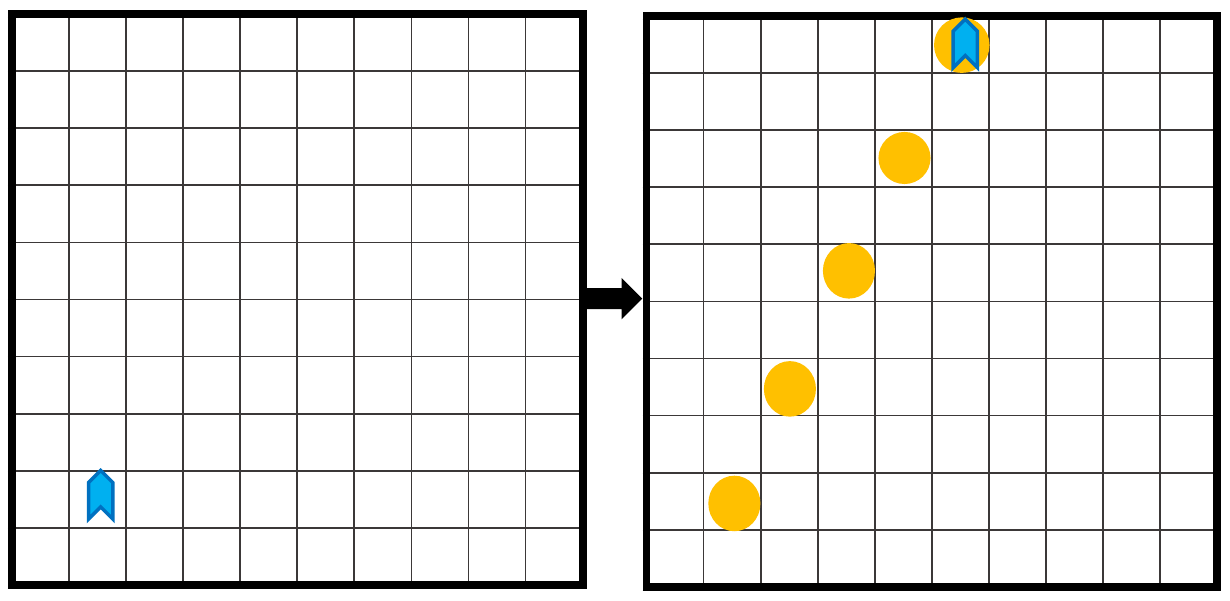}
		%\vspace{0.2mm}
		\caption{$\task^\textnormal{out}_{\textnormal{vis}, 4}$}
		 \label{fig:appendix.experiments.mcts.karel.diversity.6}
	}
	\end{subfigure}
	%%%%%%%%%%%%%%%%%
	\begin{subfigure}[b]{.32\textwidth}
	\centering
	{
		\includegraphics[height=2.00cm]{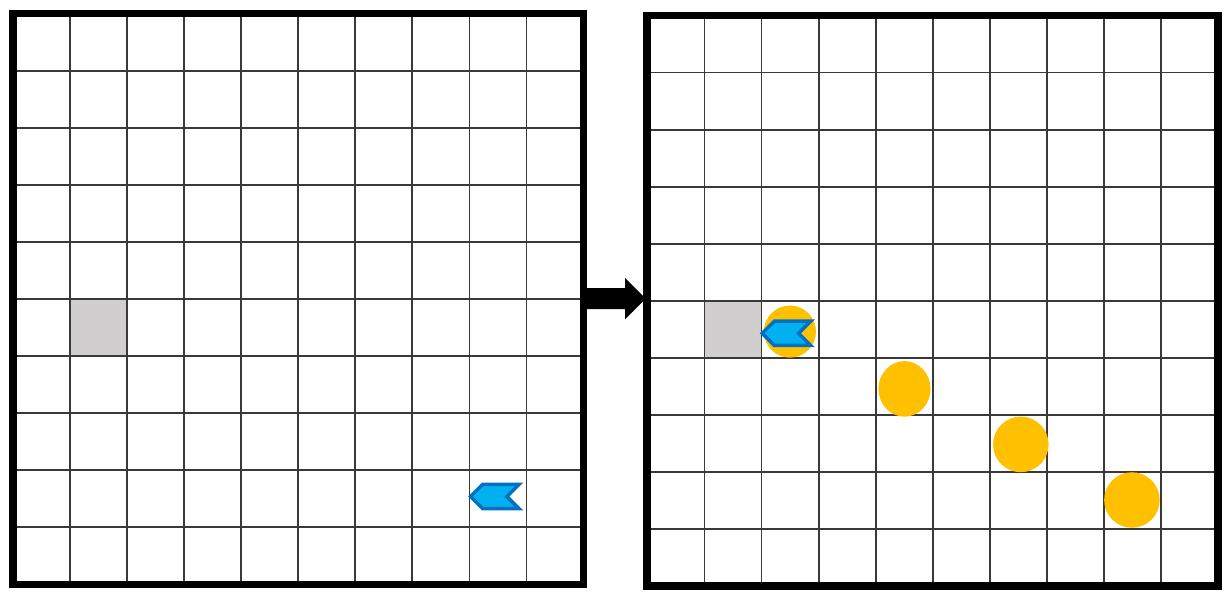}
		%\vspace{0.2mm}
		\caption{$\task^\textnormal{out}_{\textnormal{vis}, 5}$}
		 \label{fig:appendix.experiments.mcts.karel.diversity.7}
	}
	\end{subfigure}
	%%%%%%%%%%%%%%%%%
	\begin{subfigure}[b]{.32\textwidth}
	\centering
	{
		\includegraphics[height=2.00cm]{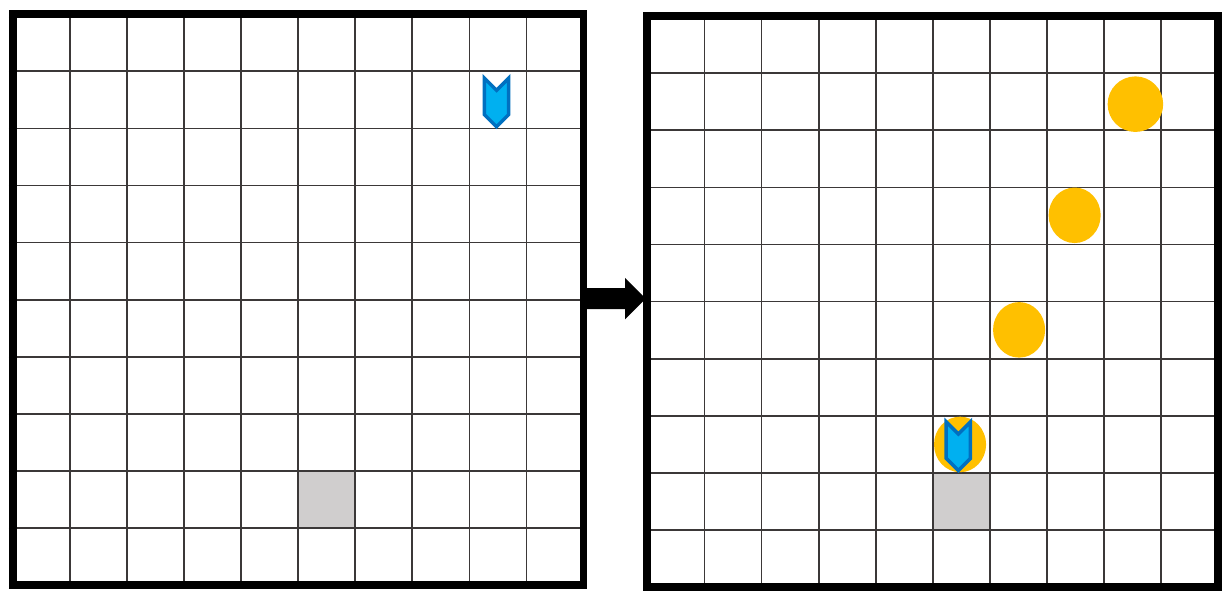}
		%\vspace{0.2mm}
		\caption{$\task^\textnormal{out}_{\textnormal{vis}, 6}$}
		 \label{fig:appendix.experiments.mcts.karel.diversity.8}
	}
	\end{subfigure}	
	%%%%%%%%%%%%%%%%%	
	%\vspace{-4mm}
	\caption{Illustration of task diversity on $\code^{\textnormal{out}}$ from \figref{fig:intro.karel.p1} which was obtained from solution code of task \karelF~by mutation. First $6$ diverse tasks are shown here.
	}
	\vspace{-4mm}
	\label{fig:appendix.experiments.mcts.karel.diversity}
\end{figure}
%%%%%%%%%%%%%%%%%%%%%%%%%%%%%%%%%%%%%
%%%%%%%%%%%%%%%%%%%%%%%%%%%%%%%%%%%%%%%%%%%%%%%%%%%%%%%%%%

\subsection{Insights into Results of \figref{fig:experiments.analysis}}
In this section, we provide further insights into the final results presented \figref{fig:experiments.analysis}. In particular, we illustrate few limitations of the codes and tasks we generate (w.r.t task-synthesis objectives defined in Section~\ref{sec.problem}):

%First, We illustrate the limitations of the codes and tasks generated by our algorithm on the $10$ reference tasks, and analyse their performance based on the  task-synthesis objectives defined in Section.\ref{sec.problem}. In particular, we illustrate three limitations of the codes and tasks we generate.

\begin{itemize}
\item \textbf{Limitations of mutated codes generated}: Column $4$ of \figref{fig:experiments.analysis} lists the number of codes generated by our mutation stage. However this set ($\#\code^\textnormal{out}_{\Delta=\textnormal{all}}$) continues to have few semantic irregularities (as discussed in Section \ref{sec.approach.mutation}). Some of these irregularities are illustrated in \figref{fig:appendix.experiments.problematic.pruned}. These codes are pruned by the symbolic execution stage of our algorithm, and the revised set of codes($\#\code^\textnormal{out}$) is listed in Column $6$ of \figref{fig:experiments.analysis}.

\item \textbf{Tasks which violate  task-synthesis objective $\textnormal{(V)}$}: Out of the final set of tasks generated (shown in Column $7$ of \figref{fig:experiments.analysis}), some of them violate objective (V). The fraction of the tasks which satisfy this particular objective are listed in Column $9$ of \figref{fig:experiments.analysis}. In \figref{fig:appendix.experiments.problematic.structure} we illustrate two examples which violate the objective, in output tasks for \hocG~and \hocH.
\item \textbf{Tasks which violate task-synthesis objective $\textnormal{(VI)}_{\delta_\textnormal{mini} = 1}$}: Column $10$ of \figref{fig:experiments.analysis} lists the fraction of the tasks generated that satisfy this particular objective, on task-minimality. In \figref{fig:appendix.experiments.problematic.minimality}, we illustrate two examples of tasks which violate this minimality constraint with $\delta_\textnormal{mini} = 1$.
\end{itemize}

%%%%%%%%%%%%%%%%%%%%%%%%%%%%%%%%%%%%%
\begin{figure}[h!]
\centering
	%%%%%%%%%%%%%%%%%	
	\begin{subfigure}[b]{.22\textwidth}
	\centering
	{
		\begin{boxcode}{3.8cm}{0.75}{0.7}
				\textcode{def }\DSLRun\textcode{()\{}\\
				\quad \DSLMove\\
				\quad \DSLTurnRight\\
				\quad \DSLMove\\
				\quad \DSLTurnRight\\
				\quad \DSLMove\\
				\textcode{\}}
				\\
				\\
				\\
				\vspace{4.3mm}
		\end{boxcode}
		\vspace{-2mm}
		\caption{\hocA: $\code^\textnormal{out}$}
		\label{fig:appendix.experiments.pruned.1}
    }
    \end{subfigure}
  	\quad  	
	%%%%%%%%%%%%%%%%%	
	\begin{subfigure}[b]{.22\textwidth}
	\centering
	{
		\begin{boxcode}{3.8cm}{0.75}{0.7}
				\textcode{def }\DSLRun\textcode{()\{}\\
				\quad \DSLTurnLeft\\				
				\quad \DSLRepeat\textcode{(5}\textcode{)\{}\\
				\quad \quad \DSLMove\\
				\quad \quad \DSLTurnRight\\
				\quad \textcode{\}}\\
				\textcode{\}}
				\\
				\\
				\\
				\\
				\vspace{0.5mm}
		\end{boxcode}
		\vspace{-2mm}
		\caption{\hocC: $\code^\textnormal{out}$}
		\label{fig:appendix.experiments.pruned.2}
    }
    \end{subfigure}
  	\quad  	
	%%%%%%%%%%%%%%%%%	
	\begin{subfigure}[b]{.22\textwidth}
	\centering
	{
		\begin{boxcode}{3.8cm}{0.75}{0.7}
				\textcode{def }\DSLRun\textcode{()\{}\\
				\quad \DSLRepeat\textcode{(4}\textcode{)\{}\\
				\quad \quad \DSLMove\\
				\quad \textcode{\}}\\
				\quad \DSLTurnLeft\\
				\quad \DSLMove\\
				\quad \DSLTurnLeft\\
				\quad \DSLRepeat\textcode{(6}\textcode{)\{}\\
				\quad \quad \DSLMove\\
				\quad \textcode{\}}\\
				\textcode{\}}
		\end{boxcode}
		\vspace{-2mm}
		\caption{\hocD: $\code^\textnormal{out}$}
		\label{fig:appendix.experiments.pruned.3}
    }
    \end{subfigure}
  	\quad  	
 	%%%%%%%%%%%%%%%%%	
	\begin{subfigure}[b]{.22\textwidth}
	\centering
	{
		\begin{boxcode}{3.8cm}{0.75}{0.7}
				\textcode{def }\DSLRun\textcode{()\{}\\
				\quad \DSLRepeatUntil\textcode{(}\DSLBoolGoal\textcode{)\{}\\
				\quad \quad \DSLMove\\
				\quad \quad \DSLTurnLeft\\
				\quad \quad \DSLMove\\
				\quad \quad \DSLTurnLeft\\
				\quad \textcode{\}}\\
				\textcode{\}}
				\\
				\\
				\\
				\vspace{1.7mm}
		\end{boxcode}
		\vspace{-2mm}
		\caption{\hocF: $\code^\textnormal{out}$}
		\label{fig:appendix.experiments.pruned.4}
    }
    \end{subfigure}
  	\\
 	%%%%%%%%%%%%%%%%%	
	\begin{subfigure}[b]{.22\textwidth}
	\centering
	{
		\begin{boxcode}{3.8cm}{0.75}{0.7}
				\textcode{def }\DSLRun\textcode{()\{}\\
				\quad \DSLRepeatUntil\textcode{(}\DSLBoolGoal\textcode{)\{}\\
				\quad \quad \DSLTurnRight\\
				\quad \quad \DSLTurnRight\\
				\quad \quad \DSLMove\\
				\quad \quad \DSLIf\textcode{(}\DSLBoolPathRight\textcode{)\{}\\
				\quad \quad \quad \DSLTurnRight\\
				\quad \quad \textcode{\}}\\
				\quad \textcode{\}}\\
				\textcode{\}}
				\\
				\vspace{1mm}
		\end{boxcode}
		\vspace{-2mm}
		\caption{\hocG: $\code^\textnormal{out}$}
		\label{fig:appendix.experiments.pruned.5}
    }
    \end{subfigure}
  	\quad  	
 	%%%%%%%%%%%%%%%%%	
	\begin{subfigure}[b]{.22\textwidth}
	\centering
	{
		\begin{boxcode}{3.8cm}{0.75}{0.7}
				\textcode{def }\DSLRun\textcode{()\{}\\
				\quad \DSLRepeatUntil\textcode{(}\DSLBoolGoal\textcode{)\{}\\
				\quad \quad \DSLIf\textcode{(}\DSLBoolPathAhead\textcode{)\{}\\
				\quad \quad \quad \DSLMove\\
				\quad \quad \quad \DSLTurnLeft\\
				\quad \quad \textcode{\}}\\
				%\quad \quad \DSLElse\textcode{(}\textcode{)\{}\\
				\quad \quad \DSLElse\textcode{\{}\\
				\quad \quad \quad \DSLTurnLeft\\
				\quad \quad \textcode{\}}\\				
				\quad \textcode{\}}\\
				\textcode{\}}
		\end{boxcode}
		\vspace{-2mm}
		\caption{\hocH: $\code^\textnormal{out}$}
		\label{fig:appendix.experiments.pruned.6}
    }
    \end{subfigure}
  	\quad  	
 	%%%%%%%%%%%%%%%%%	
	\begin{subfigure}[b]{.22\textwidth}
	\centering
	{
		\begin{boxcode}{3.8cm}{0.75}{0.70}
				\textcode{def }\DSLRun\textcode{()\{}\\
				\quad \DSLRepeat\textcode{(5}\textcode{)\{}\\
				\quad \quad \DSLPickMarker\\
				\quad \quad \DSLMove\\
				\quad \quad \DSLTurnRight\\
				\quad \quad \DSLPutMarker\\				
				\quad \textcode{\}}\\
				\textcode{\}}
				\\
				\\
				\\				
				\vspace{1.8mm}
		\end{boxcode}
		\vspace{-2mm}
		\caption{\karelC: $\code^\textnormal{out}$}
		\label{fig:appendix.experiments.pruned.7}
    }
    \end{subfigure}
  	\quad  	
 	%%%%%%%%%%%%%%%%%	
	\begin{subfigure}[b]{.22\textwidth}
	\centering
	{
		\begin{boxcode}{3.8cm}{0.75}{0.70}
			\textcode{def }\DSLRun\textcode{()\{}\\
			\quad \DSLPutMarker\\
			\quad \DSLWhile\textcode{(}\DSLBoolPathAhead\textcode{)\{}\\
			\quad \quad \DSLMove\\
			\quad \quad \DSLTurnLeft\\
			\quad \quad \DSLMove\\
			\quad \quad \DSLTurnLeft\\
			\quad \quad \DSLPutMarker\\	
			\quad \textcode{\}}\\
			\textcode{\}}
			\\
			\vspace{2mm}
		\end{boxcode}
		\vspace{-2mm}
		\caption{\karelF: $\code^\textnormal{out}$}
		\label{fig:appendix.experiments.pruned.8}
    }
    \end{subfigure}
  	\quad  	  	  	
	%%%%%%%%%%%%%%%%%		
	\vspace{-3.5mm}
    \caption{Illustration of codes with semantic irregularities in Column $\code^\textnormal{out}_{\Delta=\textnormal{all}}$ of \figref{fig:experiments.analysis}. (a)--(f) show the semantic irregularities in mutated codes generated for \hocType~tasks. All of these codes lead to circular paths in output task, and are pruned out in the symbolic execution stage, by the \FNoCut~component of the \FScore~measure. In particular, consider the semantic irregularity presented in (g). This code corresponding to the reference task \karelC, has redundant marker activity. With each iteration of \DSLRepeat, the \DSLPickMarker~and \DSLPutMarker~actions occur consecutively, leading to no change in marker activity in the output. (h) illustrates a mutated code for reference task \karelF~where if the \DSLWhile~is executed more than once, \DSLPutMarker~activity occurs in the same location consecutively leading to a crash (as only one marker is allowed per grid-cell); if \DSLWhile~is executed only once, the corresponding task generated has a short-cut to the goal.
	}
	%(h) illustrates a mutated code for reference task \karelF~where if the 'while-loop' is executed more than once, \DSLPutMarker~activity occurs in the same location consecutively leading to a crash (as only one marker is allowed per grid-cell). (ii) if the 'while-loop' is executed only once, the corresponding task generated has a short-cut to the goal.
	\vspace{-2.5mm}
	\label{fig:appendix.experiments.problematic.pruned}
\end{figure}
%%%%%%%%%%%%%%%%%%%%%%%%%%%%%%%%%%%%%
%%%%%%%%%%%%%%%%%%%%%%%%%%%%%%%%%%%%%%%%%%%%%%%%%%%%%%%%%%

%%%%%%%%%%%%%%%%%%%%%%%%%%%%%%%%%%%%%
\begin{figure}[h!]
\centering
	%%%%%%%%%%%%%%%%%
	\begin{subfigure}[b]{.22\textwidth}
	\centering
	{
		\includegraphics[height=2.39cm]{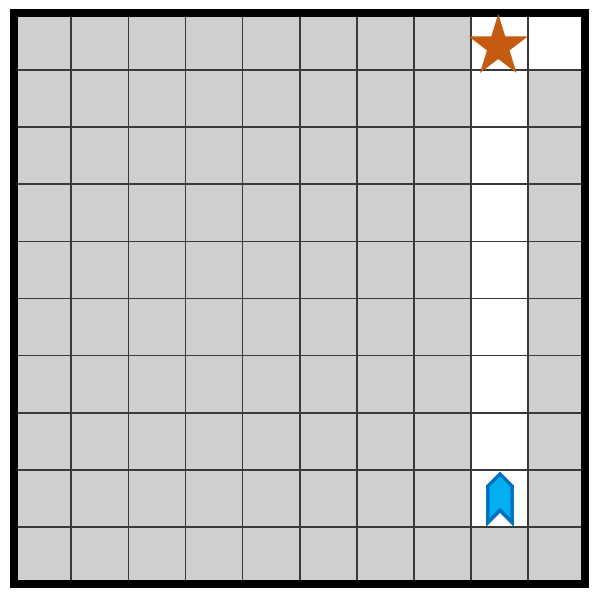}
		%\vspace{0.2mm}
		\caption{\hocG: $\task^\textnormal{out}_\textnormal{vis}$}
		 \label{fig:appendix.experiments.problematic.structure.1}
	}
	\end{subfigure}
	%%%%%%%%%%%%%%%%%	
	\begin{subfigure}[b]{.22\textwidth}
	\centering
	{
		\begin{boxcode}{4.0cm}{0.75}{0.7}
				\textcode{def }\DSLRun\textcode{()\{}\\
				\quad \DSLRepeatUntil\textcode{(}\DSLBoolGoal\textcode{)\{}\\
				\quad \quad \DSLMove\\
				\quad \quad \DSLIf\textcode{(}\DSLBoolPathRight\textcode{)\{}\\
				\quad \quad \quad \DSLTurnRight\\
				\quad \quad \quad \DSLTurnRight\\
				\quad \quad \textcode{\}}\\
				\quad \textcode{\}}\\
				\textcode{\}}
				\\
				\\
				\\
				\vspace{1mm}
		\end{boxcode}
		\vspace{-1mm}
		\caption{\hocG: $\code^\textnormal{out}$}
		\label{fig:appendix.experiments.problematic.structure.2}
    }
    \end{subfigure}
  	\quad \quad 
	%%%%%%%%%%%%%%%%%
	\begin{subfigure}[b]{.22\textwidth}
	\centering
	{
		\includegraphics[height=2.39cm]{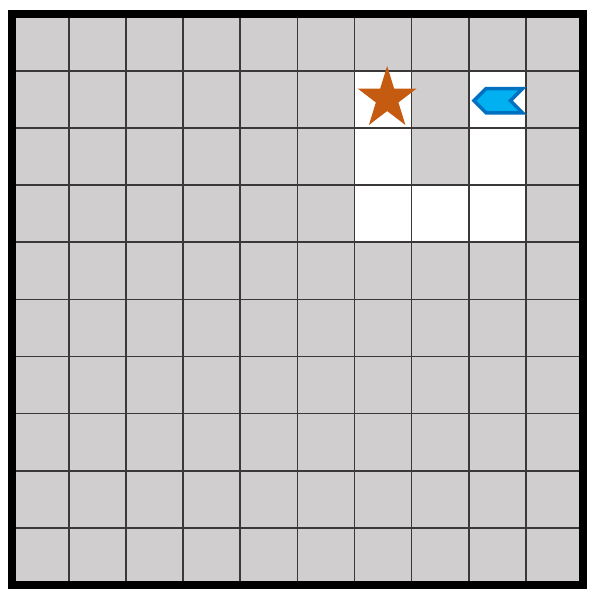}
		%\vspace{0.2mm}
		\caption{\hocH: $\task^\textnormal{out}_\textnormal{vis}$}
		 \label{fig:appendix.experiments.problematic.structure.3}
	}
	\end{subfigure}
	%%%%%%%%%%%%%%%%%	
	\begin{subfigure}[b]{.22\textwidth}
	\centering
	{
		\begin{boxcode}{4.0cm}{0.75}{0.7}
				\textcode{def }\DSLRun\textcode{()\{}\\
				\quad \DSLRepeatUntil\textcode{(}\DSLBoolGoal\textcode{)\{}\\
				\quad \quad \DSLIf\textcode{(}\DSLBoolPathAhead\textcode{)\{}\\
				\quad \quad \quad \DSLMove\\
				\quad \quad \quad \DSLMove\\				
				\quad \quad \quad \DSLTurnRight\\
				\quad \quad \textcode{\}}\\
				\quad \quad \DSLElse\textcode{\{}\\
				\quad \quad \quad \DSLTurnRight\\
				\quad \quad \textcode{\}}\\				
				\quad \textcode{\}}\\
				\textcode{\}}
				%\vspace{-1mm}
		\end{boxcode}
		\vspace{-1mm}
		\caption{\hocH: $\code^\textnormal{out}$}
		\label{fig:appendix.experiments.problematic.structure.4}
    }
    \end{subfigure}
  	\quad  	
	%%%%%%%%%%%%%%%%%		
	\vspace{-1.5mm}
\caption{Illustration of violation of task-synthesis objective (V) by generated output tasks for reference tasks \hocG~and \hocH. (a, b) show the irregularity in generated task ($\task^\textnormal{out}_\textnormal{vis}, \code^\textnormal{out}$) for \hocG~which leads to only a straight path in the visual-puzzle. The corresponding code renders the \DSLIf~construct redundant. (c, d) illustrate a similar irregularity in the generated task ($\task^\textnormal{out}_\textnormal{vis}, \code^\textnormal{out}$) for \hocH. Here, \DSLIfElse~construct is not needed for the optimal solution.
	}
	%(c, d) illustrate a similar irregularity in the generated task ($\task^\textnormal{out}_\textnormal{vis}, \code^\textnormal{out}$) for \hocH. Here, the \DSLTurnRight~action within the \DSLIf~construct is rendered redundant.
	\vspace{-2.5mm}
	\label{fig:appendix.experiments.problematic.structure}
\end{figure}
%%%%%%%%%%%%%%%%%%%%%%%%%%%%%%%%%%%%%
%%%%%%%%%%%%%%%%%%%%%%%%%%%%%%%%%%%%%%%%%%%%%%%%%%%%%%%%%%

%%%%%%%%%%%%%%%%%%%%%%%%%%%%%%%%%%%%%
\begin{figure}[h!]
\centering
	%%%%%%%%%%%%%%%%%
	\begin{subfigure}[b]{.22\textwidth}
	\centering
	{
		\includegraphics[height=2.39cm]{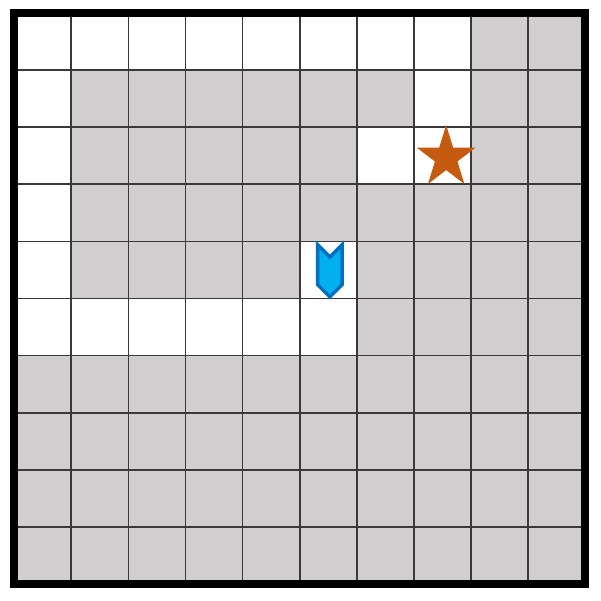}
		%\vspace{0.2mm}
		\caption{\hocG: $\task^\textnormal{out}_\textnormal{vis}$}
		 \label{fig:appendix.experiments.problematic.minimality.1}
	}
	\end{subfigure}
	%%%%%%%%%%%%%%%%%	
	\begin{subfigure}[b]{.22\textwidth}
	\centering
	{
		\begin{boxcode}{4.0cm}{0.75}{0.7}
				\textcode{def }\DSLRun\textcode{()\{}\\
				\quad \DSLMove\\
				\quad \DSLTurnRight\\
				\quad \DSLRepeatUntil\textcode{(}\DSLBoolGoal\textcode{)\{}\\
				\quad \quad \DSLMove\\
				\quad \quad \DSLIf\textcode{(}\DSLBoolPathRight\textcode{)\{}\\
				\quad \quad \quad \DSLTurnRight\\
				\quad \quad \textcode{\}}\\
				\quad \textcode{\}}\\
				\textcode{\}}
				\\
				\vspace{3mm}
		\end{boxcode}
		\vspace{-1mm}
		\caption{\hocG: $\code^\textnormal{out}$}
		\label{fig:appendix.experiments.problematic.minimality.2}
    }
    \end{subfigure}
  	\quad \quad
	%%%%%%%%%%%%%%%%%
	\begin{subfigure}[b]{.22\textwidth}
	\centering
	{
		\includegraphics[height=2.39cm]{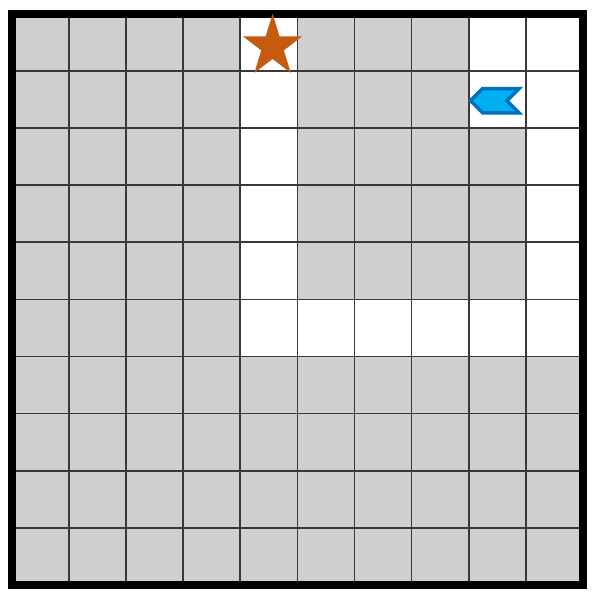}
		%\vspace{0.2mm}
		\caption{\hocH: $\task^\textnormal{out}_\textnormal{vis}$}
		 \label{fig:appendix.experiments.problematic.minimality.3}
	}
	\end{subfigure}
	%%%%%%%%%%%%%%%%%	
	\begin{subfigure}[b]{.22\textwidth}
	\centering
	{
		\begin{boxcode}{4.0cm}{0.75}{0.7}
				\textcode{def }\DSLRun\textcode{()\{}\\
				\quad \DSLTurnRight\\
				\quad \DSLMove\\
				\quad \DSLRepeatUntil\textcode{(}\DSLBoolGoal\textcode{)\{}\\
				\quad \quad \DSLIf\textcode{(}\DSLBoolPathAhead\textcode{)\{}\\
				\quad \quad \quad \DSLMove\\
				\quad \quad \textcode{\}}\\
				\quad \quad \DSLElse\textcode{\{}\\
				\quad \quad \quad \DSLTurnRight\\
				\quad \quad \textcode{\}}\\				
				\quad \textcode{\}}\\
				\textcode{\}}
				%\vspace{-1mm}
		\end{boxcode}
		\vspace{-1mm}
		\caption{\hocH: $\code^\textnormal{out}$}
		\label{fig:appendix.experiments.problematic.minimality.4}
    }
    \end{subfigure}
  	\quad  	
	%%%%%%%%%%%%%%%%%		
	\vspace{-1.5mm}
	\caption{\looseness-1 Illustration of violation of task-synthesis objective (V1), with $\delta_\textnormal{mini} = 1$ by generated output tasks for reference tasks \hocG~and \hocH. (a, b) show the violation of task minimality, for the task ($\task^\textnormal{out}_\textnormal{vis}, \code^\textnormal{out}$) generated for \hocG. In particular, the first two actions in the code shown in (b) are unnecessary. Similarly, (c, d) show the violation of task minimality, for the task ($\task^\textnormal{out}_\textnormal{vis}, \code^\textnormal{out}$) generated for \hocH.
	}
	\vspace{-2.5mm}
	\label{fig:appendix.experiments.problematic.minimality}
\end{figure}
\subsection{Adding More Variability to Output Tasks}\label{appendix.sec.simulations.more_variability}
Here, we propose a few simple extensions to our task generation algorithm allowing us to add more variability in the visual puzzles. We show that high variability can be achieved through suitable post-processing and pre-processing of the generated tasks (happening after and before the symbolic execution process, respectively). We describe each of these strategies next.

\textbf{Different grid-sizes.}
A simple strategy to enhance task variability is by altering the grid-size parameter $n$. \figref{fig:appendix.gridsz} illustrates the tasks generated with different values of the grid-size parameter $n$. For instance, this strategy is employed in generating multiple input-output pairs for Karel tasks in \textit{Intro to Programming with Karel} course by \textit{CodeHS.com}~\cite{intro_to_karel_codehs}.

%%%%%%%%%%%%%%%%%%%%%%%%%%%%%%%%%%%%%
\begin{figure}[h!]
\centering
	%%%%%%%%%%%%%%%%%
	\begin{subfigure}[b]{.22\textwidth}
	\centering
	{
		\includegraphics[height=2.39cm]{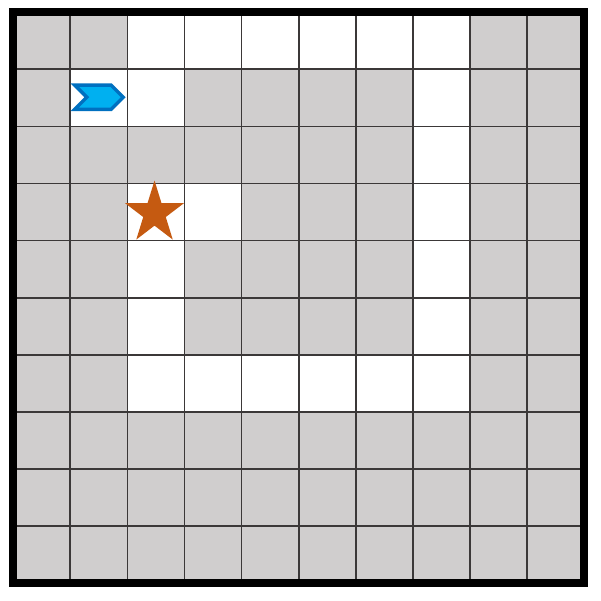}
		%\vspace{0.2mm}
		\caption{$\task^{\textnormal{out}}_{\textnormal{vis}}$; $n = 10$}
		 \label{fig:appendix.samegridsz}
	}
	\end{subfigure}
	%%%%%%%%%%%%%%%%%
	\begin{subfigure}[b]{.22\textwidth}
	\centering
	{
		\includegraphics[height=2.39cm]{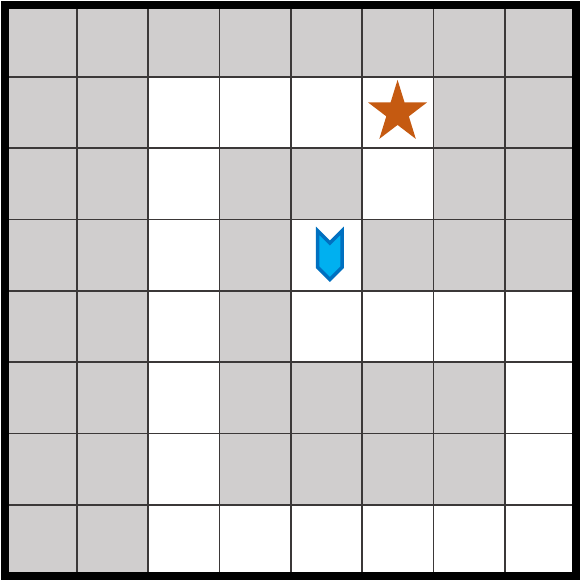}
		%\vspace{0.2mm}
		\caption{$\task^{\textnormal{out}}_{\textnormal{vis}}$; $n = 8$}
		 \label{fig:appendix.gridsz8}
	}
	\end{subfigure}
	%%%%%%%%%%%%%%%%%
	\begin{subfigure}[b]{.22\textwidth}
	\centering
	{
		\includegraphics[height=2.39cm]{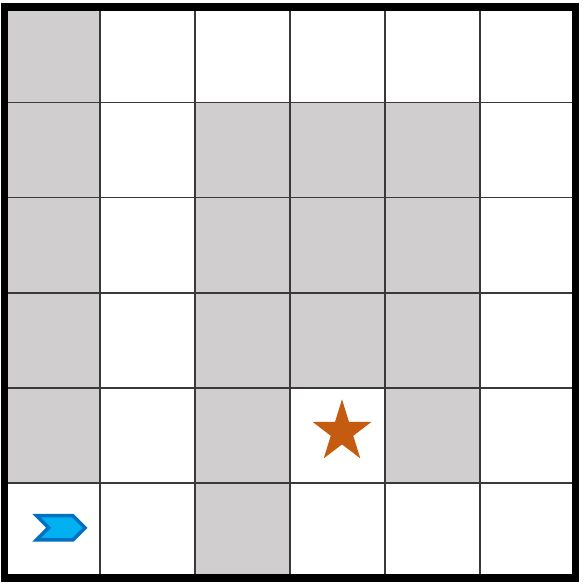}
		%\vspace{0.2mm}
		\caption{$\task^{\textnormal{out}}_{\textnormal{vis}}$; $n = 6$}
		 \label{fig:appendix.gridsz6}
	}
	\end{subfigure}
	%%%%%%%%%%%%%%%%%	
	%\vspace{-4mm}
	\caption{
	Tasks generated for $\code^{\textnormal{out}}$ from \figref{fig:intro.hoc.p1}  when varying grid-size ($n$).
	% obtained from solution code of task \hocG~by mutation
	}
	%\vspace{-4mm}
	\label{fig:appendix.gridsz}
\end{figure}
%%%%%%%%%%%%%%%%%%%%%%%%%%%%%%%%%%%%%
%%%%%%%%%%%%%%%%%%%%%%%%%%%%%%%%%%%%%%%%%%%%%%%%%%%%%%%%%%

\textbf{Post-processing: Using distractor paths.}
One of the ways to add variability to the tasks is by adding \emph{distractor-paths}, to the generated tasks. These paths are added after the symbolic execution stage is complete, and a basic task has been generated. \figref{fig:appendix.nopostprocess} shows the task generated  without any post-processing of the output of symbolic execution, carried out on $\code^{\textnormal{out}}$ from \figref{fig:intro.hoc.p1}. \figref{fig:appendix.postprocess1}, and \figref{fig:appendix.postprocess2} illustrate two different post-processing patterns that yield tasks with greater variability, for the same code.
% which was obtained from solution code of task \hocG~by mutation

\textbf{Pre-processing: Using different initializations of grid cells.}
We could also add task variability by initializing the grid-cells differently, before they are subjected to symbolic execution for task synthesis. We have a set of fixed grid-patterns, which when chosen as initializations of the grid-world yield very different looking output tasks. \figref{fig:appendix.nopreprocessing} shows the task generated for $\code^{\textnormal{out}}$ from \figref{fig:intro.hoc.p1}~without any pre-processing on the grid-cells. \figref{fig:appendix.preprocessing1} and \figref{fig:appendix.preprocessing2} show two different tasks obtained using different grid-initializations, for the same code. 

\clearpage

%%%%%%%%%%%%%%%%%%%%%%%%%%%%%%%%%%%%%
\begin{figure}[t!]
\centering
	%%%%%%%%%%%%%%%%%
	\begin{subfigure}[b]{.465\textwidth}
	\centering
	{
		\includegraphics[trim={0 0.55cm 0 0},clip,height=2.45cm]{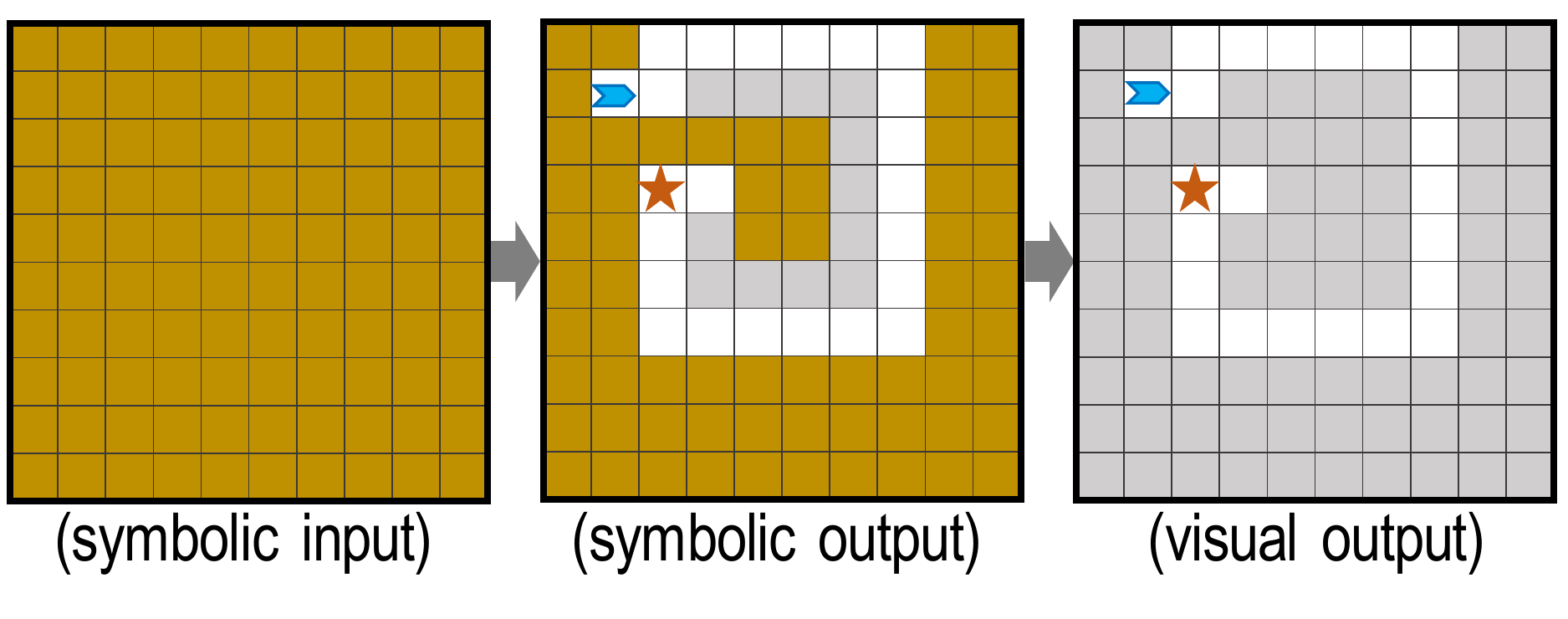}
		\vspace{-4mm}
		\caption{No post processing}
		 \label{fig:appendix.nopostprocess}
	}
	\end{subfigure}
	\\
	%%%%%%%%%%%%%%%%%
	\begin{subfigure}[b]{.465\textwidth}
	\centering
	{
		\includegraphics[trim={0 0.55cm 0 0},clip,height=2.45cm]{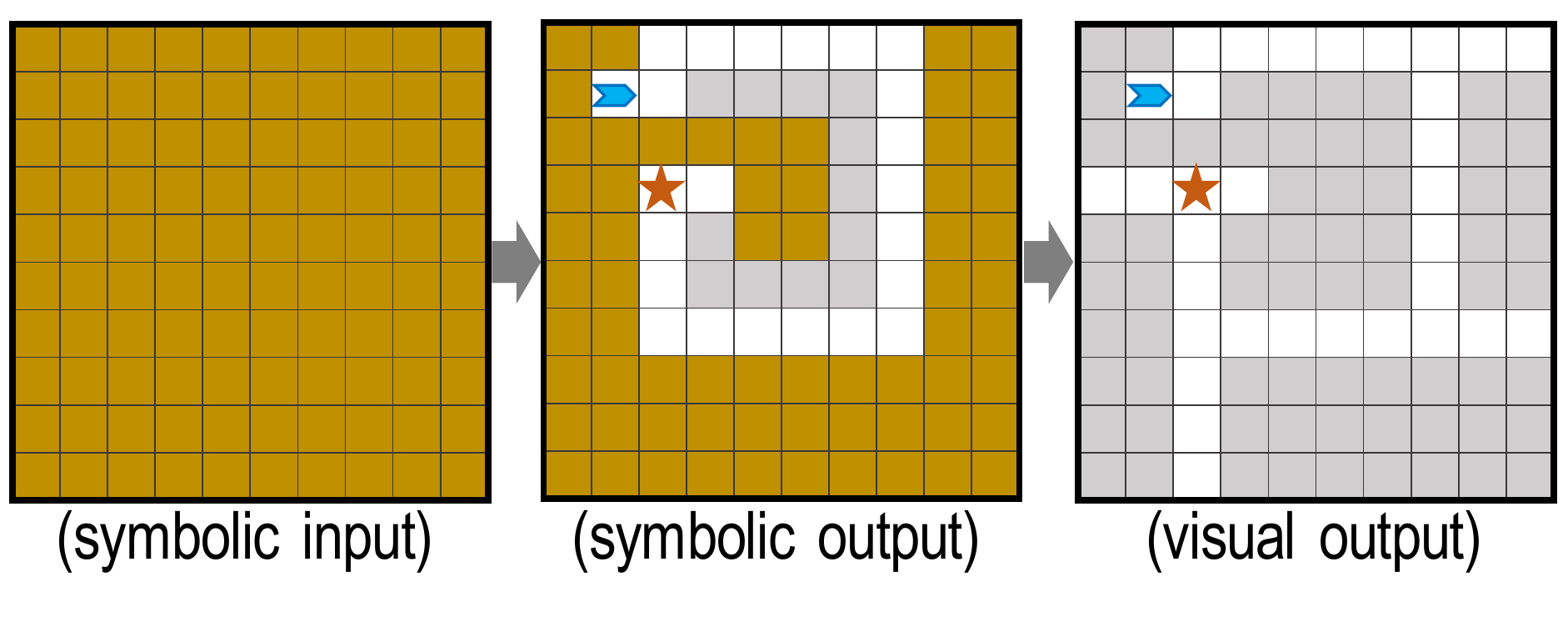}
		\vspace{-4mm}
		\caption{Post-processing 1}
		 \label{fig:appendix.postprocess1}
	}
	\end{subfigure}
	\quad \ \ 
	%%%%%%%%%%%%%%%%%
	\begin{subfigure}[b]{.465\textwidth}
	\centering
	{
		\includegraphics[trim={0 0.55cm 0 0},clip,height=2.45cm]{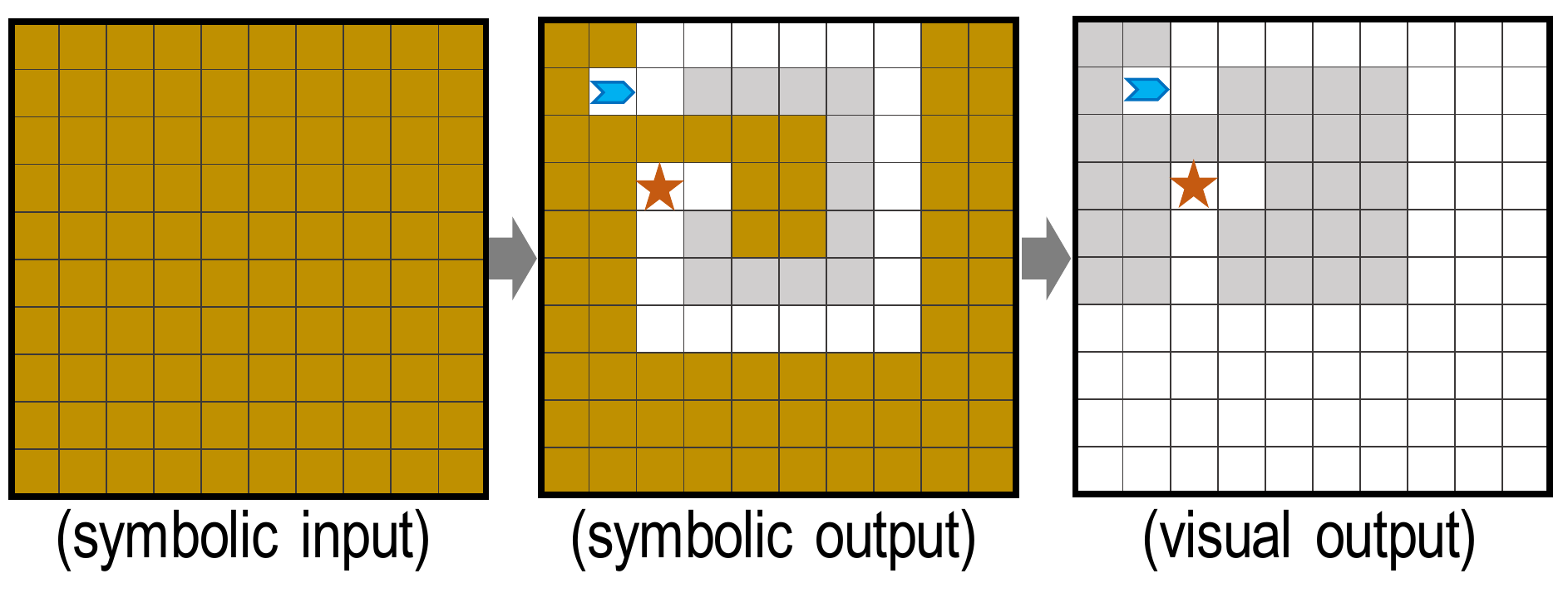}
		\vspace{-4mm}
		\caption{Post-processing 2}
		 \label{fig:appendix.postprocess2}
	}
	\end{subfigure}
	%%%%%%%%%%%%%%%%%	
	%\vspace{-2mm}
	\caption{Illustration of the (post-processing) distractor path strategy to increase task variability on tasks generated from $\code^{\textnormal{out}}$ in \figref{fig:intro.hoc.p1} which was obtained from solution code of task \hocG~by mutation. (a) illustrates the basic task obtained after symbolic execution. (b, c) show two different distractor paths added after the symbolic execution stage, yielding visually very different tasks.
	}
	%\vspace{-4mm}
	\label{fig:appendix.postprocess}
\end{figure}
%%%%%%%%%%%%%%%%%%%%%%%%%%%%%%%%%%%%%
%%%%%%%%%%%%%%%%%%%%%%%%%%%%%%%%%%%%%%%%%%%%%%%%%%%%%%%%%%

%%%%%%%%%%%%%%%%%%%%%%%%%%%%%%%%%%%%%
\begin{figure}[t!]
\centering
	%%%%%%%%%%%%%%%%%
	\begin{subfigure}[b]{.465\textwidth}
	\centering
	{
		\includegraphics[trim={0 0.55cm 0 0},clip,height=2.45cm]{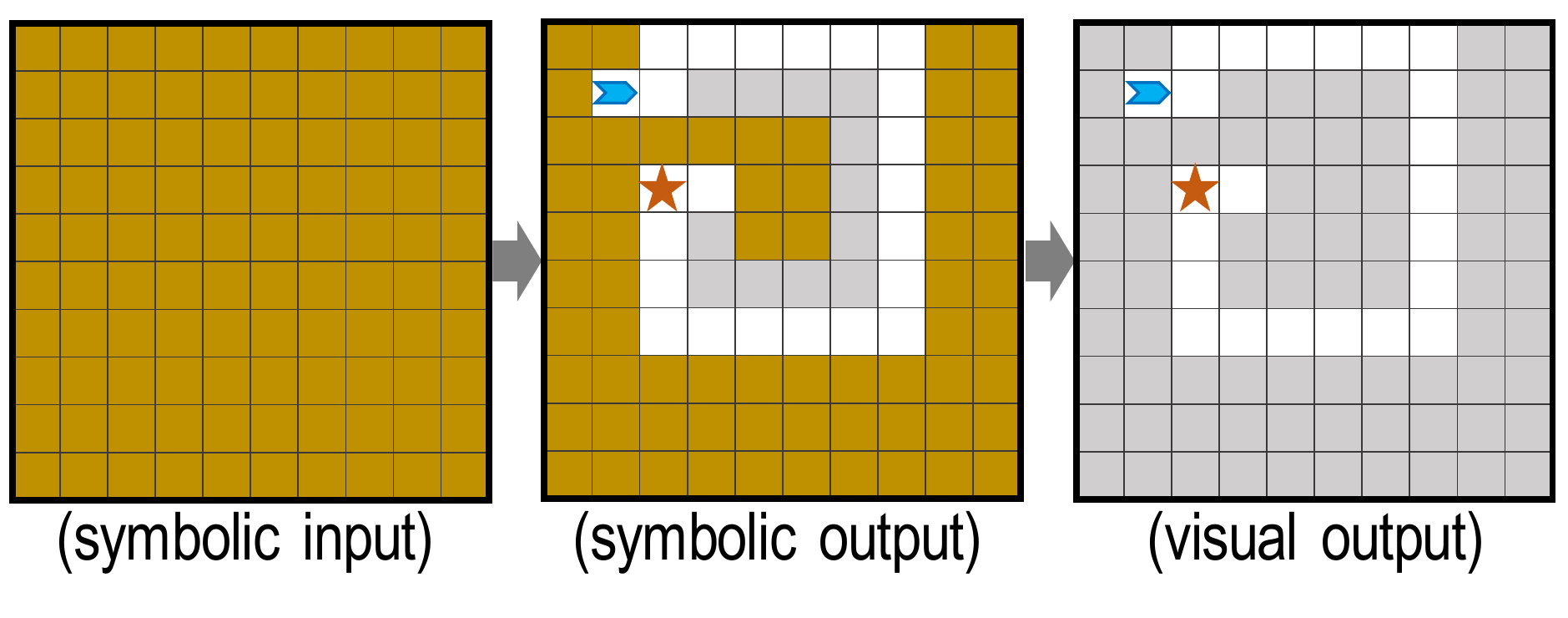}
		\vspace{-4mm}
		\caption{No grid-cell initialization}
		 \label{fig:appendix.nopreprocessing}
	}
	\end{subfigure}
	\\
	%%%%%%%%%%%%%%%%%
	\begin{subfigure}[b]{.465\textwidth}
	\centering
	{
		\includegraphics[trim={0 0.55cm 0 0},clip,height=2.45cm]{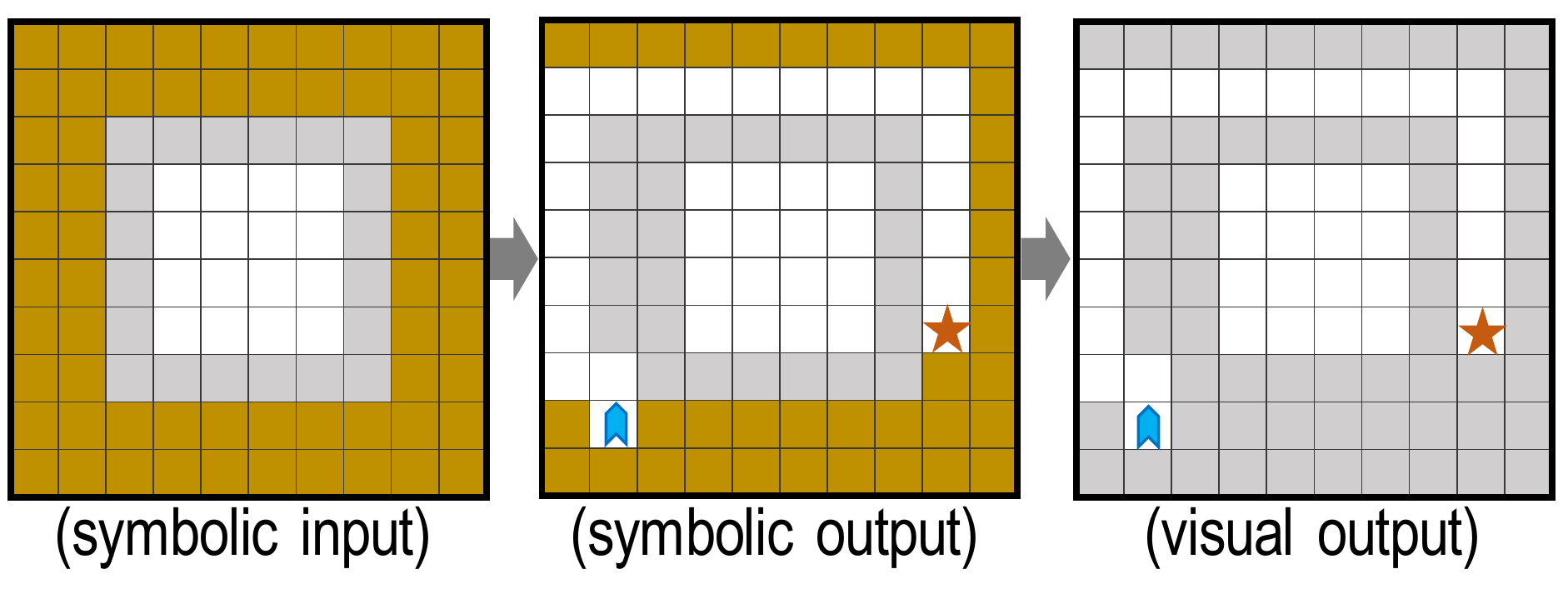}
		\vspace{-4mm}
		\caption{Grid-cell initialization 1}
		 \label{fig:appendix.preprocessing1}
	}
	\end{subfigure}
	\quad \ \ 
	%%%%%%%%%%%%%%%%%
	\begin{subfigure}[b]{.465\textwidth}
	\centering
	{
		\includegraphics[trim={0 0.55cm 0 0},clip,height=2.45cm]{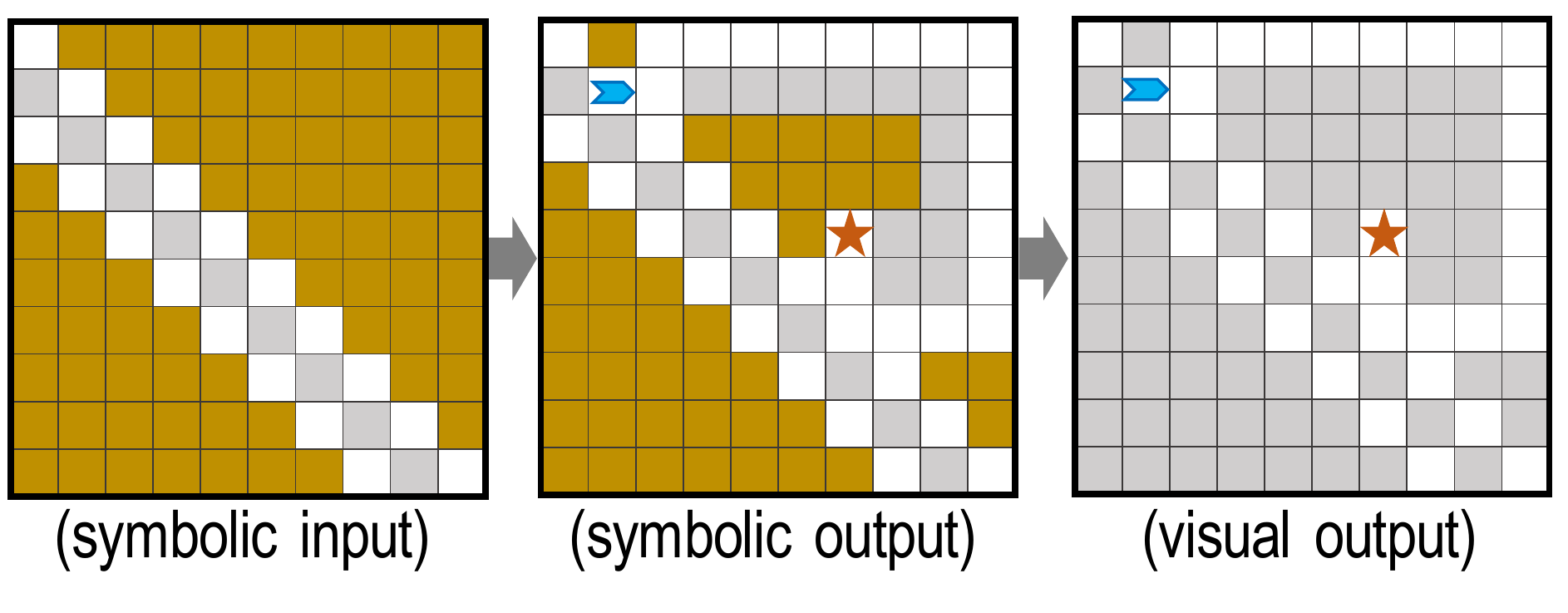}
		\vspace{-4mm}
		\caption{Grid-cell initialization 2}
		 \label{fig:appendix.preprocessing2}
	}
	\end{subfigure}
	%%%%%%%%%%%%%%%%%	
	%\vspace{-2mm}
	\caption{Illustration of the (pre-processing) grid-cell initialization strategy to increase task variability on tasks generated from $\code^{\textnormal{out}}$ in  \figref{fig:intro.hoc.p1} which was obtained from solution code of task \hocG~by mutation. (a) illustrates the basic task obtained after symbolic execution. (b, c) show two different grid-cell initializations, yielding visually very different tasks.
	}
	%\vspace{-4mm}
	\label{fig:appendix.preprocessing}
\end{figure}
%%%%%%%%%%%%%%%%%%%%%%%%%%%%%%%%%%%%%
%%%%%%%%%%%%%%%%%%%%%%%%%%%%%%%%%%%%%%%%%%%%%%%%%%%%%%%%%%

%\clearpage
%\input{9.9_appendix_misc}
}
}
{}
%%%%%%%%%%%%%%%%%%%%%%%%%%%%%%%%%%%%%%%%%%%%%%%%%%%%%%%%%%
\end{document}